\documentclass[12pt,epsf]{article}
\usepackage{amsmath}
\usepackage{amsfonts}
\usepackage{amssymb}
\usepackage{graphicx}

\newcommand {\dfn} {\stackrel{\Delta} {=}}
\newcommand{\dd}{\mbox{d}}
\newcommand {\exe} {\stackrel{\cdot} {=}}
\newcommand {\eqas} {\stackrel{\mbox{a.s.}} {=}}
\newcommand {\lexe} {\stackrel{\cdot} {\le}}

\newcommand {\reals} {{\rm I\!R}}

\newcommand {\bi} {\mbox{\boldmath $i$}}

\newcommand {\bn} {\mbox{\boldmath $n$}}

\newcommand {\br} {\mbox{\boldmath $r$}}
\newcommand {\bs} {\mbox{\boldmath $s$}}

\newcommand {\bu} {\mbox{\boldmath $u$}}
\newcommand {\bv} {\mbox{\boldmath $v$}}
\newcommand {\bw} {\mbox{\boldmath $w$}}
\newcommand {\bx} {\mbox{\boldmath $x$}}
\newcommand {\by} {\mbox{\boldmath $y$}}
\newcommand {\bz} {\mbox{\boldmath $z$}}

\newcommand {\bE} {\mbox{\boldmath $E$}}

\newcommand {\bJ} {\mbox{\boldmath $J$}}

\newcommand {\bS} {\mbox{\boldmath $S$}}

\newcommand {\bU} {\mbox{\boldmath $U$}}
\newcommand {\bV} {\mbox{\boldmath $V$}}

\newcommand {\bX} {\mbox{\boldmath $X$}}
\newcommand {\bY} {\mbox{\boldmath $Y$}}
\newcommand {\bZ} {\mbox{\boldmath $Z$}}

\newcommand{\calC}{{\cal C}}

\newcommand{\calE}{{\cal E}}

\newcommand{\calG}{{\cal G}}

\newcommand{\calI}{{\cal I}}

\newcommand{\calN}{{\cal N}}

\newcommand{\calP}{{\cal P}}

\newcommand{\calX}{{\cal X}}
\newcommand{\calY}{{\cal Y}}

\newcommand {\hbx} {\hat{\bx}}
\newcommand {\blambda} {\mbox{\boldmath $\lambda$}}
\newcommand {\bomega} {\mbox{\boldmath $\omega$}}
\begin{document}
\thispagestyle{empty}
\title{Information Theory and Statistical Physics --\\ Lecture Notes}
\author{Neri Merhav}
\date{}
\maketitle

\vspace{3cm}

\begin{figure}[ht]
\hspace*{1.5cm}\input{cover.pstex_t}
\end{figure}

\thispagestyle{empty}

\newpage

\begin{center}
{\bf \Large Information Theory and Statistical Physics --\\ Lecture Notes}

\vspace{2cm}

Neri Merhav\\
Department of Electrical Engineering \\
Technion - Israel Institute of Technology \\
Haifa 32000, ISRAEL \\
{\tt merhav@ee.technion.ac.il}
\end{center}
\vspace{1.5\baselineskip}
\setlength{\baselineskip}{1.5\baselineskip}

\begin{abstract}
This document consists of lecture notes
for a graduate course, which focuses on the relations between Information Theory and
Statistical Physics. The
course is aimed at EE graduate students in the area of
Communications and Information Theory, as well as to graduate students in
Physics who have basic background in
Information Theory.
Strong emphasis is
given to the analogy and parallelism between
Information Theory and Statistical Physics, as well as to the insights,
the analysis tools and techniques that
can be borrowed from Statistical Physics and `imported' to certain problem
areas in Information Theory. This is a research trend that has been very
active in the last few decades, and the hope is that
by exposing the student to the meeting points between these
two disciplines, we will enhance his/her background and perspective to carry
out research in the field.

A short outline of the course is as follows: Introduction; 
Elementary Statistical Physics and its Relation to Information
Theory;
Analysis Tools in
Statistical Physics;
Systems of Interacting Particles and Phase Transitions;
The Random Energy Model (REM) and
Random Channel Coding; 
Additional Topics (optional).
\end{abstract}

\newpage
\tableofcontents

\newpage
\section{Introduction}

This course is intended to EE graduate
students in the field of Communications and Information Theory, and also to
graduates of the Physics Department (in particular, graduates of the
EE--Physics program) who have basic background in Information Theory, which is
a prerequisite to this course. As its name suggests, this course focuses on
relationships and interplay between Information Theory and {\it Statistical
Physics} -- a branch of physics that deals with many--particle systems 
using probabilitistic/statistical methods in the microscopic level. 

The relationships between Information Theory and Statistical Physics (+
thermodynamics) are by no means new, and many researchers have been exploiting
them for many years. 
Perhaps the first relation, or analogy, that crosses our minds is that in both
fields,
there is a fundamental notion of {\it entropy}. Actually, in Information
Theory, the
term entropy was coined after the thermodynamic entropy.
The thermodynamic entropy was first introduced by Clausius (around 1850),
whereas its probabilistic--statistical interpretation is due to Boltzmann
(1872). It is virtually impossible to miss the functional resemblance between
the two notions of entropy, and indeed it was recognized by Shannon and von
Neumann. The well--known anecdote on this tells that von Neumann advised
Shannon to adopt this term because it would provide him with {\it ``... a
great
edge in debates because nobody really knows what entropy is anyway.''}

But the relationships between the two fields go far beyond the fact that both
share the notion of entropy. In fact,
these relationships have many aspects, and we will not
cover all of them in this course, but just to give the idea of their scope,
we will mention just a few.
\begin{itemize}

\item \underline{\it The Maximum Entropy (ME) Principle.} This is perhaps the oldest
concept that ties the two fields and it has attracted a great deal of
attention, not only of information theortists, but also that of researchers
in related fields like signal processing, image processing, and the like. 
It is about a philosopy, or a belief, which, in a nutshell, is the following:
If in a certain problem, the observed data comes from an unknown probability
distribution, but we do have some knowledge (that stems e.g., from
measurements) of certain moments of the underlying
quantity/signal/random--variable, then assume that the unknown underlying
probability distribution is the one with {\it maximum entropy} subject to
(s.t.) moment constraints corresponding to this knowledge. For example, if we
know the first and the second moment, then the ME distribution is Gaussian
with matching first and second order moments. Indeed, the Gaussian model
is perhaps the most widespread model for physical processes in Information
Theory as well as in signal-- and image processing. But why maximum entropy?
The answer to this philosophical question is rooted in the {\it second law of
thermodynamics}, which asserts that in an isolated system, the entropy cannot
decrease, and hence, when the system reaches equilibrium, its entropy reaches
its maximum. Of course, when it comes to problems in Information Theory and
other related fields, this principle becomes quite heuristic, and so, one may 
question its relevance, but nevertheless, this approach has had an enormous
impact on research trends throughout the last fifty years, after being
proposed by Jaynes in the late fifties of the previous century, and further
advocated by Shore and Johnson afterwards. In the book by Cover and Thomas,
there is a very nice chapter on this, but we will not delve into this any
further in this course.

\item \underline{\it Landauer's Erasure Principle.} Another aspect of these
relations has to do with a piece of theory whose underlying guiding principle
is that {\it information is a physical entity}. In every information bit in
the universe there is a certain amount of energy. Specifically, Landauer's
erasure principle (from the early sixties of the previous century), which is
based on a physical theory of information, asserts that every bit that one
erases, increases the entropy of the universe by $k\ln 2$, where $k$ is
Boltzmann's constant. It is my personal opinion that these kind of theories
should be taken with a grain of salt, but this is only my opinion. At any
rate, this is not going to be included in the course either.

\item \underline{\it Large Deviations Theory as a Bridge Between Information
Theory and Statistical Physics.}\\
Both Information Theory and Statistical Physics have an intimate relation to {\it large deviations theory}, 
a branch of probability theory which focuses on the assessment of the
exponential rates of decay of probabilities of rare events, where the most
fundamental mathematical tool is the {\it Chernoff bound.} This is a topic
that will be covered in the course and quite soon.

\item \underline{\it Random Matrix Theory.} How do the eigenvalues (or, more
generally, the
singular values) of random matrices behave when these matrices have very
large dimensions or if they result from products of many randomly selected matrices?
This is a hot area in probability theory with many applications, both in
Statistical Physics and
in Information Theory, especially in modern theories of wireless communication (e.g., MIMO
systems). This is again outside the scope of this course, but whoever
is interested to `taste' it, is invited to read the 2004 paper by Tulino and
Verd\'u in {\it Foundations and Trends in Communications and Information
Theory}, a relatively new journal for tutorial papers.

\item \underline{\it Spin Glasses and Coding Theory.}
It turns out that many problems in channel coding theory (and also to some
extent, source coding theory) can be mapped almost verbatim to parallel
problems in the field of physics of {\it spin glasses} -- amorphic magnetic
materials with a high degree of disorder and very complicated physical
behavior, which is cusomarily treated using statistical--mechanical
approaches. It has been many years that researchers have made attempts to `import'
analysis techniques rooted in statistical physics of spin glasses and to apply
them to analogous coding problems, with various degrees of success. This is
one of main subjects of this course and we will study it extensively, at least
from some aspects. 
\end{itemize}

We can go on and on with this list and add more items
in the context of these very fascinating meeting points between Information
Theory and Statistical Physics, but
for now, we stop here. We just mention that the last item will
form the main core of the course. We will see that, not only these relations
between Information Theory and Statistical Physics are 
interesting academically on their own right, but
moreover, they also prove useful and beneficial in that they provide us
with new insights and mathematical tools to deal with information--theoretic problems. These
mathematical tools sometimes prove a lot more efficient than traditional tools
used in Information Theory, and they may give either simpler expressions for performance
analsysis, or improved bounds, or both. 

At this point, let us have a brief review of the syllabus of this course, where
as can be seen, the physics and the Information Theory subjects are interlaced with each
other, rather than being given in two continuous, separate parts. This way, it
is hoped that the relations between Information Theory and Statistical Physics will be seen more readily.
The detailed structure of the remaining part of this course is as follows:
\begin{enumerate}
\item [1.] \underline{\it Elementary Statistical Physics and its Relation to
Information
Theory:} What is statistical physics?
Basic postulates and the micro--canonical ensemble; the canonical
ensemble: the Boltzmann--Gibbs law, the partition function, thermodynamical
potentials and their relations to information measures; the equipartition
theorem; generalized ensembles (optional);
Chernoff bounds and the Boltzmann--Gibbs law: rate functions in Information
Theory and thermal equilibrium; physics of the Shannon limits.

\item [2.] \underline{\it Analysis Tools in
Statistical Physics:} The Laplace method of integration; the saddle--point
method; transform methods for counting and for representing non--analytic
functions; examples; the replica method -- overview.

\item [3.] \underline{\it Systems of Interacting Particles and Phase
Transitions:}
Models of many--particle systems with interactions (general)
and examples; a qualitative
explanation for the existence of phase transitions in physics and
in information theory; ferromagnets and Ising
models: the 1D Ising model, the Curie-Weiss model;
randomized spin--glass models: annealed vs.\ quenched randomness, and
their relevance to coded communication systems.

\item [4.] \underline{\it The Random Energy Model (REM) and
Random Channel Coding:}
Basic derivation and phase transitions -- the glassy phase and the
paramagnetic phase; random channel codes and the REM:
the posterior distribution as an instance of the Boltzmann distribution,
analysis and phase diagrams, implications on code ensemble performance
analysis.

\item [5.] \underline{\it Additional Topics (optional)}:
The REM in a magnetic field and joint source--channel coding;
the generalized REM (GREM) and hierarchical ensembles of codes;
phase transitions in the
rate--distortion function;
Shannon capacity of infinite--range spin--glasses;
relation between temperature, de Bruijn's identity, and Fisher information;
the Gibbs inequality in Statistical Physics and its relation to the
log--sum inequality of Information Theory.
\end{enumerate}

As already said, there are also 
plenty of additional subjects that fall under the
umbrella of relations between Information Theory and Statistical Physics, which will not be covered in
this course. One very hot topic is that of codes on graphs, iterative
decoding, belief
propagation, and density evolution. The main reason for not including these
topics is that they are already covered in the course of Dr.\ Igal Sason:
``Codes on graphs.''

I would like to emphasize that prior basic background in Information
Theory will be
assumed, therefore, Information 
Theory is a prerequisite for this course. As for the physics
part, prior background in statistical mechanics could be helpful, but it is
not compulsory. The course is intended to be self--contained as far as the
physics background goes. The bibliographical list includes, in addition to
a few well known books in Information Theory, also several very good books in
elementary Statistical Physics,
as well as two books on the relations between these two fields.

As a final note, I feel compelled to clarify that the material of this
course is by no means intended to 
be presented from a very comprehensive
perspective and to consist of a full account of methods, problem areas and
results. Like in many advanced graduate courses in our department,
here too, the choice of topics, the approach, and the style strongly reflect
the personal bias of the lecturer and his/her perspective on research
interests in the field.
This is also the reason that a considerable fraction of the topics and results that will
be covered, are taken from articles in which I have been involved.


\newpage
\section{Elementary Stat.\ Physics and Its Relation to IT}

\subsection{What is Statistical Physics?}

Statistical physics is a branch in Physics which deals with systems with a
huge number of particles (or any other elementary units), e.g., of the order
of magnitude of {\it Avogadro's number}, that is, about $10^{23}$ particles.
Evidently, when it comes to systems with such an enormously
large number of particles, there is no hope to keep track of the physical
state (e.g., position and momentum) of each and every individual particle by
means of the classical methods in physics, that is, by solving a gigantic system of
differential equations pertaining to Newton's laws for all particles.
Moreover, even if these differential equations could have been solved (at least
approximately), the information that they would give us would be virtually
useless. What we normally really want to know about our physical system boils down to a
bunch of {\it macroscopic} parameters, such as energy, heat, pressure,
temperature, volume, magnetization, and the like. In other words, while we
continue to believe in the good old laws of physics that we have known for
some time, even the classical ones,
we no longer use them in the ordinary way that we are familar with from elementary
physics courses. Rather, we think of the state of the system, at any given
moment, as a
realization of a certain {\it probabilistic ensemble}. This is to say that we
approach the problem from a probabilistic (or a statistical) point of view. 
The beauty of statistical physics is that it derives the {\it macroscopic} theory of thermodynamics
(i.e., the relationships between thermodynamical potentials, temperature,
pressure, etc.) as {\it ensemble averages} that stem from this probabilistic
{\it microscopic} theory -- the theory of statistical physics, in the limit of
an infinite number of particles, that is, the {\it thermodynamic limit}.
As we shall see throughout this course, this thermodynamic limit is parallel
to the asymptotic regimes that we are used to in Information Theory, 
most notably, the one pertaining to a certain `block
length' that goes to infinity. 

\subsection{Basic Postulates and the Microcanonical Ensemble}

For the sake of concreteness, let us consider the example where our many--particle system is
a {\it gas}, namely, a system with a very large number $n$ of mobile particles, which
are free to move in a given volume. 
The {\it microscopic state} (or {\it microstate}, for short) of the system, at
each time instant $t$,  consists, in this example,
of the position $\vec{r}_i(t)$
and the momentum $\vec{p}_i(t)$ of each and every particle,
$1\le i\le n$. Since each one of
these is a vector of three components, the microstate is then given by a
$(6n)$--dimensional vector
$\vec{\bx}(t)=\{(\vec{r}_i(t),\vec{p}_i(t)),~i=1,2,\ldots,n\}$, whose trajectory
along the time axis,
in the {\it phase space}, $\reals^{6n}$, is called the {\it phase trajectory}.

Let us assume that the system is closed, i.e., {\it isolated} from its environment, in
the sense that no energy flows inside or out.
Imagine that the phase space $\reals^{6n}$ is partitioned into very small hypercubes
(or cells) $\Delta\vec{p}\times\Delta\vec{r}$. One of the basic postulates of statistical
mechanics is the following: In the very long range, the relative
amount of time at which $\vec{\bx}(t)$ spends at each such cell
converges to a certain number between $0$ and $1$, which can be given the
meaning of the {\it probability} of this cell. Thus, there is an underlying
assumption of equivalence between temporal averages and ensemble averages,
namely, this is the assumption of {\it ergodicity}.

What are then the probabilities of these cells? We would like to derive these
probabilities from first principles, based on as few as possible
basic postulates. Our first such postulate is that for an isolated system
(i.e., whose energy is fixed) all microscopic states $\{\vec{\bx}(t)\}$ are
equiprobable. The rationale behind this postulate is twofold:
\begin{itemize}
\item In the absence of additional information, there is no apparent reason
that certain regions in phase space would have preference relative to any
others.
\item This postulate is in harmony with a basic result in kinetic theory of
gases -- {\it the Liouville theorem}, which we will not touch upon in this
course, but in a nutshell, it asserts that the phase trajectories must lie
along hypersurfaces of constant probability density.\footnote{This is a result of the
energy conservation law along with the fact that probability mass behaves
like an incompressible fluid in the sense that whatever mass that flows into a certain
region from some direction must be equal to the outgoing flow from some other
direction. This is reflected in the so called continuity equation.}
\end{itemize}

Before we proceed, let us slightly broaden the scope of our discussion.
In a more general context, associated with our $n$--particle physical system, 
is a certain instantaneous microstate, generically denoted by
$\bx=(x_1,x_2,\ldots,x_n)$, where each 
$x_i$, $1\le i\le n$, may itself be a vector of several
physical quantities associated particle number 
$i$, e.g., its position, momentum,
angular momentum, magnetic moment, spin, and so on,
depending on the type and the nature 
of the physical system. For each possible value of
$\bx$, there is a certain
{\it Hamiltonian} (i.e., energy function) that assigns to
$\bx$ a certain energy $\calE(\bx)$.\footnote{For example, in the case of an {\it ideal gas},
$\calE(\bx)=\sum_{i=1}^n\frac{\|\vec{p}_i\|^2}{2m}$, independently of
the positions
$\{\vec{r}_i\}$, namely, it accounts for the contribution of the kinetic energies only. 
In more complicated situations, there might be additional contributions of potential
energy, which depend on the positions.} Now, let us denote by $\Omega(E)$ 
the {\it density--of--states} function, i.e., the
volume of the shell $\{\bx:\calE(\bx)=E\}$, or, slightly more precisely,
$\Omega(E)dE=\mbox{Vol}\{\bx:~E\le\calE(\bx)\le E+dE\}$, 
which will be denoted also as
$\mbox{Vol}\{\bx:~\calE(\bx)\approx E\}$,
where the dependence on $dE$ will normally
be ignored since $\Omega(E)$ is typically exponential in $n$ and $dE$ will
have virtually no effect on its exponential order as long as it is small.
Then, our above postulate concerning the ensemble of an isolated system, which
is called the {\it microcanonincal ensemble}, is that the probability density
$P(\bx)$ is given by
\begin{equation}
P(\bx)=\left\{\begin{array}{ll}
\frac{1}{\Omega(E)} & \calE(\bx)\approx E\\
0 & \mbox{elsewhere}\end{array}\right.
\end{equation}
In the discrete case, things are, of course, a lot easier: Then, $\Omega(E)$ would be
the number of microstates with $\calE(\bx)=E$ (exactly) and $P(\bx)$ would be
the uniform probability mass function across this set of states. 
In this case,
$\Omega(E)$ is analogous to the size of a {\it type class} in Information
Theory, and $P(\bx)$ is the uniform distribution across this type class.

Back to the
continuous case, note that $\Omega(E)$ is, in general, not dimensionless: 
In the above example of a gas, it has the
physical units of $[\mbox{length}\times\mbox{momentum}]^{3n}$, but we must get
rid of these physical units because very soon we are going to apply non--linear
functions on $\Omega(E)$, like the logarithmic function. Thus, we must
normalize this volume by an elementary reference volume. In the gas example, this reference volume
is taken to be $h^{3n}$, where $h$ is {\it Planck's constant} $\approx 6.62\times 10^{-34}$ 
Joules$\cdot$sec. Informally, the
intuition comes from the fact that $h$ is our best available ``resolution'' in the plane
spanned by each component of $\vec{r}_i$ and the corresponding component of
$\vec{p}_i$, owing to the {\it uncertainty principle} in quantum mechanics,
which tells us that the product of the standard deviations $\Delta p_a\cdot\Delta
r_a$ of each component $a$ ($a=x,y,z$) is lower bounded by $\hbar/2$, where
$\hbar=h/(2\pi)$. More
formally, this reference volume is obtained in a natural manner 
from quantum statistical mechanics: by changing
the integration variable $\vec{p}$ to $\vec{k}$ by using
$\vec{p}=\hbar\vec{k}$, where $\vec{k}$ is the wave vector.
This is a well--known relationship pertaining to particle--wave duality. Now, 
having redefined $\Omega(E)$ in units of
this reference volume, which makes it then a dimensionless quantity, the {\it entropy} is
defined as
\begin{equation}
S(E)=k\ln\Omega(E),
\end{equation}
where $k$ is {\it Boltzmann's constant} $\approx 1.38\times 10^{-23}$
Joule/degree. We will soon see what is the relationship between $S(E)$
and the information--theoretic entropy.

To get some feeling of this, it should be noted that normally,
$\Omega(E)$ behaves as an exponential function of $n$ (at least
asymptotically), and so, $S(E)$ is roughly linear in $n$. For example,
if $\calE(\bx)=\sum_{i=1}^n\frac{\|\vec{p}_i\|^2}{2m}$, then $\Omega(E)$
is the volume of a shell or surface of a $(3n)$--dimensional sphere with
radius $\sqrt{2mE}$, which is proportional 
to $(2mE)^{3n/2}V^n$, but we should divide
this by $n!$ to account for the fact that the particles are indistinguishable
and we don't count permutations as distinct physical states in this
case.\footnote{Since the particles are mobile and since they have no colors and no
identity certficiates, there is no distinction between a state where particle
no.\ 15 has position $\vec{r}$ and momentum $\vec{p}$ while particle no.\ 437
has position $\vec{r}'$ and momentum $\vec{p}'$ and a state where these two
particles are swapped.} More precisely, one obtains:
\begin{equation}
S(E)=k\ln\left[\left(\frac{4\pi
mE}{3n}\right)^{3n/2}\cdot\frac{V^n}{n!h^{3n}}\right]+\frac{3}{2}nk\approx
nk\ln\left[\left(\frac{4\pi
mE}{3n}\right)^{3/2}\cdot\frac{V}{nh^{3}}\right]+\frac{5}{2}nk.
\end{equation}
Assuming $E \propto n$ and $V\propto n$, we get $S(E)\propto n$.
A physical quantity like this, that has a linear scaling with the size of the
system $n$, is called an {\it extensive quantity}. So, energy, volume and
entropy are 
extensive quantities. 
Other quantities, which are not extensive, i.e., independent of the
system size, like temperature and pressure, are called {\it intensive}.

It is interesting to point out that from the function $S(E)$, or actually,
the function $S(E,V,n)$,
one can obtain the entire information about the relevant
macroscopic physical quantities of the system, e.g., temperature,
pressure, and so on.
The {\it temperature} $T$ of the system is defined according to:
\begin{equation}
\frac{1}{T}=\left(\frac{\partial S(E)}{\partial E}\right)_V
\end{equation}
where $(\cdot)_V$ means that the derivative is taken in constant
volume.\footnote{
This definition of temperature is related to the classical thermodynamical
definition of entropy 
as $\dd S=\dd Q/T$, where $Q$ is heat, as in the absence of
external work, when the volume $V$ is fixed, all the energy comes from heat
and so, $\dd E=\dd Q$.} Intuitively, in most situations, 
we expect that $S(E)$ would be an increasing
function of $E$ (although this is not strictly always the case), which means
$T \ge 0$. But $T$ is also expected to be increasing with $E$ (or
equivalently, $E$ is increasing with $T$, as otherwise, the heat capacity
$\dd E/\dd T < 0$). Thus, $1/T$ should decrease with $E$, which means that the
increase of $S$ in $E$ slows down as $E$ grows. In other words, we expect
$S(E)$ to be a concave function of $E$. In the above example, indeed, $S(E)$
is logarithmic in $E$ and we get $1/T\equiv\partial S/\partial E=
3nk/(2E)$, which means $E=3nkT/2$. Pressure is obtained by $P=T\cdot\partial
S/\partial V$, which in our example, gives rise to
the state equation of the ideal gas, $P=nkT/V$.

How can we also see {\it mathematically} 
that under ``conceivable conditions'', $S(E)$ is
a concave function?  We know that the Shannon entropy is also a concave
functional of the probability distribution. Is this related?

As both $E$ and $S$ are extensive quantities, let
us define $E=n\epsilon$ and 
\begin{equation}
s(\epsilon)=\lim_{n\to\infty}\frac{S(n\epsilon)}{n},
\end{equation}
i.e., the per--particle entropy as a function of the per--particle energy.
Consider the case where the Hamiltonian is additive, i.e.,
\begin{equation}
\calE(\bx)=\sum_{i=1}^n\calE(x_i)
\end{equation}
just like in the above example 
where $\calE(\bx)=\sum_{i=1}^n\frac{\|\vec{p}_i\|^2}{2m}$. Then, obviously,
\begin{equation}
\Omega(n_1\epsilon_1+n_2\epsilon_2)\ge
\Omega(n_1\epsilon_1)\cdot\Omega(n_2\epsilon_2),
\end{equation}
and so, we get:
\begin{eqnarray}
\frac{ k\ln \Omega(n_1\epsilon_1+n_2\epsilon_2)}{n_1+n_2}&\ge&
\frac{k\ln \Omega(n_1\epsilon_1)}{n_1+n_2}+\frac{k\ln
\Omega(n_2\epsilon_2)}{n_1+n_2}\nonumber\\
&=&\frac{n_1}{n_1+n_2}\cdot\frac{k\ln \Omega(n_1\epsilon_1)}{n_1}+
\frac{n_2}{n_1+n_2}\cdot\frac{k\ln
\Omega(n_2\epsilon_2)}{n_2}.
\end{eqnarray}
and so, by taking $n_1$ and $n_2$ to $\infty$, with
$n_1/(n_1+n_2)\to\lambda\in(0,1)$, we get:
\begin{equation}
s(\lambda\epsilon_1+(1-\lambda)\epsilon_2)\ge
\lambda s(\epsilon_1)+(1-\lambda)s(\epsilon_2),
\end{equation}
which establishes the concavity of $s(\cdot)$ at least in the case of an
additive Hamiltonian, which means that the entropy of mixing two systems of
particles is greater than the total entropy before they are mixed (the second
law). A similar proof
can be generalized to the
case where $\calE(\bx)$ includes also a limited degree of 
interactions (short range interactions), e.g.,
$\calE(\bx)=\sum_{i=1}^n\calE(x_i,x_{i+1})$, but this requires somewhat more
caution. In general, however, concavity may no longer hold when there are long range
interactions, e.g., where some terms of $\calE(\bx)$ depend on a linear
subset of particles. Simple examples can be found in: H.~Touchette, ``Methods for calculating
nonconcave entropies,'' 
arXiv:1003.0382v1 [cond-mat.stat-mech] 1 Mar 2010.

\vspace{0.5cm}

\begin{figure}[ht]
\hspace*{5cm}\input{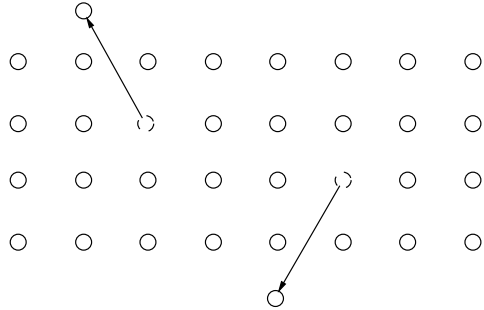}
\caption{\small Schottky defects in a crystal lattice.}
\label{schottky}
\end{figure}

\noindent
{\it Example -- Schottky defects.}
In a certain crystal, the atoms are located in a lattice, and at any positive
temperature there may be defects, where some of the atoms are dislocated (see
Fig.\ \ref{schottky}).
Assuming that defects are sparse enough, such that around each dislocated atom
all neighors are in place, the activation energy, $\epsilon_0$, required for dislocation
is fixed. Denoting the total number of atoms by $N$ and the
number of defected ones by $n$, the total energy is then $E=n\epsilon_0$, and
so,
\begin{equation}
\Omega(E)=\left(\begin{array}{cc} N \\ n
\end{array}\right)=\frac{N!}{n!(N-n)!},
\end{equation}
or, equivalently,
\begin{eqnarray}
S(E)&=&k\ln\Omega(E)=k\ln\left[\frac{N!}{n!(N-n)!}\right]\nonumber\\
&\approx& k[N\ln N-n\ln n -(N-n)\ln(N-n)] ~~~~\mbox{by the Stirling
approximation}\nonumber
\end{eqnarray}
Thus,
\begin{equation}
\frac{1}{T}=\frac{\partial S}{\partial E}= \frac{\dd S}{\dd n}\cdot\frac{\dd
n}{\dd E}=
\frac{1}{\epsilon_0}\cdot k\ln\frac{N-n}{n},
\end{equation}
which gives the number of defects as
\begin{equation}
n=\frac{N}{\exp(\epsilon_0/kT)+1}.
\end{equation}
At $T=0$, there are no defects, but their number increases gradually with $T$,
approximately according to $\exp(-\epsilon_0/kT)$. Note that from a slighly
more information--theoretic point of view,
\begin{equation}
S(E)=k\ln \left(\begin{array}{cc} N \\ n
\end{array}\right)\approx kN h_2\left(\frac{n}{N}\right)=kN
h_2\left(\frac{E}{N\epsilon_0}\right)=
kN h_2\left(\frac{\epsilon}{\epsilon_0}\right),
\end{equation}
where $$h_2(x)\dfn -x\ln x-(1-x)\ln(1-x).$$
Thus, the thermodynamical entropy is intimately related to the Shannon
entropy. We will see shortly that this is no coincidence. Note also that
$S(E)$ is indeed concave in this example. $\Box$

What happens if we have two independent systems with total energy $E$, which
lie in equilibrium with each other.
What is the temperature $T$? How does the energy split between them?
The number of combined microstates where system no.\ 1 has energy $E_1$ and
system no.\ 2 has energy $E_2=E-E_1$ is $\Omega_1(E_1)\cdot\Omega_2(E-E_1)$.
If the combined system is isolated, then the probability of such a combined
microstate is proportional to $\Omega_1(E_1)\cdot\Omega_2(E-E_1)$.
Keeping in mind that normally, $\Omega_1$ and $\Omega_2$ are exponential in
$n$, then for large $n$, this product is dominated by the value of $E_1$
for which it is maximum, or equivalently, the sum of logarithms,
$S_1(E_1)+S_2(E-E_1)$, is maximum, i.e., it is a {\bf maximum entropy} 
situation, which is {\bf the second law of thermodynamics}.
This maximum is normally achieved at the
value of $E_1$ for which the derivative vanishes, i.e.,
\begin{equation}
S_1'(E_1)-S_2'(E-E_1)=0
\end{equation}
or
\begin{equation}
S_1'(E_1)-S_2'(E_2)=0
\end{equation}
which means
\begin{equation}
\frac{1}{T_1}\equiv S_1'(E_1)=S_2'(E_2)\equiv\frac{1}{T_2}.
\end{equation}
Thus, in equilibrium, which is the maximum entropy situation, 
the energy splits in a way that temperatures are the same.

\subsection{The Canonical Ensemble}

So far we have assumed that our system is isolated, and therefore has a strictly
fixed energy $E$. Let us now relax this assumption and assume that our system
is free to exchange energy with its large environment (heat bath) and that the total energy of
the heat bath $E_0$ is by far larger than the typical energy of the system. The
combined system, composed of our original system plus the heat bath, is now an
isolated system at temperature $T$.
So what happens now?

Similarly as before, since the combined 
system is isolated, it is governed by the microcanonical
ensemble. The only difference is that now we assume that one of the systems
(the heat bath) is very large compared to the other (our test system).
This means that if our small system is in microstate $\bx$
(for whatever definition of the microstate vector) 
with energy $\calE(\bx)$, then the heat bath must have energy $E_0-\calE(\bx)$
to complement the total energy to $E_0$. The number of ways that the heat bath
may have energy $E_0-\calE(\bx)$ is $\Omega_{HB}(E_0-\calE(\bx))$,
where $\Omega_{HB}(\cdot)$ is the density--of--states function pertaining to the
heat bath. In other words, the number of microstates of the {\it combined}
system for which the small subsystem is in microstate $\bx$ 
is $\Omega_{HB}(E_0-\calE(\bx))$. Since the combined system is governed by the
microcanonical ensemble, the probability of this is proportional to 
$\Omega_{HB}(E_0-\calE(\bx))$. More precisely:
\begin{equation}
P(\bx)=\frac{\Omega_{HB}(E_0-\calE(\bx))}
{\sum_{\bx'}\Omega_{HB}(E_0-\calE(\bx'))}.
\end{equation}
Let us focus on the numerator for now, and  normalize the result at the end.
Then,
\begin{eqnarray}
P(\bx) &\propto& \Omega_{HB}(E_0-\calE(\bx))\nonumber\\
&=& \exp\{S_{HB}(E_0-\calE(\bx))/k\}\nonumber\\
&\approx&\exp\left\{\frac{S_{HB}(E_0)}{k}-\frac{1}{k}\frac{\partial
S_{HB}(E)}{\partial E}\bigg|_{E=E_0}\cdot\calE(\bx)\right\}\nonumber\\
&=&\exp\left\{\frac{S_{HB}(E_0)}{k}-\frac{1}{kT}
\cdot\calE(\bx)\right\}\nonumber\\
&\propto& \exp\{-\calE(\bx)/(kT)\}.
\end{eqnarray}
It is customary to work with the so called {\it inverse temperature}:
\begin{equation}
\beta=\frac{1}{kT}
\end{equation}
and so,
\begin{equation}
P(\bx)\propto e^{-\beta\calE(\bx)}.
\end{equation}
Thus, all that remains to do is to normalize, and 
we then obtain the {\it Boltzmann--Gibbs} (B--G) distribution, 
or the {\it canonical ensemble}, which
describes the underlying probability law in equilibrium:
\begin{center}
\fbox{
$P(\bx)=\frac{\exp\{-\beta\calE(\bx)\}}{Z(\beta)}$}
\end{center}
where $Z(\beta)$ is the normalization factor:
\begin{equation}
Z(\beta)=\sum_{\bx}\exp\{-\beta \calE(\bx)\}
\end{equation}
in the discrete case, or
\begin{equation}
Z(\beta)=\int \dd \bx\exp\{-\beta \calE(\bx)\}
\end{equation}
in the continuous case.

This is one of the most fundamental results in statistical mechanics, which
was obtained solely from the energy conservation law and the postulate that
in an isolated system the distribution is uniform. The function $Z(\beta)$
is called the {\it partition function}, and as we shall see, its meaning is by
far deeper than just being a normalization constant. Interestingly, a great
deal of the macroscopic physical quantities, like the internal energy, the
free energy, the entropy, the heat capacity, the pressure, etc., can be
obtained from the partition function.

The B--G distribution tells us then that the system ``prefers'' to visit its
low energy states more than the high energy states. And what counts
is only energy differences, not absolute energies: If we add to all states
a fixed amount of energy $E_0$, this will result in an extra factor of
$e^{-\beta E_0}$ both in the numerator and in the denominator of the B--G
distribution, which will, of course, cancel out. Another obvious observation
is that whenever the Hamiltonian is additive, that is,
$\calE(\bx)=\sum_{i=1}^n\calE(x_i)$, the various particles are statistically
independent: Additive Hamiltonians correspond to
non--interacting particles. In other words, the $\{x_i\}$'s
behave as if they were drawn from a memoryless source. And so, by the law of
large numbers $\frac{1}{n}\sum_{i=1}^n\calE(x_i)$ will tend (almost surely)
to $\epsilon=\bE\{\calE(X_i)\}$. Nonetheless, this is different from the
microcanonical ensemble where $\frac{1}{n}\sum_{i=1}^n\calE(x_i)$ was held
strictly at the value of $\epsilon$. The parallelism to 
Information Theory is as follows:
The microcanonical ensemble is parallel to the uniform distribution over
a type class and the
canonical ensemble is parallel to a memoryless source.

The two ensembles are asymptotically equivalent as far as
expectations go. They continue to be such even in cases of interactions,
as long as these are short range.
It is instructive to point out that the B--G distribution could
have been obtained also in a different manner, owing to the maximum--entropy
principle that we mentioned in the Introduction. Specifically, consider the
following optimization problem:
\begin{eqnarray}
& &\max~H(\bX)\nonumber\\
& &\mbox{s.t.}~\sum_{\bx}P(\bx)\calE(\bx)=E~~~~[\mbox{or in physicists'
notation:}~~
\langle\calE(\bX)\rangle=E]
\end{eqnarray}
By formalizing the equivalent Lagrange problem, where $\beta$ now plays the
role of a Lagrange multiplier:
\begin{equation}
\max~\left\{H(\bX)+\beta\left[E-\sum_{\bx}P(\bx)\calE(\bx)\right]\right\},
\end{equation}
or equivalently,
\begin{equation}
\min~\left\{\sum_{\bx}P(\bx)\calE(\bx)-\frac{H(\bX)}{\beta}\right\}
\end{equation}
one readily verifies that the solution to this problem is the B-G distribution
where the choice of $\beta$ {\bf controls} the average energy $E$. In many
physical systems, the Hamiltonian is a quadratic (or ``harmonic'') function,
e.g., $\frac{1}{2}mv^2$, $\frac{1}{2}kx^2$,
$\frac{1}{2}CV^2$, $\frac{1}{2}LI^2$, $\frac{1}{2}I\omega^2$, etc., in which
case the resulting B--G distribution turns out to be Gaussian. This is at
least part of the explanation why the Gaussian distribution is so frequently
encountered in Nature. Note also that indeed, we have
already seen in the Information Theory course that the Gaussian density maximizes the
(differential) entropy s.t.\ a second order moment constraint, which is
equivalent to our average energy constraint.

\subsection{Properties of the Partition Function and the Free Energy}

Let us now examine more closely the partition function and make a few
observations about its basic properties. For simplicity, we shall assume
that $\bx$ is discrete. First, let's look at the limits:
Obviously, $Z(0)$ is equal to the size of the entire set of microstates, which is also
$\sum_E\Omega(E)$, This is the high temperature limit, where
all microstates are equiprobable. At the other extreme, we have:
\begin{equation}
\lim_{\beta\to\infty} \frac{\ln Z(\beta)}{\beta}=-\min_{\bx} \calE(\bx)\dfn
-E_{GS}
\end{equation}
which describes the situation where the system is frozen to the absolute zero.
Only states with minimum energy -- the {\it ground--state energy}, prevail.

Another important property of $Z(\beta)$, or more precisely, of $\ln
Z(\beta)$,
is that it is a log--moment generating function: By taking derivatives of $\ln
Z(\beta)$, we can obtain moments (or cumulants) of $\calE(\bX)$. 
For the first moment, we have
\begin{equation}
\bE\{\calE(\bX)\}\equiv\langle\calE(\bX)\rangle
=\frac{\sum_{\bx}\calE(\bx)e^{-\beta\calE(\bx)}}{\sum_{\bx}e^{-\beta\calE(\bx)}}=
-\frac{\dd \ln Z(\beta)}{\dd \beta}.
\end{equation}
Similarly, it is easy to show (exercise) that
\begin{equation}
\mbox{Var}\{\calE(\bX)\}=\langle\calE^2(\bX)\rangle -
\langle\calE(\bX)\rangle^2=\frac{\dd^2\ln Z(\beta)}{\dd\beta^2}.
\end{equation}
This in turn implies that $\frac{\dd^2\ln Z(\beta)}{\dd\beta^2}\ge 0$,
which means that $\ln Z(\beta)$ must always be a convex function.
Higher order derivatives provide higher order moments.

Next, we look at $Z$ slightly differently than before. Instead of summing
$e^{-\beta \calE(\bx)}$ across all states, we go by energy levels (similarly
as in the method of types). 
This amounts to:
\begin{eqnarray}
Z(\beta)&=&\sum_{\bx}e^{-\beta\calE(\bx)}\nonumber\\
&=&\sum_{E}\Omega(E)e^{-\beta E}
\nonumber\\
&\approx&\sum_\epsilon e^{ns(\epsilon)/k}\cdot e^{-\beta
n\epsilon}~~~~~\mbox{recall that}~S(n\epsilon)\approx ns(\epsilon)\nonumber\\
&=&\sum_\epsilon \exp\{-n\beta[\epsilon-Ts(\epsilon)]\}\nonumber\\
&\exe&\max_\epsilon \exp\{-n\beta[\epsilon-Ts(\epsilon)]\}\nonumber\\
&=&\exp\{-n\beta\min_\epsilon[\epsilon-Ts(\epsilon)]\}\nonumber\\
&\dfn&\exp\{-n\beta[\epsilon^*-Ts(\epsilon^*)]\}\nonumber\\
&\dfn&e^{-\beta F}
\end{eqnarray}
The quantity $f\dfn \epsilon-Ts(\epsilon)$ is the (per--particle) {\it free
energy}. Similarly, the entire free energy, $F$, is defined as
\begin{equation}
F=E-TS=-\frac{\ln Z(\beta)}{\beta}.
\end{equation}
The physical meaning of the free energy is this: A change, or a difference,
$\Delta F=F_2-F_1$, in the free energy means the minimum amount of work 
it takes to transfer the system from equilibrium state 1 to another
equilibrium state 2 in an isothermal (fixed temperature) process. And this
minimum is achieved when the process is {\it quasistatic}, i.e., so slow that
the system is always almost in equilibrium. Equivalently, $-\Delta F$ is the
maximum amount of work that that can be exploited from the system, namely, the
part of the energy that
is {\it free} for doing work (i.e., not dissipated as heat) in fixed temperature.
Again, this maximum is attained by a quasistatic process.

We see that the value $\epsilon^*$ of $\epsilon$ that minimizes $f$, dominates
the partition function and hence captures most of the probability. 
As $n$ grows without bound, the energy probability distribution
becomes sharper and sharper around
$n\epsilon^*$. Thus, we see that
equilibrium in the canonical ensemble amounts to {\bf minimum free energy.}
This extends the second law of thermodynamics from the microcanonical ensemble
of isolated systems, whose equilibrium obeys the maximum entropy principle.
The maximum entropy principle is replaced, more generally, by the minimum free
energy principle.
Note that the Lagrange minimization problem that we
formalized before, i.e.,
\begin{equation}
\min~\left\{\sum_{\bx}P(\bx)\calE(\bx)-\frac{H(\bX)}{\beta}\right\},
\end{equation}
is nothing but minimization of the free energy, provided that we identify
$H$ with the physical entropy $S$ (to be done very soon) and the Lagrange
multiplier $1/\beta$ with $kT$. Thus, the B--G distribution minimizes the
free energy for a given temperature.

Although we have not yet seen this explicitly, but there were already hints
and terminology suggests that the thermodynamical entropy $S(E)$ is intimately
related to the Shannon entropy $H(\bX)$. 
We will also see it shortly in a more formal manner. But what is the
information--theoretic analogue of the free energy? 

Here is a preliminary guess based on a very rough consideration: The last chain of
equalities reminds us what happens when we sum over probabilities
type--by--type in IT problems: The exponentials $\exp\{-\beta\calE(\bx)\}$ are analoguous
(up to a normalization factor) to probabilities, which in the memoryless
case, are given by $P(\bx)=\exp\{-n[\hat{H}+D(\hat{P}\|P)]\}$. Each such
probability is weighted by the size of the type class, which as is known
from the method of types, is exponentially $e^{n\hat{H}}$, whose physical
analogue is $\Omega(E)=e^{ns(\epsilon)/k}$. The product gives
$\exp\{-nD(\hat{P}\|P)\}$ in IT and $\exp\{-n\beta f\}$ in statistical physics.
This suggests that perhaps the free energy has some analogy with the
divergence. Is this true? We will see shortly a somewhat more rigorous
argument.

More formally, let us define
\begin{equation}
\phi(\beta)=\lim_{n\to\infty}\frac{\ln Z(\beta)}{n}
\end{equation}
and, in order to avoid dragging the constant $k$, let us define
$\Sigma(\epsilon)=\lim_{n\to\infty}\frac{1}{n}\ln\Omega(n\epsilon)=s(\epsilon)/k$. Then,
the above chain of equalities, written slighlty differently, gives
\begin{eqnarray}
\phi(\beta)&=&\lim_{n\to\infty}\frac{\ln Z(\beta)}{n}\nonumber\\
&=&\lim_{n\to\infty}\frac{1}{n}\ln\left\{\sum_\epsilon 
e^{n[\Sigma(\epsilon)-\beta \epsilon]}\right\}\nonumber\\
&=&\max_\epsilon[\Sigma(\epsilon)-\beta\epsilon].\nonumber
\end{eqnarray}
Thus, $\phi(\beta)$ is (a certain variant of) the {\it Legendre
transform}\footnote{More precisely, the 1D Legendre transform of a real function
$f(x)$ is defined as $g(y)=\sup_x[xy-f(x)]$. If $f$ is convex, it can readily
be shown that: (i) The inverse transform has the very same form, i.e.,
$f(x)=\sup_y[xy-g(y)]$, and (ii) The derivatives $f'(x)$ and $g'(y)$ are inverses of each other.}
of $\Sigma(\epsilon)$. As $\Sigma(\epsilon)$ is (normally) a concave function,
then it can readily be shown (execrise) that the inverse transform is:
\begin{equation}
\Sigma(\epsilon)=\min_\beta[\beta\epsilon+\phi(\beta)].
\end{equation}
The achiever, $\epsilon^*(\beta)$, of $\phi(\beta)$ in the forward transform is obtained by
equating the derivative to zero, i.e., it is the solution to the equation
\begin{equation}
\beta=\Sigma'(\epsilon),
\end{equation}
or in other words, the inverse function of $\Sigma'(\cdot)$. By the same
token, the achiever, $\beta^*(\epsilon)$, of $\Sigma(\epsilon)$ in the backward transform
is obtained by
equating the other derivative to zero, i.e., it is the solution to the equation
\begin{equation}
\epsilon=-\phi'(\beta)
\end{equation}
or in other words, the inverse function of $-\phi'(\cdot)$.\\
{\bf Exercise}: Show that the functions $\Sigma'(\cdot)$ and $-\phi'(\cdot)$
are inverses of one another. $\Box$\\
This establishes a relationship between the
typical per--particle energy $\epsilon$ and the inverse temperature $\beta$
that gives rise to $\epsilon$ (cf.\ the Lagrange interpretation above, where we said
that $\beta$ controls the average energy).
Now, obersve that whenever $\beta$ and $\epsilon$ are related as explained
above, we have:
\begin{equation}
\Sigma(\epsilon)=\beta\epsilon+\phi(\beta)=\phi(\beta)-\beta\cdot\phi'(\beta).
\end{equation}
On the other hand, if we look at the Shannon entropy pertaining to the B--G
distribution, we get:
\begin{eqnarray}
\bar{H}(\bX)&=&\lim_{n\to\infty}\frac{1}{n}\bE\left\{\ln\frac{Z(\beta)}{e^{-\beta\calE(\bX)}}
\right\}\nonumber\\
&=&\lim_{n\to\infty}\left[\frac{\ln
Z(\beta)}{n}+\frac{\beta\bE\{\calE(\bX)\}}{n}\right]\nonumber\\
&=&\phi(\beta)-\beta\cdot\phi'(\beta).\nonumber
\end{eqnarray}
which is exactly the same expression as before,
and so, $\Sigma(\epsilon)$ and $\bar{H}$ are identical whenever $\beta$ and
$\epsilon$ are related accordingly. The former, as we recall, we defined as
the normalized logarithm of the number of microstates with per--particle 
energy $\epsilon$. Thus, we have learned that the number of such microstates
is exponentially $e^{n\bar{H}}$, a result that looks familar to what we
learned from the method of types in IT, using combinatorial arguments
for finite--alphabet sequences.
Here we got the same result from
substantially different considerations, which are applicable in situations
far more general than those of finite alphabets (continuous alphabets included).
Another look at this relation is
the following:
\begin{eqnarray}
1&\geq&\sum_{\bx:~\calE(\bx)\approx n\epsilon} P(\bx)=
\sum_{\bx:~\calE(\bx)\approx n\epsilon} \frac{\exp\{-\beta\sum_i\calE(x_i)\}}
{Z^n(\beta)}\nonumber\\
&\approx&\sum_{\bx:~\calE(\bx)\approx n\epsilon} 
\exp\{-\beta n\epsilon-n\phi(\beta)\}
=\Omega(n\epsilon)\cdot\exp\{-n[\beta\epsilon+\phi(\beta)]\}
\end{eqnarray}
which means that
$\Omega(n\epsilon)\le 
\exp\{n[\beta\epsilon+\phi(\beta)]\}$ for all $\beta$, and so,
\begin{equation}
\Omega(n\epsilon)\le 
\exp\{n\min_\beta[\beta\epsilon+\phi(\beta)]\}=
e^{n\Sigma(\epsilon)}=e^{n\bar{H}}.
\end{equation}
A compatible lower bound is obtained by observing that the minimizing $\beta$
gives rise to $\left<\calE(X_1)\right>=\epsilon$, which makes the event
$\{\bx:~\calE(\bx)\approx n\epsilon\}$ a high--probability event, by the
weak law of large numbers.
A good reference for further study and
from a more general perspective is:\\
M.~J.~W.~Hall, ``Universal geometric
approach to uncertainty, entropy, 
and information,'' {\it Phys.\ Rev.\ A}, vol.\ 59, no.\ 4, pp.\ 2602--2615,
April 1999.

Having established the identity between
the Shannon--theoretic entropy and the thermodynamical entropy, we now
move on,
as promised, to the free energy and seek its information--theoretic
counterpart. More precisely, we will look at the difference between the free energies
of two different probability distributions, one of which is the B--G
distibution. Consider first, the following chain of equalities concerning
the B--G distribution:
\begin{eqnarray}
P(\bx)&=&\frac{\exp\{-\beta\calE(\bx)\}}{Z(\beta)}\nonumber\\
&=&\exp\{-\ln Z(\beta)-\beta\calE(\bx)\}\nonumber\\
&=&\exp\{\beta[F(\beta)-\calE(\bx)]\}.
\end{eqnarray}
Consider next another probability distribution $Q$, different in general
from $P$ and hence corresponding to non--equilibrium. Let us now look at the
divergence:
\begin{eqnarray}
D(Q\|P)&=&\sum_{\bx}Q(\bx)\ln\frac{Q(\bx)}{P(\bx)}\nonumber\\
&=&-H_Q-\sum_{\bx}Q(\bx)\ln P(\bx)\nonumber\\
&=&-H_Q-\beta\sum_{\bx}Q(\bx)[F_P-\calE(\bx)]\nonumber\\
&=&-H_Q-\beta F_P+\beta\langle\calE\rangle_Q\nonumber\\
&=&\beta(F_Q-F_P)\nonumber
\end{eqnarray}
or equivalently,
\begin{center}
\fbox{
$F_Q=F_P+kT\cdot D(Q\|P)$}
\end{center}
Thus, the free energy difference is indeed related to the the divergence.
For a given temperature,
the free energy away from equilibrium is always larger than the free energy
at equilibrium. Since the system ``wants'' to minimize the free energy, it
eventually converges to the B--G distribution. More details on this can be
found in:
\begin{enumerate}
\item H.~Qian, ``Relative entropy: free energy ...,'' {\it Phys.\ Rev.\
E}, vol.\ 63, 042103, 2001.
\item G.~B.~Ba\'gci, arXiv:cond-mat/070300v1, 1 Mar.\ 2007.
\end{enumerate}
Another interesting relation between the divergence and physical quantities is
that the divergence is
proportional to the dissipated work ($=$average work $-$ free energy
difference) between two equilibrium states at the same temperature but
corresponding to two different values of some external control parameter.
Details can be found in: R.~Kawai, J.~M.~R.~Parrondo, and C.~Van den Broeck,
``Dissipation: 
the phase--space perspective,'' {\it Phys.\ Rev.\ Lett.}, vol.\ 98, 080602,
2007.

Let us now summarize the main properties of the partition function that we
have seen thus far:
\begin{enumerate}
\item $Z(\beta)$ is a continuous function. $Z(0)=|\calX^n|$ and
$\lim_{\beta\to\infty}\frac{\ln Z(\beta)}{\beta}=-E_{GS}$.
\item Generating moments: $\langle\calE\rangle =-\dd\ln Z/\dd\beta$,
$\mbox{Var}\{\calE(\bX)\}=\dd^2\ln Z/\dd\beta^2$ $\rightarrow$ convexity of
$\ln Z$, and hence also of $\phi(\beta)$.
\item $\phi$ and $\Sigma$ are a Legendre--transform pair. $\Sigma$ is
concave.
\item $\Sigma(\epsilon)$ coincides with the Shannon entropy of the B-G
distribution.
\item $F_Q=F_P+kT\cdot D(Q\|P).$
\end{enumerate}

\noindent
{\bf Exercise:} Consider $Z(\beta)$ for an {\it imaginary temperature}
$\beta=j\omega$, where $j=\sqrt{-1}$, and define $z(E)$ as the inverse
Fourier transform of $Z(j\omega)$. Show that $z(E)=\Omega(E)$ is the density of states,
i.e., for $E_1 < E_2$, the number of states with energy between $E_1$ and
$E_2$ is given by $\int_{E_1}^{E_2}z(E)\dd E$.$~~~~\Box$\\
Thus, $Z(\cdot)$ can be related to energy enumeration in two different ways:
one is by the Legendre transform of $\ln Z$ for real $\beta$, and the other
is by the inverse Fourier transform of $Z$ for imaginary $\beta$.
This double connection between $Z$ and $\Omega$ is no coincidence, as we shall
see later on.

\vspace{0.25cm}

\noindent
{\it Example -- A two level system.} Similarly to the
earlier example of Schottky defets, which was previously given in the context
of the microcanonical ensemble,
consider now a system of $n$ independent
particles, each having two possible states: state $0$ of zero energy and
state $1$, whose energy is $\epsilon_0$, i.e., $\calE(x)=\epsilon_0x$,
$x\in\{0,1\}$. The $x_i$'s are independent, each having a marginal:
\begin{equation}
P(x)=\frac{e^{-\beta\epsilon_0x}}{1+e^{-\beta\epsilon_0}}~~~x\in\{0,1\}.
\end{equation}
In this case,
\begin{equation}
\phi(\beta )=\ln(1+e^{-\beta\epsilon_0})
\end{equation}
and 
\begin{equation}
\Sigma(\epsilon)=\min_{\beta\ge
0}[\beta\epsilon+\ln(1+e^{-\beta\epsilon_0})].
\end{equation}
To find $\beta^*(\epsilon)$, we take the derivative and equate to zero:
\begin{equation}
\epsilon-\frac{\epsilon_0e^{-\beta\epsilon_0}}{1+e^{-\beta\epsilon_0}}=0
\end{equation}
which gives
\begin{equation}
\beta^*(\epsilon)=\frac{\ln(\epsilon/\epsilon_0-1)}{\epsilon_0}.
\end{equation}
On substituting this back into the above expression of $\Sigma(\epsilon)$, we
get:
\begin{equation}
\Sigma(\epsilon)=\frac{\epsilon}{\epsilon_0}\ln\left(\frac{\epsilon}{\epsilon_0}-1\right)
+\ln\left[1+\exp\left\{-\ln\left(\frac{\epsilon}{\epsilon_0}-1\right)\right\}\right],
\end{equation}
which after a short algebraic manipulation, becomes
\begin{equation}
\Sigma(\epsilon)=h_2\left(\frac{\epsilon}{\epsilon_0}\right),
\end{equation}
just like in the Schottky example. In the other direction:
\begin{equation}
\phi(\beta)=\max_\epsilon\left[h_2\left(\frac{\epsilon}{\epsilon_0}\right)-
\beta\epsilon\right],
\end{equation}
whose achiever $\epsilon^*(\beta)$ solves the zero--derivative equation:
\begin{equation}
\frac{1}{\epsilon_0}\ln\left[\frac{1-\epsilon/\epsilon_0}{\epsilon/\epsilon_0}\right]=\beta
\end{equation}
or equivalently,
\begin{equation}
\epsilon^*(\beta)=\frac{\epsilon_0}{1+e^{-\beta\epsilon_0}},
\end{equation}
which is exactly the inverse function of $\beta^*(\epsilon)$ above, and
which when plugged back into the expression of $\phi(\beta)$, indeed gives
\begin{equation}
\phi(\beta)=\ln(1+e^{-\beta\epsilon_0}).~~~~\Box
\end{equation}

\vspace{0.25cm}

\noindent
{\it Comment:} A very similar model 
(and hence with similar results)
pertains to non--interacting spins
(magnetic moments), where the only difference is that $x\in\{-1,+1\}$ rather
than $x\in\{0,1\}$. Here, the meaning of the parameter $\epsilon_0$
becomes that of a magnetic field, which is more customarily denoted by 
$B$ (or $H$), and which is either parallel or antiparallel to
that of the spin, and so the potential energy 
(in the appropriate physical units), $\vec{B}\cdot\vec{x}$, is either
$Bx$ or $-Bx$. Thus,
\begin{equation}
P(x)=\frac{e^{\beta Bx}}{2\cosh(\beta B)};~~~Z(\beta)=2\cosh(\beta B).
\end{equation}
The net {\it magnetization} per--spin is defined as
\begin{equation}
m\dfn \left< \frac{1}{n}\sum_{i=1}^nX_i\right> = \langle X_1\rangle=
\frac{\partial \phi}{\partial (\beta
B)}=\tanh(\beta B).
\end{equation}
This is the paramagnetic characteristic of the magnetization as a function of
the magnetic field: As $B\to\pm\infty$, the magnetization $m\to\pm 1$
accordingly. When the magnetic field is removed ($B=0$), the magnetization
vanishes too.
We will get back to this model and its extensions in the sequel. $\Box$

\vspace{0.25cm}

\noindent
{\bf Exercise:} Consider a system of $n$ non--interacting particles,
each having a quadratic Hamiltonian, $\calE(x)=\frac{1}{2}\alpha x^2$,
$x\in\reals$. Show that here, 
\begin{equation}
\Sigma(\epsilon)=\frac{1}{2}\ln\left(\frac{4\pi e\epsilon}{\alpha}\right)
\end{equation}
and 
\begin{equation}
\phi(\beta)=\frac{1}{2}\ln\left(\frac{2\pi}{\alpha\beta}\right).
\end{equation}
Show that $\beta^*(\epsilon)=1/(2\epsilon)$ and hence
$\epsilon^*(\beta)=1/(2\beta)$. 

\subsection{The Energy Equipartition Theorem}

From the last exercise, we have learned that 
for a quadratic Hamiltonian, $\calE(x)=\frac{1}{2}\alpha x^2$, we have
$\epsilon^*(\beta)$, namely,
the average per--particle energy, is given $1/(2\beta)=kT/2$,
independently of $\alpha$. If we have $n$ such quadratic terms, then of course, we
end up with $nkT/2$. In the case of the ideal gas, we have 3 such terms (one
for each dimension) per particle, thus a total of $3n$ terms, and so,
$E=3nkT/2$, which is exactly what we obtained also in the microcanonical
ensemble, which is equivalent (recall that this was obtained then 
by equating $1/T$ to the derivative of $S(E)=k\ln[\mbox{const}\times
E^{3n/2}]$).
In fact, we observe that in the canonical ensemble,
whenever we have an Hamiltonian of the form $\frac{\alpha}{2}x_i^2+$
some arbitrary terms that do not depend on $x_i$, 
then $x_i$ is Gaussian
(with variance $kT/\alpha$) and independent
of the other guys, i.e.,
$p(x_i)\propto
e^{-\alpha x_i^2/(2kT)}$.
Hence it contributes an amount of 
\begin{equation}
\left<\frac{1}{2}\alpha X_i^2\right> =\frac{1}{2}\alpha\cdot
\frac{kT}{\alpha}=
\frac{kT}{2}
\end{equation}
to the total
average energy, independently of $\alpha$. It is more precise to refer
to this $x_i$ as a
{\it degree of freedom} rather than a particle. This is because in
the 3D world, the kinetic energy, for example, is given by
$p_x^2/(2m)+p_y^2/(2m)+p_z^2/(2m)$, that is, each particle contributes {\it three}
additive quadratic terms rather than one (just like three independent one--dimensional
particles) and so, it contributes $3kT/2$. This principle is called the
{\it the energy equipartition theorem}. In the sequel, we will see that it is
quite intimately related to rate--distortion theory for quadratic distortion
measures. 

Below is a direct derivation of the equipartition theorem:
\begin{eqnarray}
\left<\frac{1}{2}aX^2\right> &=& \frac{\int_{-\infty}^{\infty}\dd
x(\alpha x^2/2) e^{-\beta\alpha x^2/2}}{\int_{-\infty}^{\infty}\dd
xe^{-\beta\alpha x^2/2)}}~~~\mbox{num.\ \& den.\ have closed
forms, but we use another way:}\nonumber\\
&=&-\frac{\partial}{\partial \beta}\ln\left[\int_{-\infty}^{\infty}\dd xe^{-\beta\alpha x^2/2}
\right]\nonumber\\
&=&-\frac{\partial}{\partial \beta}\ln\left[\frac{1}{\sqrt{\beta}}
\int_{-\infty}^{\infty}\dd(\sqrt{\beta}x)e^{-\alpha(\sqrt{\beta}x)^2/2}\right]\nonumber\\
&=&-\frac{\partial}{\partial \beta}\ln\left[\frac{1}{\sqrt{\beta}}
\int_{-\infty}^{\infty}\dd ue^{-\alpha u^2/2}\right]~~~\mbox{The integral is
now a constant, independent of $\beta$.}\nonumber\\
&=&\frac{1}{2}\frac{\dd
\ln\beta}{\dd\beta}=\frac{1}{2\beta}=\frac{kT}{2}.\nonumber
\end{eqnarray}
This simple trick, that bypasses the need to calculate integrals,
can easily be extended in two directions at least (exercise):
\begin{itemize}
\item Let $\bx\in\reals^n$ and let $\calE(\bx)=\frac{1}{2}\bx^TA\bx$, where $A$ is
a $n\times n$ positive definite matrix. This corresponds to a physical system
with a quadratic Hamiltonian, which includes also interactions between pairs (e.g.,
Harmonic oscillators or springs, which are coupled because they are tied
to one another). It turns out that here, regardless of $A$, we get:
\begin{equation}
\langle \calE(\bX)\rangle = \left<\frac{1}{2}\bX^TA\bX\right> = n\cdot\frac{kT}{2}.
\end{equation}
\item Back to the case of a scalar $x$, but suppose now a more general power--law
Hamiltoinan, $\calE(x)=\alpha|x|^\theta$. In this case, we get
\begin{equation}
\langle \calE(X)\rangle = \left<\alpha|X|^\theta\right> = \frac{kT}{\theta}.
\end{equation}
Moreover, if $\lim_{x\to\pm\infty} xe^{-\beta\calE(x)}=0$ for all $\beta > 0$,
and we denote $\calE'(x)\dfn \dd\calE(x)/\dd x$, then
\begin{equation}
\langle X\cdot\calE'(X)\rangle =kT.
\end{equation}
It is easy to see that the earlier power--law result is obtained as a special case of
this, as $\calE'(x)=\alpha\theta|x|^{\theta-1}\mbox{sgn}(x)$ in this case.
\end{itemize}

\vspace{0.25cm}

\noindent
{\it Example/Exercise -- Ideal gas with gravitation:} Let
\begin{equation}
\calE(x)= \frac{p_x^2+p_y^2+p_z^2}{2m}+mgz.
\end{equation}
The average kinetic energy of each particle is $3kT/2$, as said before.
The contribution of the average potential energy is $kT$ (one degree of
freedom with $\theta=1$). Thus, the total is $5kT/2$, where $60\%$ come from
kinetic energy and $40\%$ come from potential energy, universally,
that is, independent of $T$, $m$, and $g$. $\Box$

\subsection{The Grand--Canonical Ensemble (Optional)}

Looking a bit back, then a brief summary of what we have
done thus far, is the following: we started off with the microcanonical
ensemble, which was very restricitve in the sense that the energy was held
strictly fixed to the value of $E$, the number of particles was held strictly
fixed to the value of $n$, and at least in the example of a gas, the volume
was also held strictly fixed to a certain value $V$. In the passage from the
microcanonical ensemble to the canonical one, we slightly relaxed the first of
these parameters -- $E$: Rather than insisting on a fixed value of $E$, we
allowed
energy to be exchanged back and forth with the environment, and thereby to
slightly
fluctuate (for large $n$)
around a certain average value, which was controlled by temperature,
or equivalently, by the choice of $\beta$. This was done while keeping in mind
that the total energy of both system and heat bath must be kept fixed, by the
law of energy conservation, which allowed us to look at the combined system as
an isolated one, thus obeying the microcanonical ensemble.
We then had a one--to--one
correspondence between the extensive quantity $E$ and the intensive variable
$\beta$, that adjusted its average value. But the other extensive
variables, like $n$ and $V$ were still kept strictly fixed.

It turns out, that we can continue in this spirit, and `relax' also either one
of the
other variables $n$ or $V$ (but not both at the same time),
allowing it to fluctuate around a typical average
value, and controlling it by a corresponding intensive variable.
Like $E$, both $n$ and $V$ are also subjected to conservation laws
when the combined system is considered.
Each one of these relaxations, leads to a new ensemble in addition to the
microcanonical
and the canonical ensembles that we have already seen.
In the case where it is the variable $n$ that is allowed to be flexible, this
ensemble is called the {\it grand--canonical ensemble}. In the case where it
is the variable $V$, this is called the {\it Gibbs ensemble}.
And there are, of course, additional ensembles based on this principle,
depending
on what kind of the physical sytem is under discussion. We will not delve into
all of them here because this not a course in physics, after all.
We will describe, however, in some level of detail the grand--canonical
ensemble.

The fundamental idea is essentially the very same as the one we used to derive
the canonical
ensemble, we just extend it a little bit:
Let us get back to our (relatively small) subsystem, which is in contact
with a heat bath, and this time, let us allow this subsystem to exchange
with the heat bath,
not only energy, but also matter, i.e., particles. The heat bath consists of
a huge reservoir of energy and particles. The total energy is $E_0$ and the
total number of particles is $n_0$. Suppose that we can calculate the density
of
states of the heat bath as function of both its energy $E'$ and amount of
particles $n'$, call it $\Omega_{HB}(E',n')$. A microstate now is a
combnination $(\bx,n)$, where $n$ is the (variable) number of particles
in our subsystem and $\bx$ is as before for a given $n$.
From the same considerations as before, whenever our subsystem
is in state $(\bx,n)$, the heat bath can be in any one of
$\Omega_{HB}(E_0-\calE(\bx),n_0-n)$ microstates of its own. Thus,
owing to the microcanonical ensemble,
\begin{eqnarray}
P(\bx,n)&\propto& \Omega_{HB}(E_0-\calE(\bx),n_0-n)\nonumber\\
&=& \exp\{S_{HB}(E_0-\calE(\bx),n_0-n)/k\}\nonumber\\
&\approx& \exp\left\{\frac{S_{HB}(E_0,n_0)}{k}-\frac{1}{k}
\frac{\partial S_{HB}}{\partial E}\cdot\calE(\bx)-
\frac{1}{k}\frac{\partial S_{HB}}{\partial n}\cdot n
\right\}\nonumber\\
&\propto& \exp\left\{-\frac{\calE(\bx)}{kT}+\frac{\mu n}{kT}\right\}
\end{eqnarray}
where we have now defined the {\it chemical potential} $\mu$ (of the heat
bath) as:
\begin{equation}
\mu\dfn -T\cdot\frac{\partial S_{HB}(E',n')}{\partial
n'}\bigg|_{E'=E_0,n'=n_0}.
\end{equation}
Thus, we now have the grand--canonical distribution:
\begin{equation}
P(\bx,n)=\frac{e^{\beta[\mu n-\calE(\bx)]}}{\Xi(\beta,\mu)},
\end{equation}
where the denominator is called the {\it grand partition function}:
\begin{equation}
\Xi(\beta,\mu)\dfn \sum_{n=0}^\infty e^{\beta\mu
n}\sum_{\bx}e^{-\beta\calE(\bx)}\dfn \sum_{n=0}^\infty e^{\beta\mu
n}Z_n(\beta).
\end{equation}
It is sometimes convenient to change variables and to define $z=e^{\beta\mu}$
(which is called the {\it fugacity}) and then, define
\begin{equation}
\tilde{\Xi}(\beta,z)=\sum_{n=0}^\infty z^n Z_n(\beta).
\end{equation}
This notation emphasizes the fact that for a given $\beta$, $\tilde{\Xi}(z)$
is actually the $z$--transform of the sequence $Z_n$.
A natural way to think about $P(\bx,n)$ is as $P(n)\cdot P(\bx|n)$, where
$P(n)$ is proportional to $z^n Z_n(\beta)$ and $P(\bx|n)$ corresponds to the
canonical ensemble as before.

Using the grand partition function, it is now easy to obtain moments of the RV
$n$. For
example, the first moment is:
\begin{equation}
\langle n \rangle =\frac{\sum_n nz^n Z_n(\beta)}{\sum_n z^n Z_n(\beta)}=
z\cdot \frac{\partial \ln \tilde{\Xi}(\beta,z)}{\partial z}.
\end{equation}
Thus, we have replaced the fixed number of particles $n$ by a random number of
particles, which concentrates around an average controlled by the parameter
$\mu$, or
equivalently, $z$. The dominant value of $n$ is the one that maximizes the
product $z^nZ_n(\beta)$, or equivalently, $\beta\mu n+\ln Z_n(\beta)$. Thus,
$\ln\tilde{\Xi}$ is related to $\ln Z_n$ by another kind of a Legendre
transform.

When two systems, with total energy $E_0$
and a total number of particles $n_0$, are brought into contact,
allowing both energy and matter exchange, then the dominant combined states
are those for which $\Omega_1(E_1,n_1)\cdot\Omega_2(E_0-E_1,n_0-n_1)$, or
equivalently,
$S_1(E_1,n_1)+S_2(E_0-E_1,n_0-n_1)$, is maximum. By equating to zero the
partial derivatives w.r.t.\ both $E_1$ and $n_1$, we find that in equilibrium
both the temperatures $T_1$ and $T_2$ are the same and the
chemical potentials $\mu_1$ and $\mu_2$ are the same.

Finally, I would like to point out that beyond the obvious physical
significance of the grand--canonical ensemble, sometimes it proves
useful to work with it from the reason of pure mathematical convenience. This
is shown
in the following example.

\vspace{0.1cm}

\noindent
{\it Example -- Quantum Statistics.} Consider an ensemble of indistinguishable
particles, each one of which may be in a certain quantum state labeled by
$1,2,\ldots,r,\ldots$. Associated with quantum state number $r$, there is an
energy
$\epsilon_r$. Thus, if there are
$n_r$ particles in each state $r$,
the total energy is $\sum_r n_r\epsilon_r$, and so, the canonical
partition function is:
\begin{equation}
Z_n(\beta)=\sum_{\bn:~\sum_rn_r=n} \exp\{-\beta\sum_rn_r\epsilon_r\}.
\end{equation}
The constraint $\sum_rn_r=n$, which accounts for the fact that the total
number of
particles must be $n$, causes an extremely severe headache in the calculation.
However, if we pass to the grand--canonical
ensemble, things becomes extremely easy:
\begin{eqnarray}
\tilde{\Xi}(\beta,z)&=&\sum_{n\ge 0}z^n \sum_{\bn:~\sum_rn_r=n}
\exp\{-\beta\sum_rn_r\epsilon_r\}\nonumber\\
&=&\sum_{n_1\ge 0}\sum_{n_2\ge 0}\ldots z^{\sum_rn_r}
\exp\{-\beta\sum_rn_r\epsilon_r\}\nonumber\\
&=&\sum_{n_1\ge 0}\sum_{n_2\ge 0}\ldots \prod_{r\ge 1}z^{n_r}
\exp\{-\beta n_r\epsilon_r\}\nonumber\\
&=& \prod_{r\ge 1}\sum_{n_r\ge 0} [ze^{-\beta\epsilon_r}]^{n_r}
\end{eqnarray}
In the case where $n_r$ is unlimited ({\it Bose--Einstein} particles, or {\it
Bosons}),
each factor indexed by $r$ is clearly a geometric series, resulting in
$\tilde{\Xi}=\prod_r[1/(1-ze^{-\beta\epsilon_r})]$.
In the case where no quantum state can be populated by more than one particle,
owing to Pauli's exclusion principle ({\it Fermi--Dirac} particles, or {\it
Fermions}), each factor in the product
contains two terms only, pertaining to $n_r=0,1$,
and the result is $\tilde{\Xi}=\prod_r(1+ze^{-\beta\epsilon_r})$.
In both cases, this is fairly simple. Having computed
$\tilde{\Xi}(\beta,z)$, we can in principle, return to $Z_n(\beta)$ by
applying the
inverse $z$--transform. We will get back to this in the sequel.

\subsection{Gibbs' Inequality, the 2nd Law, and the Data Processing
Thm}

While the laws of physics draw the boundaries between the possible and the
impossible in Nature, the coding theorems of information theory, or
more precisely, their converses, draw the
boundaries between the possible and the impossible in coded communication
systems and data processing. Are there any relationships between these two
facts?

We are now going to demonstrate that there are some indications that
the answer to this question is affirmative. In particular, we are going to
see that there is an intimate relationship between the second law of
thermodynamics and the data processing theorem (DPT), asserting that
if $X\to U\to V$ is a Markov chain, then $I(X;U)\ge I(X;V)$. The reason for
focusing our attention on the DPT is that it is actually the most fundamental
inequality that supports most (if not all) proofs of converse theorems in IT.
Here are just a few points that make this quite clear.
\begin{enumerate}
\item {\it Lossy/lossless source coding}: Consider a source vector
$U^N=(U_1,\ldots U_N)$
compressed into a bitstream $X^n=(X_1,\ldots,X_n)$ from which the decoder
generates a reproduction $V^N=(V_1,\ldots,V_N)$ with distortion
$\sum_{i=1}^N \bE\{d(U_i,V_i)\}\le ND$. Then, by the DPT,
$I(U^N;V^N)\le I(X^n;X^n)=H(X^n)$, where $I(U^N;V^N)$ is further lower bounded
by
$NR(D)$ and $H(X^n)\le n$, which together lead to the converse to the lossy
data
compression theorem, asserting that the compression ratio $n/N$ cannot be
less than $R(D)$. The case of lossless compression is obtained as a special
case where $D=0$.
\item {\it Channel coding under bit error probability}: Let $U^N=(U_1,\ldots
U_N)$
be drawn from the binary symmetric course (BSS), designating $M=2^N$
equiprobable
messages of length $N$. The encoder maps $U^N$ into a channel input vector
$X^n$, which in turn, is sent across the channel. The receiver observes $Y^n$,
a noisy version of $X^n$, and decodes the message as $V^N$. Let
$P_b=\frac{1}{N}\sum_{i=1}^N\mbox{Pr}\{V_i\ne U_i\}$ designate
the bit error probability. Then, by the DPT, $I(U^N;V^N)\le I(X^n;Y^n)$, where
$I(X^n;Y^n)$ is further upper bounded by $nC$, $C$ being the channel capacity,
and $I(U^N;V^N)=H(U^N)-H(U^N|V^N)\ge N-\sum_{i=1}^N H(U_i|V_i)\ge
N-\sum_ih_2(\mbox{Pr}\{V_i\ne U_i\})\ge N[1-h_2(P_b)]$.
Thus, for $P_b$ to vanish, the coding rate, $N/n$ should not exceed $C$.
\item {\it Channel coding under block error probability -- Fano's inequality}:
Same as in the previous item, except that the error performance is the block
error probability $P_B=\mbox{Pr}\{V^N\ne U^N\}$. This, time $H(U^N|V^N)$,
which is identical to $H(U^N,E|V^N)$, with $E\equiv \calI\{V^N\ne U^N\}$, is
decomposed as $H(E|V^N)+H(U^N|V^N,E)$, where the first term is upper bounded
by 1 and the second term is upper bounded by $P_B\log(2^N-1) < NP_B$, owing to
the fact that the maximum of $H(U^N|V^N,E=1)$ is obtained when $U^N$ is
distributed uniformly over all $V^N\ne U^N$. Putting these facts all together,
we obtain Fano's inequality $P_B\ge 1-1/n-C/R$, where $R=N/n$ is the coding
rate. Thus, the DPT directly supports Fano's inequality,
which in turn is the main tool for proving converses to channel
coding theorems in a large variety of communication situations, including
network configurations.
\item {\it Joint source--channel coding and the separation principle}: In a
joint
source--channel situation, where the source vector $U^N$ is mapped to a
channel input vector $X^n$ and the channel output vector $Y^n$ is decoded into
a reconstruction $V^N$, the DPT gives rise to the chain
of inequalities $NR(D)\le I(U^N;V^N)\le I(X^n;Y^n)\le nC$, which is the
converse to the joint source--channel coding theorem, whose direct part can
be achieved by separate source- and channel coding. Items 1 and 2 above are
special
cases of this.
\item {\it Conditioning reduces entropy}: Perhaps even more often than the
term ``data processing theorem'' can be found as part of a proof of
a converse theorem,  one encounters an equivalent of this theorem
under the slogan ``conditioning reduces entropy''. This in turn is part of
virtually every
converse proof in the literature. Indeed, if $(X,U,V)$ is a triple of RV's,
then this statement means that $H(X|V)\ge H(X|U,V)$. If, in addition, $X\to
U\to V$
is a Markov chain, then $H(X|U,V)=H(X|U)$, and so, $H(X|V)\ge H(X|U)$, which
in turn is equivalent to the more customary form of the DPT, $I(X;U)\ge
I(X;V)$, obtained by subtracting $H(X)$ from both sides of the entropy
inequality. In fact, as we shall see shortly, it is this entropy inequality
that lends itself more naturally to a
physical interpretation. Moreover, we can think of the
conditioning--reduces--entropy inequality as another form of the DPT even in
the absence of the aforementioned Markov condition, because $X\to(U,V)\to V$
is always a
Markov chain.
\end{enumerate}

Turning now to the physics point of view,
consider a system
which may have two possibile Hamiltonians -- $\calE_0(\bx)$ and
$\calE_1(\bx)$.
Let $Z_i(\beta)$, denote the partition function pertaining to
$\calE_i(\cdot)$,
that is
\begin{equation}
Z_i(\beta)=\sum_{\bx} e^{-\beta\calE_i(\bx)},~~~i=0,1.
\end{equation}
The {\it Gibbs' inequality} asserts that
\begin{equation}
\ln Z_1(\beta)\ge \ln
Z_0(\beta)+\beta\langle\calE_0(\bX)-\calE_1(\bX)\rangle_0
\end{equation}
where $\langle\cdot\rangle_0$ denotes averaging w.r.t.\ $P_0$ -- the canonical
distribution
pertaining the Hamiltonian $\calE_0(\cdot)$. Equivalently, this inequality can
be presented as follows:
\begin{equation}
\langle\calE_1(\bX)-\calE_0(\bX)\rangle_0\ge
\left[-\frac{\ln Z_1(\beta)}{\beta}\right]-
\left[-\frac{\ln Z_0(\beta)}{\beta}\right]\equiv F_1-F_0, ~~~~~~~~\mbox{(*)}
\end{equation}
where $F_i$ is the free energy pertaining to the canonical ensemble of
$\calE_i$, $i=0,1$.

\begin{figure}[ht]
\hspace*{3cm}\input{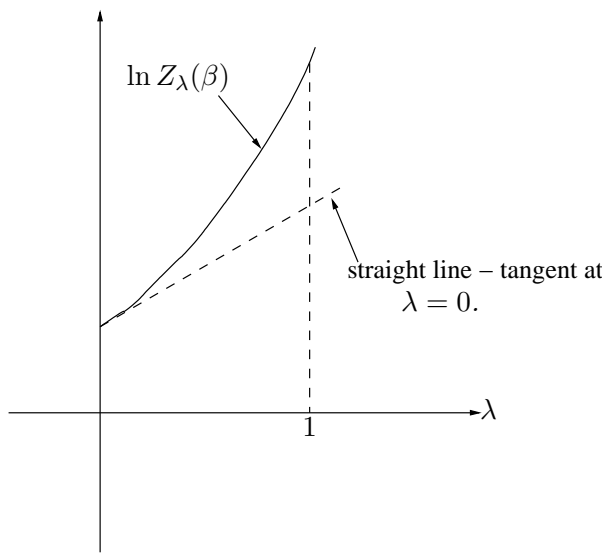}
\caption{\small The function $\ln Z_\lambda(\beta)$ is convex in $\lambda$ and
hence lies above its tangent at the origin.}
\label{lnzconvex}
\end{figure}

This inequality is easily proved by defining
an Hamiltoinan $\calE_\lambda(\bx)=(1-\lambda)\calE_0(\bx)+
\lambda\calE_1(\bx)=\calE_0(\bx)+\lambda[\calE_1(\bx)-\calE_0(\bx)]$
and using the convexity of the corresponding
log--partition function w.r.t.\ $\lambda$. Specifically, let us define the
partition
function:
\begin{equation}
Z_\lambda(\beta)=\sum_{\bx} e^{-\beta\calE_\lambda(\bx)}.
\end{equation}
Now, since $\calE_{\lambda}(\bx)$ is affine in $\lambda$, then it is easy
to show that $\dd^2\ln Z_\lambda/\dd\lambda^2\ge 0$ (just like this
was done with $\dd^2\ln Z(\beta)/\dd\beta^2\ge 0$ before) and so
$\ln Z_\lambda(\beta)$ is convex in $\lambda$ for fixed $\beta$.
It follows then that the curve of $\ln Z_\lambda(\beta)$, as a function of
$\lambda$, must lie above the straight line that is tangent to this curve
at $\lambda=0$ (see Fig.\ \ref{lnzconvex}), that is, the graph corresponding
to the affine function $\ln Z_0(\beta)+\lambda\cdot\left[\frac{\partial\ln
Z_\lambda(\beta)}{\partial\lambda}\right]_{\lambda=0}$.
In particular,
setting $\lambda=1$, we get:
\begin{equation}
\ln Z_1(\lambda)\ge \ln
Z_0(\beta)+\frac{\partial\ln
Z_\lambda(\beta)}{\partial\lambda}\bigg|_{\lambda=0}.
\end{equation}
and the second term is:
\begin{equation}
\frac{\partial\ln Z_\lambda(\beta)}{\partial\lambda}\bigg|_{\lambda=0}=
\frac{\beta\sum_{\bx}[\calE_0(\bx)-\calE_1(\bx)] e^{-\beta\calE_0(\bx)}}{
\sum_{\bx}e^{-\beta\calE_0(\bx)}}\dfn
\beta\left<\calE_0(\bX)-\calE_1(\bX)\right>_0,
\end{equation}
Thus, we have obtained
\begin{equation}
\ln \left[\sum_{\bx}e^{-\beta\calE_1(\bx)}\right]\ge
\ln \left[\sum_{\bx}e^{-\beta\calE_0(\bx)}\right]+
\beta\left<\calE_0(\bX)-\calE_1(\bX)\right>_0,
\end{equation}
and the proof is complete. In fact, the l.h.s.\ minus the r.h.s.\ is nothing
but
$D(P_0\|P_1)$, where $P_i$ is the B--G distribution pertaining to
$\calE_i(\cdot)$, $i=0,1$.

We now offer a possible physical interpretation to the Gibbs' inequality:
Imagine that a
system with Hamiltoinan $\calE_0(\bx)$ is in equilibrium for all $t < 0$, but
then, at time $t=0$, the Hamitonian changes {\it abruptly} from the
$\calE_0(\bx)$ to $\calE_1(\bx)$ (e.g., by suddenly applying a force on the
system),
which means that if the system is found at
state
$\bx$ at time $t=0$, additional energy of $W=\calE_1(\bx)-\calE_0(\bx)$ is
suddenly
`injected' into the system. This additional energy can be thought of as work
performed on the system, or as supplementary potential energy.
Since this passage between $\calE_0$ and $\calE_1$ is abrupt,
the average of $W$ should be taken
w.r.t.\ $P_0$,
as the state $\bx$ does not change instantaneously.
This average is exactly what we have at the left--hand side eq.\ (*).
The Gibbs inequality tells us then that
this average work is at least as large as
$\Delta F=F_1-F_0$, the increase in free energy.\footnote{This is
related to the interpretation
of the free--energy difference $\Delta F=F_1-F_0$
as being the maximum amount of work in an isothermal process.}
The difference $\langle W\rangle_0-\Delta F$ is due to the irreversible nature
of the abrupt energy injection, and this irreversibility
means an increase of the total entropy of the system and its environment, and
so, the Gibbs'
inequality is, in fact, a version of the second law of
thermodynamics.\footnote{
From a more general physical perspective, the Jarzynski equality tells
that under certain conditions on the test system and the heat bath,
and given any protocol $\{\lambda(t)\}$ of
changing the control variable $\lambda$ (of $\calE_\lambda(\bx)$), the work
$W$ applied to the system
is a RV which satisfies $\langle e^{-\beta W}\rangle = e^{-\beta\Delta F}$.
By Jensen's inequality, $\langle e^{-\beta W}\rangle$ is lower bounded by
$e^{-\beta\langle W\rangle}$,
and so, we obtain
$\langle W\rangle \ge \Delta F$ (which is known as the
minimum work principle), now in more generality than in the Gibbs'
inequality, which is limited to the
case where $\lambda(t)$ is a step function.
At the other extreme, when
$\lambda(t)$ changes very slowly, corresponding to a reversible process, $W$
approaches determinism, and then Jensen's inequality becomes tight, which
then gives (in the limit) $W=\Delta F$ with no increase in entropy.}
This excess work beyond the free--energy increase, $\langle W\rangle_0-\Delta
F$,
which can be thought of as the ``dissipated work,''
can easily shown (exercise) to be equal to $kT\cdot D(P_0\|P_1)$, where
$P_0$ and $P_1$ are the canonical distributions pertaining to $\calE_0$ and
$\calE_1$, respectively. Thus, the divergence is given yet another physical
significance.

Now, let us see how the Gibbs' inequality is related to the DPT.
Consider a triple of random variables $(\bX,\bU,\bV)$ which
form a Markov chain $\bX\to \bU\to \bV$.
The DPT asserts that $I(\bX;\bU)\ge I(\bX;\bV)$.
We can obtain the DPT as a special case of the Gibbs inequality as follows:
For a given realization $(\bu,\bv)$ of the random variables $(\bU,\bV)$,
consider
the Hamiltonians
\begin{equation}
\calE_0(\bx)=-\ln P(\bx|\bu)=-\ln P(\bx|\bu,\bv)
\end{equation}
and
\begin{equation}
\calE_1(\bx)=-\ln P(\bx|\bv).
\end{equation}
Let us also set $\beta=1$. Thus, for a given $(\bu,\bv)$:
\begin{equation}
\langle W\rangle_0=\langle\calE_1(\bX)-\calE_0(\bX)\rangle_0=
\sum_{\bx} P(\bx|\bu,\bv)[\ln P(\bx|\bu)-\ln P(\bx|\bv)]
=H(\bX|\bV=\bv)-H(\bX|\bU=\bu)
\end{equation}
and after further averaging w.r.t.\ $(\bU,\bV)$,
the average work becomes $H(\bX|\bV)-H(\bX|\bU)=I(\bX;\bU)-I(\bX;\bV)$.
Concerning the free energies, we have
\begin{equation}
Z_0(\beta=1)=\sum_{\bx} \exp\{-1\cdot[-\ln P(\bx|\bu,\bv)]\}=\sum_{\bx}
P(\bx|\bu,\bv)=1
\end{equation}
and similarly,
\begin{equation}
Z_1(\beta=1)=\sum_{\bx} P(\bx|\bv)=1
\end{equation}
which means that $F_0=F_1=0$, and so $\Delta F=0$ as well. So by the Gibbs
inequality, the
average work $I(\bX;\bU)-I(\bX;\bV)$ cannot be smaller than the free--energy
difference, which
in this case vanishes, namely, $I(\bX;\bU)-I(\bX;\bV)\ge 0$, which is the DPT.
Note that in this case, there is a maximum degree of irreversibility:
The identity $I(\bX;\bU)-I(\bX;\bV)=H(\bX|\bV)-H(\bX|\bU)$ means that
whole work $W=I(\bX;\bU)-I(\bX;\bV)$ goes for entropy increase
$S_1T-S_0T=H(\bX|\bV)\cdot
1-H(\bX|\bU)\cdot 1$, whereas the free energy remains unchanged, as mentioned
earlier. Note that the Jarzynski formula (cf.\ last footnote) holds in this
special case, i.e., $\langle e^{-1\cdot W}\rangle = e^{-1\cdot\Delta F}=1$.

The difference between $I(\bX;\bU)$ and $I(\bX;\bV)$,
which accounts for the rate loss in any suboptimal coded communication system,
is then given the meaning of irreversibility and entropy production in the
corresponding physical system. Optimum
(or nearly optimum) communication systems are corresponding to quasistatic
isothermal processes,
where the full free energy is exploited and no work is dissipated (or no work
is carried out at all, in the first place). In other
words, had there been a communication system that violated the converse to the
source/channel coding theorem, one could have created a corresponding physical
system that
violates the second law of thermodynamics, and this, of course, cannot be
true.

\subsection{Large Deviations Theory and Physics of Information Measures}

As I said in the Intro, large deviations theory, the branch of probability
theory that deals with exponential decay rates of probabilities of rare events, has strong
relations to IT, which we have already seen in the IT course through the
eye glasses of the method of types and Sanov's theorem. On the other hand,
large deviations theory has also a strong connection to statistical mechanics,
as we are going to see shortly. Therefore, one of the links between IT and
statistical mechanics goes through rate functions of large deviations theory, or more
concretely, Chernoff bounds. This topic is based on the paper:
N.~Merhav,
``An identity of Chernoff bounds
with an interpretation in statistical physics and
applications in information theory,''
{\it IEEE Trans.\ Inform.\ Theory}, vol.\ 54, no.\ 8, pp.\
3710--3721, August 2008.

Let us begin with a very simple question: We have a bunch of i.i.d.\ RV's\
$X_1,X_2,\ldots$ and a certain real function $\calE(x)$. How fast does the
probability of the event
$$\sum_{i=1}^n\calE(X_i)\le nE_0$$
decay as $n$ grows without bound, assuming that $E_0 <
\langle\calE(X)\rangle$ (so that this would be a rare event)?
One way to handle this problem, at least in the finite alphabet case, is the
method of types. Another method is the Chernoff bound:
\begin{eqnarray}
\mbox{Pr}\left\{\sum_{i=1}^n\calE(X_i)\le nE_0\right\}&=&
\bE \calI\left\{\sum_{i=1}^n\calE(X_i)\le nE_0\right\}~~~~\calI(\cdot)~\mbox{denoting
the indicator function}\nonumber\\
&\le&\bE \exp\left\{\beta\left[nE_0-\sum_{i=1}^n\calE(X_i)
\right]\right\}~~~~~\leftarrow~~\forall~\beta\ge 0:~\calI\{Z< a\}\le e^{\beta(a-Z)} \nonumber\\
&=&e^{\beta nE_0}\bE\exp\left\{-\beta\sum_{i=1}^n\calE(X_i)\right\}\nonumber\\
&=&e^{\beta nE_0}\bE\left\{\prod_{i=1}^n\exp\{-\beta\calE(X_i)\}\right\}\nonumber\\
&=&e^{\beta nE_0}\left[\bE\exp\{-\beta\calE(X_1)\}\right]^n\nonumber\\
&=&\exp\left\{n\left[\beta E_0+\ln\bE\exp\{-\beta\calE(X_1)\}\right]\right\}\nonumber
\end{eqnarray}
As this bound applies for every $\beta\ge 0$, the tightest bound of this
family is obtained by minimizing the r.h.s.\ over $\beta$, which yields
the exponential rate function:
\begin{equation}
\Sigma(E_0)=\min_{\beta\ge 0}[\beta E_0+\phi(\beta)],
\end{equation}
where
\begin{equation}
\phi(\beta)=\ln Z(\beta)
\end{equation}
and
\begin{equation}
Z(\beta)=\bE e^{-\beta\calE(X)}=\sum_x p(x)e^{-\beta\calE(x)}.
\end{equation}
Rings a bell? Note that $Z(\beta)$ here differs from the partition function that
we have encountered thus far only slighlty: the Boltzmann exponentials are
weighed by $\{p(x)\}$ which are independent of $\beta$. But this is not a
crucial difference: one can imagine a physical system where each microstate $x$
is actually a representative of a bunch of more refined microstates $\{x'\}$, whose
number is proportional to $p(x)$ and which all have the same energy as $x$,
that is, $\calE(x')=\calE(x)$. In the domain of the more refined system,
$Z(\beta)$ is (up to a constant) a non--weighted sum of exponentials, as it
should be. More precisely, if $p(x)$ is (or can be approximated by) a rational number
$N(x)/N$, where $N$ is independent of $x$, then imagine that each $x$ gives
rise to $N(x)$ microstates $\{x'\}$ with the same energy as $x$, so that
\begin{equation}
Z(\beta)=\frac{1}{N}\sum_xN(x)e^{-\beta\calE(x)}=\frac{1}{N}\sum_{x'}e^{-\beta\calE(x')},
\end{equation}
and we are back to an ordinary, non--weighted partition function, upto the
constant $1/N$, which is absolutely immaterial.

To summarize what we have seen thus far: the exponential rate function is given by the
Legendre transform of the log--moment generating function. The Chernoff
parameter $\beta$ to be optimized plays the role of the equilibrium temperature
pertaining to energy $E_0$.

\begin{figure}[ht]
\hspace*{5cm}\input{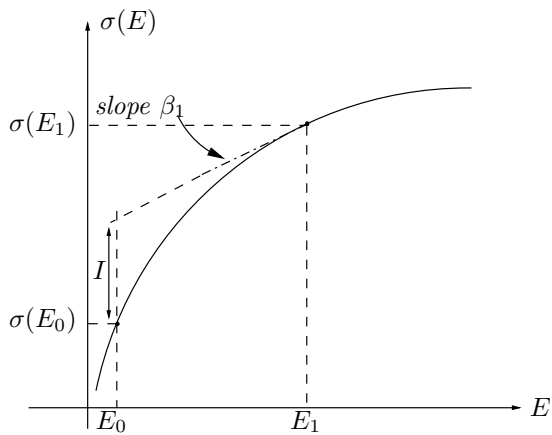}
\caption{\small Graphical interpretation of the LD rate function $I$.}
\label{chernoff}
\end{figure}

Consider next what happens when $p(x)$ is itself a B--G distribution with
Hamiltonian $\calE(x)$ at a certain inverse temperature $\beta_1$, that is
\begin{equation}
p(x)=\frac{e^{-\beta_1\calE(x)}}{\zeta(\beta_1)}
\end{equation}
with
\begin{equation}
\zeta(\beta_1)\dfn\sum_xe^{-\beta_1\calE(x)}.
\end{equation}
In this case, we have
\begin{equation}
Z(\beta)=\sum_x p(x)e^{-\beta\calE(x)}=\frac{\sum_x
e^{-(\beta_1+\beta)\calE(x)}}{\zeta(\beta_1)}=\frac{\zeta(\beta_1+\beta)}{\zeta(\beta_1)}.
\end{equation}
Thus,
\begin{eqnarray}
\Sigma(E_0)&=&\min_{\beta\ge 0}[\beta
E_0+\ln\zeta(\beta_1+\beta)]-\ln\zeta(\beta_1)\nonumber\\
&=&\min_{\beta\ge 0}[(\beta+\beta_1)
E_0+\ln\zeta(\beta_1+\beta)]-\ln\zeta(\beta_1)-\beta_1E_0\nonumber\\
&=&\min_{\beta\ge \beta_1}[\beta
E_0+\ln\zeta(\beta)]-\ln\zeta(\beta_1)-\beta_1E_0\nonumber\\
&=&\min_{\beta\ge \beta_1}[\beta
E_0+\ln\zeta(\beta)]-[\ln\zeta(\beta_1)+\beta_1E_1]+\beta_1(E_1-E_0)\nonumber
\end{eqnarray}
where $E_1$ is the energy corresponding to $\beta_1$, i.e., $E_1$ is such that
\begin{equation}
\sigma(E_1)\dfn \min_{\beta\ge 0}[\beta E_1+\ln\zeta(\beta)]
\end{equation}
is achieved  by $\beta=\beta_1$. Thus, the second bracketted term of the
right--most side of the last chain is exactly $\sigma(E_1)$, as defined.
If we now assume that $E_0 < E_1$, which is reasonable, because $E_1$ is the
average of $\calE(X)$ under $\beta_1$, and we are assuming that we are dealing
with a rare event where $E_0 < \langle \calE(X)\rangle$. In this case, the
achiever $\beta_0$ of $\sigma(E_0)$ must be larger than $\beta_1$ anyway, and
so, the first bracketted term on the right--most side of the last chain
agrees with $\sigma(E_0)$. We have obtained then that the exponential decay
rate (the rate function) is given by
\begin{equation}
I=-\Sigma(E_0)=\sigma(E_1)-\sigma(E_0)-\beta_1(E_1-E_0).
\end{equation}
Note that $I\ge 0$ thanks to the fact that $\sigma(\cdot)$ is concave. 
It has a simple graphical intepretation as the height difference,
as seen at the point $E=E_0$, between
the tangent to the curve $\sigma(E)$ at $E=E_1$ and the function $\sigma(E)$
itself (see Fig.\ \ref{chernoff}).

Another look is the following: 
\begin{eqnarray}
I&=&\beta_1\left[\left(E_0-\frac{\sigma(E_0)}{\beta_1}\right)-
\left(E_1-\frac{\sigma(E_1)}{\beta_1}\right)\right]\nonumber\\
&=&\beta_1(F_0-F_1)\nonumber\\
&=&D(P_{\beta_0}\|P_{\beta_1})\nonumber\\
&=&\min\{D(Q\|P_{\beta_1}):~\bE_Q\calE(X)\le
E_0\}~~~~\leftarrow~\mbox{exercise}\nonumber
\end{eqnarray}
The last line is exactly what we would have obtained using the method of
types. This means that the dominant instance of the large deviations event
under discussion pertains to thermal equilibrium (minimum free energy)
complying with the constraint(s) dictated by this event. This will also be the
motive of the forthcoming results.

\vspace{0.2cm}

\noindent
{\bf Exercise:} What happens if $p(x)$ is B--G with an Hamiltonian
$\hat{\calE}(\cdot)$, different from the one of the LD event? $\Box$

Let us now see how this discussion relates to very fundamental information
measures, like the rate--distortion function and channel capacity.
To this end, let us first slightly extend the above Chernoff bound.
Assume that in addition to the RV's $X_1,\ldots,X_n$, there is
also a deterministic sequence of the same length, $y_1,\ldots,y_n$, where
each $y_i$ takes on values in a finite alphabet $\calY$. Suppose also
that the asymptotic regime is such that as $n$ grows without bound, the
relative frequencies $\{\frac{1}{n}\sum_{i=1}^n1\{y_i=y\}\}_{y\in\calY}$ converge to
certain probabilities $\{q(y)\}_{y\in\calY}$. Furthermore, the $X_i$'s
are still independent, but they are no longer necessarily identically
distributed: each one of them is governed by $p(x_i|y_i)$, that is,
$p(\bx|\by)=\prod_{i=1}^n p(x_i|y_i)$. Now, the question is how does
the exponential rate function behave if we look at the event
\begin{equation}
\sum_{i=1}^n\calE(X_i,y_i)\le nE_0
\end{equation}
where $\calE(x,y)$ is a given `Hamiltonian'. What is the motivation for this
question? Where and when do we encounter such a problem?

Well, there are many examples (cf.\ the above mentioned paper), but 
here are two very classical ones, where rate functions of LD events
are directly related to very important information measures.
In both examples, the distributions $p(\cdot|y)$ are
actually the same for all $y\in\calY$ (namely, $\{X_i\}$ are again i.i.d.).
\begin{itemize}
\item {\it Rate--distortion coding.} 
Consider the good old problem of lossy compression with a randomly selected code.
Let $\by=(y_1,\ldots,y_n)$ 
be a given source sequence, typical to $Q=\{q(y),~y\in\calY\}$
(non--typical sequences are not important). Now,
let us randomly select $e^{nR}$ codebook vectors $\{\bX(i)\}$ according to
$p(\bx)=\prod_{i=1}^np(x_i)$. Here is how the direct part of the source coding
theorem essentially works: We first ask ourselves what is the probability that 
a single randomly selected codeword $\bX=(X_1,\ldots,X_n)$ would happen to
fall at distance $\le nD$ from $\by$, i.e., what is the exponential rate of
the probability of the event 
\begin{equation}
\sum_{i=1}^nd(X_i,y_i)\le nD\mbox{?}
\end{equation}
The answer is that it is exponentially about $e^{-nR(D)}$, and that's why we
need slightly more than one over this number, namely, $e^{+nR(D)}$ times to
repeat this `experiment' in order to see at least one `success',
which means being able to encode $\by$ within distortion $D$. So this is
clearly an instance of the above problem, where $\calE=d$ and $E_0=D$.
\item {\it Channel coding.} 
In complete duality, consider the classical channel coding problem,
for a discrete memoryless channel (DMC),
using a randomly selected code. Again, we have a code of size $e^{nR}$,
where each codeword is chosen independently according to
$p(\bx)=\prod_{i=1}^np(x_i)$. Let $\by$ the channel output vector, which is
(with very high probabaility), typical to $Q=\{q(y),~y\in\calY\}$, where
$q(y)=\sum_xp(x)W(y|x)$, $W$ being the single--letter transition probability
matrix of the DMC. Consider a (capacity--achieving) 
threshold decoder which selects the {\it unique}
codeword that obeys
\begin{equation}
\sum_{i=1}^n[-\ln W(y_i|X_i)]\le n[H(Y|X)+\epsilon]~~~~\epsilon>0
\end{equation}
and declares an error whenever no such codeword exists or when there is more
than one such codeword. Now, in the classical proof of the direct part of the
channel coding problem, we first ask ourselves: what is the probability that
an independently selected codeword (and hence not the one transmitted) $\bX$
will pass this threshold? The answer turns out to be exponentially $e^{-nC}$,
and hence we can randomly select up to slightly less than one over this
number, namely, $e^{+nC}$ codewords, before we start to see incorrect
codewords that pass the threshold. Again, this is clearly an instance of our
problem with $\calE(x,y)=-\ln W(y|x)$ and $E_0=H(Y|X)+\epsilon$.
\end{itemize}
Equipped with these two motivating examples, let us get back to the generic
problem we formalized, and see what happens. Once this has been done, we shall return to the
examples. There are (at least) two different ways to address the problem using Chernoff
bounds, and they lead to
two {\it seemingly} different expressions, but since the Chernoff bounding
technique gives the correct exponential behavior, these two expressions must
agree. This identity between the two expressions will have a physical
intepretation, as we shall see. 

The first approach is a direct extension of what we did before: 
\begin{eqnarray}
& &\mbox{Pr}\left\{\sum_{i=1}^n\calE(X_i,y_i)\le nE_0\right\}\nonumber\\
&=&
\bE \calI\left\{\sum_{i=1}^n\calE(X_i,y_i)\le nE_0\right\}\nonumber\\
&\le&\bE \exp\left\{\beta\left[nE_0-\sum_{i=1}^n\calE(X_i,y_i)
\right]\right\}\nonumber\\
&=&e^{n\beta E_0}\prod_{y\in\calY}\bE_y\exp\left\{-\beta
\sum_{i:y_i=y}\calE(X_i,y)\right\}~~~~
\bE_y\dfn\mbox{expectation under}~p(\cdot|y)\nonumber\\
&=&e^{\beta
nE_0}\prod_{y\in\calY}\left[\bE_y\exp\{-\beta\calE(X,y)\}\right]^{n(y)}~~~~n(y)\dfn\mbox{num.\
of}~\{y_i=y\}\nonumber\\
&=&\exp\left\{n\left[\beta
E_0+\sum_{y\in\calY}q(y)\ln\sum_{x\in\calX}p(x|y)\exp\{-\beta\calE(x,y)\}\right]\right\}\nonumber
\end{eqnarray}
and so, the resulting rate function is given by
\begin{equation}
\Sigma(E_0)=\min_{\beta\ge
0}\left[\beta E_0+\sum_{y\in\calY}q(y)\ln Z_y(\beta)\right]
\end{equation}
where
\begin{equation}
Z_y(\beta)\dfn\sum_{x\in\calX}p(x|y)\exp\{-\beta\calE(x,y)\}.
\end{equation}
In the rate--distortion example, this tells us that
\begin{equation}
R(D)=-\min_{\beta\ge 0}\left[\beta
D+\sum_{y\in\calY}q(y)\ln\sum_{x\in\calX}p(x)e^{-\beta
d(x,y)}\right].
\end{equation}
This is a well--known parametric representation of $R(D)$, which can be
obtained via a different route (see, e.g., Gray's book {\it Source Coding
Theory}), where the minimizing $\beta$ is known to have the graphical
interpretation of the local negative slope (or derivative) of the curve
of $R(D)$.
In the case of channel capacity, we obtain in a similar manner:
\begin{eqnarray}
C&=&-\min_{\beta\ge 0}\left[\beta
H(Y|X)+\sum_{y\in\calY}q(y)\ln\sum_{x\in\calX}p(x)e^{-\beta[-\ln
W(y|x)]}\right]\nonumber\\
&=&-\min_{\beta\ge 0}\left[\beta
H(Y|X)+\sum_{y\in\calY}q(y)\ln\sum_{x\in\calX}p(x)W^\beta(y|x)
\right].\nonumber
\end{eqnarray}
{\bf Exercise:} Show that for channel capacity, the minimizing $\beta$ is
always $\beta^*=1$. $\Box$

The other route is to handle each $y\in\calY$ separately: First, observe that
\begin{equation}
\sum_{i=1}^n\calE(X_i,y_i)=\sum_{y\in\calY}\sum_{i:~y_i=y}\calE(X_i,y),
\end{equation}
where now, in each partial sum over $\{i:~y_i=y\}$, we have i.i.d.\ RV's.
The event $\sum_{i=1}^n\calE(X_i,y_i)\le nE_0$ can then be thought of as the
union of all intersections
\begin{equation}
\bigcap_{y\in\calY}\left\{\sum_{i:~y_i=y}\calE(X_i,y)\le n(y)E_y\right\}
\end{equation}
where the union is across all ``possible partial energy allocations''
$\{E_y\}$ which satisfy $\sum_y q(y)E_y\le E_0$. Note that at least when the
$X_i$'s take values on a finite alphabet, each partial sum $\sum_{i:~y_i=y}\calE(X_i,y)$
can take only a polynomial number of values in $n(y)$ (why?), and so, it is
sufficient to `sample' the space of $\{E_y\}$ by polynomially many vectors
in order to cover all possible instances of the event under discussion (see
more details in the paper).
Thus,
\begin{eqnarray}
&&\mbox{Pr}\left\{\sum_{i=1}^n\calE(X_i,y_i)\le nE_0\right\}\nonumber\\
&=&\mbox{Pr}\bigcup_{\{E_y:~\sum_y q(y)E_y\le E_0\}}
\bigcap_{y\in\calY}\left\{\sum_{i:~y_i=y}\calE(X_i,y)\le
n(y)E_y\right\}\nonumber\\
&\exe&\max_{\{E_y:~\sum_y q(y)E_y\le E_0\}}\prod_{y\in\calY}\mbox{Pr}\left\{
\sum_{i:~y_i=y}\calE(X_i,y)\le n(y)E_y\right\}\nonumber\\
&\exe&\max_{\{E_y:~\sum_y q(y)E_y\le
E_0\}}\prod_{y\in\calY}\exp\left\{n(y)\min_{\beta_y\ge 0}[\beta_yE_y+\ln
Z_y(\beta)]\right\}\nonumber\\
&=&\exp\left\{n\cdot\max_{\{E_y:~\sum_y q(y)E_y\le
E_0\}}\sum_{y\in\calY}q(y)\Sigma_y(E_y)\right\}\nonumber
\end{eqnarray}
where we have defined
\begin{equation}
\Sigma_y(E_y)\dfn \min_{\beta_y\ge 0}\left[\beta_yE_y+\ln
Z_y(\beta_y)\right].
\end{equation}
We therefore arrived at an alternative expression of the rate function,
which is 
\begin{equation}
\max_{\{E_y:~\sum_y q(y)E_y\le
E_0\}}\sum_{y\in\calY}q(y)\Sigma_y(E_y).
\end{equation}
Since the two expressions must
agree, we got the following identity:
\begin{center}
\fbox{
$\Sigma(E_0)=\max_{\{E_y:~\sum_y q(y)E_y\le
E_0\}}\sum_{y\in\calY}q(y)\Sigma_y(E_y)$}
\end{center}
A few comments:\\

\noindent
1. In the paper there is also a direct proof of this identity, without
relying on Chernoff bound considerations.\\

\noindent
2. This identity accounts for a certain generalized concavity property of
the entropy function. Had all the $\Sigma_y(\cdot)$'s been the same function,
then this would have been the ordinary concavity property. What makes it
interesting is that it continues to hold for different $\Sigma_y(\cdot)$'s
too.\\

\noindent
3. The l.h.s.\ of this identity is defined by minimization over one
parameter only -- the inverse temperature $\beta$. On the other hand, on the
r.h.s.\ we have a separate inverse temperature for every $y$, because each
$\Sigma_y(\cdot)$ is defined as a separate minimization problem with its own
$\beta_y$. Stated differently, the l.h.s.\ is the minimum of a sum, whereas
in the r.h.s., for given $\{E_y\}$, we have the sum of minima. When do these
two  things agree? The answer is that it happens if all minimizers
$\{\beta_y^*\}$ happen to be the {\it same}. But $\beta_y^*$ depends on $E_y$.
So what happens is that the $\{E_y\}$ (of the outer maximization problem)
are such that the $\beta_y^*$ would all be the same, and would agree also
with the $\beta^*$ of $\Sigma(E_0)$. 
To see why this is true, consider the following chain of inequalities:
\begin{eqnarray}
& &\max_{\{E_y:~\sum_y q(y)E_y\le E_0\}}\sum_yq(y)\Sigma_y(E_y)\nonumber\\
&=&
\max_{\{E_y:~\sum_y q(y)E_y\le E_0\}}\sum_yq(y)\min_{\beta_y} [\beta_yE_y+\ln
Z_y(\beta_y)]\nonumber\\
&\le& \max_{\{E_y:~\sum_y q(y)E_y\le E_0\}}\sum_yq(y)[\beta^*E_y+\ln
Z_y(\beta^*)]~~~~~~~~~\mbox{where $\beta^*$ achieves
$\Sigma(E_0)$}\nonumber\\
&\le& \max_{\{E_y:~\sum_y q(y)E_y\le E_0\}}[\beta^*E_0+\sum_yq(y)\ln
Z_y(\beta^*)]~~~~~~~~\mbox{because $\sum_yq(y)E_y\le E_0$}\nonumber\\
&=& \beta^*E_0+\sum_yq(y)\ln
Z_y(\beta^*)~~~~~~~~\mbox{the bracketted expression no longer depends on $\{E_y\}$}\nonumber\\
&=&\Sigma(E_0).\nonumber
\end{eqnarray}
Both inequalities become equalities 
if $\{E_y\}$ would be allocated such that:\footnote{Exercise: show that there
exists an energy allocation $\{E_y\}$ that satisfies both (i) and (ii)
at the same time.}
(i) $\sum_yq(y)E_y=E_0$ and (ii) $\beta_y^*(E_y)=\beta^*$ for all $y$.
Since the $\beta$'s have the meaning of
inverse temperatures, what we have here is {\bf thermal equilibrium}:
Consider a bunch of $|\calY|$ subsystems, each one of $n(y)$ particles and
Hamiltonian $\calE(x,y)$ indexed by $y$. If all these subsystems are thermally
separated, each one with energy $E_y$, then the total entropy per particle is
$\sum_yq(y)\Sigma_y(E_y)$. The above identity tells us then what happens when
all these systems are brought into thermal contact with one another: The total
energy per particle $E_0$ is split among the different subsystems in a way
that all temperatures become the same -- thermal equilibrium. It follows then
that the dominant instance of the LD event is the one where the contributions
of each $y$, to the partial sum of energies, would correspond to equilibrium.
In the rate--distortion example, this characterizes how much distortion each
source symbol contributes typically.

Now, let us look a bit more closely on the rate--distortion
function:
\begin{equation}
R(D)=-\min_{\beta\ge 0}\left[\beta
D+\sum_{y\in\calY}q(y)\ln\sum_{x\in\calX}p(x)e^{-\beta
d(x,y)}\right].
\end{equation}
As said, the Chernoff parameter $\beta$ has
the meaning of inverse temperature.
The inverse temperature $\beta$ required to `tune' the expected
distortion (internal energy) to $D$, is the solution to the equation
\begin{equation}
D=-\frac{\partial}{\partial \beta}\sum_y q(y)\ln \sum_x p(x)e^{-\beta d(x,y)}
\end{equation}
or equivalently,
\begin{equation}
D=\sum_y q(y)\cdot \frac{\sum_xp(x)d(x,y)e^{-\beta d(x,y)}}{\sum_x p(x)\cdot e^{-\beta d(x,y)}}.
\end{equation}
The Legendre transform relation between
the log--partition function and $R(D)$ induces a one--to--one mapping between
$D$ and $\beta$
which is defined by the above equation. To emphasize this dependency, we
henceforth denote the value of $D$, corresponding to a given $\beta$, by
$D_\beta$. This expected distortion is defined w.r.t.\ the probability
distribution:
\begin{equation}
P_\beta(x,y)=q(y)\cdot P_\beta(x|y)=q(y)\cdot\frac{p(x)e^{-\beta d(x,y)}}{\sum_{x'}p(x')e^{-\beta
d(x',y)}}.
\end{equation}
On substituting
$D_\beta$ instead of $D$ in the
expression of $R(D)$, we have
\begin{equation}
-R(D_\beta)=\beta D_\beta+\sum_yq(y)\ln\sum_x
p(x)e^{-\beta d(x,y)}.
\end{equation}
Note that $R(D_\beta)$ can be represented in an integral form as follows:
\begin{eqnarray}
R(D_\beta)&=&-\int_{0}^{\beta}
\mbox{d}\hat{\beta}\cdot\left(D_{\hat{\beta}}+
\hat{\beta}\cdot\frac{\mbox{d}D_{\hat{\beta}}}
{\mbox{d}\hat{\beta}}-D_{\hat{\beta}}\right)\nonumber\\
&=&-\int_{D_{0}}^{D_\beta}\hat{\beta}\cdot\mbox{d}D_{\hat{\beta}},
\end{eqnarray}
where $D_0=\sum_{x,y}p(x)q(y)d(x,y)$ 
is the value of $D$ corresponsing to $\beta=0$, and for which $R_Q(D)=0$,
This is exactly analogous 
to the thermodynamic equation $S=\int\mbox{d}Q/T$ (following
from $1/T=\mbox{d}S/\mbox{d}Q$), that builds up the entropy from the cumulative
heat. Note that the last equation, in its differential form, reads
$\mbox{d}R(D_\beta)=-\beta\mbox{d}D_\beta$, or $\beta=-R'(D_\beta)$, which means that
$\beta$ is indeed the negative
local slope of the rate--distortion curve $R(D)$.
Returning to the integration variable $\hat{\beta}$, we have:
\begin{eqnarray}
R(D_\beta)&=&-\int_{0}^{\beta}\mbox{d}\hat{\beta}\cdot \hat{\beta}
\cdot\frac{\mbox{d}D_{\hat{\beta}}}{\mbox{d}\hat{\beta}}\nonumber\\
&=&\sum_y q(y)\int_{0}^{\beta}\mbox{d}\hat{\beta}\cdot \hat{\beta}\cdot
\mbox{Var}_{\hat{\beta}}\{d(X,y)|Y=y\}\nonumber\\
&=&\int_{0}^{\beta}\mbox{d}\hat{\beta}\cdot \hat{\beta}\cdot
\mbox{mmse}_{\hat{\beta}}\{d(X,Y)|Y\}\nonumber
\end{eqnarray}
where $\mbox{Var}_{\hat{\beta}}\{\cdot\}$ and
$\mbox{mmse}_{\hat{\beta}}\{\cdot|\cdot\}$ are taken w.r.t.
$P_{\hat{\beta}}(x,y)$. We have 
therefore introduced an integral representation for
$R(D)$ based on the MMSE in estimating the distortion variable $d(X,Y)$ based
on $Y$. In those cases where an exact expression for $R(D)$ is hard to obtain,
this opens the door to upper and 
lower bounds on $R(D)$, which are based on upper and lower bounds
on the MMSE, offered by the plethora 
of bounds available in estimation theory.\\
{\it Exercise}: Show that
$D_\beta=D_{0}-\int_{0}^\beta\mbox{d}\hat{\beta}\cdot\mbox{mmse}_{\hat{\beta}}\{d(X,Y)|Y\}$.

Finally, a word about the high--resolution regime.
The partition function of each $y$ is
\begin{equation}
Z_y(\beta)=\sum_xp(x)e^{-\beta d(x,y)},
\end{equation}
or, in the continuous case,
\begin{equation}
Z_y(\beta)=\int_{\reals}\dd x p(x)e^{-\beta d(x,y)}.
\end{equation}
Consider the $L^\theta$ distortion measure $d(x,y)=|x-y|^\theta$, where
$\theta > 0$ and consider a uniform random coding distribution over
the interval $[-A,A]$, supposing that it is the optimal (or close to optimal)
one. Suppose further that we wish to work at a very small distortion level $D$
(high res), which means a large value of $\beta$ (why?). Then,
\begin{eqnarray}
Z_y(\beta)&=&\frac{1}{2A}\int_{-A}^{+A}\dd x e^{-\beta |x-y|^\theta}\nonumber\\
&\approx&\frac{1}{2A}\int_{-\infty}^{+\infty}\dd x e^{-\beta |x-y|^\theta}
~~~(\mbox{large}~\beta) \nonumber\\
&=&\frac{1}{2A}\int_{-\infty}^{+\infty}\dd x e^{-\beta |x|^\theta}
~~~(\mbox{the integral is independent of $y$}) \nonumber
\end{eqnarray}
Thus, returning to the expression of $R(D)$, let us minimize over $\beta$ by
writing the zero--derivative equation, which yields:
\begin{equation}
D=-\frac{\partial}{\partial
\beta}\ln\left[\frac{1}{2A}\int_{-\infty}^{+\infty}\dd x e^{-\beta
|x|^\theta}\right]
\end{equation}
but this is exactly the calculation of the (generalized) equipartition
theorem, which gives $1/(\beta\theta)=kT/\theta$. Now, we already said that
$\beta=-R'(D)$, and so, $1/\beta=-D'(R)$. It follows then that the function
$D(R)$, at this high res.\ limit, obeys a simple differential equation:
\begin{equation}
D(R)=-\frac{D'(R)}{\theta}
\end{equation}
whose solution is
\begin{equation}
D(R)=D_0e^{-\theta R}.
\end{equation}
In the case where $\theta=2$ (squared error distortion),
we get that $D(R)$ is proportional to $e^{-2R}$, which is a well--known result
in high res.\ quantization theory. For the Gaussian source, this is true for
all $R$. 


\newpage
\section{Analysis Tools and Asymptotic Methods}
\subsection{Introduction}

So far we have dealt with relatively simple situations where the Hamiltonian
is additive, the resulting B--G distribution is then i.i.d., and everything is very
nice, easy,
and simple. But this is seldom the case in reality. Most models in physics,
including those that will prove relevant for IT, as we shall see in the
sequel, are way more complicated,
more difficult, but also more interesting. More often than not, they are so
complicated and difficult, that they do not lend themselves to closed--form
analysis at all. In some other cases, analysis is possible, but it 
requires some more powerful mathematical
tools and techniques, which suggest at least some asymptotic approximations.
These are tools and techniques that we must acquaint ourselves with. So the
purpose of this part of the course is to prepare these tools, before we can go on
to the more challenging settings that are waiting for us. 

Before diving into the technical stuff, I'll first try to give the flavor of
the things I am going to talk about, and I believe the best way to do this is through an example.
In quantum mechanics, as its name suggests, several physical quantites do not
really take on values in the continuum of real numbers, but only values in a discrete set,
depending on the conditions of the system. One such quantized physical quantity is
energy (for example, the energy of light comes in quanta of $h\nu$, where
$\nu$ is frequency). Suppose we have a system of $n$ mobile particles (gas), whose
energies take on discrete values, denoted $\epsilon_0 < \epsilon_1 < \epsilon_2
< \ldots$. If the particles were not interacting, then the partition function
would have been given by
\begin{equation}
\left[\sum_{r\ge 0} e^{-\beta\epsilon_r}\right]^n=\sum_{r_1\ge
0}\sum_{r_2\ge 0}\ldots\sum_{r_n\ge 0}
\exp\left\{-\beta\sum_{i=1}^n\epsilon_{r_i}\right\}=\sum_{\bn:~\sum_r
n_r=n}\frac{n!}{\prod_r n_r!}\cdot\exp\left\{-\beta\sum_r
n_r\epsilon_r\right\}.
\end{equation}
However, since the particles are indistinguishable, then
permutations among them are not considered distinct physical states (see earlier
discussion on the ideal gas), and so, the
combinatorial factor $n!/\prod_rn_r!$, that counts these permutations, should be
eliminated. In other words, the correct partition function should be
\begin{equation}
Z_n(\beta)=\sum_{\bn:~\sum_rn_r=n}\exp\left\{-\beta\sum_r
n_r\epsilon_r\right\}.
\end{equation}
The problem is that this partition function is hard to calculate in closed
form: the headache is caused mostly 
because of the constraint $\sum_rn_r=n$. However, if we define
a corresponding generating function
\begin{equation}
\Xi(\beta,z)=\sum_{n\ge 0}z^n Z_n(\beta),
\end{equation}
which is like the $z$--transform of $\{Z_n(\beta)\}$, this is easy to work with, because
\begin{equation}
\Xi(\beta,z)=\sum_{n_1\ge 0}\sum_{n_2\ge 0}\ldots z^{\sum_rn_r}\exp\left\{-\beta
\sum_rn_r\epsilon_r\right\}=\prod_r\left[\sum_{n_r}
\left(ze^{-\beta\epsilon_r}\right)^{n_r}\right].
\end{equation}
Splendid, but we still want to obtain $Z_n(\beta)$...

The idea is to apply the
inverse $z$--transform:
\begin{equation}
Z_n(\beta)=\frac{1}{2\pi j}\oint_{\calC}\frac{\Xi(\beta,z)\dd
z}{z^{n+1}}=\frac{1}{2\pi
j}\oint_{\calC}\Xi(\beta,z)e^{-(n+1)\ln z}\dd z,
\end{equation}
where $z$ is a complex variable, $j=\sqrt{-1}$, and
$\calC$ is any clockwise closed path encircling the origin and entirely in
the region of convergence.
An exact calculation of integrals of this type might be difficult, in general,
but often, we would be happy enough if at least we could identify how they behave
in the thermodynamic limit of large $n$. 

Similar needs are frequently encountered in
information--theoretic problems. One example is in universal source coding:
Suppose we have a family of sources indexed by some parameter $\theta$,
say, Bernoulli with parameter $\theta\in[0,1]$, i.e.,
\begin{equation}
P_\theta(\bx)=(1-\theta)^{N-n}\theta^n,~~~\bx\in\{0,1\}^N;~~~n=\mbox{\# of 1's}
\end{equation}
When $\theta$ is unknown, it is customary to construct a universal code as the
Shannon code w.r.t.\ a certain mixture of these sources
\begin{equation}
P(\bx)=\int_0^1\mbox{d}\theta w(\theta)P_\theta(\bx)=
\int_0^1\mbox{d}\theta w(\theta)e^{Nh(\theta)}
\end{equation}
where
\begin{equation}
h(\theta)=\ln(1-\theta)+q\ln\left(\frac{\theta}{1-\theta}\right);~~~~q=\frac{n}{N}.
\end{equation}
So here again, we need to evaluate an integral of
an exponential function of $n$ (this time, on the real line),
in order to assess the performance of this universal code.

This is exactly the point where the
first tool that we are going to study, namely, the {\it saddle point method}
(a.k.a.\ the {\it steepest descent method})
enters into the picture: it gives us a way to assess how integrals of this
kind scale as exponential functions of $n$, for large $n$.
More generally, the saddle point method is a tool for evaluating
the exponential order (plus 2nd order behavior) of an integral of the form
\begin{equation}
\int_{\calP} g(z)e^{nf(z)}\dd z~~~~\calP~\mbox{is a path in the complex
plane.}
\end{equation}
We begin with the simpler case where the integration is over the real line (or
a subset of the real line),
whose corresponding asymptotic approximation method is called the {\it Laplace
method}. The material here is taken mostly from de Bruijn's book, which appears in the
bibliographical list.

\subsection{The Laplace Method}

Consider first an integral
of the form:
\begin{equation}
F_n\dfn \int_{-\infty}^{+\infty} e^{nh(x)}\dd x,
\end{equation}
where the function $h(\cdot)$ is independent of $n$. 
How does this integral  behave exponentially for large $n$? Clearly, if it was
a sum,
like $\sum_i e^{nh_i}$, rather than an integral, and the number of terms was
finite and independent of $n$, then the dominant term, $e^{n\max_ih_i}$, would
have dictated the exponential behavior. This continues to be true even if the
sum contains even infinitely many terms provided that the tail of this series
decays sufficiently rapidly. Since the integral is, after all, a limit of sums, it
is conceivable to expect, at least when $h(\cdot)$ is ``sufficiently nice'',
that something of the same spirit would happen with
$F_n$, namely, that its exponential order would be, in analogy, 
$e^{n\max h(x)}$. In what follows, we are going to show this
more rigorously, and as a bonus, we will also be able to say
something about the second order behavior. In the above example of universal
coding, this gives rise to redundancy analysis.

We will make the following assumptions on $h$: 
\begin{enumerate}
\item $h$ is real and continuous.
\item $h$ is maximum at $x=0$ and $h(0)=0$ (w.l.o.g).
\item $h(x) < 0~~\forall x\ne 0$, and $\exists b > 0,~c> 0$ s.t.\ $|x| \ge c$
implies $h(x) \le -b$. 
\item The integral defining $F_n$ converges for all sufficiently large $n$.
W.l.o.g., let this sufficiently large $n$ be $n=1$, i.e.,
$\int_{-\infty}^{+\infty} e^{h(x)}\dd x < \infty$.
\item The derivative $h'(x)$ exists at a certain neighborhood of $x=0$, and
$h''(0) < 0$. Thus, $h'(0)=0$. 
\end{enumerate}
From these assumptions, it follows that for all $\delta > 0$, there is a
positive number $\eta(\delta)$ s.t.\ for all $|x|\ge \delta$, we have $h(x)\le
-\eta(\delta)$. For $\delta \ge c$, this is obvious from assumption 3.  
If $\delta < c$, then the maximum of the continuous function $h$ across
the interval $[\delta,c]$ is strictly negative. A similar argument applies to
the interval $[-c,-\delta]$. Consider first the tails of the integral
under discussion:
\begin{eqnarray}
\int_{|x|\ge\delta}e^{nh(x)}\dd x
&=&\int_{|x|\ge\delta}\dd
xe^{(n-1)h(x)+h(x)}\nonumber\\
&\le& \int_{|x|\ge\delta}\dd x
e^{-(n-1)\eta(\delta)+h(x)}\nonumber\\
&\le& e^{-(n-1)\eta(\delta)}\cdot\int_{-\infty}^{+\infty}e^{h(x)}\dd x
\to 0~~~~\mbox{exponentially fast}\nonumber
\end{eqnarray}
In other words, the tails' contribution is vanishingly small.
It remains to examine the integral from $-\delta$ to $+\delta$, that is,
the neighborhood of $x=0$. In this neighborhood, we shall take the Taylor
series expansion of $h$. Since $h(0)=h'(0)=0$, then
$h(x)\approx\frac{1}{2}h''(0)x^2$. More precisely, for all $\epsilon > 0$, 
there is $\delta > 0$ s.t.\ 
\begin{equation}
\bigg| h(x)-\frac{1}{2}h''(0)x^2\bigg| \le \epsilon x^2~~~\forall
|x|\le\delta .
\end{equation}
Thus, this integral is sandwiched as follows:
\begin{equation}
\int_{-\delta}^{+\delta}
\exp\left\{\frac{n}{2}(h''(0)-\epsilon)x^2\right\}\dd x\le
\int_{-\delta}^{+\delta}
e^{nh(x)}\dd x
\le \int_{-\delta}^{+\delta}
\exp\left\{\frac{n}{2}(h''(0)+\epsilon)x^2\right\}\dd x.
\end{equation}
The right--most side is further upper bounded by
\begin{equation}
\int_{-\infty}^{+\infty}
\exp\left\{\frac{n}{2}(h''(0)+\epsilon)x^2\right\}\dd x
\end{equation}
and since $h''(0) < 0$, then $h''(0)+\epsilon=-(|h''(0)|-\epsilon)$, and so,
the latter is a Gaussian integral given by
\begin{equation}
\sqrt{\frac{2\pi}{(|h''(0)|-\epsilon)n}}.
\end{equation}
The left--most side of the earlier sandwich is further lower
bounded by
\begin{eqnarray}
&&\int_{-\delta}^{+\delta}
\exp\left\{-\frac{n}{2}(|h''(0)|+\epsilon)x^2\right\}\dd x\nonumber\\
&=&
\int_{-\infty}^{+\infty}\exp\left\{-\frac{n}{2}(|h''(0)|+\epsilon)x^2\right\}\dd
x-\int_{|x|\ge \delta}
\exp\left\{-\frac{n}{2}(|h''(0)|+\epsilon)x^2\right\}\dd x\nonumber\\
&=&\sqrt{\frac{2\pi}{(|h''(0)|+\epsilon)n}}
-2Q(\delta\sqrt{n(|h''(0)|+\epsilon)})\nonumber\\
&\ge&\sqrt{\frac{2\pi}{(|h''(0)|+\epsilon)n}}
-O\left(\exp\left\{-\frac{n}{2}(|h''(0)|+\epsilon)\delta^2\right\}\right)\nonumber\\
&\sim& \sqrt{\frac{2\pi}{(|h''(0)|+\epsilon)n}}\nonumber
\end{eqnarray}
where the notation $A_n\sim B_n$ means that $\lim_{n\to\infty}A_n/B_n=1$.
Since $\epsilon$ and hence $\delta$ can be made arbitrary small, we find that
\begin{equation}
\int_{-\delta}^{+\delta}
e^{nh(x)}\dd x \sim \sqrt{\frac{2\pi}{|h''(0)|n}}.
\end{equation}
Finally, since the tails contribute an exponentially small term, which
is negligible compared to the contribution of $O(1/\sqrt{n})$ order
of the integral across $[-\delta,+\delta]$, we get:
\begin{equation}
\int_{-\infty}^{+\infty}
e^{nh(x)}\dd x \sim \sqrt{\frac{2\pi}{|h''(0)|n}}.
\end{equation}
Slightly more generally, if $h$ is maximized at an arbitrary point $x=x_0$
this is completely immaterial because an integral over the entire real line
is invariant under translation of the integration variable. If, furthermore,
the maximum $h(x_0)$ is not necessarily zero, we can make it zero by
decomposing $h$ according to
$h(x)=h(x_0)+[h(x)-h(x_0)]$ and moving the first term as a constant factor of $e^{nh(x_0)}$ 
outside of the integral. The result would then be
\begin{equation}
\int_{-\infty}^{+\infty}
e^{nh(x)}\dd x \sim e^{nh(x_0)}\cdot\sqrt{\frac{2\pi}{|h''(x_0)|n}}
\end{equation}
Of course, the same considerations continue to apply if $F_n$ is
defined over any finite or half--infinite interval that contains the maximizer
$x=0$, or more generally $x=x_0$ as an internal point. 
It should be noted, however, that if $F_n$ is defined over a finite or
semi--infinite interval and the maximum of $h$ is obtained at an edge of this
interval, then the derivative of $h$ at that point does not necessarily
vanish, and the Gaussian integration would not apply anymore. In this case,
the local behavior around the maximum would be approximated by an exponential
$\exp\{-n|h'(0)|x\}$ or
$\exp\{-n|h'(x_0)|x\}$ instead, which gives a somewhat different expression.
However, the factor $e^{nh(x_0)}$, which is the most important factor, would
continue to appear. Normally, this will be the only term that will
interest us, whereas the other factor, which provides the second order behavior
will not be important for us.
A further extension in the case where the
maximizer is an internal point at which the derivative vanishes, is this:
\begin{center}
\fbox{
$\int_{-\infty}^{+\infty}g(x)
e^{nh(x)}\dd x \sim g(x_0)e^{nh(x_0)}\cdot\sqrt{\frac{2\pi}{|h''(x_0)|n}}$}
\end{center}
where $g$ is another function that does not depend on $n$.
This technique, of approximating an integral of a function,
which is exponential in some large parameter $n$, by neglecting the
tails and approximating it by a Gaussian integral around the maximum,
is called the {\it Laplace method of integration}.

\subsection{The Saddle Point Method}

We now expand the scope to integrals along paths in the complex plane, which
are also encountered and even more often than one would expect (cf.\ the earlier example).
As said, the extension of the Laplace integration technique to the complex
case is called the saddle--point method or the steepest descent method, for
reasons that will become apparent shortly. Specifically,
we are now interested in an integral of the form
\begin{equation}
F_n=\int_{\calP} e^{nh(z)}\dd z~~\mbox{or more generally}~~
F_n=\int_{\calP} g(z)e^{nh(z)}\dd z
\end{equation}
where $z=x+jy$ is a complex variable ($j=\sqrt{-1}$), and $\calP$ is a certain
path (or curve) in the complex plane, starting at some point $A$ and ending at
point $B$. We will focus first on the former integral, without the factor $g$.
We will assume that $\calP$ is fully contained in a region where
$h$ is analytic (differentiable as many times as we want). 

The first
observation, in this case, is that the value
of the integral depends actually only on $A$ and $B$, and not on the details
of $\calP$: 
Consider any alternate path $\calP'$  from $A$ to $B$
such that $h$ has no singularities in the
region surrounded by $\calP\bigcup\calP'$. Then, the integral of $e^{nh(z)}$
over the closed path $\calP\bigcup\calP'$ (going from $A$ to $B$ via $\calP$
and returning to $A$ via $\calP'$) vanishes, which means that 
the integrals from $A$ to $B$ via $\calP$ and via $\calP'$ are the same.
This means that we actually have the {\it freedom} to select the integration path,
as long as we do not go too far, to the other side of some singularity point,
if there is any. This point will be important in our forthcoming
considerations.

An additional important observation has to do with yet another basic property of analytic
functions: the {\it maximum modulus theorem}, which basically
tells that the modulus of an analytic function has no maxima. We will not
prove here this theorem, but in a nutshell, the point is this: Let
\begin{equation}
h(z)=u(z)+jv(z)=u(x,y)+jv(x,y),
\end{equation}
where $u$ and $v$ are real functions. If $h$ is analytic, the following
relationships (a.k.a.\ the Cauchy--Riemann conditions)\footnote{This is related
to the fact that for the derivative $f'(z)$ to exist, it should be independent
of the direction at which $z$ is perturbed, whether it is, e.g., the
horizontal or the vertical direction, i.e., $f'(z)=\lim_{\delta\to
0}[f(z+\delta)-f(z)]/\delta=\lim_{\delta\to 0}[f(z+j\delta)-f(z)]/(j\delta)$,
where $\delta$ goes to zero along the reals.}
between the partial derivatives of $u$ and $v$ must hold:
\begin{equation}
\frac{\partial u}{\partial x}=\frac{\partial v}{\partial y}; ~~~~~~
\frac{\partial u}{\partial y}=-\frac{\partial v}{\partial x}.
\end{equation}
Taking the second order partial derivative of $u$:
\begin{equation}
\frac{\partial^2 u}{\partial x^2}=\frac{\partial^2 v}{\partial x\partial y}=
\frac{\partial^2 v}{\partial y\partial x}=-\frac{\partial^2 u}{\partial
y^2}
\end{equation}
where the first equality is due to the first Cauchy--Riemann condition and the
third equality is due to the second Cauchy--Riemann condition. Equivalently,
\begin{equation}
\frac{\partial^2 u}{\partial x^2}+\frac{\partial^2 u}{\partial
y^2}=0,
\end{equation}
which is the {\it Laplace equation}. This means, among other things, that no
point at which $\partial u/\partial x=\partial u/\partial y=0$ can be a local
maximum (or a local minimum) of $u$, because if it is a local maximum in the
$x$--direction, in which case, $\partial^2 u/\partial x^2 < 0$, then
$\partial^2 u/\partial y^2$ must be positive, which makes it a local minimum
in the $y$--direction, and vice versa. In other words, every point of zero
partial derivatives of $u$ must be a {\it saddle point}.
This discussion applies now to the modulus of the integrand $e^{nh(z)}$
because
\begin{equation}
\bigg|\exp\{nh(z)\}\bigg|= \exp[n\mbox{Re}\{h(z)\}]=e^{nu(z)}.
\end{equation}
Of course, if $h'(z)=0$ at some $z=z_0$, then $u'(z_0)=0$ too, and then
$z_0$ is a saddle point of $|e^{nh(z)}|$. Thus, 
zero--derivative points of $h$ are saddle points.

Another way to see this is the following: Given a complex analytic function
$f(z)$, we argue that the average of $f$ over a circle always agrees
with its value at the center of this circle. Specifically, consider the circle
of radius $R$ centered at $z_0$, i.e., $z=z_0+Re^{j\theta}$. Then,
\begin{eqnarray}
\frac{1}{2\pi}\int_{-\pi}^{\pi} f\left(z_0+Re^{j\theta}\right)\mbox{d}\theta
&=&\frac{1}{2\pi j}\int_{-\pi}^{\pi}
\frac{f\left(z_0+Re^{j\theta}\right)jRe^{j\theta}\mbox{d}\theta}{Re^{j\theta}}\nonumber\\
&=&\frac{1}{2\pi j}\oint_{z=z_0+Re^{j\theta}}
\frac{f\left(z_0+Re^{j\theta}\right)\mbox{d}\left(z_0+Re^{j\theta}\right)}{Re^{j\theta}}\nonumber\\
&=&\frac{1}{2\pi j}\oint_{z=z_0+Re^{j\theta}}
\frac{f(z)\mbox{d}z}{z-z_0}=f(z_0).
\end{eqnarray}
and so,
\begin{equation}
|f(z_0)|\le \frac{1}{2\pi}\int_{-\pi}^{\pi} \bigg|
f\left(z_0+Re^{j\theta}\right)\bigg|\mbox{d}\theta
\end{equation}
which means that $|f(z_0)|$ cannot be strictly larger than {\it all} $|f(z)|$
in any neighborhood (an arbitrary radius $R$) of $z_0$. Now, apply this fact to
$f(z)=e^{nh(z)}$. 

Equipped with this background, let us return to our integral $F_n$. Since we
have the freedom to choose the path $\calP$, suppose that we can find one
which passes through a saddle point $z_0$ (hence the name of the method) 
and that $\max_{z\in\calP}|e^{nh(z)}|$
is attained at $z_0$. We expect then, that similarly as in the Laplace method,
the integral would be dominated by $e^{nh(z_0)}$.
Of course, such a path would be fine only if it
crosses the saddle point $z_0$ at a direction w.r.t.\ which $z_0$ is a local maximum of
$|e^{nh(z)}|$, or equivalently, of $u(z)$. Moreover, in order to apply our
earlier results of the Laplace method, we will find it convenient to draw
$\calP$ such that any point $z$ in the vicinity of $z_0$, where in the Taylor
expansion is:
\begin{equation}
h(z)\approx h(z_0)+\frac{1}{2}h''(z_0)(z-z_0)^2~~~~\mbox{(recall that
$h'(z_0)=0$.)}
\end{equation}
the second term, $\frac{1}{2}h''(z_0)(z-z_0)^2$ is purely {\bf real and
negative},
and then it behaves locally as a negative parabola, just like in the Laplace
case. This means that
\begin{equation}
\mbox{arg}\{h''(z_0)\}+ 2\mbox{arg}(z-z_0)=\pi
\end{equation}
or equivalently
\begin{equation}
\mbox{arg}(z-z_0)=\frac{\pi-\mbox{arg}\{h''(z_0)\}}{2}\dfn\theta.
\end{equation}
Namely, $\calP$ should cross $z_0$ in the direction $\theta$.
This direction is called the {\it axis} of $z_0$, and it can be shown to be
the direction of {\bf steepest descent} from the peak at $z_0$ (hence
the name).\footnote{Note that in the direction $\theta-\pi/2$, which is
perpendicular to the axis, $\mbox{arg}[h''(z_0)(z-z_0)^2]=\pi-\pi=0$, which
means that $h''(z_0)(z-z_0)^2$ is real and positive (i.e., it behaves like a positive
parabola). Therefore, in this direction,
$z_0$ is a local minimum.}

\begin{figure}[ht]
\hspace*{5cm}\input{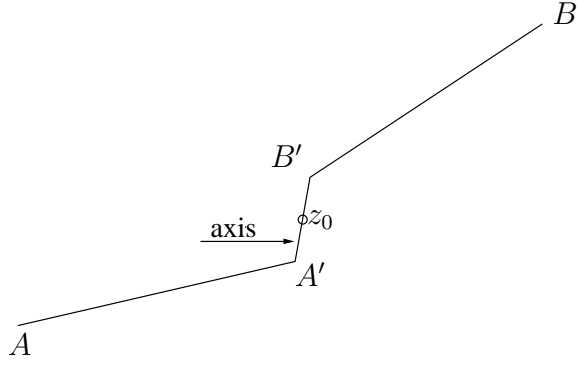}
\caption{\small A path $\calP$ from $A$ to $B$, passing via $z_0$ along the
axis.}
\label{saddlepoint}
\end{figure}

So pictorially, what we are going to do is choose a path $\calP$ from $A$ to
$B$, which will be
composed of three parts (see Fig.\ \ref{saddlepoint}): The parts $A\to A'$ and
$B'\to B$ are quite arbitrary as they constitute the tail of the integral.
The part from $A'$ to $B'$, in the vicinity of $z_0$, is a straight line on the axis of $z_0$.

Now, let us decompose $F_n$ into its three parts:
\begin{equation}
F_n=\int_A^{A'}e^{nh(z)}\dd z+\int_{A'}^{B'}e^{nh(z)}\dd
z+\int_{B'}^{B}e^{nh(z)}\dd z.
\end{equation}
As for the first and the third terms,
\begin{equation}
\bigg|\left(\int_A^{A'}+\int_{B'}^{B}\right)\dd z e^{nh(z)}\bigg|\le
\left(\int_A^{A'}+\int_{B'}^{B}\right)\dd z |e^{nh(z)}|=
\left(\int_A^{A'}+\int_{B'}^{B}\right)\dd z e^{nu(z)}
\end{equation}
whose contribution is negligible compared to $e^{nu(z_0)}$, just like the
tails in the Laplace method. As for the middle integral,
\begin{equation}
\int_{A'}^{B'}e^{nh(z)}\dd z\approx e^{nh(z_0)}\int_{A'}^{B'}
\exp\{nh''(z_0)(z-z_0)^2/2\}\dd z.
\end{equation}
By changing from the complex integration variable $z$ to the real variable
$x$, running from $-\delta$ to $+\delta$,
with $z=z_0+xe^{j\theta}$ (motion along the axis), we get exactly the Gaussian
integral of the Laplace method, leading to
\begin{equation}
\int_{A'}^{B'}\exp\{nh''(z_0)(z-z_0)^2/2\}\dd
z=e^{j\theta}\sqrt{\frac{2\pi}{n|h''(z_0)|}}
\end{equation}
where the factor $e^{j\theta}$ is due to the change of variable
($\mbox{d}z=e^{j\theta}\mbox{d}x$). Thus,
\begin{equation}
F_n\sim e^{j\theta}\cdot e^{nh(z_0)}\sqrt{\frac{2\pi}{n|h''(z_0)|}},
\end{equation}
and slightly more generally,
\begin{center}
\fbox{
$ 
\int_{\calP} g(z)e^{nh(z)}\dd z\sim e^{j\theta}g(z_0)
e^{nh(z_0)}\sqrt{\frac{2\pi}{n|h''(z_0)|}}
$}
\end{center}
The idea of integration along the axis is that along this direction, the
`phase' of $e^{nh(z)}$ is locally constant, and only the modulus varies. Had
the integration been along another direction with an imaginary component
$j\phi(z)$, the
function $e^{nh(z)}$ would have undergone `modulation', i.e., it would have
oscillated with a complex exponential $e^{nj\phi(z)}$ of a very high
`frequency' (proportional
to $n$) and then $e^{nu(z_0)}$ would not have guaranteed to dictate the modulus
and to dominate the integral.

Now, an important comment is in order:
What happens if there is more than one saddle point?
Suppose we have two saddle points, $z_1$ and $z_2$. 
On a first thought, one may be
concerned by the following consideration: We can construct two paths
from $A$ to $B$, path $\calP_1$ crossing
$z_1$, and path $\calP_2$ crossing $z_2$.
Now, if $z_i$ is the highest point along $\calP_i$ for both $i=1$ and
$i=2$, then $F_n$ is exponentially both $e^{nh(z_1)}$ and
$e^{nh(z_2)}$ at the same time. If $h(z_1)\ne h(z_2)$, this is a contradiction.
But the following consideration shows that this cannot happen as long
as $h(z)$ is analytic within the region $\calC$ 
surround by $\calP_1\cup\calP_2$.
Suppose conversely, that the scenario described above happens.
Then either $z_1$ or $z_2$ maximize $|e^{nh(z)}|$ along the closed
path $\calP_1\cup\calP_2$. Let us say that it is $z_1$. We claim that
then $z_1$ cannot be a saddle point, for the following reason: No point in
the interior of $\calC$ can be higher than $z_1$, because if there was
such a point, say, $z_3$, then we had
\begin{equation}
\max_{z\in\calC}|e^{nh(z)}|\ge
|e^{nh(z_3)}|> 
|e^{nh(z_1)}|=\max_{z\in \calP_1\cup\calP_2}|e^{nh(z)}|
\end{equation}
which contradicts the maximum modulus principle. This then means, among other
things, that in every neighborhood of $z_1$, all points in $\calC$ are lower
than $z_1$, including points found in a direction perpendicular to the
direction of the axis through $z_1$. But this contradicts the fact that
$z_1$ is a saddle point: 
Had it been a saddle point, it would be a local maximum along the axis and a
local minimum along the perpendicular direction. Since $z_1$ was assumed
a saddle point, then it cannot be the highest point on $\calP_1$, which
means that it doesn't dominate the integral. 

One might now be concerned
by the thought that the integral along $\calP_1$ is then dominated by
an even higher contribution, which still seems to contradict the lower
exponential order of $e^{nh(z_2)}$ attained by the path $\calP_2$. 
However, this is not the case. The highest point on the path is guaranteed
to dominate the integral only if it is a saddlepoint. Consider, for example,
the integral $F_n=\int_{a+j0}^{a+j2\pi}e^{nz}
\mbox{d}z$. Along the vertical line from $a+j0$ to $a+j2\pi$, the modulus
(or attitude) is $e^{na}$ everywhere. If the attitude alone had been whatever
counts (regardless of whether it is a saddle point or not), 
the exponential order of (the modulus of) this integral would be $e^{na}$.
However, the true value of this integral is zero! The reason for this
disagreement is that there is no saddle point along this path.

What about a path $\calP$ that crosses both $z_1$ and $z_2$? This cannot
be a good path for the saddle point method, for the following reason:
Consider two slightly perturbed versions of $\calP$: 
path $\calP_1$, which
is very close to $\calP$, it crosses $z_1$, but it makes 
a tiny detour that bypasses $z_2$,
and similarly path $\calP_2$, passing via $z_2$, but
with a small deformation near $z_1$. Path $\calP_2$
includes $z_2$ as saddle point, 
but it is not the highest point on the path, since
$\calP_2$ passes near $z_1$, which is higher. Path $\calP_1$ includes $z_1$
as saddle point, but it cannot be the highest point on the path because we
are back to the same situation we were 
two paragraphs ago. Since both $\calP_1$
and $\calP_2$ are bad choices, and since they are both 
arbitrarily close to $\calP$,
then $\calP$ cannot be good either.

To summarize: if we have multiple saddle points, 
we should find the one with the
{\it lowest} attitude and then we have a chance to find a path through 
this saddlepoint (and only this one) along which this saddle point is dominant.

Let us look now at a few examples.\\

\noindent
{\it Example 1 -- relation between $\Omega(E)$ and $Z(\beta)$ revisited.}
Assuming, without essential loss of generality, that the ground--state energy 
of the system is zero,
we have seen before the relation $Z(\beta)=\int_0^\infty
\mbox{d}E\Omega(E)e^{-\beta E}$, which actually means that $Z(\beta)$ is the
Laplace transform of $\Omega(E)$. Consequently, this means that $\Omega(E)$ is the inverse
Laplace transform of $Z(\beta)$, i.e.,
\begin{equation}
\Omega(E)=\frac{1}{2\pi j}\int_{\gamma-j\infty}^{\gamma+j\infty} e^{\beta E}
Z(\beta)\mbox{d}\beta,
\end{equation}
where the integration in the complex plane is along the vertical line
$\mbox{Re}(\beta)=\gamma$, which is chosen to the right of all singularity
points of $Z(\beta)$. In the large $n$ limit, this becomes
\begin{equation}
\Omega(E)=\frac{1}{2\pi j}\int_{\gamma-j\infty}^{\gamma+j\infty}
e^{n[\beta\epsilon+\phi(\beta)]}
\mbox{d}\beta,
\end{equation}
which can now be assessed using the saddle point method. The derivative of the
bracketed term at the exponent vanishes at the value of $\beta$ that solves
the equation $\phi'(\beta)=-\epsilon$, which is $\beta^*(\epsilon)\in\reals$,
thus we will choose $\gamma=\beta^*(\epsilon)$ (assuming that this is a
possible choice) and thereby let the integration path pass through
this saddle point. At $\beta=\beta^*(\epsilon)$,
$|\exp\{n[\beta\epsilon+\phi(\beta)]\}|$ has its maximum along the vertical
direction, $\beta=\beta^*(\epsilon)+j\omega$, $-\infty <\omega< +\infty$
(and hence it dominates the integral), but since it is a saddle point,
it {\it minimizes} $|\exp\{n[\beta\epsilon+\phi(\beta)]\}|=
\exp\{n[\beta\epsilon+\phi(\beta)]\}$, in the horizontal direction (the real
line). Thus,
$\Omega(E)\exe
\exp\{n\min_{\beta\in\reals}[\beta\epsilon+\phi(\beta)]\}=e^{n\Sigma(\epsilon)}$, as we
have seen before.

\noindent
{\it Example 2 -- size of a type class.}
Here is a question which we know how to answer using the method of types. 
Among all binary sequences of length $N$, how many have $n$ 1's and $(N-n)$ 0's?
\begin{eqnarray}
M_n&=&\sum_{\bx\in\{0,1\}^N}\calI\left\{\sum_{i=1}^Nx_i=n\right\}\nonumber\\
&=&\sum_{x_1=0}^1\ldots\sum_{x_N=0}^1\calI\left\{\sum_{i=1}^Nx_i=n\right\}\nonumber\\
&=&\sum_{x_1=0}^1\ldots\sum_{x_N=0}^1\frac{1}{2\pi}\int_0^{2\pi}\dd\omega
\exp\left\{j\omega\left(n-\sum_{i=1}^Nx_i\right)\right\}\nonumber\\
&=& \int_0^{2\pi}\frac{\dd \omega}{2\pi}\sum_{x_1=0}^1\ldots\sum_{x_N=0}^1
\exp\left\{j\omega\left(n-\sum_{i=1}^Nx_i\right)\right\}\nonumber\\
&=& \int_0^{2\pi}\frac{\dd\omega}{2\pi}e^{j\omega n}
\prod_{i=1}^N\left[ \sum_{x_i=0}^1
e^{-j\omega x_i}\right]\nonumber\\
&=& \int_0^{2\pi}\frac{\dd \omega}{2\pi}
e^{j\omega n}(1+e^{-j\omega})^N\nonumber\\
&=& \int_0^{2\pi}\frac{\dd \omega}{2\pi}\exp\{N[j\omega\alpha+\ln
(1+e^{-j\omega})]\}~~~~\alpha\dfn\frac{n}{N}\nonumber\\
&=& \int_0^{2\pi j}\frac{\dd z}{2\pi j}\exp\{N[z\alpha+\ln
(1+e^{-z})]\} ~~~~~j\omega\longrightarrow z
\end{eqnarray}
This is an integral with a starting point $A$ at the origin and an ending point
$B$ at $2\pi j$. Here, $h(z)=z\alpha+\ln(1+e^{-z})$, and the saddle point,
where $h'(z)=0$, is
on the real axis: $z_0=\ln\frac{1-\alpha}{\alpha}$,
where $h(z_0)$ gives the binary entropy of $\alpha$, as expected. Thus, the
integration path must be deformed to pass through this point on the real
axis, and then to approach back the imaginary axis, so as to arrive at $B$.
There is one serious caveat here, however: 
The points $A$ and $B$ are both higher than
$z_0$: While $u(z_0)=-\alpha\ln(1-\alpha)-(1-\alpha)\ln(1-\alpha)$,
at the edges we have $u(A)=u(B)=\ln 2$. So this is not a good saddle--point
integral to work with. 

Two small modifications can, however, fix the problem:
The first is to define the integration interval of $\omega$  to be
$[-\pi,\pi]$ rather than $[0,2\pi]$ (which is, of course, legitimate), and
then $z$ would run from $-j\pi$ to $+j\pi$. The second is the following:
Consider again the first line of the expression of $M_n$ above, but before
we do anything else, let us multiply the 
whole expression (outside the summation)
by $e^{\theta n}$ ($\theta$ an aribtrary real), whereas the
summand will be multiplied by $e^{-\theta\sum_ix_i}$, which exactly cancels
the factor of $e^{\theta n}$ for every non--zero term of this sum. We can
now repeat exactly the same calculation as above (exercise), 
but this time we get:
\begin{equation}
M_n=
\int_{\theta-j\pi}^{\theta+j \pi}\frac{\dd z}{2\pi j}\exp\{N[z\alpha+\ln
(1+e^{-z})]\},
\end{equation}
namely, we moved the integration path to a parallel vertical line and shifted it
by the amount of $\pi$ to the south.
Now, we have the freedom
to choose $\theta$. The obvious choice is to set
$\theta=\ln\frac{1-\alpha}{\alpha}$, so that we cross the saddle point
$z_0$. Now $z_0$ is the highest point on the path (exercise: please verify).
Moreover, the vertical direction of the integration is also the direction
of the axis of $z_0$ (exercise: verify this too), so now everything is fine.
Also, the second order factor of $O(1/\sqrt{n})$ 
of the saddle point integration agrees with the same
factor that we can see from the Stirling approximation in the more refined
formula.

A slightly different look at this example 
is as follows. Consider the Schottky example and the partition
function
\begin{equation}
Z(\beta)=\sum_{\bx} e^{-\beta\epsilon_0\sum_ix_i},
\end{equation}
which, on the one hand, is given by $\sum_{n=0}^N M_ne^{-\beta\epsilon_0n}$,
and on the other hand, is given also by $(1+e^{-\beta\epsilon_0})^N$. 
Thus, defining $s=e^{-\beta\epsilon_0}$, we have $Z(s)=\sum_{n=0}^N
M_ns^n$, and so, $Z(s)=(1+s)^N$ is the $z$--transform of the finite sequence
$\{M_n\}_{n=0}^N$. Consequently,
$M_n$ is given by the inverse $z$--transform of $Z(s)=(1+s)^N$, i.e.,
\begin{eqnarray}
M_n&=&\frac{1}{2\pi j}\oint (1+s)^Ns^{-n-1}\dd s\nonumber\\
&=& \frac{1}{2\pi j}\oint \exp\{N[\ln(1+s)-\alpha\ln s]\}\dd s
\end{eqnarray}
This time, the integration path is any closed path
that surrounds the origin, the saddle point is $s_0=\alpha/(1-\alpha)$, so we take 
the path to be a circle whose radius is
$r=\frac{\alpha}{1-\alpha}$. 
The rest of the calculation is essentially the same as before, and of
course, so is the result. Note that this is actually the very same integral as before
up to a change of the integration variable from $z$ to $s$, according to $s=e^{-z}$,
which maps the vertical straight line between $\theta-\pi j$ and $\theta+\pi
j$ onto a circle 
of radius $e^{-\theta}$, centered at the origin.
$\Box$

\vspace{0.2cm}

\noindent
{\it Example 3 -- surface area of a sphere.}
Let us compute the surface area of an $n$--dimensional sphere with radius
$nR$:
\begin{eqnarray}
S_n&=&\int_{\reals^n}\dd \bx \delta\left(nR-\sum_{i=1}^nx_i^2\right)\nonumber\\
&=&e^{n\alpha R}\int_{\reals^n}\dd \bx e^{-\alpha\sum_ix_i^2}\cdot
\delta\left(nR-\sum_{i=1}^nx_i^2\right)~~~\mbox{($\alpha > 0$ to be chosen
later.)}\nonumber\\
&=&e^{n\alpha R}\int_{\reals^n}\dd \bx
e^{-\alpha\sum_ix_i^2}\int_{-\infty}^{+\infty}\frac{\dd \theta}{2\pi}
e^{j\theta(nR-\sum_ix_i^2)}\nonumber\\
&=&e^{n\alpha R}\int_{-\infty}^{+\infty}\frac{\dd \theta}{2\pi}e^{j\theta
nR}\int_{\reals^n}\dd \bx
e^{-(\alpha+j\theta)\sum_ix_i^2}\nonumber\\
&=&e^{n\alpha R}\int_{-\infty}^{+\infty}\frac{\dd \theta}{2\pi}e^{j\theta
nR}\left[\int_{\reals}\dd x
e^{-(\alpha+j\theta)x^2}\right]^n\nonumber\\
&=&e^{n\alpha R}\int_{-\infty}^{+\infty}\frac{\dd \theta}{2\pi}e^{j\theta
nR}\left(\frac{\pi}{\alpha+j\theta}\right)^{n/2}\nonumber\\
&=&\frac{\pi^{n/2}}{2\pi}\int_{-\infty}^{+\infty}\dd \theta
\exp\left\{n\left[(\alpha+j\theta)
R-\frac{1}{2}\ln(\alpha+j\theta)\right]\right\}\nonumber\\
&=&\frac{\pi^{n/2}}{2\pi}\int_{\alpha-j\infty}^{\alpha+j\infty}\dd z
\exp\left\{n\left[zR-\frac{1}{2}\ln z\right]\right\}.
\end{eqnarray}
So here $h(z)=zR-\frac{1}{2}\ln z$ and the integration is along an arbitrary
vertical straight line parametrized by $\alpha$. We will choose this straight line to pass thru the
saddle point $z_0=\frac{1}{2R}$ (exercise: show that this is indeed the
highest point on the path). Now, $h(z_0)=\frac{1}{2}\ln(2\pi e R)$, just
like the differential entropy of a Gaussian RV (is this a coincidence?).
$\Box$

\vspace{0.2cm}

\noindent
{\bf Comment:} In these examples, we used an additional trick: whenever we had
to deal with an `ugly' function like the $\delta$ function, we presented it as
an inverse transform of a `nice' function, and then changed the order of
integrations/summations. This idea will be repeated in the sequel. It is used
very frequently by physicists.

\subsection{The Replica Method}

The replica method is one of the most useful tools, which originally comes
from statistical physics, but it finds its use in a variety of other fields,
with Communications and Information Theory included (e.g., multiuser
detection). As we shall see, there are many models in statistical physics, where
the partition function $Z$ depends, among other things, on a bunch of {\it
random} parameters (to model disorder), and then $Z$, or $\ln Z$, becomes, of
course, a random variable as well. Further, it turns out that more often than not, 
the RV $\frac{1}{n}\ln Z$ exhibits a concentration property, or in the
jargon of physicists, a {\it self--averaging} property: in the thermodynamic limit
of $n\to\infty$, it falls in the vicinity of its expectation $\frac{1}{n}\langle\ln
Z\rangle$, with very high probability. Therefore, the computation of the 
per--particle free energy (and hence also many other physical quantities), for a
typical realization of these random parameters, is associated with the
computation of $\langle\ln Z\rangle$. The problem is that in most of the
interesting cases, the exact closed form calculation of this expectation is
extremely difficult if not altogether impossible. This is the point where the
replica method enters into the picture.

Before diving into the description of the replica method, 
it is important to make a certain digression: This is a non--rigorous,
heuristic method, and it is not quite clear (yet) what are exactly the
conditions under which it gives the correct result.
Physicists tend to believe in it very strongly, because in many situations
it gives results that make sense, live in harmony with intuition, or make
good fit to experimental results and/or simulation results. The problem is
that when there are no other means to test its validity, there is no 
certainty that it is credible and reliable. In such cases, I believe that
the correct approach would be to refer to the results it provides, as a
certain educated guess or as a conjecture, rather than a solid scientific truth. 
As we shall see shortly, the problematics of the replica method is not
just that it depends on a certain interchangeability between a limit and
an integral, but more severely, that the procedure that it proposes, is 
actually not even well--defined. 
In spite of all this, since this method is so widely used, it would be
inappropriate to completely ignore it in a course of this kind, and therefore,
we will devote to the replica method 
at least a short period of time, presenting it
in the general level, up to a certain point. 
However, we will not use the replica method elsewhere in this course.

Consider then the calculation of $\bE\ln Z$. The problem is that $Z$ is a sum,
and it is not easy to say something intelligent on the logarithm of a sum of
many terms, let alone the expectation of this log--sum. If, instead, we had to
deal with integer moments of $Z$, $\bE Z^m$, life would have been much easier, because
integer moments of sums, are sums of products. Is there a way then that we can relate moments
$\bE Z^m$ to $\bE\ln Z$? The answer is, in principle, affirmative if {\bf real},
rather than just integer, moments are allowed. These could be related via the
simple relation
\begin{equation}
\bE\ln Z=
\lim_{m\to 0}\frac{\bE
Z^m-1}{m}=
\lim_{m\to 0}\frac{\ln \bE Z^m}{m}
\end{equation}
provided that the expectation operator and the limit over $m$ can be
interchanged. But we know how to deal only with integer moments of $m$.
The first courageous idea of the replica method, at this point, is to offer
the following recipe: Compute $\bE Z^m$, for positive integer $m$, and obtain
an expression which is a function of $m$. Once this has been done, now {\it
forget} that $m$ is an
integer, and think of it as a {\it real} variable. Finally, use the
above identity, taking the limit of $m\to 0$.

Beyond the technicality of interchanging the expectation operator with the
limit, which is, after all, OK in most conceivable cases, there is a more serious concern
here, and this is that the above procedure is not well--defined, as mentioned
earlier: We derive an expression $f(m)\dfn \bE Z^m$, which is
originally meant for $m$ integer only, and then `interpolate' in between
integers by using the same expression, in other words, we take the analytic
continuation. Actually, the right--most side of the above identity is $f'(0)$
where $f'$ is the derivative of $f$.
But there are infinitely many functions of a continuous variable $m$ that
pass through given points at integer values of $m$: If $f(m)$ is such, then
$\tilde{f}(m)=f(m)+g(m)$ is good as well, for every $g$ that vanishes on the integers, for
example, take $g(m)=A\sin(\pi m)$. Nonetheless, $\tilde{f}'(0)$ might be different from
$f'(0)$, and this is indeed the case with the example where $g$ is sinusoidal.
So in this step of the procedure there is some weakness, but this is simply
ignored...

After this introduction, let us now present the replica method on a concrete
example, which is
essentially taken
from the book by M\'ezard and Montanari.
In this example, $Z=\sum_{i=1}^{2^n} e^{-\beta E_i}$, where
$\{E_i\}_{i=1}^{2^n}$ are i.i.d.\ RV's. In the sequel, we will work with this
model quite a lot, after we see why, when and where it is relevant. It is called the {\it
random energy model} (REM). But for now, this is just a technical example on which
we demonstrate the replica method. As the replica method suggests, let's first
look at the integer moments. First, what we have is:
\begin{equation}
Z^m=\left[\sum_{i=1}^{2^n} e^{-\beta E_i}\right]^m=
\sum_{i_1=1}^{2^n}\ldots \sum_{i_m=1}^{2^n} \exp\{-\beta\sum_{a=1}^m
E_{i_a}\}.
\end{equation}
The right--most side can be thought of as the partition function pertaining to
a new system, consisting of $m$ independent replicas (hence the name of the
method) of the original system. Each configuration of the new system is
indexed by an $m$--tuple $\bi=(i_1,\ldots,i_m)$, where each $i_a$ runs from $1$ to
$2^n$, and the energy is $\sum_a E_{i_a}$. Let us now rewrite $Z^m$ slightly
differently:
\begin{eqnarray}
Z^m&=&\sum_{i_1=1}^{2^n}\ldots \sum_{i_m=1}^{2^n} \exp\left\{-\beta\sum_{a=1}^m
E_{i_a}\right\}\nonumber\\
&=&\sum_{\bi}\exp\left\{-\beta\sum_{a=1}^m\sum_{j=1}^{2^n}\calI(i_a=j)E_j\right\}~~~~~~\mbox{$\calI(\cdot)=$
indicator function}\nonumber\\
&=&\sum_{\bi}\exp\left\{-\beta\sum_{j=1}^{2^n}\sum_{a=1}^m\calI(i_a=j)E_j\right\}\nonumber\\
&=&\sum_{\bi}\prod_{j=1}^{2^n}\exp\left\{-\beta\sum_{a=1}^m\calI(i_a=j)E_j\right\}\nonumber
\end{eqnarray}
Let us now further suppose that each $E_j$ is $\calN(0,nJ^2/2)$, as is
customary in the REM, for reasons that we shall see later on. Then, taking
expecations w.r.t.\ this distribution, we get:
\begin{eqnarray}
\bE
Z^m&=&\sum_{\bi}\bE\prod_{j=1}^{2^n}\exp\left\{-\beta\sum_{a=1}^m\calI(i_a=j)E_j\right\}\nonumber\\
&=&\sum_{\bi}\prod_{j=1}^{2^n}\exp\left\{\frac{\beta^2nJ^2}{4}
\sum_{a,b=1}^m\calI(i_a=j)\calI(i_b=j)\right\} ~~~~\mbox{using independence and
Gaussianity}\nonumber\\
&=&\sum_{\bi}\exp\left\{\frac{\beta^2nJ^2}{4}
\sum_{a,b=1}^m\sum_{j=1}^{2^n}\calI(i_a=j)\calI(i_b=j)\right\}\nonumber\\
&=&\sum_{\bi}\exp\left\{\frac{\beta^2nJ^2}{4}
\sum_{a,b=1}^m\calI(i_a=i_b)\right\}.\nonumber
\end{eqnarray}
We now define an $m\times m$ binary matrix $Q$, called the {\it overlap
matrix}, whose entries are
$Q_{ab}=\calI(i_a=i_b)$. Note that the summand in the last expression depends on $\bi$
only via $Q$. Let $N_n(Q)$ denote the number of configurations $\{\bi\}$ whose
overlap matrix is $Q$. We have to exhaust all possible overlap matrices, which are all
binary symmetric matrices with 1's on the main diagonal. Observe that the
number of such matrices is $2^{m(m-1)/2}$ whereas the number of configurations
is $2^{nm}$. Thus we are dividing the exponentially large number of
configurations into a relatively small number (independent of $n$) of
equivalence classes, something that rings the bell of the method of types. 
Let us suppose, for now, that there is some function $s(Q)$ such that
$N_n(Q)\exe e^{ns(Q)}$, and so
\begin{equation}
\bE Z^m\exe \sum_Q e^{ng(Q)}
\end{equation}
with:
\begin{equation}
g(Q)=\frac{\beta^2J^2}{4}\sum_{a,b=1}^mQ_{ab}+s(Q).
\end{equation}
From this point onward, the strategy is to use the saddle point method.
Note that the function $g(Q)$ is symmetric under replica permutations: let
$\pi$ be a permutation operator of $m$ objects and let $Q^\pi$ be the overlap
matrix with entries $Q_{ab}^\pi=Q_{\pi(a)\pi(b)}$. Then, $g(Q^\pi)=g(Q)$.  
This property is called {\it replica symmetry} (RS), and this property is
inherent to the replica method. In light of this, the first natural idea that
comes to our mind is to postulate that the saddle point is symmetric too, in
other words, to assume that the saddle--point $Q$ has 1's on its main
diagonal and all other entries are taken to be the same (binary) value, call it $q_0$. 
Now, there are only two possibilities:
\begin{itemize}
\item $q_0=0$ and then $N_n(Q)=2^n(2^n-1)\cdot\cdot\cdot(2^n-m+1)$, which implies that
$s(Q)=m\ln 2$, and then $g(Q)=g_0(Q)\dfn m(\beta^2J^2/4+\ln 2)$, thus
$(\ln \bE Z^m)/m=\beta^2J^2/4+\ln 2$, and so is the limit as $m\to 0$. 
Later on, we will compare this with the
result obtained from a more rigorous derivation.
\item $q_0=1$, which means that all components of $\bi$ are the same, and then
$N_n(Q)=2^n$, which means 
that $s(Q)=\ln 2$ and so, $g(Q)=g_1(Q)\dfn m^2\beta^2J^2/4+\ln 2$.
\end{itemize}
Now, one should check which one of these saddle points is the dominant one,
depending on $\beta$ and $m$. For $m\ge 1$, the behavior is dominated by
$\max\{g_0(Q),g_1(Q)\}$, which is $g_1(Q)$ for $\beta \ge \beta_c(m)\dfn
\frac{2}{J}\sqrt{\ln 2/m}$, and $g_0(Q)$ otherwise. 
For $m < 1$ (which is, in fact, the relevant case for $m\to 0$), one should
look at $\min\{g_0(Q),g_1(Q)\}$ (!), which is $g_0(Q)$ in the
high--temperature range.
As it turns out, in certain regions in the
$\beta$--$m$ plane, we must back off from the `belief' that dominant
configurations are {\it purely} symmetric, 
and resort to the quest for dominant configurations
with a lower level of symmetry. The first step, after having exploited the
purely symmetric case above, is called {\it one--step replica symmetry
breaking} (1RSB), and this means some partition of the set $\{1,2,\ldots,m\}$
into two complementary subsets (say, of equal size) and postulating a saddle point
$Q$ of the following structure: 
\begin{equation}
Q_{ab}=\left\{\begin{array}{ll}
1 & a=b\\
q_0 & \mbox{$a$ and $b$ are in the same subset}\\
q_1 & \mbox{$a$ and $b$ are in different subsets}
\end{array}\right.
\end{equation}
In further steps of symmetry breaking, one may split $\{1,2,\ldots,m\}$ to a larger number of
subsets 
or even introduce certain hierarchical structures. The replica method
includes a variety of heuristic guidelines in this context. We will not delve
into them any further in the framework of this course, but the interested
student/reader can easily find more details 
in the literature, specifically, in the book by M\'ezard and
Montanari. 


\newpage
\section{Interacting Particles and Phase Transitions}

\subsection{Introduction -- Origins of Interactions}

As I said already in the introductory part on the analysis tools and 
asymptotic methods, until now, we have dealt almost exclusively with
systems that have additive Hamiltonians, $\calE(\bx)=\sum_i\calE(x_i)$,
which means that the particles are i.i.d.\ and there is no interaction:
each particle behaves as if it was alone in the world. In Nature, of course,
this is seldom really the case. Sometimes this is still a reasonably good
approximation, but in many others the interactions 
are appreciably strong and cannot be neglected.
Among the different particles there could be
many sorts of mutual forces, e.g., mechanical, electrical, magnetic, etc.
There could also be interactions that stem from quantum--mechanical effects:
Pauli's exclusion principle asserts that for a certain type of particles,
called Fermions (e.g., electrons), no quantum state can be populated by more
than one particle. This gives rise to a certain mutal influence between
particles. Another type of interaction stems from the fact that the particles
are indistinguishable, so permutations between them are not considered as
distinct states. We have already seen this as an example at the beginning of
the previous set of lecture notes: In a quantum gas, as we eliminated the
combinatorial factor (that counted indistinguishable
states as distinguishable ones), we created statistical 
dependence, which physically means interactions.\footnote{Indeed, in the case
of the boson gas, there is a well--known effect referred to as {\it
Bose--Einstein condensation}, which is actually a phase transition, but phase
transitions can occur only in systems of interacting particles, as will be
discussed in this set of lectures.}


\subsection{A Few Models That Will be Discussed in This Subsection Only}

\begin{figure}[ht]
\hspace*{4cm}\input{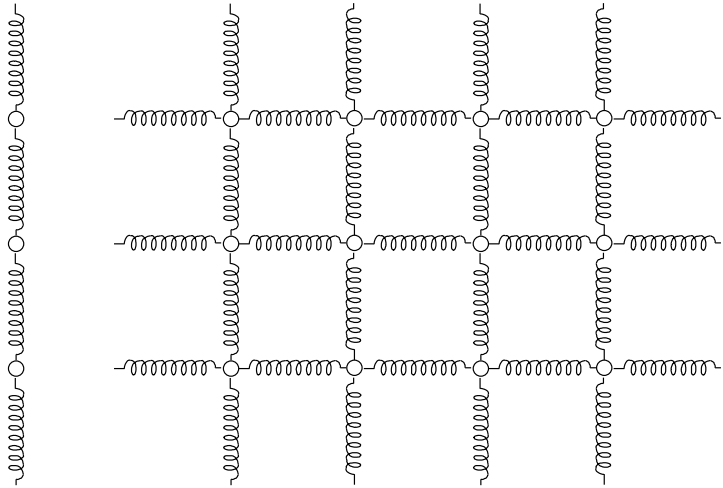}
\caption{\small Elastic interaction forces between adjacent atoms in a
one--dimensional lattice (left part of the figure) and in a two--dimensional
lattice (right part).}
\label{springs2d}
\end{figure}

\begin{figure}[ht]
\hspace*{5cm}\input{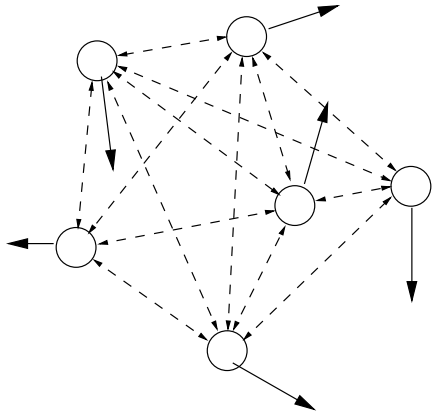}
\caption{\small Mobile particles and mutual forces between them.}
\label{spheres}
\end{figure}

The simplest forms of deviation from the purely 
additive Hamiltonian structure 
are those that consists, in addition to the individual energy
terms $\{\calE(x_i)\}$, also terms that depend on pairs, and/or triples,
and/or even larger cliques of particles. 
In the case of purely pairwise interactions, this
means a structure like the following:
\begin{equation}
\calE(\bx)=\sum_{i=1}^n\calE(x_i)+\sum_{(i,j)}\varepsilon(x_i,x_j)
\end{equation}
where the summation over pairs can be defined over all pairs $i\ne j$, 
or over some of the pairs, according to a given rule, e.g., depending
on the distance between particle $i$ and particle $j$, and according to
the geometry of the system, or according to a certain graph whose edges
connect the relevant pairs of variables (that in turn, are designated as nodes).
For example, in a one--dimensional array 
(a lattice) of
particles, a customary model accounts for interactions between neighboring
pairs only, neglecting more remote ones, thus the second term above would be
$\sum_{i}\varepsilon(x_i,x_{i+1})$. A well known special case of this is
that of a solid, i.e., a crystal lattice, where 
in the one--dimensional version of the model, atoms are thought of as
a chain of masses connected by springs (see left part of Fig.\ \ref{springs2d}),
i.e., an array of coupled harmonic oscillators. In
this case,
$\varepsilon(x_i,x_{i+1})=\frac{1}{2}K(u_{i+1}-u_i)^2$, where $K$ is a
constant and $u_i$ is the displacement of the $i$-th atom from its
equilibrium location, i.e., the potential energies of the springs.
This model has an easy analytical solution (by applying a Fourier transform
on the sequence $\{u_i\}$), where by ``solution'', we mean a closed--form,
computable formula for the log--partition function, at least in the
thermodynamic limit.
In higher dimensional arrays (or lattices), similar interactions apply,
there are just more neighbors to each site,
from the various directions (see right part of Fig.\ \ref{springs2d}). In a system where the
particles are mobile and hence their locations vary 
and have no geometrical structure,
like in a gas, the interaction terms are also
potential energies pertaining to the mutual forces (see Fig.\ \ref{spheres}), and these normally
depend solely on the distances $\|\vec{r}_i-\vec{r}_j\|$.
For example, in a non--ideal gas,
\begin{equation}
\calE(\bx)=\sum_{i=1}^n\frac{\|\vec{p}_i\|^2}{2m}+
\sum_{i\ne j}V(\|\vec{r}_i-\vec{r}_j\|).
\end{equation}
A very simple special case is that of 
hard spheres (Billiard balls), without any
forces, where
\begin{equation}
V(\|\vec{r}_i-\vec{r}_j\|)=\left\{\begin{array}{ll}
\infty & \|\vec{r}_i-\vec{r}_j\| < 2R\\
0 & \|\vec{r}_i-\vec{r}_j\| \ge 2R
\end{array}\right.
\end{equation}
which expresses the simple fact that balls cannot physcially overlap.
This model can (and indeed is) being used to obtain bounds on sphere--packing
problems, which are very relevant to channel coding theory.
This model is also solvable, but this is beyond the scope of this course.

\subsection{Models of Magnetic Materials -- General}

Yet another example of a model, or more precisely, a very large class of
models with interactions, are those of magnetic materials. These models
will closely accompany our dicussions from this point onward, because some
of them lend themselves to mathematical formalisms that are analogous
to those of coding problems, as we shall see. Few of these models are
solvable, but most of them are not. For the purpose of our discussion, a
magnetic material is one for which the important property of each particle
is its {\it magnetic moment}. The magnetic moment is a vector proportional to
the angular momentum of a revolving charged particle (like a rotating electron,
or a current loop), or the {\it spin}, and it designates the intensity of its
response to the net magnetic field that this particle `feels'. This magnetic
field may be 
the superposition of an externally applied magnetic field and the magnetic
fields generated by the neighboring spins. 

\begin{figure}[ht]
\hspace*{5cm}\input{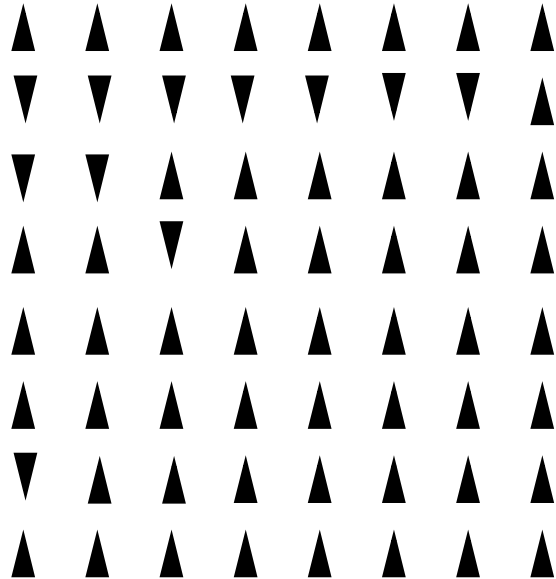}
\caption{\small Ilustration of a spin array on a square lattice.}
\label{spinglass}
\end{figure}

Quantum mechanical considerations
dictate that each spin, which will be denoted by $s_i$,
is quantized -- it may take only one out of finitely many 
values. In the simplest case to be adopted in our study -- only two values.
These will be designated by $s_i=+1$ (``spin up'') and $s_i=-1$ (``spin
down''), corresponding to the same intensity, but in two opposite directions,
one parallel to the magnetic field, and the other -- antiparallel (see Fig.\
\ref{spinglass}).
The Hamiltonian associated with an array of spins $\bs=(s_1,\ldots,s_n)$
is customarily modeled (up to certain constants that, among other things, accommodate for the
physical units) with a structure like this:
\begin{equation}
\calE(\bs)=-B\cdot\sum_{i=1}^ns_i-\sum_{(i,j)} J_{ij}s_is_j,
\end{equation}
where $B$ is the externally applied magnetic field 
and $\{J_{ij}\}$
are the coupling constants that designate the levels of interaction between
spin pairs, and they depend on properties of the magnetic material and on the
geometry of the system. The first term accounts for the 
contributions of potential energies 
of all spins due to the magnetic field, which in general, are 
given by the inner
product $\vec{B}\cdot\vec{s}_i$, but since each $\vec{s}_i$ is either
parallel or antiparallel to $\vec{B}$, as said, these boil down to simple
products, where only the sign of each $s_i$ counts. Since $P(\bs)$ is
proportional to $e^{-\beta\calE(\bs)}$, the spins `prefer' to be parallel,
rather than antiparallel to the magnetic field. 
The second term in the above Hamiltonian accounts for the interaction energy.
If $J_{ij}$ are all positive,
they also prefer to be parallel to one another (the probability for this is
larger), which is the case where
the material is called {\it ferromagnetic} (like iron and nickel). If they are all
negative, the material is {\it antiferromagnetic}. In the mixed case, it is
called a {\it spin glass}. In the latter, the behavior is rather complicated,
as we shall see later on.

Of course, the above model for the Hamiltonian can (and, in fact, is being)
generalized to include interactions formed also, by triples, quadruples, or
any fixed size $p$ (that does not grow with $n$) of spin--cliques.
At this point, it is instructive to see the relation between spin--array
models
(especially, those that involve large cliques of spins) to channel codes, in
particular, linear codes.
Consider a linear code defined by a set of $m$ partiy--check equations (in
$GF(2)$),
each involving the modulo--2 sum of some subset of the components of the
codeword $\bx$.
I.e., the $\ell$--th equation is: $x_{i_1^\ell}\oplus
x_{i_2^\ell}\oplus\cdot\cdot\cdot\oplus 
x_{i_{k_\ell}}^\ell=0$, $\ell=1,\ldots,m$. Transforming from
$x_i\in\{0,1\}$ to $s_i\in\{-1,+1\}$ via $s_i=1-2x_i$, this is
equivalent to $s_{i_1^\ell}s_{i_2^\ell}\cdot\cdot\cdot
s_{i_{k_{\ell}}^\ell}=1$. The MAP decoder would estimate $\bs$ based on
the posterior
\begin{equation}
P(\bs|\by)=\frac{P(\bs)P(\by|\bs)}{Z(\by)};~~~Z(\by)=\sum_{\bs}P(\bs)P(\by|\bs)=P(\by),
\end{equation}
where $P(\bs)$ is normally assumed uniform over the codewords (we will
elaborate on this posterior later).
Assuming, e.g., a BSC or a Gaussian channel $P(\by|\bs)$,
the relevant distance between the codeword $\bs=(s_1,\ldots,s_n)$
and the channel output $\by=(y_1,\ldots,y_n)$ is proportional to
$\|\bs-\by\|^2=\mbox{const.}-2\sum_is_iy_i$.
Thus, $P(\bs|\by)$ can be thought of as a B--G distribution with Hamiltonian
\begin{equation}
\calE(\bs|\by)= -J\sum_{i=1}^ns_iy_i+\sum_{\ell=1}^m
\phi(s_{i_1^\ell}s_{i_2^\ell}\cdot\cdot\cdot s_{i_{k_{\ell}}^\ell})
\end{equation}
where $J$ is some constant (depending on the channel parameters),
the function $\phi(u)$ vanishes for $u=1$ and becomes infinite for $u\ne
1$, and the partition function given by the denominator of $P(\bs|\by)$.
The first term plays the analogous role to that of the contribution of the
magnetic field in a spin system model,
where each `spin' $s_i$ `feels' a different magnetic field proportional to $y_i$, and the
second term accounts for the interactions among cliques of spins. In the case
of LDPC
codes, where each parity check equation involves only a small number of bits
$\{s_i\}$, these
interaction terms amount to cliques of relatively small sizes.\footnote{Error
correction codes can be represented by bipartite graphs with two types of
nodes: variable nodes corresponding to the various $s_i$ and function nodes
corresponding to cliques. There is an edge between variable node $i$ and
function node $j$ if $s_i$ is a member in clique $j$. Of course each $s_i$ may
belong to more than one clique. When all cliques are of size 2, there is no
need for the function nodes, as edges between nodes $i$ and $j$ simply
correspond to 
partity check equations involving $s_i$ and $s_j$.}
For a
general code, the second term is replaced by $\phi_{\calC}(\bs)$, which is
zero for $\bs\in\calC$ and infinite otherwise.

Another aspect of this model of a coded communication system pertains to
calculations of mutual information and capacity. The mutual information
between $\bS$ and $\bY$ is, of course, given by
\begin{equation}
I(\bS;\bY)=H(\bY)-H(\bY|\bS).
\end{equation}
The second term is easy to calculate for every additive channel -- it is
simply the entropy of the additive noise. The first term is harder to
calculate: 
\begin{equation}
H(\bY)=-\bE\{\ln P(\bY)\}=
-\bE\{\ln Z(\bY)\}.
\end{equation}
Thus, we are facing a problem of calculating the free energy of a spin system
with random magnetic fields designated by the components of $\bY$. This is 
the kind of calculations we mentioned earlier in the context of the replica
method. Indeed, the replica method is used extensively in this context.

As we will see in the sequel, it is also customary to introduce 
an inverse temperature parameter $\beta$, by defining
\begin{equation}
P_\beta(\bs|\by)=\frac{P^\beta(\bs)P^\beta(\by|\bs)}{Z(\beta|\by)}=
\frac{e^{-\beta\calE(\bs|\by)}}{Z(\beta|\by)}
\end{equation}
where $\beta$ controls the sharpness of the posterior distribution and
\begin{equation}
Z(\beta|\by)=\sum_{\bs} e^{-\beta\calE(\bs|\by)}.
\end{equation}
The motivations of this will be discussed extensively later on.

We will get back
to this important class of models, as well as its many extensions, shortly.
But before that, we discuss a very important effect that exists in some
systems with strong interactions (both in magnetic
materials and in other models): the effect of {\it phase transitions}.

\subsection{Phase Transitions -- A Qualitative Discussion}

Loosely speaking, a phase transition means an abrupt change in the
collective behavior of a physical system, as we change gradually
one of the externally controlled parameters, like the temperature, pressure,
or magnetic field, and so on. The most common example of a phase transition in our everyday
life is the water that we boil in the kettle when we make coffee, or when it
turns into ice as we put it in the freezer. 
What exactly are these phase transitions?
Before we refer to this question, it should be noted that there are also ``phase
transitions'' in the behavior of communication systems: As the SNR
passes a certain limit (for which capacity crosses the coding rate), there is
a sharp transition between reliable and unreliable communication, where the
error probability (almost) `jumps' from $0$ to $1$ or vice versa.
We also know about certain threshold effects in highly non--linear communication systems.
Are there any relationships between these phase transitions
and those of physics? We will see shortly that the answer is generally
affirmative.

In physics, phase transitions can occur only if the system has interactions. Consider,
the above example of an array of spins with $B=0$, and let us
suppose that all $J_{ij}> 0$ are equal, and thus will be denoted commonly by $J$. Then,
\begin{equation}
P(\bs)=\frac{\exp\left\{\beta J\sum_{(i,j)}s_is_j\right\}}{Z(\beta)}
\end{equation}
and, as mentioned earlier, this is a ferromagnetic model, where all spins
`like' to be in the same direction, especially when $\beta$ and/or $J$ is
large. In other words, the interactions, in this case, tend to introduce {\it
order} into the system. On the other hand, the second law talks about maximum
entropy, which tends to increase the {\it disorder}. So there are two
conflicting effects here. Which one of them prevails? 

\begin{figure}[ht]
\hspace*{5cm}\input{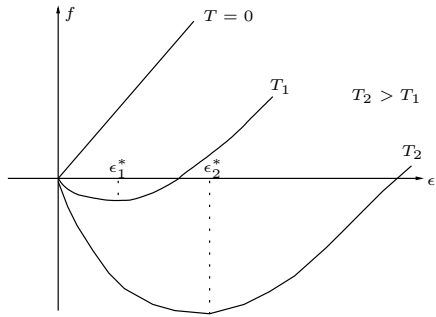}
\caption{\small Qualitative graphs of $f(\epsilon)$ at various temperatures.
The minimizing $\epsilon$ increases with $T$.}
\label{fgraph}
\end{figure}

The answer turns out to depend on
temperature. Recall that in the canonical ensemble, equilibrium is attained at
the point of minimum free energy $f=\epsilon-Ts(\epsilon)$. Now, $T$ plays
the role of a weighting factor for the entropy. At low temperatures, the
weight of the second term of $f$ is small, and minimiizing $f$ is approximately
(and for $T=0$, this is exact)
equivalent to minimizing $\epsilon$, which is obtained by states with a high
level of order, as $\calE(\bs)=-J\sum_{(i,j)}s_is_j$, in this example. 
As $T$ grows, however, the weight of the term $-Ts(\epsilon)$ increases, and 
$\min f$, becomes more and more equivalent to $\max s(\epsilon)$, which is achieved by
states with a high level of disorder (see Fig.\ \ref{fgraph}). 
Thus, the order--disorder characteristics depend
primarily on temperature. 
It turns out that for some magnetic systems of this kind, this transition between
order and disorder may be abrupt, in which case, we call it a {\it phase
transition}.
At a certain critical temperature, called
the {\it Curie temperature}, there
is a sudden transition between order and disorder. In the ordered phase,
a considerable fraction of the spins align in the same
direction, which means that the system is spontaneously magnetized (even
without an external magnetic field), whereas in
the disordered phase, about half of the spins are in either direction, and
then the
net magnetization vanishes.
This happens if the
interactions, or more precisely, their dimension in some sense, is strong
enough. 

What is the mathematical significance of a phase transition? If we look at
the partition function, $Z(\beta)$, which is the key to all physical
quantities of interest, then
for every finite $n$, this is simply the sum of a bunch of exponentials in
$\beta$ and therefore it is continuous and differentiable as many times as we
want. So what kind of abrupt changes could there possibly be in the behavior
of this function?

It turns out that while this is true for all finite $n$, it is no longer necesarily true if we
look at the thermodynamical limit, i.e., if we look at the
behavior of $\phi(\beta)=\lim_{n\to\infty}\frac{\ln
Z(\beta)}{n}$. While $\phi(\beta)$ must be continuous for all $\beta >0$
(since it is convex), it need not necessarily have continuous derivatives.
Thus, a phase transition, if exists, is fundamentally an asymptotic property, it
may exist in the thermodynamical limit only. While a physical system is, after
all finite, it is nevertheless well approximated by the thermodynamical limit
when it is very large.
By the same token, if we look at the analogy with a coded communication
system: for any finite block--length $n$, the error probability is a `nice'
and smooth
function of the SNR, but in the limit of large $n$, it behaves like a step
function that jumps between $0$ and $1$ at the critical SNR. 
We will see that the two things are
related.

Back to the physical aspects, the above discussion explains also why a system without interactions,
where all $\{x_i\}$ are i.i.d., cannot have phase transitions. In this case,
$Z_n(\beta)=[Z_1(\beta)]^n$, and so, $\phi(\beta)=\ln Z_1(\beta)$, which is
always a `nice' function without any irregularities. For a phase transition to
occur, the particles must behave in some collective manner, which is the case
only if interactions take place.

There is a distinction between two types of phase transitions:
\begin{itemize}
\item If $\phi(\beta)$ has a discontinuous first order derivative, then this
is called a {\it first order phase transition}. 
\item If $\phi(\beta)$ has a continuous first order derivative, 
but a discontinuous second order derivative then this
is called a {\it second order phase transition}, or a {\it continuous phase
transition}.
\end{itemize}

We can talk, of course, about phase transitions w.r.t.\ additional parameters other than
temperature. In the above magnetic example, if we introduce back the magnetic
field $B$ into the picture, then $Z$, and hence also $\phi$, become functions of $B$
too. If we then look at derivative of
\begin{equation}
\phi(\beta,B)=\lim_{n\to\infty}\frac{\ln Z(\beta,B)}{n}=
\lim_{n\to\infty}\frac{1}{n}\ln\left[\sum_{\bs}\exp\left\{\beta
B\sum_{i=1}^ns_i+\beta J\sum_{(i,j)}s_is_j\right\}\right]
\end{equation}
w.r.t.\ the product $(\beta B)$, which multiplies the magnetization,
$\sum_is_i$, at the exponent,
this would give exactly the average magnetization per spin
\begin{equation}
m(\beta,B)=\left<\frac{1}{n}\sum_{i=1}^n S_i\right>,
\end{equation}
and this quantity might not always be continuous. 
Indeed, as I mentioned earlier, below the Curie
temperature there might be a spontaneous magnetization. If $B\downarrow
0$, then this magnetization is positive, and if $B\uparrow 0$, it is negative,
so there is a discontinuity at $B=0$.
We will see this more concretely later on.  
We next discuss a few solvable models of spin arrays, with and without
phase transitions.

\subsection{The One--Dimensional Ising Model}

According to this model,
\begin{equation}
\calE(\bs)=-B\sum_{i=1}^ns_i -J\sum_{i=1}^n s_is_{i+1}
\end{equation}
with the periodic boundary condition $s_{n+1}=s_1$. Thus, 
\begin{eqnarray}
Z(\beta,B)&=&\sum_{\bs}\exp\left\{\beta B\sum_{i=1}^ns_i+\beta J\sum_{i=1}^n
s_is_{i+1}\right\}~~~\mbox{Note: the kind of sums encountered in Markov
chains}\nonumber\\
&=&\sum_{\bs}\exp\left\{h\sum_{i=1}^ns_i+K\sum_{i=1}^n
s_is_{i+1}\right\}~~~~~h\dfn\beta B,~~K\dfn \beta J\nonumber\\
&=&\sum_{\bs}\exp\left\{\frac{h}{2}\sum_{i=1}^n(s_i+s_{i+1})+K\sum_{i=1}^n
s_is_{i+1}\right\}~~~~~\mbox{(just to symmetrize the expression)}\nonumber
\end{eqnarray}
Consider now the $2\times 2$ matrix $P$ whose entries are
$\exp\{\frac{h}{2}(s+s')+Kss'\}$, $s,s\in\{-1,+1\}$, i.e., 
\begin{equation}
P=\left(\begin{array}{cc}
e^{K+h} & e^{-K} \\
e^{-K} & e^{K-h}\end{array}\right).
\end{equation}
Also, $s_i=+1$ will be represented by the column vector $\sigma_i=(1,0)^T$ and
$s_i=-1$ will be represented by $\sigma_i=(0,1)^T$. Thus,
\begin{eqnarray}
Z(\beta,B)&=&\sum_{\sigma_1}\cdot\cdot\cdot\sum_{\sigma_n} 
(\sigma_1^T
P\sigma_2)\cdot(\sigma_2^TP\sigma_2)\cdot\cdot\cdot(\sigma_n^TP\sigma_1)\nonumber\\
&=&\sum_{\sigma_1}\sigma_1^TP\left(\sum_{\sigma_2}\sigma_2\sigma_2^T\right)P
\left(\sum_{\sigma_3}\sigma_3\sigma_3^T\right)P\cdot\cdot\cdot P
\left(\sum_{\sigma_n}\sigma_n\sigma_n^T\right)P\sigma_1\nonumber\\
&=&\sum_{\sigma_1}\sigma_1^TP\cdot I\cdot P\cdot I
\cdot\cdot\cdot I\cdot P\sigma_1\nonumber\\
&=&\sum_{\sigma_1}\sigma_1^TP^n\sigma_1\nonumber\\
&=&\mbox{tr}\{P^n\}\nonumber\\
&=&\lambda_1^n+\lambda_2^n
\end{eqnarray}
where $\lambda_1$ and $\lambda_2$ are the eigenvalues of $P$, which are
\begin{equation}
\lambda_{1,2}=e^K\cosh(h)\pm\sqrt{e^{-2K}+e^{2K}\sinh^2(h)}.
\end{equation}
Letting $\lambda_1$ denote the larger (the dominant) eigenvalue, i.e.,
\begin{equation}
\lambda_{1}=e^K\cosh(h)+\sqrt{e^{-2K}+e^{2K}\sinh^2(h)},
\end{equation}
then clearly,
\begin{equation}
\phi(h,K)=\lim_{n\to\infty}\frac{\ln Z}{n}=\ln\lambda_1.
\end{equation}
The average magnetization is
\begin{eqnarray}
M(h,K)&=&\left< \sum_{i=1}^n S_i\right>\nonumber\\
&=&\frac{\sum_{\bs}(\sum_{i=1}^ns_i)\exp\{h\sum_{i=1}^ns_i+K\sum_{i=1}^ns_is_{i+1}\}}
{\sum_{\bs}\exp\{h\sum_{i=1}^ns_i+K\sum_{i=1}^ns_is_{i+1}\}}\nonumber\\
&=&\frac{\partial \ln Z(h,K)}{\partial h}
\end{eqnarray}
and so, the per--spin magnetization is:
\begin{equation}
m(h,K)\dfn\lim_{n\to\infty}\frac{M(h,K)}{n}=\frac{\partial \phi(h,K)}{\partial
h}= \frac{\sinh(h)}{\sqrt{e^{-4K}+\sinh^2(h)}}
\end{equation}
or, returning to the original parametrization:
\begin{equation}
m(\beta,B)=
\frac{\sinh(\beta B)}{\sqrt{e^{-4\beta J}+\sinh^2(\beta B)}}.
\end{equation}
For $\beta > 0$ and $B > 0$ this is a nice function, and so, there is are no
phase transitions and no spontaneous magnetization 
at any finite temperature.\footnote{Note, in particular, that for
$J=0$ (i.i.d.\ spins) we get paramagnetic characteristics
$m(\beta,B)=\tanh(\beta B)$, in agreement with
the result pointed out in the example of two--level systems, in one of our
earlier discussions.} However, at the absolute zero ($\beta\to\infty$), we get
\begin{equation}
\lim_{B\downarrow 0}\lim_{\beta\to\infty}m(\beta,B)=+1;~~
\lim_{B\uparrow 0}\lim_{\beta\to\infty}m(\beta,B)=-1,
\end{equation}
thus $m$ is discontinuous w.r.t.\ $B$ at $\beta\to\infty$, which means that
there is a phase transition at $T=0$. In other words, the Curie temperature is
$T_c=0$.

We see then that one--dimensional Ising model is easy to handle, but it is not
very interesting in the sense that there is actually no phase transition. The
extension to the two--dimensional Ising model on the square lattice is
surprisingly more difficult, but it is still solvable, albeit without a magnetic
field. It was first solved by
Onsager in 1944, who has shown that it exhibits a phase transition with
Curie temperture given by 
\begin{equation}
T_c=\frac{2J}{k\ln(\sqrt{2}+1)},
\end{equation}
where $k$ is
Boltzmann's constant. For lattice dimension $\ge 3$, the problem is still
open. 

It turns out then that whatever counts for the existence of phase transitions, is
not the intensity of the interactions (designated by the magnitude of $J$),
but rather the ``dimensionality'' of the 
structure of the pairwise interactions. If we
denote by
$n_\ell$ the number of $\ell$--th order neighbors of every given site, namely, the
number of sites that can be reached within $\ell$ steps from the given site, then
whatever counts is how fast does the sequence $\{n_\ell\}$ grow, or more
precisely, what is the value of $d\dfn \lim_{\ell\to\infty}\frac{1}{\ell}\ln
n_\ell$, which is exactly the ordinary dimensionality for hypercubic lattices. 
Loosely speaking, this dimension must be sufficiently large for a phase transition to exist.

To demonstrate this point, we next discuss an extreme case of a model where
this dimensionality
is actually infinite. In this model ``everybody is a neighbor of everybody
else'' and to the same extent, so it definitely has the highest connectivity
possible. This is not quite a physically realistic model, but
the nice thing about it is that it is easy to solve and that it exhibits a
phase transition that is fairly similar to those that exist in real systems. It is also
intimately related to a very popular approximation method in statistical
mechanics, called the {\it mean field approximation}. Hence it is sometimes
called the {\it mean field model}. It is also known as the {\it Curie--Weiss
model} or the {\it infinite range model}. 

Finally, I should comment that there are
other ``infinite--dimensional'' Ising models, like the one defined on the
Bethe lattice (an infinite tree without a root and without leaves), 
which is also easily solvable (by recursion) and it also
exhibits phase transitions (see Baxter's book), but we will not discuss it here.

\subsection{The Curie--Weiss Model}

\begin{figure}[ht]
\hspace*{1cm}\input{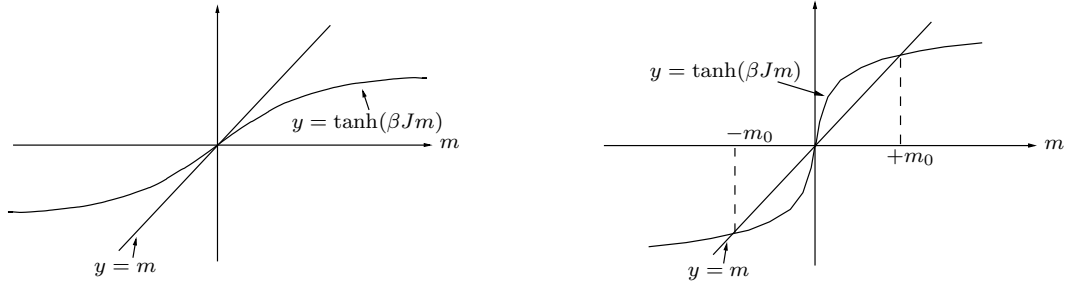}
\caption{\small Graphical solutions of equation $m=\tanh(\beta Jm)$: The
left part corresponds to the case $\beta J < 1$, where there is one
solution only, $m^*=0$. The right part corresponds to the case $\beta J > 1$, where
in addition to the zero solution, there are two non--zero solutions $m^*=\pm
m_0$.}
\label{cw1}
\end{figure}

\begin{figure}[ht]
\hspace*{1cm}\input{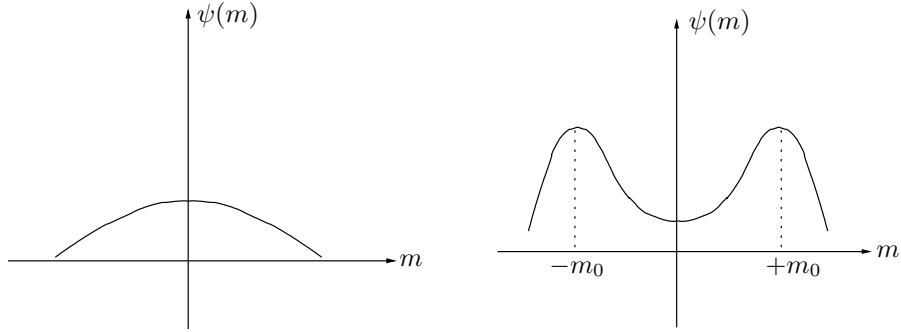}
\caption{\small The function $\psi(m)=  h_2((1+m)/2)+\beta Jm^2/2$ has a
unique maximum at $m=0$ when $\beta J < 1$ (left graph) and two local maxima
at $\pm m_0$,
in addition to a local minimum at $m=0$, when $\beta J > 1$ (right graph).}
\label{cw2}
\end{figure}

According to the Curie--Weiss (C--W) model,
\begin{equation}
\calE(\bs)=-B\sum_{i=1}^n s_i-\frac{J}{2n}\sum_{i\ne j}s_is_j.
\end{equation}
Here, all pairs $\{(s_i,s_j)\}$ ``talk to each other'' with the same
``voice intensity'', $J/(2n)$, and without any geometry. The $1/n$ factor here
is responsible for
keeping the energy of the system extensive (linear in $n$), as the number of
interaction terms is quadratic in $n$. The factor $1/2$ compensates for the
fact that the summation over $i\ne j$ counts each pair twice. The first
observation is the trivial fact that
\begin{equation}
\left(\sum_is_i\right)^2=\sum_is_i^2+\sum_{i\ne j}s_is_j=n+\sum_{i\ne j}s_is_j
\end{equation}
where the second equality holds since $s_i^2\equiv 1$. It follows then, that
our Hamiltonian is, upto a(n immaterial) constant, equivalent to
\begin{equation}
\calE(\bs)=-B\sum_{i=1}^n s_i-\frac{J}{2n}\left(\sum_{i=1}^ns_i\right)^2=
-n\left[B\cdot\left(\frac{1}{n}\sum_{i=1}^n s_i\right)+
\frac{J}{2}\left(\frac{1}{n}\sum_{i=1}^ns_i\right)^2\right],
\end{equation}
thus $\calE(\bs)$ depends on $\bs$ only via the magnetization
$m(\bs)=\frac{1}{n}\sum_is_i$. This fact makes the C--W model very easy to
handle similarly as in the method of types:
\begin{eqnarray}
Z_n(\beta,B)&=&\sum_{\bs}\exp\left\{n\beta\left[B\cdot
m(\bs)+\frac{J}{2}m^2(\bs)\right]\right\}\nonumber\\
&=&\sum_{m=-1}^{+1}\Omega(m)\cdot e^{n\beta(Bm+Jm^2/2)}\nonumber\\
&\exe&\sum_{m=-1}^{+1}e^{nh_2((1+m)/2)}\cdot e^{n\beta(Bm+Jm^2/2)}\nonumber\\
&\exe&\exp\left\{n\cdot\max_{|m|\le 1}\left[h_2\left(\frac{1+m}{2}\right)+
\beta Bm+\frac{\beta m^2J}{2}\right]\right\}\nonumber
\end{eqnarray}
and so,
\begin{equation}
\phi(\beta,B)=\max_{|m|\le 1}\left[h_2\left(\frac{1+m}{2}\right)+
\beta Bm+\frac{\beta m^2J}{2}\right].
\end{equation}
The maximum is found by equating the derivative to zero, i.e.,
\begin{equation}
0=\frac{1}{2}\ln\left(\frac{1-m}{1+m}\right)+\beta B+\beta Jm
\equiv -\tanh^{-1}(m)+\beta B+\beta Jm
\end{equation}
or equivalently, the maximizing (and hence the dominant) 
$m$ is a solution $m^*$ to the equation\footnote{Once again, for $J=0$, we are
back to non--interacting spins and then this equation gives the
paramagnetic behavior $m=\tanh(\beta B)$.}
$$m=\tanh(\beta B+\beta Jm).$$
Consider first the case
$B=0$, where the equation boils down to
\begin{equation}
m=\tanh(\beta Jm).
\end{equation}
It is instructive to look at this equation graphically. 
Referring to Fig.\ \ref{cw1}, we have to make a distinction between
two cases: If $\beta J <1$, namely, $T > T_c\dfn J/k$, the slope of the
function $y=\tanh(\beta Jm)$ at the origin, $\beta J$, is smaller than the slope of the
linear function $y=m$, which is $1$, thus these two graphs intersect only at
the origin. It is easy to check that in this case, the second derivative of
$\psi(m)\dfn h_2((1+m)/2)+\beta Jm^2/2$ at $m=0$ is negative, and therefore it is indeed the
maximum (see Fig.\ \ref{cw2}, left part). Thus,
the dominant magnetization is $m^*=0$, which means
disorder and hence no spontaneous
magnetization for $T > T_c$.
On the other hand, when $\beta J > 1$, which means temperatures lower than
$T_c$, the initial slope of the $\tanh$ function is larger than that of the
linear function, but since the $\tanh$ cannot take values outside the interval
$(-1,+1)$, the two functions must intersect also at two additional, symmetric, non--zero
points, which we denote by $+m_0$ and $-m_0$ (see Fig.\ \ref{cw1}, right part). 
In this case, it can readily be
shown that the second derivative of 
$\psi(m)$ is positive at the origin (i.e., there is a local
minimum at $m=0$) and negative at $m=\pm m_0$, which means that 
there are maxima at these two points (see Fig.\ \ref{cw2}, right part). 
Thus, the dominant magnetizations are
$\pm m_0$, each capturing about half of the probability.

\begin{figure}[ht]
\hspace*{1cm}\input{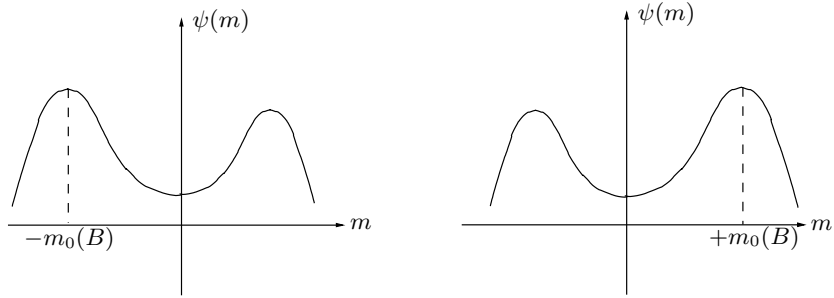}
\caption{\small The case $\beta J > 1$ with a magnetic field $B$. The left
graph corresponds to $B < 0$ and the right graph -- to $B > 0$.}
\label{cw3}
\end{figure}

Consider now the case $\beta J > 1$, where the magnetic field $B$ is brought
back into the picture. This will break the symmetry of the right graph of
Fig.\ \ref{cw2} and the corresponding graphs of $\psi(m)$ would be as in Fig.\
\ref{cw3}, where now the higher local maximum (which is also the global one)
is at $m_0(B)$ whose sign is as that of $B$. But as $B\to 0$, $m_0(B)\to m_0$
of Fig.\ \ref{cw2}.
\begin{figure}[ht]
\hspace*{1cm}\input{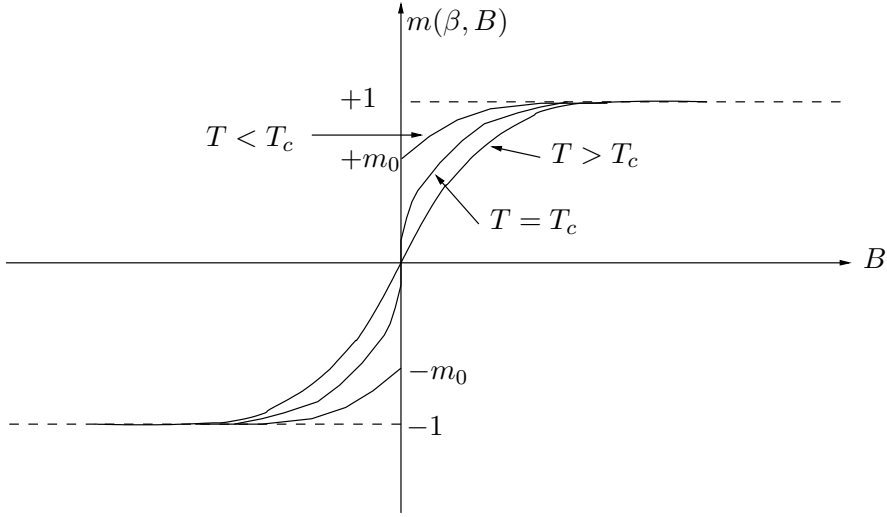}
\caption{\small
Magnetization vs.\ magnetic field: For $T < T_c$ there is spontaneous
magnetization: $\lim_{B\downarrow 0}m(\beta,B)=+m_0$ and
$\lim_{B\uparrow 0}m(\beta,B)=-m_0$, and so there is a discontinuity at
$B=0$.}
\label{cw4}
\end{figure}
Thus, we see the spontaneous magnetization here. Even after removing the
magnetic field, the system remains magnetized to the level of $m_0$, depending
on the direction (the sign) of $B$ before its removal. Obviously, the
magnetization $m(\beta,B)$ has a discontinuity at $B=0$ for $T < T_c$, which is a first
order phase transition w.r.t.\ $B$ (see Fig.\ \ref{cw4}). We note that the point $T=T_c$ is
the boundary between the region of existence and the 
region of non--existence of a phase transition
w.r.t. $B$. Such a point is called a {\it critical point}. 
The phase transition w.r.t.\ $\beta$ is of the second order.

Finally, we should mention here an alternative technique that can be used
to analyze this model, which is useful in many other contexts as well. It
is based on the idea of using a transform integral, in this case, the 
{\it Hubbard--Stratonovich transform}, and then the saddle point method.
Specifically, we have the following chain of equalities:
\begin{eqnarray}
Z(h,K)&=&\sum_{\bs} \exp\left\{h\sum_{i=1}^n
s_i+\frac{K}{2n}\left(\sum_{i=1}^ns_i\right)^2\right\}~~~~h\dfn \beta
B,~K\dfn \beta J\nonumber\\
&=&\sum_{\bs} \exp\left\{h\sum_{i=1}^n
s_i\right\}\cdot\exp\left\{\frac{K}{2n}\left(\sum_{i=1}^ns_i\right)^2\right\}\nonumber\\
&=&\sum_{\bs} \exp\left\{h\sum_{i=1}^n
s_i\right\}\cdot\sqrt{\frac{n}{2\pi K}}\int_\reals \dd z
\exp\left\{-\frac{nz^2}{2K}+z\cdot\sum_{i=1}^ns_i\right\}\nonumber\\
&=&\sqrt{\frac{n}{2\pi K}}\int_\reals \dd z e^{-nz^2/(2K)}
\sum_{\bs} \exp\left\{(h+z)\sum_{i=1}^n
s_i\right\}\nonumber\\
&=&\sqrt{\frac{n}{2\pi K}}\int_\reals \dd z e^{-nz^2/(2K)}
\left[\sum_{s=-1}^1 e^{(h+z)s}\right]^n\nonumber\\
&=&\sqrt{\frac{n}{2\pi K}}\int_\reals \dd z e^{-nz^2/(2K)}
[2\cosh(h+z)]^n\nonumber\\
&=&2^n\cdot\sqrt{\frac{n}{2\pi K}}\int_\reals \dd z
\exp\{n[\ln\cosh(h+z)-z^2/(2K)]\}\nonumber
\end{eqnarray}
Using the the saddle point method (or the Laplace method), this integral is
dominated by the maximum of the function in the square brackets at the
exponent of the integrand, or equivalently, the minimum of the function
\begin{equation}
\gamma(z)=\frac{z^2}{2K}-\ln\cosh(h+z).
\end{equation}
by equating its derivative to zero, we get the very same equation as
$m=\tanh(\beta B+\beta Jm)$ by setting $z=\beta Jm$.
The function $\gamma(z)$ is different from the function
$\psi$ that we maximized earlier, but the extremum is the same. This function
is called the {\it Landau free energy}.

\subsection{Spin Glass Models With Random Parameters and Random Code
Ensembles}

So far we discussed only models where the non--zero coupling coefficients,
$\bJ=\{J_{ij}\}$ are equal, thus they are either all positive (ferromagnetic models) or
all negative (antiferromagnetic models). As mentioned earlier, there are also
models where the signs of these coefficients are mixed, which are called {\it
spin glass} models. 

Spin glass models have a much more complicated and more interesting behavior
than ferromagnets, because there might be metastable states due to the fact
that not necessarily all spin pairs $\{(s_i,s_j)\}$ can be in their preferred
mutual polarization. It might be the case that some of these pairs are
``frustrated.'' In order to model situations of amorphism and disorder in
such systems, it is customary to model the coupling coeffcients as random
variables. 

Some models allow, in addition to the random coupling
coefficients, also random local fields, i.e., the term $-B\sum_is_i$ in the
Hamiltonian, is replaced by $-\sum_iB_is_i$, where $\{B_i\}$ are random
variables, similarly as in the representation of $P(\bs|\by)$ pertaining to a coded 
communicaion system, as discussed earlier, where $\{y_i\}$ play the role of local
magnetic fields. The difference, however, is that here the $\{B_i\}$ are
normally assumed i.i.d., whereas in the communication system model $P(\by)$
exhibits memory (even if the channel is memoryless) due to memory in $P(\bs)$.
Another difference is that in the physics model, the distribution of $\{B_i\}$
is assumed to be independent of temperature, whereas in coding, if we introduce a
temperature parameter by exponentiating (i.e., $P_\beta(\bs|\by)\propto
P^\beta(s)P^\beta(\by|\bs)$), the induced marginal of $\by$ will depend on $\beta$.

In the following discussion, let us refer to the case where only the coupling
coefficients $\bJ$ are random variables (similar things can be said in the
more general case, discussed in the last paragraph). This model with random
parameters means that there are now 
two levels of randomness:
\begin{itemize}
\item Randomness of the coupling coefficients $\bJ$.
\item Randomness of the spin configuration $\bs$ given $\bJ$, according
to the Boltzmann distribution, i.e.,
\begin{equation}
P(\bs|\bJ)=\frac{\exp\left\{\beta\left[B\sum_{i=1}^ns_i+\sum_{(i,j)}J_{ij}s_is_j\right]\right\}}
{Z(\beta,B|\bJ)}.
\end{equation}
\end{itemize}
However, these two sets of RV's have a rather different stature.
The underlying setting is normally such that $\bJ$ is considered to be
randomly drawn once and for all, and then remain fixed, whereas $\bs$ keeps
varying all the time (according to the dynamics of the system).
At any rate, the time scale along which $\bs$ varies is much smaller than that
of $\bJ$. Another difference is that $\bJ$ is normally not
assumed to depend on temperature, whereas $\bs$, of course, does. In the
terminlogy of physicists, $\bs$ is considered an {\it annealed} RV, whereas
$\bJ$ is considered a {\it quenched} RV. Accordingly, there is a
corresponding distinction between {\it annealed averages} and {\it quenched
averages}.

Actually, there is (or, more precisely, should be) a parallel distinction
when we consider ensembles of randomly chosen codes in Information Theory.
When we talk about random coding, we normally think of the randomly chosen
code as being drawn once and for all, we don't reselect it after each
transmission (unless there are security reasons to do so), and so, a random
code should be thought of us a quenched entity, whereas the source(s) and
channel(s) are more naturally thought of as annealed entities. Nonetheless,
this is not what we usually do in Information Theory. We normally take double
expectations of some performance measure w.r.t.\ both source/channel and
the randomness of the code, on the same footing.\footnote{
There are few exceptions to this rule, e.g., a paper by
Barg and Forney, IEEE Trans.\ on IT, Sept.\ 2002, and several follow--ups.}
We will elaborate on this point later on.

Returning to spin glass models, let's see what is exactly the difference
between the quenched averaging and the annealed one. If we examine, for
instance, the free energy, or the log--partition function, $\ln
Z(\beta|\bJ)$, this is now a RV, of course, because it depends on the random
$\bJ$. If we denote by $\langle\cdot\rangle_{\bJ}$ the
expectation w.r.t.\ the randomness of $\bJ$, then quenched averaging means
$\langle\ln Z(\beta|\bJ)\rangle_{\bJ}$ (with the motivation of
the self--averaging property of the RV $\ln Z(\beta|\bJ)$ in many cases),
whereas annealed averaging means $\ln\langle Z(\beta|\bJ)\rangle_{\bJ}$.
Normally, the relevant average is the quenched one, but it is typically
also much harder to calculate (and it is customary to apply the replica
method then).
Clearly, the annealed average is never smaller than the quenched one because of Jensen's
inequality, but they sometimes coincide at high temperatures.
The difference between them is that in quenched averaging, the
dominant realizations of $\bJ$ are the typical ones, whereas in annealed
averaging, this is not necessarily the case. This follows from the
following sketchy consideration. As for the annealed average, we have:
\begin{eqnarray}
\left<Z(\beta|\bJ\right>&=&\sum_{\bJ}P(\bJ)Z(\beta|\bJ)\nonumber\\
&\approx&\sum_{\alpha}\mbox{Pr}\{\bJ:~Z(\beta|\bJ)\exe e^{n\alpha}\}\cdot
e^{n\alpha}\nonumber\\
&\approx&\sum_{\alpha}e^{-nE(\alpha)}\cdot
e^{n\alpha}~~~~~~\mbox{(assuming exponential probabilities)}\nonumber\\
&\exe& e^{n\max_\alpha[\alpha-E(\alpha)]}
\end{eqnarray}
which means that the annealed average is dominated by realizations
of the system with 
\begin{equation}
\frac{\ln Z(\beta|\bJ)}{n}\approx \alpha^*\dfn
\mbox{arg}\max_\alpha[\alpha-E(\alpha)],
\end{equation}
which may differ from the typical value of $\alpha$,
which is 
\begin{equation}
\alpha=\phi(\beta)
\equiv\lim_{n\to\infty}\frac{1}{n}\left<\ln
Z(\beta|\bJ)\right>.
\end{equation}
On the other hand, when it comes to quenched averaging, 
the RV $\ln Z(\beta|\bJ)$ behaves linearly in $n$, 
and concentrates strongly around 
the typical value $n\phi(\beta)$,
whereas other values are weighted by (exponentially) decaying probabilities.

In the coded communication
setting, there is a strong parallelism.
Here, there is a distinction between the exponent of the
average error probability, $\ln\bE P_e(\calC)$ (annealed) and the average
exponent of the error probability $\bE\ln P_e(\calC)$ (quenched), where
$P_e(\calC)$ is the error probability of a randomly selected code $\calC$.
Very similar things can be said here too. 

The literature on spin glasses includes many models for the randomness of
the coupling coefficients. We end this part by listing just a few.
\begin{itemize}
\item The {\it Edwards--Anderson} (E--A) model, where $\{J_{ij}\}$ are non--zero
for nearest--neighbor pairs only (e.g., $j=i\pm 1$ in one--dimensional model).
According to this model, these $J_{ij}$'s are i.i.d.\ RV's, which are normally
modeled to have a zero--mean Gaussian pdf, or binary symmetric with levels $\pm J_0$.
It is customary to work with a zero--mean distribution if we have a pure spin
glass in mind. If the mean is nonzero, the model has either a ferromangetic
or an anti-ferromagnetic bias, according to the sign of the mean.
\item The {\it Sherrington--Kirkpatrick} (S--K) model, which is similar to the
E--A model, except that the support of $\{J_{ij}\}$ is extended to include all
$n(n-1)/2$ pairs, and not only nearest--neighbor pairs. This can be thought of
as a stochastic version of the C--W model in the sense that here too, there is
no geometry, and every spin `talks' to every other spin to the same extent,
but here the coefficients are random, as said. 
\item The {\it $p$--spin} model, which is similar to the S--K model, but now
the interaction term consists, not only of pairs, but also triples,
quadraples, and so on, up to cliques of size $p$, i.e., products 
$s_{i_1}s_{i_2}\cdot\cdot\cdot s_{i_p}$, where $(i_1,\ldots,i_p)$ exhaust all
possible subsets of $p$ spins out of $n$. Each such term has a Gaussian
coefficient $J_{i_1,\ldots,i_p}$ with an appropriate variance.
\end{itemize}
Considering the $p$--spin model, it turns out that if we look at the extreme
case of $p\to\infty$ (taken after the thermodynamic limit $n\to\infty$), the
resulting behavior turns out to be extremely erratic: all energy levels
$\{\calE(\bs)\}_{\bs\in\{-1,+1\}^n}$ become i.i.d.\ Gaussian RV's. This is,
of course, a toy model, which has very little to do with reality (if any),
but it is surprisingly interesting and easy to work with. It is called the
{\it random energy model} (REM). We have already mentioned it as an example
on which we demonstrated the replica method. We are next going to talk about
it extensively because it turns out to be very relevant for random coding
models.


\newpage
\section{The Random Energy Model and Random Coding}

\subsection{The REM in the Absence of a Magnetic Field}

The REM was proposed by the French physicist Bernard Derrida in the early
eighties of the previous century in a series of papers:
\begin{enumerate}
\item B.~Derrida, ``Random--energy model: limit of a family of disordered
models,'' 
{\it Phys.\ Rev.\ Lett.}, vol.\ 45, no.\ 2, pp.\ 79--82, July 1980.
\item B.~Derrida, ``The random energy model,'' {\it Physics Reports} (Review
Section of 
Physics Letters), vol.\ 67, no.\ 1, pp.\ 29--35, 1980.
\item B.~Derrida, ``Random--energy model: an exactly solvable model for
disordered systems,'' 
{\it Phys.\ Rev.\ B}, vol.\ 24, no.\ 5, pp.\ 2613--2626, September 1981.
\end{enumerate}
Derrida showed in one of his papers that, since the correlations
between the random energies of two configurations, $\bs$ and $\bs'$ in the
$p$--spin model are given by
\begin{equation}
\left(\frac{1}{n}\sum_{i=1}^ns_is_i'\right)^p,
\end{equation}
and since $|\frac{1}{n}\sum_{i=1}^ns_is_i'| < 1$, these correlations vanish
as $p\to\infty$. This has motivated him to propose a model according to which
the configurational energies $\{\calE(\bs)\}$, in the absence of a magnetic field, are simply
i.i.d.\ zero--mean Gaussian RV's with a variance that grows linearly with $n$
(again, for reasons of extensivity). More concretely, this variance
is taken to be $nJ^2/2$, where $J$ is a constant parameter. This means that we
forget that the spin array has any structure of the kind that
we have seen before, and we simply randomly draw an independent RV
$\calE(\bs)\sim\calN(0,nJ^2/2)$ (and other distributions are also possible)
for every configuration $\bs$. Thus,
the partition function $Z(\beta)=\sum_{\bs} e^{-\beta\calE(\bs)}$ is a random
variable as well, of course.

This is a toy model that does not 
describe faithfully any realistic physical system, but we
will devote to it some considerable time, for several reasons:
\begin{itemize}
\item It is simple and easy to analyze.
\item In spite of its simplicity, it is rich enough to exhibit phase
transitions, and therefore it is interesting.
\item Last but not least, it will prove very relevant to the analogy with
coded communication systems with randomly selected codes.
\end{itemize}
As we shall see quite shortly, there is an intimate relationship between
phase transitions of the REM and phase transitions in the behavior of coded
communication systems, most notably, transitions between reliable and
unreliable communication, but others as well. 

What is the basic idea that stands behind the analysis of the REM?
As said, 
\begin{equation}
Z(\beta)=\sum_{\bs} e^{-\beta\calE(\bs)}
\end{equation}
where $\calE(\bs)\sim\calN(0,nJ^2/2)$ are i.i.d. Consider the
density of states $\Omega(E)$, which is now a RV: $\Omega(E)\dd E$ is the
number of configurations $\{\bs\}$ whose randomly selected energy $\calE(\bs)$
happens to fall between $E$ and $E+\dd E$, and of course,
\begin{equation}
Z(\beta)=\int_{-\infty}^{+\infty}\dd E\Omega(E)e^{-\beta E}.
\end{equation}
How does the RV $\Omega(E)\dd E$ behave like?
First, observe that, ignoring non--exponential factors: 
\begin{equation}
\mbox{Pr}\{E\le\calE(\bs)\le E+\dd E\}\approx f(E)\dd E\exe e^{-E^2/(nJ^2)}\dd
E,
\end{equation}
and so,
\begin{equation}
\langle \Omega(E)\dd E\rangle \exe 2^n \cdot
e^{-E^2/(nJ^2)}=\exp\left\{n\left[\ln
2-\left(\frac{E}{nJ}\right)^2\right]\right\}.
\end{equation}
We have reached the pivotal point behind the analysis of the REM,
which is based on a fundamental principle that goes far beyond the
analysis of the first moment of $\Omega(E)\dd E$. In fact,
this principle is frequently used in random coding arguments in IT:

Suppose that we have $e^{nA}$ ($A > 0$, independent of $n$) independent events $\{\calE_i\}$, each one
with probability $\mbox{Pr}\{\calE_i\}=e^{-nB}$ ($B > 0$, independent of $n$).
What is the probability that at least one of the $\calE_i$'s would occur?
Intuitively, we expect that in order to see at least one or a few successes,
the number of experiments should be at least about $1/\mbox{Pr}\{\calE_i\}=e^{nB}$.
If $A > B$ then this is the case. On the other hand, for $A < B$, 
the number of trials is probably insufficient for 
seeing even one success. Indeed, a more rigorous argument gives:
\begin{eqnarray}
\mbox{Pr}\left\{\bigcup_{i=1}^{e^{nA}}\calE_i\right\}&=&
1-\mbox{Pr}\left\{\bigcap_{i=1}^{e^{nA}}\calE_i^c\right\}\nonumber\\
&=&1-\left(1-e^{-nB}\right)^{e^{nA}}\nonumber\\
&=&1-\left[e^{\ln(1-e^{-nB})}\right]^{e^{nA}}\nonumber\\
&=&1-\exp\{e^{nA}\ln(1-e^{-nB})\}\nonumber\\
&\approx& 1-\exp\{-e^{nA}e^{-nB}\}\nonumber\\
&=& 1-\exp\{-e^{n(A-B)}\}\nonumber\\
&\to&\left\{\begin{array}{ll}
1 & A > B\\
0 & A < B\end{array}\right.
\end{eqnarray}
BTW, the 2nd line could have been shown also by the union bound, as $\sum_i
\mbox{Pr}\{\calE_i\}=e^{nA}e^{-nB}\to 0$. Exercise: What happens when $A=B$?

Now, to another question: For $A > B$, how many of the $\calE_i$'s would occur in
a typical realization of this set of experiments? The number $\Omega_n$ of `successes'
is given by $\sum_{i=1}^{e^{nA}}\calI\{\calE_i\}$, namely, it is the sum of $e^{nA}$ i.i.d.\ binary
RV's whose expectation is $\bE\{\Omega_n\}=e^{n(A-B)}$. Therefore, its probability distribution
concentrates very rapidly around its mean. In fact, the events
$\{\Omega_n \ge e^{n(A-B+\epsilon)}\}$ ($\epsilon > 0$, independent
of $n$) and
$\{\Omega_n \le e^{n(A-B-\epsilon)}\}$ are large deviations events
whose probabilities decay exponentially in the number of experiments, $e^{nA}$,
i.e., {\it double--exponentially} (!) in $n$.\footnote{This will be shown
rigorously later on.}
Thus, for $A > B$, the number of successes is ``almost deterministically''
about $e^{n(A-B)}$.

\begin{figure}[ht]
\hspace*{0cm}\input{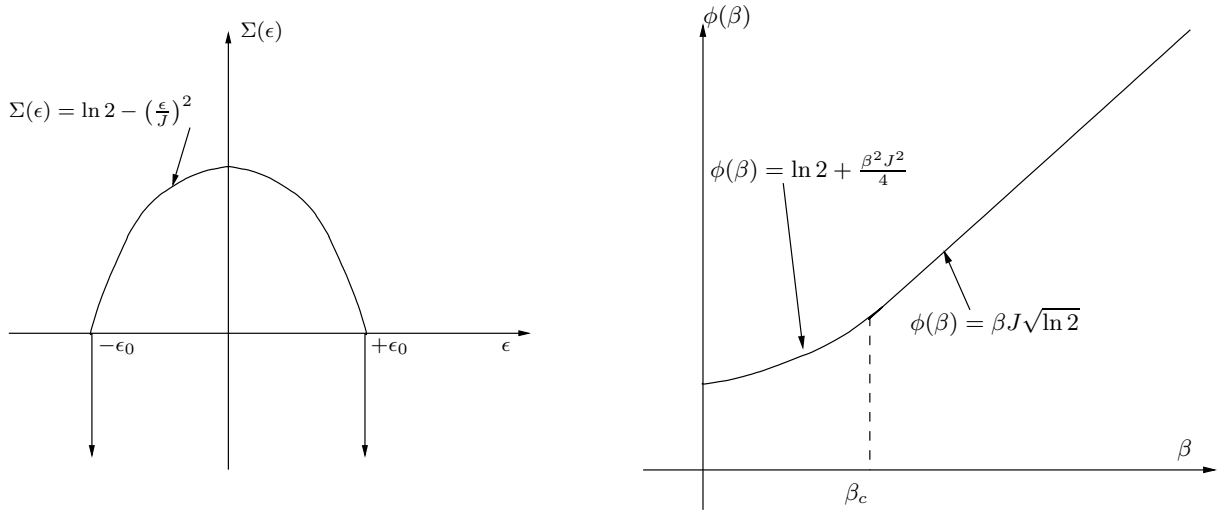}
\caption{\small The entropy function and the normalized 
log--partition function of the REM.}
\label{freentrem}
\end{figure}

Now, back to the REM:
For $E$ whose absolute value is less than
\begin{equation}
E_0\dfn nJ\sqrt{\ln 2}
\end{equation}
the exponential increase rate, $A=\ln 2$, of the number $2^n=e^{n\ln 2}$ 
of configurations, $=$ the number
of independent trials 
in randomly drawing energies $\{\calE(\bs)\}$, is faster
than the exponential decay rate of the probability,
$e^{-n[E/(nJ)]^2)}=e^{-n(\epsilon/J)^2}$ (i.e., $B=(\epsilon/J)^2$)
that $\calE(\bs)$ would happen to fall around $E$. In other words, the number
of these trials is way larger than one over this probability and
in view of the earlier discussion, the probability that
\begin{equation}
\Omega(E)\dd E=\sum_{\bs}\calI\{E\le\calE(\bs)\le E+\dd E\}.
\end{equation}
would deviate from its mean $\exe 
\exp\{n[\ln 2-(E/(nJ))^2]\}$, by a multiplicative factor that falls out of the
interval $[e^{- n\epsilon},e^{+ n\epsilon}]$, decays
double--exponentially with $n$. 
In other words, we argue that for $-E_0 < E < +E_0$,
the event
\begin{equation}
e^{-n\epsilon}\cdot 
\exp\left\{n\left[\ln
2-\left(\frac{E}{nJ}\right)^2\right]\right\}\le \Omega(E)\dd E\le
e^{+n\epsilon}\cdot 
\exp\left\{n\left[\ln
2-\left(\frac{E}{nJ}\right)^2\right]\right\}
\end{equation}
happens with probability that tends to unity in a double--exponential rate.
As discussed, $-E_0 < E < +E_0$ is exactly the condition for the expression in
the square brackets at the exponent $[\ln 2 -(\frac{E}{nJ})^2]$ to be
positive, thus $\Omega(E)\dd E$ is exponentially large.
On the other hand, if $|E|> E_0$, the number of trials $2^n$ is way smaller
than one over the probability of falling around $E$, and so, most of the
chances are that we will see no configurations at all with energy about $E$.
In other words, for these large values of $|E|$, $\Omega(E)=0$ for typical
realizations of the REM. It follows then that for such a typical realization,
\begin{eqnarray}
Z(\beta)&\approx&\int_{-E_0}^{+E_0}\left<\dd E\cdot\Omega(E)\right>e^{-\beta
E}\nonumber\\
&\exe&\int_{-E_0}^{+E_0}\dd E\cdot\exp\left\{n\left[\ln
2-\left(\frac{E}{nJ}\right)^2\right]\right\}\cdot e^{-\beta E}\nonumber\\
&=&\int_{-E_0}^{+E_0}\dd E\cdot\exp\left\{n\left[\ln
2-\left(\frac{E}{nJ}\right)^2-\beta\cdot\left(\frac{E}{n}\right)\right]\right\}\nonumber\\
&=&n\cdot\int_{-\epsilon_0}^{+\epsilon_0}\dd\epsilon\cdot\exp\left\{n\left[\ln
2-\left(\frac{\epsilon}{J}\right)^2-\beta\epsilon\right]\right\}~~~
\epsilon\dfn\frac{E}{n},~
\epsilon_0\dfn\frac{E_0}{n}=J\sqrt{\ln 2},~\nonumber\\
&\exe&\exp\left\{n\cdot\max_{|\epsilon|\le\epsilon_0}\left[\ln
2-\left(\frac{\epsilon}{J}\right)^2-\beta\epsilon\right]\right\}~~~\mbox{by
Laplace integration} \nonumber
\end{eqnarray}
The maximization problem at the exponent is very simple: it is that of a
quadratic function across an interval. The solution is of either one of two
types, depending on whether the maximum is attained at a zero--derivative
internal point in $(-\epsilon_0,+\epsilon_0)$ or at an edgepoint. The choice
between the two depends on $\beta$. Specifically, we obtain the following:
\begin{equation}
\phi(\beta)=\lim_{n\to\infty}\frac{\ln Z(\beta)}{n}=\left\{\begin{array}{ll}
\ln 2+\frac{\beta^2J^2}{4} & \beta\le \beta_c\\
\beta J\sqrt{\ln 2} & \beta> \beta_c
\end{array}\right.
\end{equation}
where $\beta_c=\frac{2}{J}\sqrt{\ln 2}$. What we see here is a phase
transition. The function $\phi(\beta)$ changes its behavior abruptly at
$\beta=\beta_c$, from being quadratic in $\beta$ to being linear in $\beta$
(see also Fig.\ \ref{freentrem}, right part).
The function $\phi$ is continuous (as always), and so is its first derivative,
but the second derivative is not. Thus, it is a second order phase transition.
Note that in the quadratic range, this expression is precisely the same as we
got using the replica method, when we hypothesized that the dominant
configuration is fully symmetric and is given by $Q=I_{m\times m}$. Thus, the replica
symmetric solution indeed gives the correct result in the high temperature
regime, but the low temperature regime seems to require symmetry breaking.
Thus, the condition $R >\ln 2-h_2(\delta)$ is equivalent to

What is the significance of each one of these phases?
Let's begin with the second line of the above expression of $\phi(\beta)$,
which is $\phi(\beta)=\beta J\sqrt{\ln 2}\equiv \beta\epsilon_0$ for $\beta >
\beta_c$. What is the meaning of linear dependency of $\phi$ in $\beta$?
Recall that the entropy $\Sigma$ is given by
$$\Sigma(\beta)=\phi(\beta)-\beta\cdot\phi'(\beta),$$ 
which in the case where $\phi$ is linear, simply vanishes. 
Zero entropy means
that the partition function is dominated by a subexponential number of ground--state
configurations (with per--particle energy about $\epsilon_0$), 
just like when it is frozen
(see also Fig.\
\ref{freentrem}, left part: $\Sigma(-\epsilon_0)=0$).
This is why we will refer to this
phase as the {\it frozen phase} or the {\it glassy phase}.\footnote{In this
phase, the system
behaves like a glass: on the one hand, it is frozen (so it consolidates), but
on the other hand, it remains disordered and amorphous, like a liquid.}
In the high--temperature range, on the other hand, the entropy is strictly
positive and the dominant per--particle energy level is
$\epsilon^*=-\frac{1}{2}\beta J^2$, which is the point of zero--derivative of
the function $[\ln 2-(\epsilon/J)^2-\beta\epsilon]$. Here the partition
is dominated by exponentially many (exercise: what is the exponent?)
configurations whose energy is $E^*=n\epsilon^*=-\frac{n}{2}\beta J^2$.
As we shall see later on, in this range the behavior of the system is
essentially paramagnetic (like in a system of i.i.d.\ spins), and so it is
called the {\it paramagnetic phase}. 

We therefore observe that the type of
phase transition here is different than in the Curie--Weiss model. We are not
talking here about spontaneous magnetization transition, but rather on a glass
transition. In fact, we will not see here a spontaneous magnetization even
if we add a magnetic field (time permits, this will be seen later on).

From $\phi(\beta)$, one can go ahead and calculate other physical quantities,
but we will not do this now. As a final note in this context, I wish to
emphasize that since the calculation of $Z$ was carried out for the typical
realizations of the quenched RV's $\{\calE(\bs)\}$, we have actually
calculated the quenched average of $\lim_n(\ln Z)/n$. As for the annealed
average, we have 
\begin{eqnarray}
\lim_{n\to\infty}\frac{\ln\langle Z(\beta)\rangle}{n}&=&
\lim_{n\to\infty}\frac{1}{n}\ln\left[\int_{\reals}
\langle\Omega(E)\dd \epsilon\rangle e^{-\beta n\epsilon}\right]\nonumber\\
&=&\lim_{n\to\infty}\frac{1}{n}\ln\left[\int_{\reals}
\exp\left\{n\left[\ln 2-\left(\frac{\epsilon}{J}\right)^2-\beta
\epsilon\right]\right\}\right]\nonumber\\
&=&\max_{\epsilon\in\reals}\left[\ln 2-\left(\frac{\epsilon}{J}\right)^2-\beta
\epsilon\right]~~~~\mbox{Laplace integration}\nonumber\\
&=&\ln 2+\frac{\beta^2J^2}{4},
\end{eqnarray}
which is the paramagnetic expression, without any phase transition since the
maximization over $\epsilon$ is not constrained.

\subsection{The Random Code Ensemble and its Relation to the REM}

Let us now see how does the REM relate to random code ensembles. The
discussion in this part is based on M\'ezard and Montanari's book, as well
as on the paper: N.~Merhav, ``Relations between random coding exponents and
the statistical physics of random codes,''
{\it IEEE Trans.\ Inform.\ Theory}, vol.\ 55, no.\ 1, pp.\ 83--92, January
2009. Another relevant paper is: A.~Barg and G.~D.~Forney, Jr., ``Random
codes: minimum
distances and error exponents,''
{\it IEEE Trans.\ Inform.\ Theory},
vol.\ 48, no.\ 9, pp.\ 2568--2573, September 2002.

Consider a DMC, $P(\by|\bx)=\prod_{i=1}^n p(y_i|x_i)$, fed by an input
$n$--vector that belongs to
a codebook $\calC=\{\bx_1,\bx_2,\ldots,\bx_M\}$, $M=e^{nR}$,
with uniform priors, where $R$ is the coding rate in nats per channel use.
The induced posterior, for $\bx\in\calC$,
is then:
\begin{eqnarray}
P(\bx|\by)&=&\frac{P(\by|\bx)}{\sum_{\bx'\in\calC}P(\by|\bx')}\nonumber\\
&=&\frac{e^{-\ln[1/P(\by|\bx)]}}
{\sum_{\bx'\in\calC}e^{-\ln[1/P(\by|\bx')]}}.
\end{eqnarray}
Here, the second line is deliberately written in a form that resembles the
Boltzmann distribution, which
naturally suggests to
consider, more generally, the posterior distribution parametrized
by $\beta$, that is
\begin{eqnarray}
\label{pbeta}
P_\beta(\bx|\by)&=&\frac{P^\beta(\by|\bx)}{\sum_{\bx'\in\calC}
P^\beta(\by|\bx')}\nonumber\\
&=&\frac{e^{-\beta\ln[1/P(\by|\bx)]}}{\sum_{\bx'\in\calC}
e^{-\beta\ln[1/P(\by|\bx')]}}\nonumber\\
&\dfn&\frac{e^{-\beta\ln[1/P(\by|\bx)]}}{Z(\beta|\by)}\nonumber
\end{eqnarray}
There are a few motivations for introducing the temperature parameter:
\begin{itemize}
\item It allows a degree of freedom
in case there is some uncertainty regarding the channel
noise level (small $\beta$ corresponds to high noise level). 
\item It is
inspired by the ideas behind simulated annealing techniques: by
sampling from $P_\beta$ while gradually increasing $\beta$
(cooling the system), the minima of the energy function
(ground states) can be found.
\item By applying symbolwise
maximum a-posteriori (MAP) decoding, i.e., decoding the $\ell$--th
symbol of $\bx$ as
$\mbox{arg}\max_a P_\beta(x_\ell=a|\by)$, where
\begin{equation}
P_\beta(x_\ell=a|\by)=
\sum_{\bx\in\calC:~x_\ell=a}P_\beta(\bx|\by),
\end{equation}
we obtain
a family of
{\it finite--temperature decoders}
(originally proposed by Ruj\'an in 1993)
parametrized by $\beta$, where $\beta=1$ corresponds
to minimum symbol error probability (with respect to the
real underlying channel $P(\by|\bx)$)
and $\beta\to\infty$ corresponds to minimum block error probability.
\item
This is one of our main motivations:
the corresponding partition function, $Z(\beta|\by)$, namely,
the sum of (conditional) probabilities raised to some power $\beta$, is
an expression frequently encountered in
R\'enyi information measures as well as
in the analysis of random
coding exponents using Gallager's techniques.
Since the partition function plays a key role in statistical mechanics,
as many physical quantities can be derived from it,
then it is natural to ask if it can also be used to
gain some insights regarding the behavior of random codes at various
temperatures and coding rates.
\end{itemize}

For the sake of simplicity, let us suppose further now that we are dealing
with the binary symmetric channel (BSC) with crossover probability $p$,
and so,
\begin{equation}
P(\by|\bx)=p^{d(\bx,\by)}(1-p)^{n-d(\bx,\by)}=(1-p)^ne^{-Jd(\bx,\by)},
\end{equation}
where $J=\ln\frac{1-p}{p}$ and $d(\bx,\by)$ is the Hamming distance.
Thus, the partition function can be presented as follows:
\begin{equation}
Z(\beta|\by)=(1-p)^{\beta n}\sum_{\bx\in\calC} e^{-\beta
Jd(\bx,\by)}.
\end{equation}

Now consider the fact that the codebook $\calC$ is selected at random:
Every codeword is randomly chosen independently of all other codewords.
At this point, the analogy to the REM, and hence also its relevance, become
apparent: If each codeword is selected independently, then the `energies'
$\{Jd(\bx,\by)\}$ pertaining to the partition function
\begin{equation}
Z(\beta|\by)=(1-p)^{\beta n}\sum_{\bx\in\calC} e^{-\beta
Jd(\bx,\by)},
\end{equation}
(or, in the case of a more general channel, the energies
$\{-\ln[1/P(\by|\bx)]\}$ pertaining to the partition function
$Z(\beta|\by)=\sum_{\bx\in\calC}e^{-\beta\ln[1/P(\by|\bx)]}$),
are i.i.d.\ random variables for all codewords in
$\calC$, with the exception of the codeword $\bx_0$ that was actually
transmitted and generated $\by$.\footnote{This one is still independent, but
it has a different distribution, and hence will be handled separately.}
Since we have seen phase transitions in the REM, it is conceivable to expect
them also in the statistical physics of the random code ensemble, and indeed
we will see them shortly.

Further, we assume that each symbol of each codeword is drawn by fair coin
tossing, i.e., independently and with equal probabilities for `0' and `1'.
As said, we have to distinguish now between the contribution of the correct
codeword $\bx_0$, which is
\begin{equation}
Z_c(\beta|\by)\dfn (1-p)^{\beta n}e^{-Jd(\bx_0,\by)}
\end{equation}
and the contribution of all other (incorrect) codewords:
\begin{equation}
Z_e(\beta|\by)\dfn(1-p)^{\beta
n}\sum_{\bx\in\calC\setminus\{\bx_0\}}e^{-Jd(\bx,\by)}.
\end{equation}
Concerning the former, things are very simple: Typically, the channel
flips about $np$ bits out the $n$ transmissions, which means that
with high probability, $d(\bx_0,\by)$ is about $np$, and so
$Z_c(\beta|\by)$ is expected to take values around $(1-p)^{\beta n}e^{-\beta Jnp}$.
The more complicated and more interesting question is how does
$Z_e(\beta|\by)$ behave, and here the treatment will be very similar to 
that of the REM.

\begin{figure}[ht]
\hspace*{5cm}\input{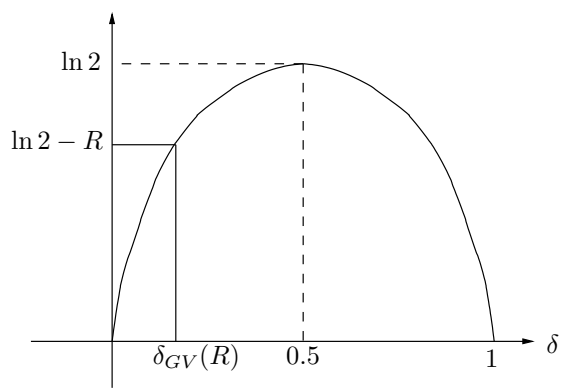}
\caption{\small The Gilbert--Varshamov distance as the smaller solution to the
equation $R+h_2(\delta)-\ln 2=0$.}
\label{binentropy}
\end{figure}

Given $\by$, define $\Omega_{\by}(d)$ as the number of incorrect codewords
whose Hamming distance from $\by$ is exactly $d$. Thus,
\begin{equation}
Z_e(\beta|\by)=(1-p)^{\beta n} \sum_{d=0}^n\Omega_{\by}(d)\cdot e^{-\beta J
d}.
\end{equation}
Just like in the REM, here too the enumerator $\Omega_{\by}(d)$ is the sum of
an exponential number, $e^{nR}$, of binary i.i.d.\ RV's:
\begin{equation}
\Omega_{\by}(d)=\sum_{\bx\in\calC\setminus\{\bx_0\}}\calI\{d(\bx,\by)=d\}.
\end{equation}
According to the method of types, the probability of a single `success'
$\{d(\bX,\by)=n\delta\}$ is given by
\begin{equation}
\mbox{Pr}\{d(\bX,\by)=n\delta\}\exe\frac{e^{nh_2(\delta)}}{2^n}=
\exp\{-n[\ln 2-h_2(\delta)]\}.
\end{equation}
So, just like in the REM, we have an exponential number of trials, $e^{nR}$,
each one with an exponentially decaying probability of success, $e^{-n[\ln
2-h_2(\delta)]}$. We already
know how does this experiment behave: It depends which exponent is faster. If
$R >\ln 2-h_2(\delta)$, we will typically see about $\exp\{n[R+h_2(\delta)-\ln 2]\}$
codewords at distance $d=n\delta$ from $\by$. Otherwise, we see none.
So the critical value of $\delta$ is the solution to the equation
\begin{equation}
R+h_2(\delta)-\ln 2=0.
\end{equation}
There are two solutions to this equation, which are symmetric about $1/2$.
The smaller one is called the Gilbert--Varshamov (G--V) distance\footnote{The
G--V distance was originally defined and used in coding theory for the BSC.} and it
will be denoted by $\delta_{GV}(R)$ (see Fig.\ \ref{binentropy}).
The other solution is, of course, $\delta=1-\delta_{GV}(R)$. 
Thus, the condition $R >\ln 2-h_2(\delta)$ is equivalent to
$\delta_{GV}(R)<\delta<1-\delta_{GV}(R)$, and so, for a typical
code in the ensemble:
\begin{eqnarray}
Z_e(\beta|\by)&\approx&(1-p)^{\beta
n}\sum_{\delta=\delta_{GV}(R)}^{1-\delta_{GV}(R)}
\exp\{n[R+h_2(\delta)-\ln 2]\}\cdot e^{-\beta Jn\delta}\nonumber\\
&=&(1-p)^{\beta
n}e^{n(R-\ln 2)}\cdot\sum_{\delta=\delta_{GV}(R)}^{1-\delta_{GV}(R)}
\exp\{n[h_2(\delta)-\beta J\delta]\}\nonumber\\
&=&(1-p)^{\beta
n}e^{n(R-\ln 2)}\cdot
\exp\left\{n\cdot\max_{\delta_{GV}(R)\le\delta\le1-\delta_{GV}(R)}
[h_2(\delta)-\beta J\delta]\right\}\nonumber
\end{eqnarray}
Now, similarly as in the REM, we have to maximize a certain function
within a limited interval. And again, there are two phases, corresponding
to whether the maximizer falls at an edgepoint (glassy phase) or at an internal point with
zero derivative (paramagnetic phase). It is easy to show (exercise: fill in
the details) that in the paramagnetic phase, the maximum is attained at 
\begin{equation}
\delta^* =p_\beta\dfn\frac{p^\beta}{p^\beta+(1-p)^\beta}
\end{equation}
and then
\begin{equation}
\phi(\beta)=R-\ln 2+\ln[p^\beta+(1-p)^\beta].
\end{equation}
In the glassy phase, $\delta^*=\delta_{GV}(R)$ and then
\begin{equation}
\phi(\beta)=\beta[\delta_{GV}(R)\ln p+(1-\delta_{GV}(R))\ln(1-p)],
\end{equation}
which is again, linear in $\beta$ and hence corresponds to zero entropy.
The boundary between the two phases occurs when $\beta$ is such that
$\delta_{GV}(R)=p_\beta$, which is equivalent to
\begin{equation}
\beta=\beta_c(R)\dfn\frac{\ln[(1-\delta_{GV}(R))/\delta_{GV}(R)]}{\ln[(1-p)/p]}.
\end{equation}

So $\beta < \beta_c(R)$ is the paramagnetic phase of $Z_e$ and
$\beta > \beta_c(R)$ is its glassy phase.

\begin{figure}[ht]
\hspace*{4cm}\input{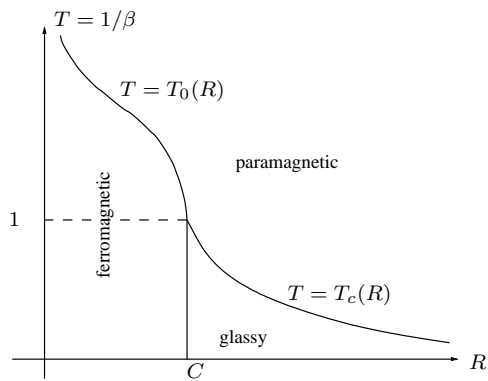}
\caption{Phase diagram of the finite--temperature MAP decoder.}
\label{phasediag}
\end{figure}

But now we should remember that $Z_e$ is only part of the partition function
and it is time to put the contribution of $Z_c$ back into the picture.
Checking the dominant contribution of $Z=Z_e+Z_c$ as a function of $\beta$ and $R$, we can
draw a phase diagram, where we find that there are actually three phases, two
contributed by $Z_e$, as we have already seen (paramagnetic and glassy), plus
a third phase -- contributed by $Z_c$, namely, the {\it ordered} or the {\it
ferromagnetic} phase, where $Z_c$ dominates (cf.\ Fig.\ \ref{phasediag}),
which means reliable communication, as the correct codeword $\bx_0$ dominates
the partition function and hence the posterior distribution.
The boundaries of the ferromagnetic phase
designate phase transitions from reliable to unreliable decoding.

Both the glassy phase and the paramagnetic phase correspond to unreliable
communication. What is the essential difference between them? As in the REM,
the difference is that in the glassy phase, $Z$ is dominated by a
subexponential number of codewords at the `ground--state energy', namely, that
minimum seen distance of $n\delta_{GV}(R)$, whereas in the paramagnetic phase,
the dominant contribution comes from an exponential number of codewords at
distance $np_\beta$. In the glassy phase, there is {\it seemingly} a smaller
degree of uncertainty since $H(\bX|\bY)$ that is induced from the
finite--temperature posterior has zero entropy. But this is fictitious since
the main support of the posterior belongs to incorrect codewords. This is to
say that we may have the illusion that we know quite a lot about the
transmitted codeword, but what we know is wrong! This is like an event of 
an undetected error. In both glassy and paramagnetic phases, above capacity, the 
ranking of the correct codword, in the list of
decreasing $P_\beta(\bx|\by)$, is about $e^{n(R-C)}$. 

\noindent
{\bf Exercise}: convince yourself that the phase diagram is as depicted in
Fig.\ \ref{phasediag} and find the equations of the boundaries between phases.
Note that the triple point is $(C,1)$ where $C=\ln 2-h_2(p)$ is the  channel capacity.
Also, the ferro--glassy boundary is the vertical straight line $R=C$. What does
this mean? $\Box$

\subsection{Random Coding Exponents}

It turns out that these findings are relevant to ensemble performance analysis of
codes. This is because many of the bounds on code performance include
summations of $P^\beta(\by|\bx)$ (for some $\beta$), which are exactly
the partition functions that we work with in the foregoing discussion.
These considerations can sometimes even help to get tighter bounds.
We will now demonstrate this point in the context of the analysis of
the probability of correct decoding above capacity. 

First, we have
\begin{eqnarray}
P_c&=&\frac{1}{M}\sum_{\by}\max_{\bx\in\calC} P(\by|\bx)~~~~M\dfn e^{nR}\nonumber\\
&=&\lim_{\beta\to\infty}\frac{1}{M}\sum_{\by}\left[\sum_{\bx\in\calC}
P^\beta(\by|\bx)\right]^{1/\beta}\nonumber
\end{eqnarray}
The expression in the square brackets is readily identified with the partition
function, and we
note that the combination of $R > C$ and $\beta\to\infty$ takes us deep into
the glassy phase. Taking the ensemble average, we get:
\begin{equation}
\bar{P}_c=\lim_{\beta\to\infty}\frac{1}{M}\sum_{\by}\bE\left\{\left[\sum_{\bx\in\calC}
P^\beta(\by|\bx)\right]^{1/\beta}\right\}.
\end{equation}
At this point, the traditional approach would be to insert the expectation into
the square brackets by applying Jensen's inequality (for $\beta > 1$), 
which would give us an upper
bound. Instead, our previous treatment of random code ensembles as a REM--like 
model can give us a hand on
exponentially tight evaluation of the last expression, with Jensen's
inequality being avoided. Consider the following chain:
\begin{eqnarray}
\bE\left\{\left[\sum_{\bx\in\calC}P^\beta(\by|\bx)\right]^{1/\beta}\right\}&=&
(1-p)^n\bE\left\{\left[\sum_{d=0}^n\Omega_{\by}(d)e^{-\beta
Jd}\right]^{1/\beta}\right\}\nonumber\\
&\exe&(1-p)^n\bE\left\{\left[\max_{0\le d\le n}\Omega_{\by}(d)e^{-\beta
Jd}\right]^{1/\beta}\right\}\nonumber\\
&=&(1-p)^n\bE\left\{\max_{0\le d\le n}[\Omega_{\by}(d)]^{1/\beta}\cdot e^{-Jd}
\right\}\nonumber\\
&\exe&(1-p)^n\bE\left\{\sum_{d=0}^n[\Omega_{\by}(d)]^{1/\beta}\cdot e^{-Jd}
\right\}\nonumber\\
&=&(1-p)^n\sum_{d=0}^n\bE\left\{[\Omega_{\by}(d)]^{1/\beta}\right\}\cdot e^{-Jd}
\nonumber\\
&\exe&(1-p)^n\max_{0\le d\le n}\bE\left\{[\Omega_{\by}(d)]^{1/\beta}\right\}\cdot e^{-Jd}
\nonumber
\end{eqnarray}
Thus, it boils down to the calculation of (non--integer) moments of
$\Omega_{\by}(d)$. At this point, we adopt the main ideas of the
treatment of the REM, distinguishing between the values of $\delta$
below the G--V distance, and those that are above it. 
Before we actually assess the moments of $\Omega_{\by}(d)$, we take a closer
look at the asymptotic behavior of these RV's. This will also rigorize our
earlier discussion on the Gaussian REM. 

For two numbers $a$ and $b$ in $[0,1]$, 
let us define the binary divergence as
\begin{equation}
D(a\|b)=a\ln\frac{a}{b}+(1-a)\ln\frac{1-a}{1-b}.
\end{equation}
Using the inequality
$$\ln(1+x)=-\ln\left(1-\frac{x}{1+x}\right)\ge \frac{x}{1+x},$$
we get the following lower bound to $D(a\|b)$:
\begin{eqnarray}
D(a\|b)&=&a\ln\frac{a}{b}+(1-a)\ln\frac{1-a}{1-b}\nonumber\\
&=&a\ln\frac{a}{b}+(1-a)\ln\left(1+\frac{b-a}{1-b}\right)\nonumber\\
&\ge&a\ln\frac{a}{b}+(1-a)\cdot\frac{(b-a)/(1-b)}{1+(b-a)/(1-b)}\nonumber\\
&=&a\ln\frac{a}{b}+b-a\nonumber\\
&>&a\left(\ln\frac{a}{b}-1\right)\nonumber
\end{eqnarray}
Now, as mentioned earlier, $\Omega_{\by}(d)$ is the sum of $e^{nR}$ i.i.d.\
binary RV's, i.e., Bernoulli RV's with parameter $e^{-n[\ln 2-h_2(\delta)]}$.
Consider the event $\Omega_{\by}(d) \ge e^{nA}$, $A\ge 0$, which means that
the relative frequency of `successes' exceeds
$\frac{e^{nA}}{e^{nR}}=e^{-n(R-A)}$. Then this is a large deviations event if
$e^{-n(R-A)} > e^{-n[\ln 2-h_2(\delta)]}$, that is,
\begin{equation}
A > R+h_2(\delta)-\ln 2.
\end{equation}
Using the {\bf Chernoff bound} (exercise: fill in the details), one can easily show
that
\begin{equation}
\mbox{Pr}\{\Omega_{\by}(d) \ge e^{nA}\}\le
\exp\{-e^{nR}D(e^{-n(R-A)}\|e^{-n[\ln 2-h_2(\delta)]})\}.
\end{equation}
Note: we have emphasized the use of the Chernoff bound as opposed to the method of
types since the method of types would introduce the factor of the number of
type classes, which is in this case $(e^{nR}+1)$. 
Now, by applying the above lower bound to the binary divergence, we can
further upper bound the last expression as
\begin{eqnarray}
\mbox{Pr}\{\Omega_{\by}(d) \ge e^{nA}\}&\le&
\exp\{-e^{nR}\cdot e^{-n(R-A)}\cdot(n[\ln 2-R-h_2(\delta)+A]-1)\}\nonumber\\
&=&\exp\{-e^{nA}\cdot(n[\ln 2-R-h_2(\delta)+A]-1)\}\nonumber
\end{eqnarray}
Now, suppose first that $\delta_{GV}(R)<\delta<1-\delta_{GV}(R)$, and take
$A=R+h_2(\delta)-\ln 2+\epsilon$, where $\epsilon > 0$ may not necessarily be
small. In this case, the term in the square brackets is $\epsilon$, 
which means that the right--most side decays doubly--exponentially rapidly. 
Thus, for $\delta_{GV}(R)<\delta<1-\delta_{GV}(R)$, 
the probability that $\Omega_{\by}(d)$ exceeds $\bE\{\Omega_{\by}(d)\}\cdot
e^{n\epsilon}$ decays double--exponentially fast with $n$.
One can show in a similar manner (exercise: please do)\footnote
{This requires a slighly different lower bound to the binary divergence.} that
$\mbox{Pr}\{\Omega_{\by}(d)< \bE\{\Omega_{\by}(d)\}\cdot e^{-n\epsilon}\}$
decays in a double exponential rate as well. Finally, consider the case
where $\delta<\delta_{GV}(R)$ or $\delta>1-\delta_{GV}(R)$, and let $A=0$.
This is also a large deviations event, and hence the above bound continues to
be valid. Here, by setting $A=0$, we get an ordinary exponential decay:
\begin{equation}
\mbox{Pr}\{\Omega_{\by}(d)\ge 1\}\lexe e^{-n[\ln 2-R-h_2(\delta)]}.
\end{equation}

Now, after having prepared these results, let's get back to the evaluation of
the moments of $\Omega_{\by}(d)$. Once again, we separate between the two
ranges of $\delta$. For $\delta<\delta_{GV}(R)$ or $\delta>1-\delta_{GV}(R)$,
we have the following:
\begin{eqnarray}
\bE\{[\Omega_{\by}(d)]^{1/\beta}\}&\exe&
0^{1/\beta}\cdot\mbox{Pr}\{\Omega_{\by}(d)=0\}+e^{n\cdot 0/\beta}\cdot
\mbox{Pr}\{1\le \Omega_{\by}(d)\le e^{n\epsilon}\}+\mbox{double--exp.\
terms}\nonumber\\
&\exe&e^{n\cdot 0/\beta}\cdot\mbox{Pr}\{\Omega_{\by}(d)\ge 1\}\nonumber\\
&\exe&e^{-n[\ln 2-R-h_2(\delta)]}\nonumber
\end{eqnarray}
Thus, in this range, $\bE\{[\Omega_{\by}(d)]^{1/\beta}\}\exe e^{-n[\ln
2-R-h_2(\delta)]}$ independently of $\beta$.
On the other hand in the range $\delta_{GV}(R)<\delta<1-\delta_{GV}(R)$,
\begin{eqnarray}
\bE\{[\Omega_{\by}(d)]^{1/\beta}\}&\exe&
(e^{n[R+h_2(\delta)-\ln 2]})^{1/\beta}\cdot\mbox{Pr}\{
e^{n[R+h_2(\delta)-\ln 2-\epsilon]}\le \Omega_{\by}(d)\le 
e^{n[R+h_2(\delta)-\ln 2+\epsilon]}\}+\nonumber\\
& &+\mbox{double--exp.\ terms}\nonumber\\
&\exe&
e^{n[R+h_2(\delta)-\ln 2]/\beta}\nonumber
\end{eqnarray}
since the probability $\mbox{Pr}\{
e^{n[R+h_2(\delta)-\ln 2-\epsilon]}\le \Omega_{\by}(d)\le 
e^{n[R+h_2(\delta)-\ln 2+\epsilon]}\}$ tends to unity double--exponentially
rapidly. So to summarize, we have shown that the moment of $\Omega_{\by}(d)$
undergoes a phase transition, as it behaves as follows:
\begin{equation}
\bE\{[\Omega_{\by}(d)]^{1/\beta}\}\exe\left\{\begin{array}{cc}
e^{n[R+h_2(\delta)-\ln 2]} & \delta<\delta_{GV}(R)~~\mbox{or}~~
\delta>1-\delta_{GV}(R)\\
e^{n[R+h_2(\delta)-\ln 2]/\beta} & \delta_{GV}(R)<\delta<1-\delta_{GV}(R)
\end{array}\right.
\end{equation}

Finally, by plugging these moments back into the expression of $\bar{P}_c$
(exercise: fill in the details), and taking 
the limit $\beta\to\infty$, we eventually get:
\begin{equation}
\lim_{\beta\to\infty}\bE\left\{\left[\sum_{\bx\in\calC}P^\beta(\by|\bx)\right]^{1/\beta}\right\}
\exe e^{-nF_g}
\end{equation}
where $F_g$ is the free energy of the glassy phase, i.e.,
\begin{equation}
F_g=\delta_{GV}(R)\ln\frac{1}{p}+(1-\delta_{GV}(R))\ln\frac{1}{1-p}
\end{equation}
and so, we obtain a very simple relation between the exponent of $\bar{P}_c$
and the free energy of the glassy phase:
\begin{eqnarray}
\bar{P}_c&\exe&\frac{1}{M}\sum_{\by} e^{-nF_g}\nonumber\\
&=&\exp\{n(\ln 2-R-F_g)\}\nonumber\\
&=&\exp\{n[\ln
2-R+\delta_{GV}(R)\ln p+(1-\delta_{GV}(R))\ln(1-p)]\}
\nonumber\\
&=&\exp\{n[h_2(\delta_{GV}(R))+
\delta_{GV}(R)\ln p+(1-\delta_{GV}(R))\ln(1-p)]\}
\nonumber\\
&=& e^{-nD(\delta_{GV}(R)\|p)}\nonumber
\end{eqnarray}
The last expression has an intuitive interpretation. It answers the following
question: what is the probability that the channel would flip less than
$n\delta_{GV}(R)$ bits although $p>\delta_{GV}(R)$? This is exactly the
relevant question for correct decoding in the glassy phase, because in that
phase, there is a ``belt'' of codewords ``surrounding'' $\by$ at radius
$n\delta_{GV}(R)$ -- these are the codewords that dominate the partition
function in the glassy phase and there are no codewords closer to $\by$. The event of correct
decoding happens if the channel flips less than $n\delta_{GV}(R)$ bits and
then $\bx_0$ is closer to $\by$ more than all belt--codewords. Thus, $\bx_0$ is
decoded correctly.

One can also derive an upper bound on the error
probability at $R< C$. The partition
function $Z(\beta|\by)$ plays a role there too according to Gallager's
classical bounds. We will not delve now into it, but we only comment that
in that case, the calculation is performed in the paramagnetic regime rather
than the glassy regime that we have seen in the calculation of $\bar{P}_c$.
The basic technique, however, is essentially the same.

We will now demonstrate the usefulness of this technique of 
assessing moments of distance enumerators
in a certain problem of decoding with an erasure option.
Consider the BSC with a crossover probability $p< 1/2$,
which is unknown and one employs a universal detector that operates
according
to the following decision rule: Select the message $m$ if
\begin{equation}
\label{univdcn}
\frac{e^{-n\beta\hat{h}(\bx_m\oplus\by)}}
{\sum_{m'\ne m}e^{-n\beta\hat{h}(\bx_{m'}\oplus\by)}} \ge e^{nT}
\end{equation}
where $\beta > 0$ is an inverse temperature parameter and
$\hat{h}(\bx\oplus\by)$ is the binary entropy pertaining to the relative
number of
1's in the vector resulting from
bit--by--bit XOR of $\bx$ and $\by$,
namely, the binary entropy function computed
at the normalized Hamming distance between $\bx$ and $\by$. If no message $m$
satisfies
(\ref{univdcn}), then an erasure is declared.

We have no optimality claims regarding this decision rule, but arguably, it is
a reasonable
decision rule (and hence there is motivation to analyze its performance): 
It is a universal version of the optimum decision rule: 
\begin{equation}
\mbox{Decide on $m$ if}~~\frac{P(\by|\bx_m)}{\sum_{m'\ne m}P(\by|\bx_{m'})}\ge e^{nT}
~~\mbox{and erase otherwise.}
\end{equation}
The minimization
of $\hat{h}(\bx_m\oplus\by)$ among all codevectors $\{\bx_m\}$,
namely, the {\it minimum conditional entropy decoder} is a well--known
universal decoding rule
in the ordinary decoding regime, without erasures,
which in the simple case of the BSC, is equivalent
to the {\it maximum mutual information} (MMI) decoder 
and to the {\it generalized likelihood ratio test} (GLRT) decoder,
which jointly maximizes the likelihood over both the message and the unknown
parameter.
Here we adapt the minimum conditional entropy decoder to the structure
proposed by
the optimum decoder with erasures, where the (unknown) likelihood of each
codeword $\bx_m$
is basically replaced by its maximum $e^{-n\hat{h}(\bx_m\oplus\by)}$, but with
an additional
degree of freedom of scaling the exponent by $\beta$. 
The parameter $\beta$ controls the relative importance of the codeword with the
second highest score.
For example, when $\beta\to \infty$,\footnote{As $\beta$ varies it is
plausible to let $T$ scale linearly with $\beta$.}
only the first and the second highest scores count in the
decision, whereas if $\beta\to 0$, the
differences between the scores of all codewords are washed out.

To demonstrate the advantage of the proposed analysis technique, we will now
apply it in comparison to the traditional approach of using Jensen's inequality and
supplementing an additional
parameter $\rho$ in the bound so as to 
monitor the loss of tightness due to the use of Jensen's inequality. 
Let us analyze the probability
of the event $\calE_1$ that the transmitted codeword $\bx_m$ does not satisfy
(\ref{univdcn}).
We then have the following chain of inequalities,
where the first few steps are common to the two analysis
methods to be compared:
\begin{eqnarray}
\mbox{Pr}\{\calE_1\}&=&\frac{1}{M}\sum_{m=1}^M
\sum_{\by}P(\by|\bx_m)\cdot 1\left\{\frac{e^{nT}\sum_{m'\ne m}
e^{-n\beta\hat{h}(\bx_{m'}\oplus\by)}}{e^{-n\beta\hat{h}(\bx_{m}\oplus\by)}}\ge
1\right\}\nonumber\\
&\le&\frac{1}{M}\sum_{m=1}^M\sum_{\by}P(\by|\bx_m)\cdot
\left[\frac{e^{nT}\sum_{m'\ne m}
e^{-n\beta\hat{h}(\bx_{m'}\oplus\by)}}{e^{-n\beta\hat{h}(\bx_{m}\oplus\by)}}\right]^s\nonumber\\
&=&\frac{e^{nsT}}{M}\sum_{m=1}^M \sum_{\by}P(\by|\bx_m)\cdot e^{n\beta
s\hat{h}(\bx_{m}\oplus\by)}
\cdot\left[\sum_{m'\ne m}e^{-n\beta\hat{h}(\bx_{m'}\oplus\by)}\right]^s
\end{eqnarray}
Considering now the ensemble of codewords drawn indepedently by fair coin
tossing, we have:
\begin{eqnarray}
\overline{\mbox{Pr}}\{\calE_1\}&\le&
e^{nsT}\sum_{\by}
\bE\left\{P(\by|\bX_1)\cdot \exp[n\beta
s\hat{h}(\bX_{1}\oplus\by)]\right\}\cdot
\bE\left\{\left[\sum_{m >
1}\exp[-n\beta\hat{h}(\bX_m\oplus\by)]\right]^s\right\}\nonumber\\
&\dfn&e^{nsT}\sum_{\by}A(\by)\cdot B(\by)
\end{eqnarray}
The computation of $A(\by)$ is as follows:
Denoting the Hamming weight of a binary sequence $\bz$ by $w(\bz)$, we have:
\begin{eqnarray}
A(\by)&=&\sum_{\bx}2^{-n}(1-p)^n\cdot\left(\frac{p}{1-p}\right)^{w(\bx\oplus\by)}
\exp[n\beta s\hat{h}(\bx\oplus\by)]\nonumber\\
&=&\left(\frac{1-p}{2}\right)^n\sum_{\bz}\exp\left[n\left(w(\bz)\ln\frac{p}{1-p}+\beta
s\hat{h}(\bz)
\right)\right]\nonumber\\
&\exe&\left(\frac{1-p}{2}\right)^n\sum_\delta
e^{nh(\delta)}\cdot\exp\left[n\left(\beta s h(\delta)
-\delta\ln\frac{1-p}{p}\right)\right]\nonumber\\
&\exe&\left(\frac{1-p}{2}\right)^n\exp\left[n\max_\delta\left((1+\beta
s)h(\delta)
-\delta\ln\frac{1-p}{p}\right)\right].
\end{eqnarray}
It is readily seen by ordinary optimization that
\begin{equation}
\max_\delta\left[(1+\beta s)h(\delta)-\delta\ln\frac{1-p}{p}\right]
=(1+\beta s)\ln\left[p^{1/(1+\beta s)}+(1-p)^{1/(1+\beta s)}\right]-\ln(1-p)
\end{equation}
and so upon substituting back into the the bound on
$\overline{\mbox{Pr}}\{\calE_1\}$, we get:
\begin{equation}
\overline{\mbox{Pr}}\{\calE_1\}\le
\exp\left[n\left(sT+(1+\beta s)\ln\left[p^{1/(1+\beta s)}+(1-p)^{1/(1+\beta
s)}\right]-\ln 2
\right)\right]\cdot\sum_{\by}B(\by).
\end{equation}
It remains then to assess the exponential order of $B(\by)$ and this will now
be done
in two different ways. The first is Forney's way of using Jensen's inequality
and
introducing the additional parameter $\rho$, i.e.,
\begin{eqnarray}
B(\by)&=&\bE\left\{\left(\left[\sum_{m>1}\exp[-n\beta
\hat{h}(\bX_m\oplus\by)\right]^{s/\rho}\right)^\rho\right\}\nonumber\\
&\le&\bE\left\{\left(\sum_{m>1}
\exp[-n\beta s\hat{h}(\bX_m\oplus\by)/\rho]\right)^\rho\right\}~~~~~0\le s/\rho\le
1\nonumber\\
&\le&e^{n\rho R}\left(\bE\left\{\exp[-n\beta s\hat{h}(\bX_m\oplus\by)/\rho]
\right\}\right)^\rho,~~~~~\rho\le 1
\end{eqnarray}
where in the second line we have used the 
following inequality\footnote{To see why this is true, think of $p_i=a_i/(\sum_i a_i)$ as probabilities,
and then $p_i^\theta\ge p_i$, which implies $\sum_i p_i^\theta \ge \sum_i
p_i=1$. The idea behind the introduction of the new parameter $\rho$ is to
monitor the possible loss of exponential tightness due to the use of Jensen's inequality.
If $\rho=1$, there is no loss at all due to Jensen, but there is maximum loss in the
second line of the chain. If $\rho=s$, it is the other way around. Hopefully,
after optimization over $\rho$, the overall loss in tightness is minimized.}
for non--negative $\{a_i\}$ and $\theta\in[0,1]$:
\begin{equation}
\left(\sum_i a_i\right)^\theta \le \sum_ia_i^\theta.
\end{equation}
Now,
\begin{eqnarray}
\bE\left\{\exp[-n\beta s\hat{h}(\bX_m\oplus\by)/\rho]\right\}&=&
2^{-n}\sum_{\bz}\exp[-n\beta s\hat{h}(\bz)/\rho]\nonumber\\
&\exe&2^{-n}\sum_{\delta}e^{nh(\delta)}\cdot e^{-n\beta
sh(\delta)/\rho}\nonumber\\
&=&\exp[n([1-\beta s/\rho]_+-1)\ln 2],
\end{eqnarray}
where $[u]_+\dfn\max\{u,0\}$. Thus, we get
\begin{equation}
B(\by)\le \exp(n[\rho(R-\ln 2)+[\rho-\beta s]_+]),
\end{equation}
which when substituted back into the bound on
$\overline{\mbox{Pr}}\{\calE_1\}$, yields
an exponential rate of
\begin{eqnarray}
\tilde{E}_1(R,T)&=&\max_{0\le s\le\rho\le 1}
\left\{(\rho-[\rho-\beta s]_+)\ln 2-\right.\nonumber\\
& &\left.-(1+\beta s)\ln\left[p^{1/(1+\beta s)}+
(1-p)^{1/(1+\beta s)}\right]-\rho R-sT\right\}.
\end{eqnarray}
On the other hand, estimating $B(\by)$ by the new method, we have:
\begin{eqnarray}
B(\by)&=&\bE\left\{\left[\sum_{m>1}\exp[-n\beta\hat{h}(\bX_m\oplus\by)]\right]^s\right\}\nonumber\\
&=&\bE\left\{\left[\sum_{\delta}\Omega_{\by}(n\delta)\exp[-n\beta
h(\delta)]\right]^s\right\}\nonumber\\
&\exe&\sum_{\delta}\bE\{\Omega_{\by}^s(n\delta)\}\cdot\exp(-n\beta
sh(\delta))\nonumber\\
&\exe&\sum_{\delta\in\calG_R^c}e^{n[R+h(\delta)-\ln 2]}\cdot\exp[-n\beta
sh(\delta)]+
\sum_{\delta\in\calG_R}e^{ns[R+h(\delta)-\ln 2]}\cdot\exp[-n\beta
sh(\delta)]\nonumber\\
&\dfn& U+V,
\end{eqnarray}
where $\calG_R=\{\delta:~\delta_{GV}(R)\le \delta\le 1-\delta_{GV}(R)\}$.
Now, $U$ is dominated by the term $\delta=0$ if $\beta s> 1$ and
$\delta=\delta_{GV}(R)$ if $\beta s < 1$. It is then easy to see
that $U\exe\exp[-n(\ln 2-R)(1-[1-\beta s]_+)]$. Similarly,
$V$ is dominated by the term $\delta=1/2$ if $\beta < 1$ and
$\delta=\delta_{GV}(R)$ if $\beta\ge 1$. Thus, $V\exe \exp[-ns(\beta[\ln
2-R]-R[1-\beta]_+)]$.
Therefore, defining
\begin{equation}
\phi(R,\beta,s)=\min\{(\ln 2-R)(1-[1-\beta s]_+),s(\beta[\ln
2-R]-R[1-\beta]_+)\},
\end{equation}
the resulting exponent is
\begin{equation}
\hat{E}_1(R,T)=\max_{s\ge 0}\left\{\phi(R,\beta,s)
-(1+\beta s)\ln\left[p^{1/(1+\beta s)}+(1-p)^{1/(1+\beta
s)}\right]-sT\right\}.
\end{equation}

Numerical comparisons show
that while there are many quadruples $(p,\beta,R,T)$ for which the two exponents
coincide, there are also situations where $\hat{E}_1(R,T)$ exceeds
$\tilde{E}_1(R,T)$.
To demonstrate these situations,
consider the values $p=0.1$, $\beta=0.5$,
$T=0.001$, and let $R$ vary from $0$ to $0.06$ in steps of $0.01$. Table 1
summarizes numerical values of both exponents, where the optimizations over
$\rho$ and $s$
were conducted by an exhaustive search with a step size of $0.005$ in each
parameter. In the case
of $\hat{E}_1(R,T)$, where $s\ge 0$ is not limited
to the interval $[0,1]$ (since Jensen's inequality
is not used), the numerical search over $s$ was limited to the interval
$[0,5]$.\footnote{
It is interesting to note that for some values of $R$, the optimum value $s^*$
of the parameter
$s$ was indeed larger than $1$. For example, at rate $R=0$, we have
$s^*=2$ in the above search resolution.}

\begin{table}
\begin{center}
\begin{tabular}{||l|c|c|c|c|c|c|c||} \hline
$$ & $R=0.00$ & $R=0.01$ & $R=0.02$ & $R=0.03$ & $R=0.04$ & $R=0.05$ &
$R=0.06$\\ \hline
$\tilde{E}_1(R,T)$ & 0.1390 & 0.1290 & 0.1190 & 0.1090 & 0.0990 & 0.0890 &
0.0790 \\ \hline
$\hat{E}_1(R,T)$ & 0.2211 & 0.2027 & 0.1838 & 0.1642 & 0.1441 & 0.1231 &
0.1015 \\ \hline
\end{tabular}
\caption{Numerical values of $\tilde{E}_1(R,T)$ and $\hat{E}_1(R,T)$
as functions of $R$ for $p=0.1$, $\beta=0.5$, and $T=0.001$.}
\end{center}
\end{table}

\begin{figure}[h!t!b!]
\centering
\includegraphics[width=8.5cm, height=8.5cm]{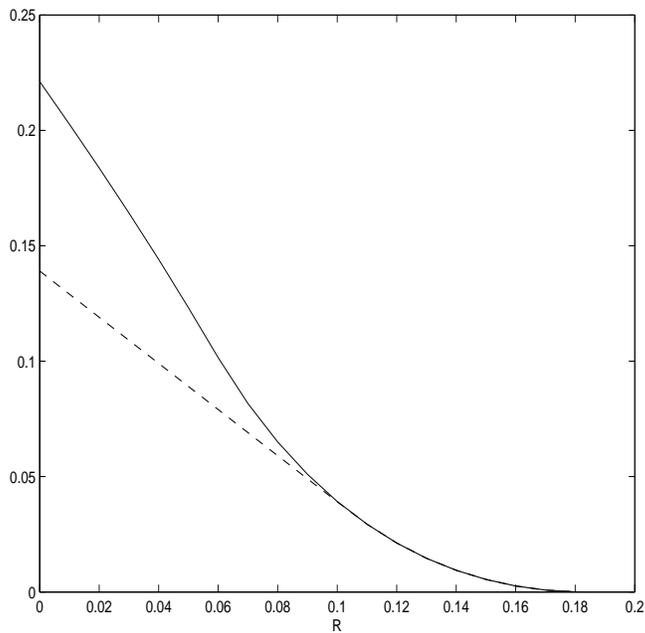}
\caption{Graphs of $\hat{E}_1(R,T)$ (solid line) and $\tilde{E}_1(R,T)$
(dashed line) 
as functions of $R$
for $p=0.1$, $T=0.001$ and $\beta=0.5$.}
\label{bounds}
\end{figure}

As can be seen (see also Fig.\ \ref{bounds}), the numerical values of the exponent
$\hat{E}_1(R,T)$
are considerably larger than those of $\tilde{E}_1(R,T)$ in this example,
which means that the analysis technique proposed here, not only
simplifies exponential error bounds, but sometimes leads also to significantly
tighter bounds.


There are other examples where these techniques are used in more involved
situations, and in some of them they yield better performance bounds
compared to traditional methods. Here is a partial list of papers:
\begin{itemize}
\item 
R.~Etkin, N.~Merhav and E.~Ordentlich, ``Error exponents of optimum
decoding for the
interference channel,''
{\it IEEE Trans.\ Inform.\ Theory}, vol.\ 56, no.\ 1, pp.\ 40--56, January
2010.
\item Y.~Kaspi and N.~Merhav, ``Error exponents of optimum decoding
for the degraded broadcast channel using moments of type class enumerators,''
{\it Proc.\ ISIT 2009}, pp.\ 2507--2511, Seoul, South Korea, June--July 2009.
Full version: available in
arXiv:0906.1339.
\item A.~Somekh--Baruch and N.~Merhav, ``Exact random coding exponents
for erasure decoding,''
to appear in {\it Proc.\ ISIT 2010}, June 2010, Austin, Texas, U.S.A.
\end{itemize}



\newpage
\section{Additional Topics (Optional)}
\subsection{The REM With a Magnetic Field and Joint
Source--Channel Coding}
\subsubsection{Magnetic Properties of the REM}

Earlier, we studied the REM in the absence of an external magnetic field.
The Gaussian randomly drawn energies that we discussed 
were a caricature of the interaction energies in the $p$--spin glass 
model for an extremely large
level of disorder, in the absence of a magnetic field. 

\begin{figure}[ht]
\hspace*{5cm}\input{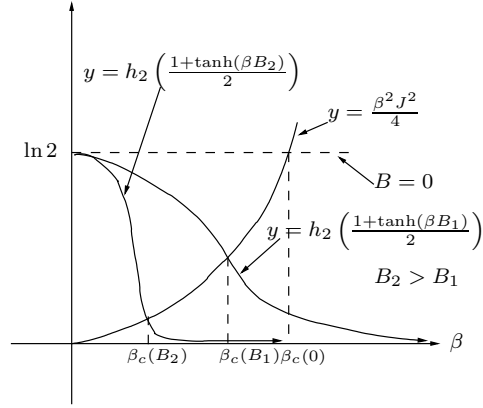}
\caption{\small Graphical presentation of the solution $\beta_c(B)$ to the
equation $\frac{1}{4}\beta^2J^2=h_2((1+\tanh(\beta B))/2)$ for various values
of $B$.}
\label{betacb}
\end{figure}

\begin{figure}[ht]
\hspace*{5cm}\input{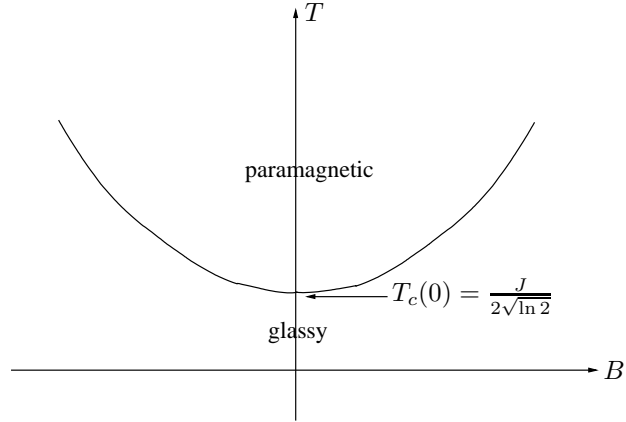}
\caption{\small Phase diagram in the $B$--$T$ plane.}
\label{tvsb}
\end{figure}

We are now going to expand the analysis of the REM so as to incorporate
also an external magnetic field $B$. This will turn out to be relevant
to a more general communication setting, namely, that of joint
source--channel coding, where as we shall see, the possible skewedness of
the probability disitribution of the source (when it is not symmetric) plays a role that 
is analogous to that of a magnetic field.
The Hamiltonian in the presence of the magnetic field is
\begin{equation}
\calE(\bs)=-B\sum_{i=1}^ns_i+\calE_I(\bs)
\end{equation}
where $\calE_I(\bs)$ stands for the interaction energy, previously modeled
to be $\calN(0,\frac{1}{2}nJ^2)$ according to the REM.
Thus, the partition function is now
\begin{eqnarray}
Z(\beta,B)&=&\sum_{\bs}e^{-\beta\calE(\bs)}\nonumber\\
&=&\sum_{\bs}e^{-\beta\calE_I(\bs)+\beta B\sum_{i=1}^ns_i}\nonumber\\
&=&\sum_{\bs}e^{-\beta\calE_I(\bs)+n\beta
Bm(\bs)}~~~~~m(\bs)=\frac{1}{n}\sum_is_i\nonumber\\
&=&\sum_m\left[\sum_{\bs:~m(\bs)=m}e^{-\beta\calE_I(\bs)}\right]\cdot
e^{+n\beta Bm}\nonumber\\
&\dfn& \sum_m Z_0(\beta,m)\cdot e^{+n\beta Bm}\nonumber
\end{eqnarray}
where $Z_0(\beta,m)$ is the {\it partial partition function}, defined to be
the expression in the square brackets in the second to the last line.\footnote{Note
that the relation between $Z_0(\beta,m)$ to $Z(\beta,B)$ is similar to the
relation between $\Omega(E)$ of the microcanonical ensemble to $Z(\beta)$
of the canonical one (a Legendre relation in the log domain): we are replacing
the fixed magnetization $m$, which is an extensive quantity, by an intensive
variable $B$ that controls its average.}
Now, observe
that $Z_0(\beta,m)$ is just like the partition function of the REM without
magnetic field, except that it has a smaller number of configurations -- only
those with magnetization $m$, namely, about $\exp\{nh_2((1+m)/2)\}$
configurations. Thus, the analysis of $Z_0(\beta,m)$ is precisely the same as
in the REM except that every occurrence of the term $\ln 2$
should be replaced by $h_2((1+m)/2)$. Accordingly, 
\begin{equation}
Z_0(\beta,m)\exe e^{n\psi(\beta,m)}
\end{equation}
with
\begin{eqnarray}
\psi(\beta,m)&=&\max_{|\epsilon|\le J\sqrt{h_2((1+m)/2)}}\left[
h_2\left(\frac{1+m}{2}\right)-\left(\frac{\epsilon}{J}\right)^2-\beta\epsilon\right]\nonumber\\
&=&\left\{\begin{array}{ll}
h_2\left(\frac{1+m}{2}\right)+\frac{\beta^2J^2}{4} & \beta\le\beta_m\dfn
\frac{2}{J}\sqrt{h_2\left(\frac{1+m}{2}\right)}\\
\beta J\sqrt{h_2\left(\frac{1+m}{2}\right)} &\beta
>\beta_m
\end{array}\right.\nonumber
\end{eqnarray}
and from the above relation between $Z$ and $Z_0$, we readily have the
Legendre relation
\begin{equation}
\phi(\beta,B)=\max_m[\psi(\beta,m)+\beta m B].
\end{equation}
For small $\beta$ (high temperature), the maximizing (dominant) $m$ is
attained with zero--derivative:
\begin{equation}
\frac{\partial}{\partial
m}\left[h_2\left(\frac{1+m}{2}\right)+\frac{\beta^2J^2}{4}+\beta mB\right]=0
\end{equation}
that is
\begin{equation}
\frac{1}{2}\ln\frac{1-m}{1+m}+\beta B =0
\end{equation}
which yields
\begin{equation}
m^*=m_p(\beta,B)\dfn\tanh(\beta B)
\end{equation}
which is exactly the paramagnetic characteristic of magnetization vs.\ magnetic
field (like that of i.i.d.\ spins), hence the name ``paramagnetic phase.''
Thus, plugging $m^*=\tanh(\beta B)$ back into the expression of $\phi$,
we get:
\begin{equation}
\phi(\beta,B)=h_2\left(\frac{1+\tanh(\beta
B)}{2}\right)+\frac{\beta^2J^2}{4}+\beta B \tanh(\beta B).
\end{equation}
This solution is valid as long as the condition 
\begin{equation}
\beta\le\beta_{m^*}=\frac{2}{J}\sqrt{h_2\left(\frac{1+\tanh(\beta B)}{2}\right)}
\end{equation}
holds, or equivalently, the condition
\begin{equation}
\frac{\beta^2J^2}{4}\le h_2\left(\frac{1+\tanh(\beta B)}{2}\right).
\end{equation}
Now, let us denote by $\beta_c(B)$ the solution $\beta$ to the equation:
\begin{equation}
\frac{\beta^2J^2}{4}= h_2\left(\frac{1+\tanh(\beta B)}{2}\right).
\end{equation}
As can be seen from the graphical illustration (Fig.\ \ref{betacb}),
$\beta_c(B)$ is a decreasing function and hence $T_c(B)\dfn 1/\beta_c(B)$
is increasing. Thus, the phase transition temperature is increasing with $|B|$
(see Fig.\ \ref{tvsb}).
Below $\beta=\beta_c(B)$, we are in the glassy phase, where $\phi$ is given
by:
\begin{equation}
\phi(\beta,B)=\max_m\left[\beta J\sqrt{h_2\left(\frac{1+m}{2}\right)}+\beta
mB\right]=\beta\cdot\max_m\left[J\sqrt{h_2\left(\frac{1+m}{2}\right)}+mB\right]
\end{equation}
thus, the maximizing $m$ does not depend on $\beta$, only on $B$. On the other
hand, it should be the same solution that we get on the boundary
$\beta=\beta_c(B)$, and so, it must be:
\begin{equation}
m^*=m_g(B)\dfn\tanh(B\beta_c(B)).
\end{equation}
Thus, in summary
\begin{equation}
\phi(\beta,B)=\left\{\begin{array}{ll}
h_2\left(\frac{1+m_p(\beta,B)}{2}\right)+\frac{\beta^2J^2}{4}+\beta
Bm_p(\beta,B) & \beta\le\beta_c(B)\\
\beta J\sqrt{h_2\left(\frac{1+m_g(B)}{2}\right)}+\beta Bm_g(B) & \beta >
\beta_c(B)\end{array}\right.
\end{equation}
In both phases $B\to 0$ implies $m^*\to 0$, therefore the REM does not
exhibit spontaneous magnetization, only a glass transition, as described.

\begin{figure}[ht]
\hspace*{5cm}\input{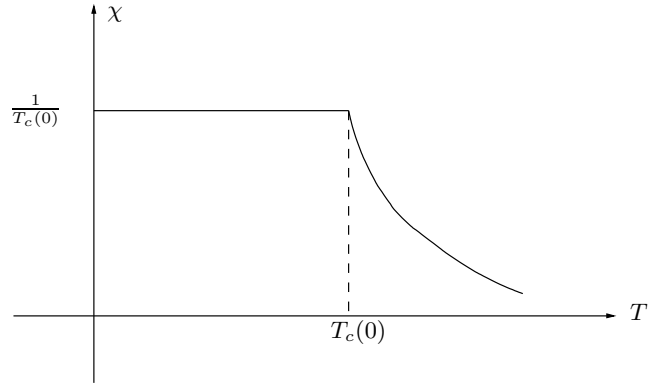}
\caption{\small $\chi$ vs.\ $T$.}
\label{chi}
\end{figure}

Finally, we mention an important parameter in the physics of magnetic
materials -- the weak--field {\it magnetic susceptibility}, which is defined as
$\chi\dfn\frac{\partial m^*}{\partial B}|_{B=0}$. It can readily be shown that
in the REM case
\begin{equation}
\chi=\left\{\begin{array}{ll}
\frac{1}{T} & T\ge T_c(0)\\
\frac{1}{T_c(0)} & T< T_c(0)\end{array}\right.
\end{equation}
The graphical illustration of this function is depicted in Fig.\ \ref{chi}.
The $1/T$ behavior for high temperature is known as {\it Curie's law}. As
we heat a magnetic material up, it becomes more and more difficult to
magnetize. The fact that here $\chi$ has an upper limit of $1/T_c(0)$
follows from the random interactions between spins, which make the
magnetization more difficult too.

\subsubsection{Relation to Joint Source--Channel Coding}

We now relate these derivations to the behavior of joint source--channel
coding systems. The full details of this part are in:
N.~Merhav, ``The random energy model in a
magnetic field and joint source--channel coding,''
{\it Physica A: Statistical
Mechanics and Its Applications}, vol.\ 387, issue 22, pp.\ 5662--5674,
September 15, 2008.

\begin{figure}[ht]
\hspace*{4cm}\input{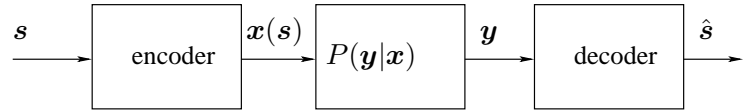}
\caption{\small Block diagram of joint source--channel communication system.}
\label{blkdiag}
\end{figure}

Consider again our coded communication system with a few slight modifications
(cf.\ Fig.\ \ref{blkdiag}).
Rather than $e^{nR}$ equiprobable messages for channel coding, we are now
talking about joint source--channel coding where the message probabilities are
skewed by the source probability distribution, which may not be symmetric. In
particular, we consider the following: Suppose we have a vector
$\bs\in\{-1,+1\}^N$ emitted from a binary memoryless source with symbol
probabilities $q=\mbox{Pr}\{S_i=+1\}=1-\mbox{Pr}\{S_i=-1\}$. The channel is
still a BSC with crossover $p$. For every $N$--tuple emitted by the source,
the channel conveys $n$ channel binary symbols, which are the components of a
codeword $\bx\in\{0,1\}^n$, such that the ratio $\theta=n/N$, the
{\it bandwidth expansion factor}, remains fixed. The mapping from $\bs$ to
$\bx$ is the encoder. As before, we shall concern ourselves with random codes,
namely, for every $\bs\in\{-1,+1\}^N$, we randomly select an independent codevector
$\bx(\bs)\in\{0,1\}^n$ by fair coin tossing, as before. Thus, we randomly
select $2^N$ codevectors, each one of length $n=N\theta$.
As in the case of pure channel coding, we consider the finite--temperature
posterior:
\begin{equation}
P_\beta(\bs|\by)=\frac{[P(\bs)P(\by|\bx(\bs))]^\beta}{Z(\beta|\by)}
\end{equation}
with
\begin{equation}
Z(\beta|\by)=\sum_{\bs}[P(\bs)P(\by|\bx(\bs))]^\beta,
\end{equation}
corresponding to the finite--temperature decoder:
\begin{equation}
\hat{s}_i=\mbox{arg}\max_{s=\pm
1}\sum_{\bs:~s_i=s}[P(\bs)P(\by|\bx(\bs))]^\beta.
\end{equation}
Once again, we separate the contributions of
$Z_c(\beta|\by)=[P(\bs_0)P(\by|\bx(\bs_0))]^\beta$, $\bs_0$ being the
true source message, and
\begin{equation}
Z_e(\beta|\by)=\sum_{\bs\ne\bs_0}[P(\bs)P(\by|\bx(\bs))]^\beta.
\end{equation}
As we shall see quite shortly, $Z_e$ behaves like the REM in a magnetic
field given by $B=\frac{1}{2}\ln\frac{q}{1-q}$. 
Accordingly, we will henceforth denote $Z_e(\beta)$
also by $Z_e(\beta,B)$, to emphasize the analogy to the REM in a magnetic
field.

\begin{figure}[ht]
\hspace*{4cm}\input{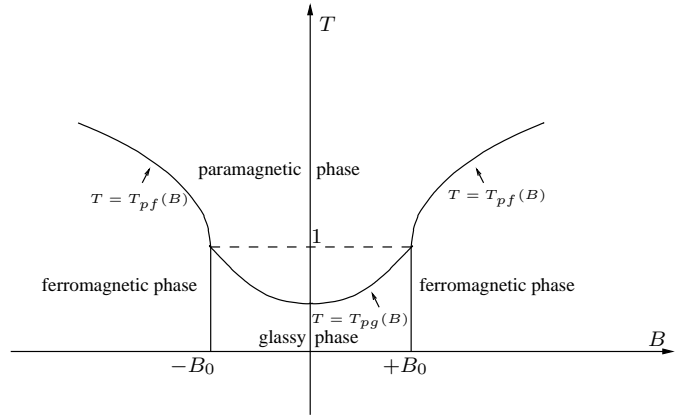}
\caption{\small Phase diagram of joint source--channel communication system.}
\label{jscphasediag}
\end{figure}

To see that $Z_e(\beta,B)$ behaves like the REM
in a magnetic field, consider
the following: first, denote by $N_1(\bs)$ the number of $+1$'s in $\bs$,
so that the magnetization,
$m(\bs)\dfn\frac{1}{N}[\sum_{i=1}^N1\{s_i=+1\}-\sum_{i=1}^N1\{s_i=-1\}]$,
pertaining to spin configuration $\bs$,
is given by $m(\bs)=2N_1(\bs)/N-1$. Equivalently,
$N_1(\bs)=N(1+m(\bs))/2$,
and then
\begin{eqnarray}
P(\bs)&=&q^{N_1(\bs)}(1-q)^{N-N_1(\bs)}\nonumber\\
&=&(1-q)^{N}\left(\frac{q}{1-q}\right)^{N(1+m(\bs))/2}\nonumber\\
&=&[q(1-q)]^{N/2}\left(\frac{q}{1-q}\right)^{Nm(\bs))/2}\nonumber\\
&=&[q(1-q)]^{N/2}e^{Nm(\bs)B}\nonumber
\end{eqnarray}
where $B$ is defined as above.
By the same token, for the binary symmetric channel we have:
\begin{equation}
P(\by|\bx)=p^{d_H(\bx,\by)}(1-p)^{n-d_H(\bx,\by)}=(1-p)^n
e^{-Jd_H(\bx,\by)}
\end{equation}
where $J=\ln\frac{1-p}{p}$ and $d_H(\bx,\by)$ is the Hamming distance,
as defined earlier.
Thus,
\begin{eqnarray}
Z_e(\beta,B)&=&[q(1-q)]^{N\beta/2}\sum_m\left[\sum_{\bx(\bs):~m(\bs)=m}
e^{-\beta\ln[1/P(\by|\bx(\bs))]}\right]e^{N\beta mB}\nonumber\\
&=&[q(1-q)]^{\beta N/2}(1-p)^{n\beta}\sum_m\left[\sum_{\bx(\bs):~m(\bs)=m}
e^{-\beta Jd_H(\bx(\bs),\by)}\right]e^{\beta NmB}\nonumber\\
&\dfn&[q(1-q)]^{N\beta/2}(1-p)^{n\beta}\sum_m Z_0(\beta,m|\by)
e^{\beta NmB}\nonumber
\end{eqnarray}
The resemblance to the REM in a magnetic field is now self--evident.
In analogy to the above analysis of the REM, $Z_0(\beta,m)$ here
behaves like in the REM without a magnetic field, namely, it contains
exponentially
$e^{Nh((1+m)/2)}=e^{nh((1+m)/2)/\theta}$ terms, with the random energy levels
of the REM being replaced now by random Hamming distances
$\{d_H(\bx(\bs),\by)\}$
that are induced by the random selection of the code
$\{\bx(\bs)\}$.
Using the same
considerations as with the REM in channel coding, 
we now get (exercise: fill in the details):
\begin{eqnarray}
\psi(\beta,m)&\dfn&\lim_{n\to\infty}\frac{\ln Z_0(\beta,m|\by)}{n}\nonumber\\
&=&\max_{\delta_m\le\delta\le
1-\delta_m}\left[\frac{1}{\theta}h_2\left(\frac{1+m}{2}\right)+h_2(\delta)-\ln
2-\beta
J\delta\right]~~~~\delta_m\dfn
\delta_{GV}\left(\frac{1}{\theta}h_2\left(\frac{1+m}{2}\right)\right)\nonumber\\
&=&\left\{\begin{array}{ll}
\frac{1}{\theta}h_2\left(\frac{1+m}{2}\right)+h_2(p_\beta)-\ln 2-\beta
Jp_\beta & p_\beta \ge \delta_m\\
-\beta J\delta_m & p_\beta < \delta_m
\end{array}\right.\nonumber
\end{eqnarray}
where again, 
\begin{equation}
p_\beta=\frac{p^\beta}{p^\beta+(1-p)^\beta}.
\end{equation}
The condition $p_\beta \ge \delta_m$ is equivalent to
\begin{equation}
\beta\le\beta_0(m)\dfn\frac{1}{J}\ln\frac{1-\delta_m}{\delta_m}.
\end{equation}
Finally, back to the full partition function:
\begin{equation}
\phi(\beta,B)=\lim_{n\to\infty}\frac{1}{N}\ln\left[\sum_mZ_0(\beta,m|\by)e^{N\beta
Bm}\right]=\max_m[\theta\psi(\beta,m)+\beta mB].
\end{equation}
For small enough $\beta$, the dominant $m$ is the one that maximizes
$[h_2((1+m)/2)+\beta m B]$, which is again the paramagnetic magnetization
\begin{equation}
m^*=m_p(\beta,B)=\tanh(\beta B).
\end{equation}
Thus, in high decoding temperatures, the source vectors $\{\bs\}$ that
dominate the posterior $P_\beta(\bs|\by)$ behave like a paramagnet under
a magentic field defined by the prior $B=\frac{1}{2}\ln\frac{q}{1-q}$. 
In the glassy regime, similarly as before, we get:
\begin{equation}
m^*=m_g(B)\dfn \tanh(B\beta_c(B))
\end{equation}
where this time, $\beta_c(B)$, the glassy--paramagnetic boundary,
is defined as the solution to the equation
\begin{equation}
\ln 2-h_2(p_\beta)=\frac{1}{\theta}h_2\left(\frac{1+\tanh(\beta
B)}{2}\right).
\end{equation}
The full details are in the paper. Taking now into account also $Z_c$,
we get a phase diagram as depicted in Fig.\ \ref{jscphasediag}.
Here,
\begin{equation}
B_0\dfn \frac{1}{2}\ln\frac{q^*}{1-q^*}
\end{equation}
where $q^*$ is the solution to the equation
\begin{equation}
h_2(q)=\theta[\ln 2-h_2(p)],
\end{equation}
namely, it is the boundary between reliable and unreliable communication.

\newpage
\subsection{The Generalized Random Energy Model (GREM) and Hierarchical
Coding}

In the mid--eighties of the previous century, Derrida extended the REM to
the generalized REM (GREM), which has an hierarchical tree sturcture to
accommodate possible correlations between 
energy levels of various configurations (and hence is somewhat closer to
reality). It turns out to have direct
relevance to performance analysis of codes with a parallel hierarchical
structure. Hierarchicial structured codes are frequently encountered in many
contexts, e.g., tree codes, multi--stage codes for progressive coding and successive 
refinement, codes for the degraded broadcast channel, codes with a binning
structure (like in G--P and W--Z coding and coding for the wiretap channel),
and so on. This part is based on the following papers:

\begin{itemize}
\item
B.~Derrida, ``A generalization of the random energy model which includes
correlations 
between energies,'' {\it J.\ de Physique -- Lettres}, vol.\ 46, L--401-107,
May 1985.
\item
B.~Derrida and E.~Gardner, ``Solution of the generalised random 
energy model,'' {\it J.\ Phys.\ C: Solid State Phys.}, vol.\ 19,
pp.\ 2253--2274, 1986.
\item
N.~Merhav, ``The generalized random energy model and its
application to the statistical physics of
ensembles of hierarchical codes,''
{\it IEEE Trans.\ Inform.\ Theory},
vol.\ 55, no.\ 3, pp.\ 1250--1268, March 2009.
\end{itemize}

\begin{figure}[ht]
\hspace*{4cm}\input{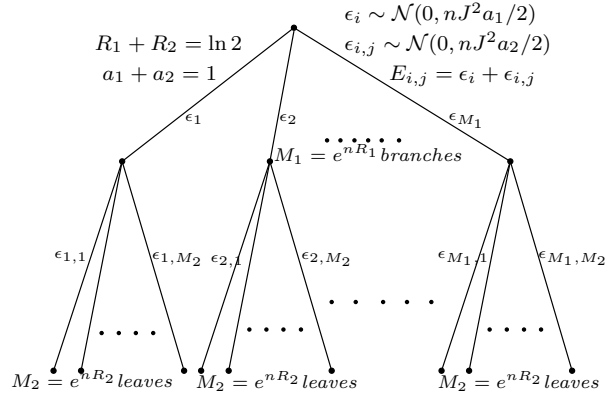}
\caption{\small The GREM with $K=2$ stages.}
\label{grem}
\end{figure}

We begin from the physics of the GREM. For simplicity, we limit ourselves to
two stages, but the discussion and the results extend to any fixed, finite 
number of stages. The GREM is defined by a few parameters: (i) a number $0 <
R_1 < \ln 2$ and $R_2=\ln 2-R_1$.
(ii) a number $0< a_1<1$ and $a_2=1-a_1$. Given these parameters, we now
partition the set of $2^n$ configurations into $e^{nR_1}$ groups, each having
$e^{nR_2}$ configurations.\footnote{
Later, we will see that in the analogy to hierarchical codes,
$R_1$ and $R_2$ will have the meaning of coding rates
at two stages of a two--stage code.} The easiest way to describe it is with a tree
(see Fig.\ \ref{grem}), each leaf of which represents one spin configuration.
Now, for each branch in this tree, we randomly draw
an independent random variable, which will be referred to as an {\it energy component}:
First, for every branch outgoing from the root, we randomly draw
$\epsilon_i\sim\calN(0,a_1nJ^2/2)$, $1\le i\le e^{nR_1}$. Then, for each
branch $1\le j\le e^{nR_2}$, emanating from node no. $i$, $1\le
i\le e^{nR_1}$, we randomly draw $\epsilon_{i,j}\sim\calN(0,a_2nJ^2/2)$.
Finally, we define the energy associated with each configuration, or
equivalently, each
leaf indexed by $(i,j)$, as $E_{i,j}=\epsilon_i+\epsilon_{i,j}$,
$1\le i\le e^{nR_1}$, $1\le j\le e^{nR_2}$.

Obviously, the marginal pdf of each $E_{i,j}$ is $\calN(0,nJ^2/2)$, just like in the
ordinary REM. However, unlike in the ordinary REM, here the configurational
energies $\{E_{i,j}\}$ are correlated: Every two leaves with a common parent
node $i$ have an energy component $\epsilon_i$ in common and
hence their total energies are correlated. 

An extension of the
GREM to $K$ stages is parametrized by $\sum_{\ell=1}^K R_\ell =\ln 2$ and
$\sum_{\ell=1}^Ka_\ell =1$, where one first divides the entirety of $2^n$
configurations into $e^{nR_1}$ groups, then each such group is subdivided 
into $e^{nR_2}$ subgroups, and so on. For each branch of generation no.\ $\ell$,
an independent energy component is drawn according to $\calN(0,a_\ell
nJ^2/2)$ and the total energy pertaining to each configuration, or a leaf, is
the sum of energy components along the path from the root to that leaf. An
extreme case of the GREM is where $K=n$, which is referred to as the {\it
directed polymer on a tree} or a {\it directed polymer in a random medium}. We
will say a few words about it later, although it has a different asymptotic
regime than the GREM, because in the GREM, $K$ is assumed fixed while $n$ grows
without bound in the thermodynamic limit.

Returning back to the case of $K=2$ stages, the analysis of the GREM is conceptually a
simple extension of that of the REM: First, we ask ourselves what is the
typical number of branches emanating from the root whose first--generation
energy component,
$\epsilon_i$, is about $\epsilon$? The answer is very similar to that of the
REM: Since we have $e^{nR_1}$ independent trials of an experiment for which
the probability of a single success is exponentially
$e^{-\epsilon^2/(nJ^2a_1)}$, then for a typical realization:
\begin{equation}
\Omega_1(\epsilon)\approx\left\{\begin{array}{ll}
0 & |\epsilon| >nJ\sqrt{a_1R_1}\\
\exp\left\{n\left[R_1-\frac{1}{a_1}\left(\frac{\epsilon}{nJ}\right)^2\right]\right\}
& |\epsilon| <nJ\sqrt{a_1R_1}\end{array}\right.
\end{equation}
Next, we ask ourselves what is the typical
number $\Omega_2(E)$ of configurations with total energy about $E$? Obviously,
each such configuration should have a first--generation energy component
$\epsilon$ and second--generation energy component $E-\epsilon$, for some
$\epsilon$. Thus,
\begin{equation}
\Omega_2(\epsilon)\approx
\int_{-nJ\sqrt{a_1R_1}}^{+nJ\sqrt{a_1R_1}} \dd\epsilon \Omega_1(\epsilon)\cdot
\exp\left\{n\left[R_2-\frac{1}{a_2}\left(\frac{E-\epsilon}{nJ}\right)^2\right]\right\}.
\end{equation}
It is important to understand here the following point: Here, we {\it no
longer} zero--out the factor 
\begin{equation}
\exp\left\{n\left[R_2-\frac{1}{a_2}\left(\frac{E-\epsilon}{nJ}\right)^2\right]\right\}
\end{equation}
when the expression in the square brackets at the exponent becomes negative,
as we did in the first stage and in the REM. The reason is simple:
Given $\epsilon$, we are conducting $\Omega_1(\epsilon)\cdot e^{nR_1}$
indepenent trials of an experiment whose success rate is
\begin{equation}
\exp\left\{-\frac{n}{a_2}\left(\frac{E-\epsilon}{nJ}\right)^2\right\}.
\end{equation}
Thus, whatever counts is whether the {\it entire} integrand has a positive
exponent or not. 

Consider next the entropy. The entropy behaves as follows:
\begin{equation}
\Sigma(E)=\lim_{n\to\infty}\frac{\ln
\Omega_2(E)}{n}=\left\{\begin{array}{ll}
\Sigma_0(E) & \Sigma_0(E)\ge 0\\
-\infty & \Sigma_0(E)< 0\end{array}\right.
\end{equation}
where $\Sigma_0(E)$ is the exponential rate of the above integral, which after
applying the Laplace method, is shown to be:
\begin{equation}
\Sigma_0(E)=\max_{|\epsilon|\le +nJ\sqrt{a_1R_1}}\left[
R_1-\frac{1}{a_1}\left(\frac{\epsilon}{nJ}\right)^2+R_2-
\frac{1}{a_2}\left(\frac{E-\epsilon}{nJ}\right)^2\right].
\end{equation}
How does the function $\Sigma(E)$ behave like? 

It turns out that to answer
this question, we will have to distinguish between two cases: (i) $R_1/a_1 <
R_2/a_2$ and (ii) $R_1/a_1\ge R_2/a_2$.\footnote{Accordingly, in coding, this
will mean a distinction between two cases of the relative coding rates at the
two stages.} First, observe that $\Sigma_0(E)$ is an even function, i.e., it
depends on $E$ only via $|E|$, and it is monotonoically non--increasing in
$|E|$. Solving the optimization problem pertaining to $\Sigma_0$, we readily
find:
\begin{equation}
\Sigma_0(E)=\left\{\begin{array}{ll}
\ln 2-\left(\frac{E}{nJ}\right)^2 & |E|\le E_1\nonumber\\
R_2-\frac{1}{a_2}\left(\frac{E}{nJ}-\sqrt{a_1R_1}\right)^2 & |E|>
E_1\end{array}\right.
\end{equation}
where $E_1\dfn nJ\sqrt{R_1/a_1}$. This is a phase transition due to the fact
that the maximizing $\epsilon$ becomes an edgepoint of its allowed interval.
Imagine now that we gradually increase $|E|$ from zero upward. Now the
question is what is encountered first: The energy level $\hat{E}$, 
where $\Sigma(E)$ jumps to $-\infty$, or $E_1$ where this phase transition
happens? In other words, is $\hat{E}< E_1$ or $\hat{E}> E_1$? In the former
case, the phase transition at $E_1$ will not be apparent because $\Sigma(E)$
jumps to $-\infty$ before, and that's it. In this case, according to the
first line of $\Sigma_0(E)$, $\ln 2-(E/nJ)^2$ vanishes at $\hat{E}=nJ\sqrt{\ln
2}$ and we get:
\begin{equation}
\Sigma(E)=\left\{\begin{array}{ll}
\ln 2-\left(\frac{E}{nJ}\right)^2 & |E|\le \hat{E}\\
-\infty & |E|> \hat{E}\end{array}\right.
\end{equation}
exactly like in the ordinary REM. It follows then that in this case,
$\phi(\beta)$ which is the Legendre transform of $\Sigma(E)$ will also be like
in the ordinary REM, that is:
\begin{equation}
\phi(\beta)=\left\{\begin{array}{ll}
\ln 2 +\frac{\beta^2J^2}{4} & \beta\le\beta_0\dfn \frac{2}{J}\sqrt{\ln 2}\\
\beta J\sqrt{\ln 2} & \beta>\beta_0\end{array}\right.
\end{equation}
As said, the condition for this is:
\begin{equation}
nJ\sqrt{\ln 2}\equiv \hat{E}\le E_1\equiv nJ\sqrt{\frac{R_1}{a_1}}
\end{equation}
or, equivalently,
\begin{equation}
\frac{R_1}{a_1}\ge \ln 2.
\end{equation}
On the other hand, in the opposite case, $\hat{E}> E_1$, the phase transition
at $E_1$ is apparent, and so, there are now {\it two} phase
transtions:
\begin{equation}
\Sigma(E)=\left\{\begin{array}{ll}
\ln 2-\left(\frac{E}{nJ}\right)^2 & |E|\le E_1\\
R_2-\frac{1}{a_2}\left(\frac{E}{nJ}-\sqrt{a_1R_1}\right)^2 & E_1< |E|\le
\hat{E}\\
-\infty & |E| > \hat{E}\end{array}\right.
\end{equation}
and accordingly (exercise: please show this):
\begin{equation}
\phi(\beta)=\left\{\begin{array}{ll}
\ln 2 +\frac{\beta^2J^2}{4} & \beta\le\beta_1\dfn
\frac{2}{J}\sqrt{\frac{R_1}{a_1}}\\
\beta J\sqrt{a_1R_1} + R_2+\frac{a_2\beta^2J^2}{4} & \beta_1\le
\beta<\beta_2\dfn \frac{2}{J}\sqrt{\frac{R_2}{a_2}}\\
\beta J(\sqrt{a_1R_1}+\sqrt{a_2R_2}) & \beta \ge \beta_2\end{array}\right.
\end{equation}
The first line is a purely paramagnetic phase. In the second line, the
first--generation
branches are glassy (there is a subexponential number of dominant
ones) but the second--generation is still paramagnetic. In the third line,
both generations are glassy, i.e., a subexponential number of
dominant first--level branches, each followed by 
a subexponential number of second--level ones, thus a total of a
subexponential number of dominant configurations overall.

Now, there is a small technical question: what is it that guarantees that
$\beta_1 <\beta_2$ whenever $R_1/a_1 < \ln 2$? We now argue that these two
inequalities are, in fact, equivalent. In a paper by Cover and Ordentlich
(IT Transactions, March 1996), the following inequality is proved for
two positive vectors $(a_1,\ldots,a_n)$ and
$(b_1,\ldots,b_n)$:
\begin{equation}
\min_i\frac{a_i}{b_i}\le \frac{\sum_{i=1}^n a_i}{\sum_{i=1}^n b_i}
\le \max_i\frac{a_i}{b_i}.
\end{equation}
Thus,
\begin{equation}
\min_{i\in\{1,2\}}\frac{R_i}{a_i}\le \frac{R_1+R_2}{a_1+a_2}\le
\max_{i\in\{1,2\}}\frac{R_i}{a_i},
\end{equation}
but in the middle expression the numerator is $R_1+R_2=\ln 2$ and 
the denominator is $a_1+a_2=1$, thus it is exactly $\ln 2$. In other words,
$\ln 2$ is always in between $R_1/a_1$ and $R_2/a_2$. So $R_1/a_1 < \ln 2$ iff
$R_1/a_1 < R_2/a_2$, which is the case where $\beta_1<\beta_2$.
To summarize our findings thus far, we have shown that:\\
{\it Case A}: $R_1/a_1 < R_2/a_2$ -- two phase transitions:
\begin{equation}
\phi(\beta)=\left\{\begin{array}{ll}
\ln 2 +\frac{\beta^2J^2}{4} & \beta\le\beta_1\\
\beta J\sqrt{a_1R_1} + R_2+\frac{a_2\beta^2J^2}{4} & \beta_1\le
\beta<\beta_2\\
\beta J(\sqrt{a_1R_1}+\sqrt{a_2R_2}) & \beta \ge \beta_2\end{array}\right.
\end{equation}
{\it Case B}: $R_1/a_1 \ge R_2/a_2$ -- one phase transition, like in the REM:
\begin{equation}
\phi(\beta)=\left\{\begin{array}{ll}
\ln 2 +\frac{\beta^2J^2}{4} & \beta\le\beta_0\\
\beta J\sqrt{\ln 2} & \beta>\beta_0\end{array}\right.
\end{equation}

We now move on to our coding problem, this time it is about source coding 
with a fidelity criterion. For simplicity, we will assume a binary symmetric
source (BSS) and the Hamming distortion. Consider the following hierarchical
structure of a code: Given a block length $n$, we break it into two segments
of lengths $n_1$ and $n_2=n-n_1$. For the first segment, we randomly select
(by fair coin tossing) a codebook $\hat{\calC}=\{\hat{\bx}_i,~1\le i\le
e^{n_1R_1}\}$. For the second segment, we do the following: For each $1\le
i\le e^{n_1R_1}$, we randomly select (again, by fair coin tossing) a codebook
$\tilde{\calC}_i=\{\tilde{\bx}_{i,j},~1\le j\le e^{n_2R_2}\}$.
Now, given a source vector $\bx\in\{0,1\}^n$, segmentized as $(\bx',\bx'')$,
the encoder seeks a pair $(i,j)$, $~1\le i\le e^{n_1R_1}$, $1\le j\le
e^{n_2R_2}$, such that $d(\bx',\hat{\bx}_i)+d(\bx'',\tilde{\bx}_{i,j})$ is
minimum, and then transmits $i$ using $n_1R_1$ nats and $j$ -- using
$n_2R_2$ nats, thus a total of $(n_1R_1+n_2R_2)$ nats, which means an
average rate of $R=\lambda R_1+(1-\lambda)R_2$ nats per symbol, where
$\lambda=n_1/n$. Now, there are a few questions that naturally arise:
\begin{itemize}
\item {\it What is the motivation for codes of this structure?}
The decoder has a reduced delay. It can decode the first $n_1$ symbols after
having received the first $n_1R_1$ nats, and does not have to wait until
the entire transmission of length $(n_1R_1+n_2R_2)$ has been received. 
Extending this idea to $K$ even segments of length $n/K$, the decoding delay
is reduced from $n$ to $n/K$. In the limit of $K=n$, in which case it is a
tree code, the decoder is actually delayless.
\item {\it What is the relation to the GREM?} The hierarchical structure of
the code is that of a tree, exactly like the GREM. The role of the energy
components at each branch is now played by the segmental
distortions $d(\bx',\hat{\bx}_i)$ and $d(\bx'',\tilde{\bx}_{i,j})$.
The parameters $R_1$ and $R_2$ here are similar to those of the GREM.
\item {\it Given an overall rate $R$, suppose
we have the freedom to choose $\lambda$, $R_1$ and $R_2$, such that
$R=\lambda R_1+(1-\lambda)R_2$, are some choice better than others in
some sense?} This is exactly what we are going to check out..
\end{itemize}

As for the performance criterion, here, 
we choose to examine performance in terms of the characteristic
function of
the overall distortion, $\bE[\exp\{-s\cdot\mbox{distortion}\}]$.
This is, of course, a much more informative figure of merit than
the average distortion, because in principle, it gives information on the
{\it entire probability
distribution} of the distortion. In particular, it generates all the moments of
the distortion
by taking derivatives at $s=0$, and it is useful in deriving Chernoff bounds on
probabilities of
large deviations events concerning the distortion. 
More formally, we make the following definitions:
Given a code $\calC$ (any block code, not necessarily of the class we defined),
and a source vector $\bx$, we define
\begin{equation}
\Delta(\bx)=\min_{\hbx\in\calC} d(\bx,\hbx),
\end{equation}
and we will be interested in the exponential rate of
\begin{equation}
\Psi(s)\dfn\bE\{\exp[-s\Delta(\bX)]\}.
\end{equation}
This quantity can be easily related to the ``partition function'':
\begin{equation}
Z(\beta|\bx)\dfn\sum_{\hbx\in\calC} e^{-\beta d(\bx,\hbx)}.
\end{equation}
In particular, 
\begin{equation}
\bE\{\exp[-s\Delta(\bX)]\}=\lim_{\theta\to
\infty}\bE\left\{[Z(s\cdot\theta|\bX)]^{1/\theta}\right\}.
\end{equation}
Thus, to analyze the characteristic function of the distortion,
we have to assess (noninteger) moments of the partition function. 

\begin{figure}[ht]
\hspace*{4cm}\input{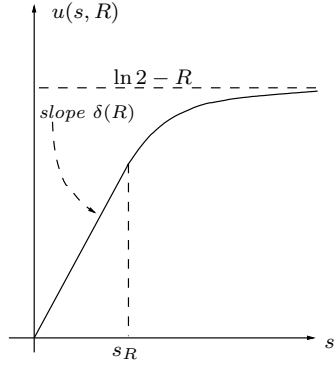}
\caption{\small Qualitative graph of the function $u(s,R)$ as a function of
$s$ for fixed $R$.}
\label{usrn}
\end{figure}

Let's first see
what happens with ordinary random block codes, without any structure. 
This calculation is very similar the one we did before in the context
of channel coding:
\begin{eqnarray}
\bE\left\{[Z(s\cdot\theta|\bX)]^{1/\theta}\right\}&=&
\bE\left\{\left[\sum_{\hbx\in\calC} e^{-s\theta
d(\bx,\hbx)}\right]^{1/\theta}\right\}\nonumber\\
&=&\bE\left\{\left[\sum_{d=0}^n\Omega(d)e^{-s\theta
d}\right]^{1/\theta}\right\}\nonumber\\
&\exe&\sum_{d=0}^n\bE\left\{\left[\Omega(d)\right]^{1/\theta}\right\}\cdot e^{-s
d}\nonumber
\end{eqnarray}
where, as we have already shown in the past:
\begin{equation}
\bE\left\{\left[\Omega(d)\right]^{1/\theta}\right\}\exe\left\{\begin{array}{ll}
e^{n[R+h_2(\delta)-\ln 2]} & \delta\le\delta_{GV}(R)~\mbox{or}~
\delta\ge 1-\delta_{GV}(R)\\
e^{n[R+h_2(\delta)-\ln 2]/\theta} & \delta_{GV}(R)\le \delta\le
1-\delta_{GV}(R)\end{array}\right.
\end{equation}
Note that $\delta_{GV}(R)$ is exactly the distortion--rate function of the BSS
w.r.t.\ the Hamming distortion. 
By plugging the expression of
$\bE\{[\Omega(d)]^{1/\theta}\}$ back into that of 
$\bE\{[Z(s\cdot\theta|\bX)]^{1/\theta}\}$ and carrying out the
maximization pertaining to the dominant contribution, we eventually (exercise:
please show that) obtain:
\begin{equation}
\Psi(s)\exe e^{-nu(s,R)}
\end{equation}
where
\begin{eqnarray}
u(s,R)&=&\ln
2-R-\max_{\delta\le\delta_{GV}(R)}[h_2(\delta)-s\delta]\nonumber\\
&=&\left\{\begin{array}{ll}
s\delta_{GV}(R) & s\le s_R\\
v(s,R) & s > s_R\end{array}\right.
\end{eqnarray}
with 
\begin{equation}
s_R\dfn \ln\left[\frac{1-\delta_{GV}(R)}{\delta_{GV}(R)}\right]
\end{equation}
and
\begin{equation}
v(s,R)\dfn\ln 2-R+s-\ln(1+e^s).
\end{equation}
The function $u(s,R)$ is depicted qualitatively in Fig.\ \ref{usrn}.

Let's now move on to the hierarchical codes. The analogy with the GREM is
fairly clear. Given $\bx$, there are about $\Omega_1(\delta_1)\exe
e^{n_1[R_1+h_2(\delta_1)-\ln 2]}$ first--segment codewords $\{\hat{\bx}_i\}$
in $\hat{\calC}$ at distance $n_1\delta_1$ from the first segment $\bx'$ of
$\bx$, provided that $R_1+h_2(\delta_1)-\ln 2 >0$ and $\Omega_1(\delta_1)=0$
otherwise. For each such first--segment codeword, there are about
$e^{n_2[R_2+h_2(\delta_2)-\ln 2]}$ second--segment codewords
$\{\tilde{\bx}_{i,j}\}$ at distance $n_2\delta_2$ from the second segment
$\bx''$ of $\bx$. Therefore, for $\delta=\lambda\delta_1+(1-\lambda)\delta_2$,
\begin{eqnarray}
\Omega_2(\delta)&=&\sum_{\delta_1=\delta_{GV}(R_1)}^{1-\delta_{GV}(R_1)}
e^{n_1[R_1+h_2(\delta_1)-\ln 2]}\cdot
e^{n_2[R_2+h_2((\delta-\lambda\delta_1)/(1-\lambda))-\ln
2]}\nonumber\\
&\exe&\exp\left\{n\cdot\max_{\delta_1\in[\delta_{GV}(R_1),1-\delta_{GV}(R_1)]}
\left[R+\lambda
h_2(\delta_1)+(1-\lambda)h_2\left(\frac{\delta-\lambda\delta_1}{1-\lambda}\right)\right]\right\}\nonumber
\end{eqnarray}
In analogy to the analysis of the GREM, here too, there is a distinction between
two cases: $R_1\ge R\ge R_2$ and $R_1 < R < R_2$. 
In the first case, the behavior is just like in the REM:
\begin{equation}
\Sigma(\delta)=\left\{\begin{array}{ll}
R+h_2(\delta)-\ln 2 & \delta\in[\delta_{GV}(R),1-\delta_{GV}(R)]\\
-\infty & \mbox{elsewhere}\end{array}\right.
\end{equation}
and then, of course, $\phi(\beta)=-u(\beta,R)$ behaves exactly like that of
a general random code, in spite of the hierarchical structure.
In the other case, we have two phase transitions:
\begin{equation}
\phi(\beta,R)=\left\{\begin{array}{ll}
-v(\beta,R) & \beta < \beta(R_1)\\
-\lambda\beta\delta_{GV}(R_1)-(1-\lambda)v(\beta,R_2) & \beta(R_1) < \beta <
\beta(R_2)\\
-\beta[\lambda\delta_{GV}(R_1)+(1-\lambda)\delta_{GV}(R_2)] & \beta >
\beta(R_2)\end{array}\right.
\end{equation}
The last line is the purely glassy phase and this is the relevant phase because of the
limit $\theta\to 0$ that we take in order to calculate $\Psi(s)$.
Note that at this phase the slope is
$\lambda\delta_{GV}(R_1)+(1-\lambda)\delta_{GV}(R_2)$
which means that code behaves as if the two segments were coded {\it
separately}, which is worse that $\delta_{GV}(R)$ due to convexity arguments. 
Let's see this more concretely on the characteristic function:
This time, it will prove convenient to define $\Omega(d_1,d_2)$ as an
enumerator of codewords whose distance is $d_1$ at the first segment
and $d_2$ -- on the second one. 
Now,
\begin{equation}
\bE\left\{Z^{1/\theta}(s\cdot\theta)\right\}=\bE\left\{\left[\sum_{d_1=0}^n\sum_{d_2=0}^n
\Omega(d_1,d_2)\cdot e^{-s\theta(d_1+d_2)}\right]^{1/\theta}\right\}\exe
\sum_{d_1=0}^n\sum_{d_2=0}^n\bE\left\{\Omega^{1/\theta}(d_1,d_2)\right\}\cdot
e^{-s(d_1+d_2)}.
\end{equation}
Here, we should distinguish between four types of terms depending on whether
or not $\delta_1\in[\delta_{GV}(R_1),1-\delta_{GV}(R_1)]$ and whether or not
$\delta_2\in[\delta_{GV}(R_2),1-\delta_{GV}(R_2)]$. In each one of these
combinations, the behavior is different (the details are in the paper). 
The final results are as follows:
\begin{itemize}
\item For $R_1 < R_2$,
\begin{equation}
\lim_{n\to\infty}\left[-\frac{1}{n}\ln\bE \exp\{-s\Delta(\bX)\}\right]=
\lambda u(s,R_1)+(1-\lambda)u(s,R_2)
\end{equation}
which means the behavior of two independent, decoupled codes for the
two segments, which is bad, of course.
\item For $R_1 \ge R_2$,
\begin{equation}
\lim_{n\to\infty}\left[-\frac{1}{n}\ln\bE \exp\{-s\Delta(\bX)\}\right]=
u(s,R)~~~~\forall s\le s_0
\end{equation}
where $s_0$ is some positive constant. This means that the code behaves like
an unstructured code (with delay) for all $s$ up to a certain $s_0$ and the
reduced decoding delay is obtained for free. Note that the domain of small $s$
is relevant for moments of the distortion. For $R_1=R_2$, $s_0$ is unlimited.
\end{itemize}
Thus, the conclusion is that if we must work at different rates, it is better
to use the higher rate first.

Finally, we discuss a related model that we mentioned earlier,
which can be thought of as an extreme case of the GREM with $K=n$. 
This is the {\it directed polymer in a
random medium} (DPRM): 
Consider a {\it Cayley tree}, namely, a full balanced tree with branching
ratio $d$ and depth $n$ (cf.\ Fig.\ \ref{tree}, where $d=2$ and $n=3$). Let us index
the branches by a pair of integers $(i,j)$, where $1\le i\le n$ describes the
generation (with $i=1$ corresponding to the $d$ branches that emanate from the
root), and $0\le j\le d^i-1$ enumerates the branches of the $i$--th generation, say,
from left to right (again, see Fig.\ \ref{tree}).
For each branch $(i,j)$, $1\le j\le d^i$, $1\le i\le n$,
we randomly draw an independent random variable $\varepsilon_{i,j}$ according
to a fixed probability function $q(\varepsilon)$ (i.e., a probability mass
function in the discrete
case, or probability density function in the continuous case). As explained
earlier, the asymptotic regime here is different from that of the GREM: In the GREM we
had a fixed number of stages $K$ that didn't grow with $n$ and exponentially
many branches emanating from each internal node. Here, we have $K=n$ and a
fixed number $d$ of branches outgoing from each note.

\begin{figure}[ht]
\hspace*{5cm}\input{tree.pstex_t}
\caption{A Cayley tree with branching factor $d=2$ and depth $n=3$.}
\label{tree}
\end{figure}

A {\it walk} $\bw$, from the root of the tree to one of its leaves, is
described
by a finite
sequence $\{(i,j_i)\}_{i=1}^n$, where $0\le j_1\le d-1$ and $dj_i\le
j_{i+1}\le dj_i+d-1$, $i=1,2,\ldots,(n-1)$.\footnote{In fact, for a given $n$,
the
number $j_n$ alone dictates the entire walk.} For a given realization of the
RV's
$\{\varepsilon_{i,j}:~i=1,2,\ldots,n,~j=0,1,\ldots,d^i-1\}$, we define the
Hamiltonian associated with $\bw$ as
$\calE(\bw)=\sum_{i=1}^n\varepsilon_{i,j_i}$,
and then the partition function as:
\begin{equation}
\label{pftree}
Z_n(\beta)=\sum_{\bw} \exp\{-\beta\calE(\bw)\}.
\end{equation}
It turns out that this model is exactly solvable (in many ways) and one can
show (see e.g., E.~Buffet, A.~Patrick, and J.~V.~Pul\'e, ``Directed polymers
on trees: a martingale approach,'' {\it J.\ Phys.\ A: Math.\ Gen.}, vol.\ 26,
pp.\ 1823--1834, 1993) that it admits a glassy phase transition:
\begin{equation}
\phi(\beta)=\lim_{n\to\infty}\frac{\ln Z_n(\beta)}{n}=\left\{\begin{array}{ll}
\phi_0(\beta) & \beta < \beta_c\\
\phi_0(\beta_c) & \beta \ge \beta_c\end{array}\right.~~~~\mbox{almost surely}
\end{equation}
where
\begin{equation}
\phi_0(\beta)\dfn \frac{\ln [d\cdot \bE e^{-\beta\rho(\epsilon)}]}{\beta}
\end{equation}
and $\beta_c$ is the value of $\beta$ that minimizes $\phi_0(\beta)$. 

In analogy to the hierachical codes inspired by the GREM, consider now an ensemble
of tree codes for encoding source $n$--tuples,
$\bx=(x_1,\ldots,x_n)$, which is
defined as follows: Given a coding rate $R$ (in nats/source--symbol), which is
assumed to be the natural logarithm of some positive integer
$d$, and given a
probability distribution on the reproduction alphabet,
$Q=\{q(y),~y\in\calY\}$,
let us draw $d=e^R$ independent copies of $Y$ under $Q$, and denote them by
$Y_1,Y_2,\ldots,Y_d$. We shall refer to the randomly chosen set,
$\calC_1=\{Y_{1},Y_{2},\ldots,Y_{d}\}$, as our `codebook' for the first
source symbol, $X_1$. Next, for each $1\le j_1\le d$, we randomly select
another
such codebook under $Q$,
$\calC_{2,j_1}=\{Y_{j_1,1},Y_{j_1,2},\ldots,Y_{j_1,d}\}$, for the second
symbol,
$X_2$. Then, for each $1\le j_1\le d$ and $1\le j_2\le d$, we again draw under
$Q$
yet another codebook
$\calC_{3,j_1,j_2}=\{Y_{j_1,j_2,1},Y_{j_1,j_2,2},\ldots,Y_{j_1,j_2,d}\}$, for
$X_3$, and so on.
In general, for each $t\le n$, we randomly draw $d^{t-1}$ codebooks under
$Q$, which are indexed by $(j_1,j_2,\ldots,j_{t-1})$, $1\le j_k\le d$, $1\le
k\le t-1$.

Once the above described random code selection process is complete, the
resulting
set of codebooks $\{\calC_1,\calC_{t,j_1,\ldots,j_{t-1}},~2\le t\le n,~1\le
j_k\le
d,~1\le k\le t-1\}$
is revealed to both the
encoder and decoder, and the encoding--decoding system works as follows:
\begin{itemize}
\item {\it Encoding:} Given a source $n$--tuple $X^n$, find a vector of
indices $(j_1^*,j_2^*,\ldots,j_n^*)$ that minimizes the overall distortion
$\sum_{t=1}^n \rho(X_t,Y_{j_1,\ldots,j_t})$. Represent each component $j_t^*$
(based on $j_{t-1}^*$)
by $R=\ln d$ nats (that is, $\log_2d$ bits), thus a total of $nR$ nats.
\item {\it Decoding:}
At each time $t$ ($1\le t\le n$), after having decoded
$(j_1^*,\ldots,j_t^*)$, output the reproduction symbol
$Y_{j_1^*,\ldots,j_t^*}$.
\end{itemize}

In order to analyze the rate--distortion performance of this ensemble of
codes, we now make the following assumption:

\vspace{0.15cm}

\noindent
{\it The random coding distribution $Q$ is such that the
distribtion of the RV $\rho(x,Y)$ is the same for all $x\in\calX$.}

It turns out that this assumption is fulfilled
quite often -- it is the case whenever the random
coding distribution together with distortion function exhibit a sufficiently
high degree of symmetry. For example, if $Q$ is the uniform distribution over
$\calY$ and the rows of the distortion matrix $\{\rho(x,y)\}$ are permutations
of each other, which is in turn the case, for example,
when $\calX=\calY$ is a group and $\rho(x,y)=\gamma(x-y)$
is a difference distortion function w.r.t.\ the group difference operation.
Somewhat more generally, this assumption
still holds when the different rows of the
distortion matrix
are formed by permutations of each other subject to the following
rule: $\rho(x,y)$ can be
swapped with $\rho(x,y')$ provided that $q(y')=q(y)$.

For a given $\bx$ and a given realization of the set of codebooks, define
the partition function in analogy to that of the DPRM:
\begin{equation}
Z_n(\beta)=\sum_{\bw} \exp\{-\beta\sum_{t=1}^n\rho(x_t,Y_{j_1,\ldots,j_t})\},
\end{equation}
where the summation extends over all $d^n$ possible walks,
$\bw=(j_1,\ldots,j_n)$,
along the Cayley tree. 
Clearly, considering our symmetry assumption,
this falls exactly under the umbrella of the DPRM, with the distortions
$\{\rho(x_t,Y_{j_1,\ldots,j_t})\}$ playing the role of the branch energies
$\{\varepsilon_{i.j}\}$. Therefore, $\frac{1}{n\beta}\ln Z_n(\beta)$ converges
almost surely, as $n$ grows without bound, to $\phi(\beta)$, now defined as
\begin{equation}
\label{physresult1}
\phi(\beta)=\left\{\begin{array}{ll}
\phi_0(\beta) & \beta\le\beta_c\\
\phi_0(\beta_c) & \beta>\beta_c
\end{array}\right.
\end{equation}
where now
\begin{eqnarray}
\phi_0(\beta)&\dfn&\frac{\ln[d\cdot\bE\{e^{-\beta\rho(x,Y)}\}]}{\beta}\nonumber\\
&=&\frac{\ln[e^R\cdot\bE\{e^{-\beta\rho(x,Y)}\}]}{\beta}\nonumber\\
&=&\frac{R+\ln[\bE\{e^{-\beta\rho(x,Y)}\}]}{\beta},\nonumber
\end{eqnarray}
Thus, for every $(x_1,x_2,\ldots)$, the distortion is
given by
\begin{eqnarray}
\limsup_{n\to\infty}\frac{1}{n}\sum_{t=1}^n\rho(x_t,Y_{j_1^*,\ldots,j_t^*})
&\dfn&\limsup_{n\to\infty}\frac{1}{n}\min_{\bw}
\left[\sum_{t=1}^n\rho(x_t,Y_{j_1,\ldots,j_t})\right]\nonumber\\
&=&\limsup_{n\to\infty}\limsup_{\ell\to\infty}\left[-\frac{\ln
Z_n(\beta_\ell)}{n\beta_\ell}\right]\nonumber\\
&\le&\limsup_{\ell\to\infty}\limsup_{n\to\infty}\left[-\frac{\ln
Z_n(\beta_\ell)}{n\beta_\ell}\right]\nonumber\\
&\eqas&-\liminf_{\ell\to\infty}\phi(\beta_\ell)\nonumber\\
&=& -\phi_0(\beta_c)\nonumber\\
&=&\max_{\beta\ge
0}\left[-\frac{\ln[\bE\{e^{-\beta\rho(x,Y)}\}]+R}{\beta}\right]\nonumber\\
&=& D(R)\nonumber,
\end{eqnarray}
where: (i) $\{\beta_\ell\}_{\ell\ge 1}$ is an arbitrary sequence tending to
infinity, (ii) the almost--sure equality in the above mentioned paper, 
and (iii)
the justification of the inequality at the third line is left as an exercise.
The last equation is easily obtained by inverting the function $R(D)$ in its
parametric representation that we have seen earlier:
\begin{equation}
R(D)=-\min_{\beta\ge 0}\min_Q\left\{\beta
D+\sum_{x\in\calX}p(x)\ln\left[\sum_{y\in\calY}q(y)e^{-\beta\rho(x,y)}\right]\right\}.
\end{equation}
Thus, the ensemble of tree codes achieves $R(D)$ almost surely.


\newpage
\subsection{Phase Transitions of the Rate--Distortion Function}

The material in this part is based on the paper:
K.~Rose, ``A mapping approach to rate-distortion computation and 
analysis,'' {\em IEEE Trans.~Inform.~Theory\/}, vol.\ 40, no.\ 6, pp.\
1939--1952, November 1994.

We have seen in one of the earlier meetings that the rate--distortion function
of a source $P=\{p(x),~x\in\calX\}$ can be expressed as
\begin{equation}
R(D)=-\min_{\beta\ge 0}\left[\beta D+\sum_xp(x)\ln\left(\sum_yq(y)e^{-\beta
d(x,y)}\right)\right]
\end{equation}
where $Q=\{q(y),y\in\calY\}$ is the output marginal of the test channel, which
is also the one that minimizes this expression.
We are now going to take a closer look at this function in the context of the
quadratic distortion function $d(x,y)=(x-y)^2$.
As said, the optimum $Q$ is the one that minimizes the above expression, or
equivalently, the free energy
\begin{equation}
f(Q)=-\frac{1}{\beta}\sum_xp(x)\ln\left(\sum_yq(y)e^{-\beta d(x,y)}\right)
\end{equation}
and in the continuous case, summations should be replaced by integrals:
\begin{equation}
f(Q)=-\frac{1}{\beta}\int_{-\infty}^{+\infty}
\dd xp(x)\ln\left(\int_{-\infty}^{+\infty} \dd yq(y)e^{-\beta d(x,y)}\right).
\end{equation}
Rose suggests to represent the RV $Y$ as a function of
$U\sim\mbox{unif}[0,1]$, and then, instead of optimizing $Q$,
one should optimize the function $y(u)$ in:
\begin{equation}
f(y(\cdot))=-\frac{1}{\beta}\int_{-\infty}^{+\infty}
\dd xp(x)\ln\left(\int_0^1 \dd \mu(u)e^{-\beta d(x,y(u))}\right),
\end{equation}
where $\mu(\cdot)$ is the Lebesgue measure (the uniform measure).
A necessary condition for optimality,\footnote{The details are in the paper,
but intuitively, 
instead of a function $y(u)$ of a continuous variable $u$,
think of a vector $\by$ whose components are indexed by $u$, which take on
values in some grid of $[0,1]$. In other words,
think of the argument of the logarithmic function as
$\sum_{u=0}^1 e^{-\beta d(x,y_u)}$.} which must hold for almost every $u$ is:
\begin{equation}
\int_{-\infty}^{+\infty}\dd xp(x)\cdot\left[
\frac{e^{-\beta d(x,y(u))}}{\int_0^1 \dd \mu(u')e^{-\beta
d(x,y(u'))}}\right]\cdot\frac{\partial d(x,y(u))}{\partial y(u)}=0.
\end{equation}
Now, let us define the {\it support} of $y$ as the set of values that $y$ may
possibly take on. Thus, this support is a subset of the set of all points
$\{y_0=y(u_0)\}$ for which:
\begin{equation}
\int_{-\infty}^{+\infty}\dd xp(x)\cdot\left[
\frac{e^{-\beta d(x,y_0)}}{\int_0^1 \dd \mu(u')e^{-\beta
d(x,y(u'))}}\right]\cdot\frac{\partial d(x,y(u))}{\partial y(u)}\bigg|_{y(u)=y_0}=0.
\end{equation}
This is because $y_0$ must be a point that is obtained as $y(u)$ for some $u$.
Let us define now the posterior:
\begin{equation}
q(u|x)=\frac{e^{-\beta d(x,y(u))}}{\int_0^1 \dd \mu(u')e^{-\beta
d(x,y(u'))}}.
\end{equation}
Then,
\begin{equation}
\int_{-\infty}^{+\infty}\dd xp(x)q(u|x)\cdot\frac{\partial
d(x,y(u))}{\partial y(u)}=0.
\end{equation}
But $p(x)q(u|x)$ is a joint distribution $p(x,u)$, which can also be thought
of as $\mu(u)p(x|u)$. So, if we divide the last equation by $\mu(u)$, we get,
for almost all $u$:
\begin{equation}
\int_{-\infty}^{+\infty}\dd x p(x|u)\frac{\partial d(x,y(u))}{\partial y(u)}=0.
\end{equation}
Now, let's see what happens in the case of the quadratic distortion,
$d(x,y)=(x-y)^2$. Let us suppose that the support of $Y$ includes some
interval $\calI_0$ as a subset. For a given $u$, $y(u)$ is nothing other than
a number, and so the optimality condition must hold for every $y\in\calI_0$.
In the case of the quadratic distortion, this optimality criterion means
\begin{equation}
\int_{-\infty}^{+\infty}\dd xp(x)\lambda(x)(x-y)
e^{-\beta(x-y)^2}=0,~~~~\forall y\in\calI_0
\end{equation}
with
\begin{equation}
\lambda(x)\dfn \frac{1}{\int_0^1 \dd \mu(u)e^{-\beta d(x,y(u))}}=
\frac{1}{\int_{-\infty}^{+\infty} \dd yq(y)e^{-\beta d(x,y)}},
\end{equation}
or, equivalently,
\begin{equation}
\int_{-\infty}^{+\infty}\dd xp(x)\lambda(x)\frac{\partial}{\partial
y}\left[e^{-\beta(x-y)^2}\right]=0.
\end{equation}
Since this must hold for all $y\in\calI_0$, then all derivatives of the
l.h.s.\ must vanish within $\calI_0$, i.e.,
\begin{equation}
\int_{-\infty}^{+\infty}\dd xp(x)\lambda(x)\frac{\partial^n}{\partial
y^n}\left[e^{-\beta(x-y)^2}\right]=0.
\end{equation}
Now, considering the Hermitian polynomials
\begin{equation}
H_n(z)\dfn e^{\beta z^2}\frac{\dd^n}{\dd z^n}(e^{-\beta z^2})
\end{equation}
this requirement means
\begin{equation}
\int_{-\infty}^{+\infty}\dd
xp(x)\lambda(x)H_n(x-y)e^{-\beta(x-y)^2}=0.
\end{equation}
In words: $\lambda(x)p(x)$ is orthogonal to all Hermitian polynomials of order
$\ge 1$ w.r.t.\ the weight function $e^{-\beta z^2}$. Now, as is argued in the
paper, since these polynomials are complete in $L^2(e^{-\beta z^2})$, we get
\begin{equation}
p(x)\lambda(x)=\mbox{const.}
\end{equation}
because $H_0(z)\equiv 1$ is the only basis function orthogonal to all $H_n(z)$,
$n\ge 1$. This
yields, after normalization:
\begin{equation}
p(x)=\sqrt{\frac{\beta}{\pi}}\int_0^1\dd\mu(u)e^{-\beta(x-y(u))^2}=
\sqrt{\frac{\beta}{\pi}}\int_{-\infty}^{+\infty}\dd
yq(y)e^{-\beta(x-y)^2}=Q\star\calN\left(0,\frac{1}{2\beta}\right).
\end{equation}
The interpretation of the last equation is simple: the marginal of $X$ is
given by the convolution between the marginal of $Y$ and the zero--mean
Gaussian distribution with variance $D=1/(2\beta)$ ($=kT/2$ of the
equipartition theorem, as we already saw). This means that $X$ must be
representable as 
\begin{equation}
X=Y+Z
\end{equation}
where $Z\sim \calN\left(0,\frac{1}{2\beta}\right)$ and independent of $Y$.
From the Information Theory course we know that this is exactly what happens
when $R(D)$ coincides with its Gaussian lower bound, a.k.a.\ the {\it Shannon
lower bound}. Here is a reminder of this:
\begin{eqnarray}
R(D)&=&h(X)-\max_{\bE(X-Y)^2\le D}h(X|Y)\nonumber\\
&=&h(X)-\max_{\bE(X-Y)^2\le D}h(X-Y|Y)\nonumber\\
&\ge&h(X)-\max_{\bE(X-Y)^2\le D}h(X-Y)~~~~\mbox{equality if}~(X-Y)\perp Y\nonumber\\
&=&h(X)-\max_{\bE Z^2\le D}h(Z)~~~~Z\dfn X-Y\nonumber\\
&\ge &h(X)-\frac{1}{2}\ln(2\pi e D)~~~~\mbox{equality
if}~Z\sim\calN(0,D)\nonumber\\
&\dfn& R_{\mbox{SLB}}(D)\nonumber
\end{eqnarray}
The conclusion then is that if the support of $Y$ includes an interval (no
matter how small) then $R(D)$ coincides with $R_{\mbox{SLB}}(D)$. This implies
that in all those cases that $R_{\mbox{SLB}}(D)$ is not attained, the support
of the optimum test channel output distribution must be singular, i.e., it
cannot contain an interval. It can be, for example, a set of isolated points. 

But we also know that whenever $R(D)$ meets the SLB for some $D=D_0$, then
it must also coincide with it for all $D < D_0$. This follows from the
following consideration: If $X$ can be represented as $Y+Z$, where
$Z\sim\calN(0,D_0)$ is independent of $Y$,  then for every $D < D_0$, 
we can always decompose $Z$ as $Z_1+Z_2$, where $Z_1$ and $Z_2$ are both
zero--mean independent Gaussian RV's with variances $D_0-D$ and $D$,
respectively. Thus,
\begin{equation}
X=Y+Z=(Y+Z_1)+Z_2\dfn Y'+Z_2
\end{equation}
and we have represented $X$ as a noisy version of $Y'$ with noise variance
$D$. Whenever $X$ can be thought of as 
a mixture of Gaussians, $R(D)$ agrees with its SLB for
all $D$ upto the variance of the narrowest Gaussian in this mixture.
Thus, in these cases:
\begin{equation}
R(D)~~\left\{\begin{array}{ll}
= R_{\mbox{SLB}}(D) & D\le D_0\\
> R_{\mbox{SLB}}(D) & D> D_0\end{array}\right.
\end{equation}
It follows then that in all these cases,
the optimum output marginal contains intervals for all $D\le D_0$ and then
becomes abruptly singular as $D$ exceeds $D_0$. From the viewpoint of
statistical mechanics, this looks like a phase transition, then. Consider first
an infinite temperature, i.e., $\beta=0$, which means unlimited distortion.
In this case, the optimum output marginal puts all its mass on one point:
$y=\bE(X)$, so it is definitely singular. This remains true even if we increase
$\beta$ to the inverse temperature that corresponds to $D_{\max}$, the
smallest distortion for which $R(D)=0$. If we further increase $\beta$, the
support of $Y$ begins to change. In the next step it can include 2 points,
then 3 points, etc. Then, if there is $D_0$ below which the SLB is met, then
the support of $Y$ abruptly becomes one that contains one interval at least. 
This point is also demonstrated numerically in the paper.

An interesting topic for research evolves around possible extensions of these
results to more general distortion measures, other than the quadratic
distortion measure.


\newpage
\subsection{Capacity of the Sherrington--Kirkpartrick Spin Glass}

This part is based on the paper:
O.~Shental and I.~Kanter, ``Shannon capacity of infinite--range
spin--glasses,''  
Technical Report, Bar Ilan University, 2005.
In this work, the authors consider the S--K model with independent Gaussian
coupling coefficients, and they count the number $N(n)$ of meta--stable states in the
absence of magnetic field. A meta-stable state means that each spin is in its
preferred polarization according to the net field that it `feels'. i.e.,
\begin{equation}
s_i=\mbox{sgn}\left(\sum_j J_{ij}s_j\right),~~~~~~i=1,\ldots,n.
\end{equation}
They refer to the limit $\lim_{n\to\infty}[\ln N(n)]/n$ as the capacity $C$
of the S--K model. However, they take an annealed rather than a quenched
average, thus the resulting capacity is somewhat optimistic. The reason that
this work is brought here is that many of the mathematical tools we have
been exposed to are used here. The main result in this work is that
\begin{equation}
C=\ln[2(1-Q(t))]-\frac{t^2}{2}
\end{equation}
where 
\begin{equation}
Q(t)\dfn\frac{1}{2\pi}\int_t^\infty \dd u\cdot e^{-u^2/2}
\end{equation}
and $t$ is the solution
to the equation
\begin{equation}
t=\frac{e^{-t^2/2}}{\sqrt{2\pi}[1-Q(t)]}.
\end{equation}

The authors even address a slighlty more general question: Quite obviously,
the metastability condition is that for every $i$ there exists $\lambda_i > 0$
such that
\begin{equation}
\lambda_is_i=\sum_j J_{ij}s_j.
\end{equation}
But they actually answer the following question:
Given a constant $K$, what is the expected number of states for which
there is $\lambda_i>K$ for each $i$ such that $\lambda_is_i=\sum_j J_{ij}s_j$?
For $K\to-\infty$, one expects $C\to\ln 2$, and for
$K\to\infty$, one expects $C\to 0$. The case of interest is exactly in the
middle, where $K=0$.

Moving on to the analysis, we first observe that for each such state,
\begin{equation}
\int_K^\infty\cdot\cdot\cdot\int_K^\infty\prod_{i=1}^n\left[\dd\lambda_i
\delta\left(\sum_\ell J_{i\ell}s_\ell-\lambda_is_i\right)\right]=1
\end{equation}
thus
\begin{equation}
N(n)=\int_K^\infty\cdot\cdot\cdot\int_K^\infty\prod_{i=1}^n\dd\lambda_i\sum_{\bs}
\left<\prod_{i=1}^n\delta\left(\sum_\ell
J_{i\ell}s_\ell-\lambda_is_i\right)\right>_{\bJ}.
\end{equation}
Now, according to the S--K model, $\{J_{i\ell}\}$ are $n(n-1)/2$ i.i.d.
zero--mean Gaussian RV's with variance $J^2/n$. Thus,
\begin{equation}
\bar{N}(n)=\left(\frac{n}{2\pi
J^2}\right)^{n(n-1)/4}\int_{\reals^{n(n-1)/2}}\dd\bJ
\exp\left\{-\frac{n}{2J^2}\sum_{i>\ell}J_{i\ell}^2\right\}\cdot\sum_{\bs}
\int_K^\infty\cdot\cdot\cdot\int_K^\infty\dd\blambda \cdot\prod_{i=1}^n
\delta\left(\sum_\ell J_{i\ell} s_\ell-\lambda_is_i\right).
\end{equation}
The next step is to represent each Dirac as an inverse Fourier transform
of an exponent
\begin{equation}
\delta(x)=\frac{1}{2\pi}\int_{-\infty}^{+\infty}\dd\omega e^{j\omega
x}~~~~~j=\sqrt{-1}
\end{equation}
which then becomes:
\begin{eqnarray}
\bar{N}(n)&=&\left(\frac{n}{2\pi
J^2}\right)^{n(n-1)/4}\int_{\reals^{n(n-1)/2}}\dd\bJ
\exp\left\{-\frac{n}{2J^2}\sum_{i>\ell}J_{i\ell}^2\right\}\cdot\sum_{\bs}
\int_K^\infty\cdot\cdot\cdot\int_K^\infty\dd\blambda \times\nonumber\\
&&\int_{\reals^n}\frac{\dd\bomega}{(2\pi)^n}
\prod_{i=1}^n\exp\left\{j\omega_i\left(\sum_\ell
J_{i\ell}s_\ell-\lambda_is_i\right)\right\}\nonumber\\
&=&\left(\frac{n}{2\pi
J^2}\right)^{n(n-1)/4}\int_{\reals^{n(n-1)/2}}\dd\bJ\sum_{\bs}
\int_K^\infty\cdot\cdot\cdot\int_K^\infty\dd\blambda \times\nonumber\\
&&\int_{\reals^n}\frac{\dd\bomega}{(2\pi)^n}
\exp\left\{-\frac{n}{2J^2}\sum_{i>\ell}J_{i\ell}^2+j\sum_{i>\ell}J_{i\ell}
(\omega_is_\ell+\omega_{\ell}s_i)-j\sum_i\omega_is_i\lambda_i\right\}
\end{eqnarray}
We now use the Hubbard--Stratonovich transform:
\begin{equation}
\int_\reals\dd x e^{ax^2+bx}\equiv \sqrt{\frac{\pi}{a}}e^{b^2/(4a)}
\end{equation}
with $a=n/(2J^2)$ and $b=\omega_is_\ell+\omega_\ell s_i$:
\begin{equation}
\bar{N}(n)=\sum_{\bs}\int_K^\infty\cdot\cdot\cdot\int_K^\infty\dd\blambda
\int_{\reals^n}\frac{\dd\bomega}{(2\pi)^n}\prod_{i=1}^n
e^{-j\omega_is_i\lambda_i}\prod_{i>\ell}\exp\{-(\omega_is_\ell+\omega_\ell
s_i)^2J^2/(2n)\}.
\end{equation}
Next observe that the summand doesn't actually depend on $\bs$ because each
$s_i$ is multiplied by an integration variable that runs over $\reals$ and
thus the sign of $s_i$ may be absorbed by this integration variable anyhow
(exercise: convince yourself). Thus, all $2^n$ contributions are the same as
that of $\bs=(+1,\ldots,+1)$:
\begin{equation}
\bar{N}(n)=2^n\int_K^\infty\cdot\cdot\cdot\int_K^\infty\dd\blambda
\int_{\reals^n}\frac{\dd\bomega}{(2\pi)^n}\prod_{i=1}^n
e^{-j\omega_i\lambda_i}\prod_{i>\ell}\exp\{-(\omega_i+\omega_\ell
)^2J^2/(2n)\}.
\end{equation}
Now, consider the following identity (exercise: prove it):
\begin{equation}
\frac{J^2}{2n}\sum_{i>\ell}(\omega_i+\omega_\ell)^2=J^2\frac{(n-1)}{2n}\sum_i\omega_i^2+
\frac{J^2}{n}\sum_{i>\ell}\omega_i\omega_\ell,
\end{equation}
and so for large $n$,
\begin{equation}
\frac{J^2}{2n}\sum_{i>\ell}(\omega_i+\omega_\ell)^2\approx
\frac{J^2}{2}\sum_i\omega_i^2+
\frac{J^2}{n}\sum_{i>\ell}\omega_i\omega_\ell\approx
\frac{J^2}{2}\sum_i\omega_i^2+\frac{J^2}{2n}\left(\sum_{i=1}^n\omega_i\right)^2.
\end{equation}
thus
\begin{equation}
\bar{N}(n)\approx 2^n\int_K^\infty\cdot\cdot\cdot\int_K^\infty\dd\blambda
\int_{\reals^n}\frac{\dd\bomega}{(2\pi)^n}\prod_{i=1}^n
\exp\left\{-j\omega_i\lambda_i-\frac{J^2}{2}\sum_{i=1}^n\omega_i^2-\frac{J^2}{2n}
\left(\sum_{i=1}^n\omega_i\right)^2\right\}.
\end{equation}
We now use again the Hubbard--Stratonovich transform
\begin{equation}
e^{a^2}\equiv\int_\reals\frac{\dd t}{2\pi} e^{j\sqrt{2}at-t^2/2}
\end{equation}
and then, after changing variables $\lambda_i\to J\lambda_i$ and 
$J\omega_i\to\omega_i$ (exercise: show that), we get:
\begin{equation}
\bar{N}(n)\approx\frac{1}{\pi^n}\cdot\frac{1}{\sqrt{2\pi}}
\int_{K/J}^\infty\cdot\cdot\cdot\int_{K/J}^\infty\dd\blambda
\int_\reals \dd t e^{-t^2/2}\prod_{i=1}^n\left[\int_\reals\dd\omega_i
\exp\left\{j\omega_i\left(-\lambda_i+\frac{t}{\sqrt{n}}\right)-\frac{1}{2}\sum_{i=1}^n
\omega_i^2\right\}\right]
\end{equation}
which after changing $t/\sqrt{n}\to t$, becomes
\begin{eqnarray}
\bar{N}(n)&\approx&\frac{1}{\pi^n}\cdot\frac{n}{\sqrt{2\pi}}
\int_\reals \dd t
e^{-nt^2/2}\left[\int_{K/\lambda}^\infty\dd\lambda\int_\reals \dd\omega
e^{j\omega(t-\lambda)-\omega^2/2}\right]^n\nonumber\\
&=&\frac{1}{\pi^n}\cdot\frac{n}{\sqrt{2\pi}}
\int_\reals \dd t
e^{-nt^2/2}\left[\sqrt{2\pi}\int_{K/\lambda}^\infty\dd\lambda
e^{-(t-\lambda)^2/2}\right]^n~~~~~\mbox{(again, the H--S identity)}\nonumber\\
&=&\frac{1}{\pi^n}\cdot\frac{n}{\sqrt{2\pi}}
\int_\reals \dd t
e^{-n(t+K/J)^2/2}\left[\sqrt{2\pi}\int_{-\infty}^t\dd\lambda
e^{-\lambda^2/2}\right]^n~~~~~~~t\to t+K/J,~\lambda\to
-\lambda+t+K/J\nonumber\\
&=&\frac{1}{\pi^n}\cdot\frac{n}{\sqrt{2\pi}}
\int_\reals \dd t
e^{-n(t+K/J)^2/2}\cdot [2\pi(1-Q(t))]^n\nonumber\\
&=&\frac{n}{\sqrt{2\pi}}
\int_\reals \dd t
\exp\left\{-\frac{n}{2}(t+K/J)^2+\ln[2(1-Q(t))]\right\}\nonumber\\
&\exe&\exp\left\{n\cdot\max_t\left[\ln(2(1-Q(t))-\frac{(t+K/J)^2}{2}\right]\right\}~~~
\mbox{Laplace integration}\nonumber
\end{eqnarray}
The maximizing $t$ zeroes out the derivative, i.e., it solves the equation
\begin{equation}
\frac{e^{-t^2/2}}{\sqrt{2\pi}[1-Q(t)]}=t+\frac{K}{J}
\end{equation}
which for $K=0$, gives exactly the asserted result about the capacity.


\newpage
\subsection{Generalized Temperature, de Bruijn's Identity, and Fisher
Information}

Earlier, we defined temperature by
\begin{equation}
\frac{1}{T}=\left(\frac{\partial S}{\partial E}\right)_V.
\end{equation}
This definition corresponds to equilibrium. We now describe a generalized
definition that is valid also for non--equilibrium situations, and see how
it relates to concepts in information theory and estimation 
theory, like the Fisher information. The derivations here follow the paper:
K.~R.~Narayanan and A.~R.~Srinivasa, ``On the thermodynamic temperature of a
general 
distribution,'' arXiv:0711.1460v2 [cond-mat.stat-mech], Nov.\ 10, 2007.

As we know, when the Hamiltonian is quadratic $\calE(x)=\frac{\alpha}{2}x^2$, 
the Boltzmann distribution is Gaussian:
\begin{equation}
P(\bx)=\frac{1}{Z}\exp\left\{-\beta\cdot\frac{\alpha}{2}\sum_{i=1}^nx_i^2\right\}
\end{equation}
and by the equipartition theorem:
\begin{equation}
\bar{E}(P)\dfn\left<\frac{\alpha}{2}\sum_{i=1}^nX_i^2\right>_P=n\frac{kT}{2}.
\end{equation}
We also computed the entropy, which is nothing but the entropy of a Gaussian
vector $S(P)=\frac{nk}{2}\ln(\frac{2\pi e}{\alpha\beta}).$
Consider now another probability density function $Q(\bx)$, which means a
non--equilibrium probability law if it differs from $P$, and let's look
also at the energy and the entropy pertaining to $Q$:
\begin{equation}
\bar{E}(Q)=\left<\frac{\alpha}{2}\sum_{i=1}^nX_i^2\right>_Q=\int \dd \bx Q(\bx)
\cdot\left[\frac{\alpha}{2}\sum_{i=1}^nx_i^2\right]
\end{equation}
\begin{equation}
S(Q)=k\cdot\langle-\ln Q(\bX)\rangle_Q=-k\int \dd \bx Q(\bx)\ln Q(\bx).
\end{equation}
In order to define a notion of generalized temperature, we have to define some
sort of derivative of $S(Q)$ w.r.t. $\bar{E}(Q)$. This definition could make
sense if it turns out that the ratio between the response of $S$ to
perturbations in $Q$ and the response of $\bar{E}$ to the same perurbations,
is independent of the ``direction'' of this perturbation, as long as it is
``small'' in some reasonable sense. It turns out the de Bruijn identity helps
us here.

Consider now the perturbation of $\bX$ by $\sqrt{\delta}\bZ$ thus
defining the perturbed version of $\bX$ as $\bX_\delta=\bX+
\sqrt{\delta}\bZ$, where $\delta > 0$ is small and $\bZ$ is an {\it arbitrary}
i.i.d.\ zero--mean 
random vector, {\it not necessarily Gaussian}, whose components all have unit
variance.
Let $Q_\delta$ denote the
density of $\bX_\delta$ (which is, of course, the convolution between $Q$ and
the density
of $Z$, scaled by $\sqrt{\delta}$). The proposed generalized
definition of temperature is:
\begin{equation}
\frac{1}{T}\dfn \lim_{\delta\to 0}
\frac{S(Q_\delta)-S(Q)}{\bar{E}(Q_\delta)-\bar{E}(Q)}.
\end{equation}
The denominator is easy since
\begin{equation}
\bE\|\bX+\sqrt{\delta}\bZ\|^2-\bE\|\bX\|^2=2\sqrt{\delta}\bE\bX^T\bZ+n\delta=n\delta
\end{equation}
and so, $\bar{E}(Q_\delta)-\bar{E}(Q)=n\alpha\delta/2$.
In view of the above, our new definition of temperature becomes:
\begin{equation}
\frac{1}{T}\dfn \frac{2k}{n\alpha}\cdot\lim_{\delta\to
0}\frac{h(\bX+\sqrt{\delta}\bZ)-h(\bX)}{\delta}=\frac{2k}{n\alpha}\cdot\frac{\partial
h(\bX+\sqrt{\delta}\bZ)}{\partial\delta}\bigg|_{\delta=0}.
\end{equation}
First, it is important to understand
that the numerator of the middle expression is positive (and hence so is $T$) since
\begin{equation}
S(Q_\delta)=kh(\bX+\sqrt{\delta}\bZ)\ge
kh(\bX+\sqrt{\delta}\bZ|\bZ)=kh(\bX)=S(Q).
\end{equation}
In order to move forward from this point, we will need a piece of background.
A well--known notion from estimation theory is the {\it Fisher information},
which is the basis for the Cram\'er--Rao bound for unbiased parameter
estimators:
Suppose we have a family of pdf's $\{Q_\theta (x)\}$ where $\theta$ is a
continuous valued parameter. The Fisher info is defined as
\begin{equation}
J(\theta)=\bE_\theta\left\{\left[\frac{\partial
\ln Q_\theta(X)}{\partial\theta}\right]^2\right\}=\int_{-\infty}^{+\infty}
\frac{\dd
x}{Q_\theta(x)}\left[\frac{\partial}{\partial\theta}Q_\theta(x)\right]^2.
\end{equation}
Consider now the special case where $\theta$ is a translation parameter,
i.e., $Q_\theta(x)=Q(x-\theta)$, then
\begin{eqnarray}
J(\theta)&=&
\int_{-\infty}^{+\infty}
\frac{\dd
x}{Q(x-\theta)}\left[\frac{\partial}{\partial\theta}Q(x-\theta)\right]^2\nonumber\\
&=&\int_{-\infty}^{+\infty}
\frac{\dd
x}{Q(x-\theta)}\left[\frac{\partial}{\partial x}Q(x-\theta)\right]^2~~~~~
\frac{\partial Q(x-\theta)}{\partial x}=-
\frac{\partial Q(x-\theta)}{\partial \theta}
\nonumber\\
&=&\int_{-\infty}^{+\infty}
\frac{\dd
x}{Q(x)}\left[\frac{\partial}{\partial x}Q(x)\right]^2\nonumber\\
&\dfn& J(Q)~~~~\mbox{with a slight abuse of notation.}\nonumber
\end{eqnarray}
independently of $\theta$.
For the vector case, we define the Fisher info matrix, whose elements are
\begin{equation}
J_{ij}(Q)=\int_{\reals^n}\frac{\dd\bx}{Q(\bx)}\left[\frac{\partial 
Q(\bx)}{\partial x_i}\cdot \frac{\partial Q(\bx)}{\partial
x_j}\right]~~~i,j=1,\ldots,n.
\end{equation}
Shortly, we will relate $T$ with the trace of this matrix.

To this end, we will need the following result, which is a variant of the
well--known {\it de Bruijn identity}, first for the scalar case: Let $Q$ be the pdf of a scalar RV $X$
of finite variance. Let $Z$ be a unit variance RV which is symmetric around
zero, and let $X_\delta=X+\sqrt{\delta}Z$. Then,
\begin{equation}
\frac{\partial h(X+\sqrt{\delta} Z)}{\partial\delta}\bigg|_{\delta=0}=\frac{J(Q)}{2}.
\end{equation}
The original de Bruijn identity allows only a Gaussian perturbation $Z$, but it
holds for any $\delta$. Here, on the other hand, we allow an 
arbitrary density $M(z)$ of $Z$,
but we insist on $\delta\to 0$. The proof of this result is essentially
similar to the proof of the original result, which can be found, for example,
in the book by Cover and Thomas: 
Consider the characteristic functions:
\begin{equation}
\Phi_X(s)=\int_{-\infty}^{+\infty}\dd x e^{sx}Q(x)
\end{equation}
and
\begin{equation}
\Phi_Z(s)=\int_{-\infty}^{+\infty}\dd z e^{sz}M(z).
\end{equation}
Due to the independence
\begin{eqnarray}
\Phi_{X_\delta}(s)&=&\Phi_X(s)\cdot\Phi_{\sqrt{\delta}Z}(s)\nonumber\\
&=&\Phi_X(s)\cdot\Phi_{Z}(\sqrt{\delta}s)\nonumber\\
&=&\Phi_X(s)\cdot\int_{-\infty}^{+\infty}\dd z
e^{\sqrt{\delta}sz}M(z)\nonumber\\
&=&\Phi_X(s)\cdot\sum_{i=0}^\infty\frac{(\sqrt{\delta}s)^i}{i!}\mu_i(M)
~~~~\mu_i(M)~\mbox{being the $i$--th moment of $Z$}\nonumber\\
&=&\Phi_X(s)\cdot\left(1+\frac{\delta
s^2}{2}+\cdot\cdot\cdot\right)~~~\mbox{odd moments vanish due to
symmetry}\nonumber
\end{eqnarray}
Applying the inverse Fourier transform, we get:
\begin{equation}
Q_\delta(x)=Q(x)+\frac{\delta}{2}\cdot\frac{\partial^2Q(x)}{\partial x^2}+o(\delta),
\end{equation}
and so,
\begin{equation}
\frac{\partial Q_\delta(x)}{\partial\delta}\bigg|_{\delta=0}=
\frac{1}{2}\cdot\frac{\partial^2Q(x)}{\partial x^2}\sim
\frac{1}{2}\cdot\frac{\partial^2Q_\delta(x)}{\partial x^2}.
\end{equation}
Now, let's look at the entropy:
\begin{equation}
h(X_\delta)=-\int_{-\infty}^{+\infty}\dd x Q_\delta(x)\ln Q_\delta(x).
\end{equation}
Taking the derivative w.r.t.\ $\delta$, we get:
\begin{eqnarray}
\frac{\partial h(X_\delta)}{\partial\delta}&=&
-\int_{-\infty}^{+\infty}\dd x\left[\frac{\partial
Q_\delta(x)}{\partial\delta}+\frac{\partial Q_\delta(x)}{\partial\delta}\cdot
\ln Q_\delta(x)\right]\nonumber\\
&=&-\frac{\partial}{\partial\delta}\int_{-\infty}^{+\infty}\dd x Q_\delta(x)-
\int_{-\infty}^{+\infty}\dd x\frac{\partial
Q_\delta(x)}{\partial\delta}\cdot\ln Q_\delta(x)\nonumber\\
&=&-\frac{\partial}{\partial\delta} 1-
\int_{-\infty}^{+\infty}\dd x\frac{\partial
Q_\delta(x)}{\partial\delta}\cdot\ln Q_\delta(x)\nonumber\\
&=&-\int_{-\infty}^{+\infty}\dd x\frac{\partial
Q_\delta(x)}{\partial\delta}\cdot\ln Q_\delta(x)
\end{eqnarray}
and so, 
\begin{equation}
\frac{\partial h(X_\delta)}{\partial\delta}\bigg|_{\delta=0}=
-\int_{-\infty}^{+\infty}\dd x\cdot\frac{\partial
Q_\delta(x)}{\partial\delta}\bigg|_{\delta=0}\cdot\ln Q(x)
=-\int_{-\infty}^{+\infty}\dd x\cdot\frac{1}{2}\frac{\dd^2 Q(x)}{\dd^2 x}\cdot\ln
Q(x).
\end{equation}
Integrating by parts, we obtain:
\begin{equation}
\frac{\partial h(X_\delta)}{\partial\delta}\bigg|_{\delta=0}=
\left[-\frac{1}{2}\cdot\frac{\dd Q(x)}{\dd x}\cdot \ln
Q(x)\right]_{-\infty}^{+\infty}+\frac{1}{2}\int_{-\infty}^{+\infty}
\frac{\dd x}{Q(x)}\left[\frac{\partial Q(x)}{\partial x}\right]^2.
\end{equation}
The first term can be shown to vanish (see paper and/or C\&T) and the
second term is exactly $J(Q)/2$. This completes the proof of the
(modified) de Bruijn identity.

\vspace{0.2cm}

\noindent
{\bf Exercise}: Extend this to the vector case, showing that
for a vector $\bZ$ with i.i.d.\ components, all symmetric around the origin:
\begin{equation}
\frac{\partial
h(\bX+\sqrt{\delta}\bZ)}{\partial\delta}=\frac{1}{2}\sum_{i=1}^n
\int_{\reals^n}\frac{\dd\bx}{Q(\bx)}\left[\frac{\partial Q(\bx)}{\partial
x_i}\right]^2=\frac{1}{2}\sum_{i=1}^n
J_{ii}(Q)=\frac{1}{2}\mbox{tr}\{J(Q)\}.~~~~\Box
\end{equation}

Putting all this together, we end up with the following generalized definition of
temperature:
\begin{equation}
\frac{1}{T}=\frac{k}{n\alpha}\cdot\mbox{tr}\{J(Q)\}.
\end{equation}
In the `stationary' case, where $Q$ is symmetric w.r.t.\ all
components of $\bx$, $\{J_{ii}\}$ are all the same quantity, call it
$J(Q)$, and then
\begin{equation}
\frac{1}{T}=\frac{k}{\alpha}\cdot J(Q)
\end{equation}
or, equivalently,
\begin{equation}
T=\frac{\alpha}{kJ(Q)}=\frac{\alpha}{k}\cdot\mbox{CRB}
\end{equation}
where CRB is the Cram\'er--Rao bound. High temperature means a lot of noise
and this in turn means that it is hard to estimate the mean of $X$.
In the Boltzmann case, $J(Q)=1/\mbox{Var}\{X\}=\alpha\beta=\alpha/(kT)$
and we are back to the ordinary definition of temperature.

Another way to look at this result is as an extension of the equipartition
theorem: As we recall, in the ordinary case of a quadratic Hamiltonian and
in equilibrium, we have:
\begin{equation}
\langle\calE(X)\rangle =\left<\frac{\alpha}{2}X^2\right>=\frac{kT}{2}
\end{equation}
or
\begin{equation}
\frac{\alpha}{2}\sigma^2\dfn \frac{\alpha}{2}\langle X^2\rangle=\frac{kT}{2}.
\end{equation}
In the passage to the more general case, $\sigma^2$ should be replaced by
$1/J(Q)=\mbox{CRB}$. Thus, the induced generalized equipartition function,
doesn't talk about average energy but about the CRB:
\begin{equation}
\frac{\alpha}{2}\cdot \mbox{CRB}=\frac{kT}{2}.
\end{equation}
Now, the CRB is a lower bound to the estimation error which, in this case, is
a transaltion parameter. For example, let $x$ denote the location of a mass
$m$ tied to a spring of strength $m\omega_0^2$ and equilibrium location
$\theta$. Then, 
\begin{equation}
\calE(x)=\frac{m\omega_0^2}{2}(x-\theta)^2.
\end{equation}
In this case, $\alpha=m\omega_0^2$, and we get:
\begin{equation}
\mbox{estimation error
energy}=\frac{m\omega_0^2}{2}\cdot\bE(\hat{\theta}(X)-\theta)^2\ge \frac{kT}{2}
\end{equation}
where $\hat{\theta}(X)$ is any unbiased estimator of $\theta$ based on a
measurement of $X$. This is to say that the generalized equipartition theorem
talks about the estimation error energy in the general case. Again, in the
Gaussian case, the best estimator is $\hat{\theta}(x)=x$ and we are back to
ordinary energy and the ordinary equipartition theorem.


\newpage
\subsection{The Gibbs Inequality and the Log--Sum Inequality}

In one of our earlier meetings, we have seen the Gibbs' inequality,
its physical significance, and related it to the second law and the DPT.
We now wish to take another look at the Gibbs' inequality, from a completely
different perspective, namely, as a tool for generating useful bounds
on the free energy, in situations where the exact calculation is difficult
(see Kardar's book, p.\ 145).
As we show in this part, this inequality is nothing else than the {\it
log--sum inequality}, which is used in Information Theory, mostly
for proving certain {\it qualitative} properties of information measures,
like the data processing theorem of the divergence, etc. But this equivalence
now suggests that the log--sum inequality can perhaps be used in a similar
way that it is used in physics, and then it could perhaps yields useful bounds
on certain information measures. We try to demonstrate this point here.

Suppose we have an
Hamiltonian $\calE(\bx)$ for which we wish to know the partition function
\begin{equation}
Z(\beta)=\sum_{\bx} e^{-\beta\calE(\bx)}
\end{equation}
but it is hard, if not impossible, to calculate in closed--form.
Suppose further that for another, somewhat different Hamiltonian,
$\calE_0(\bx)$, it is rather easy to make calculations. The Gibbs' inequality
can be presented as a lower bound
on $\ln Z(\beta)$ in terms of B--G statistics pertaining to $\calE_0$.
\begin{equation}
\ln \left[\sum_{\bx}e^{-\beta\calE(\bx)}\right]\ge
\ln \left[\sum_{\bx}e^{-\beta\calE_0(\bx)}\right]+
\beta\left<\calE_0(\bX)-\calE(\bX)\right>_0,
\end{equation}
The idea now is that we can obtain
pretty good bounds thanks to the fact that we may have some freedom in the
choice of $\calE_0$. For example, one can define a parametric family of
functions $\calE_0$ and maximize the r.h.s.\ w.r.t.\ the parameter(s) of this
family, thus obtaining the tightest lower bound
within the family. We next demonstrate this with an example:

\vspace{0.2cm}

\noindent
{\it Example -- Non--harmonic oscillator.}
Consider the potential function 
\begin{equation}
V(z)=Az^4
\end{equation}
and so
\begin{equation}
\calE(x)=\frac{p^2}{2m}+Az^4,
\end{equation}
where we approximate the second term by
\begin{equation}
V_0(z)=\left\{\begin{array}{ll}
0 & |z|\le\frac{L}{2}\\
+\infty & |z|>\frac{L}{2}\end{array}\right.
\end{equation}
where $L$ is a parameter to be optimized.
Thus,
\begin{eqnarray}
Z_0&=&\frac{1}{h}\int_{-\infty}^{+\infty}\dd p\int_{-\infty}^{+\infty}\dd z
e^{-\beta[V_0(z)+p^2/(2m)]}\nonumber\\
&=&\frac{1}{h}\int_{-\infty}^{+\infty}\dd p\cdot e^{-\beta p^2/(2m)}\int_{-L/2}^{+L/2}\dd z
\nonumber\\
&=&\frac{\sqrt{2\pi mkT}}{h}\cdot L\nonumber
\end{eqnarray}
and so, by the Gibbs inequality:
\begin{eqnarray}
\ln Z&\ge&\ln Z_0+\beta\langle\calE_0(\bX)-\calE(\bX)\rangle_0\nonumber\\
&\ge&\ln Z_0-\frac{1}{kT}\cdot\frac{1}{L}\int_{-L/2}^{+L/2}\dd z \cdot Az^4\nonumber\\
&\ge&\ln \left[\frac{L\sqrt{2\pi mkT}}{h}\right] -\frac{AL^4}{80kT}\nonumber\\
&\dfn& f(L)\nonumber
\end{eqnarray}
To maximize $f(L)$ we equate its derivative to zero:
\begin{equation}
0=\frac{\dd f}{\dd L}\equiv \frac{1}{L}-\frac{AL^3}{20kT}~\Longrightarrow~
L^*=\left(\frac{20kT}{A}\right)^{1/4}.
\end{equation}
Plugging this back into the Gibbs lower bound and comparing to
the {\it exact} value of $Z$ (which is still computable in this example), we
find that $Z_{\mbox{approx}}\approx 0.91 Z_{\mbox{exact}}$, which is not that
bad considering the fact that the infinite potential well seems to be quite a
poor approximation to the fourth order power law potential $V(z)=Az^4$.

As somewhat better approximation is the harmonic one:
\begin{equation}
V_0(z)= \frac{m\omega_0^2}{2}\cdot z^2
\end{equation}
where now $\omega_0$ is the free parameter to be optimized. This gives
\begin{equation}
Z_0=\frac{1}{h}\int_{-\infty}^{+\infty}\dd p\int_{-\infty}^{+\infty}\dd z
e^{-\beta[m\omega_0^2z^2/2+p^2/(2m)]}=\frac{kT}{\hbar\omega_0}~~~~~~~~\hbar
=\frac{h}{2\pi}
\end{equation}
and this time, we get:
\begin{eqnarray}
\ln Z&\ge& \ln\left(\frac{kT}{\hbar\omega_0}\right)+\frac{1}{kT}\left<
\frac{m\omega_0^2Z^2}{2}-AZ^2\right>_0\nonumber\\
&=&
\ln\left(\frac{kT}{\hbar\omega_0}\right)+\frac{1}{2}-\frac{3AkT}{m^2\omega_0^4}\nonumber\\
&\dfn& f(\omega_0)\nonumber
\end{eqnarray}
Maximizing $f$:
\begin{equation}
0=\frac{\dd f}{\dd
\omega_0}\equiv -\frac{1}{\omega_0}+\frac{12AkT}{m^2\omega_0^5}~
\Longrightarrow~\omega_0^*=\frac{(12AkT)^{1/4}}{\sqrt{m}}.
\end{equation}
This time, we get $Z_{\mbox{approx}}\approx 0.95 Z_{\mbox{exact}}$, i.e.,
this approximation is even better. $\Box$

So much for physics. Let's look now at the Gibbs inequality slightly
differently. What we actually did, in a nutshell, and in
different notation, is the following: Consider the function:
\begin{equation}
Z(\lambda)=\sum_{i=1}^n a_i^{1-\lambda}b_i^\lambda=\sum_{i=1}^n
a_ie^{-\lambda\ln(a_i/b_i)},
\end{equation}
where $\{a_i\}$ and $\{b_i\}$ are positive reals. 
Since $\ln Z(\lambda)$ is convex (as before), we have:
\begin{eqnarray}
\ln\left(\sum_{i=1}^n b_i\right)&\equiv&\ln Z(1)\nonumber\\
&\ge&\ln Z(0)+1\cdot \frac{\dd\ln
Z(\lambda)}{\dd\lambda}\bigg|_{\lambda=0}\nonumber\\
&=&\ln\left(\sum_{i=1}^n a_i\right)+\frac{\sum_{i=1}^n
a_i\ln(b_i/a_i)}{\sum_{i=1}^n a_i}\nonumber
\end{eqnarray}
which is nothing but the log--sum inequality, which in IT, is more customarily
written as:
\begin{equation}
\sum_{i=1}^n a_i\ln\frac{a_i}{b_i}\ge \left(\sum_{i=1}^n a_i\right)\cdot\ln
\frac{\sum_{i=1}^n a_i}{\sum_{i=1}^n b_i}.
\end{equation}
Returning to the form:
\begin{equation}
\ln\left(\sum_{i=1}^n b_i\right)\ge 
\ln\left(\sum_{i=1}^n a_i\right)+\frac{\sum_{i=1}^n
a_i\ln(b_i/a_i)}{\sum_{i=1}^n a_i},
\end{equation}
the idea now is, once again, to lower bound an expression
$\ln(\sum_{i=1}^n b_i)$ which may be hard to calculate, by
the expression on the l.h.s.\ which is hopefully easier, and
allows a degree of freedom concerning the choice of $\{a_i\}$,
at least in accordance to some structure, and depending on a limited
set of parameters.

Consider, for example, a hidden Markov model (HMM), which is the output
of a DMC $W(\by|\bx)=\prod_{t=1}^n W(y_t|x_t)$ fed by a first--order
Markov process $\bX$, governed by $Q(\bx)=\prod_{t=1}^n Q(x_t|x_{t-1})$.
The entropy rate of the hidden Markov process  $\{Y_t\}$ does not admit
a closed--form expression, so we would like to have at least good bounds. 
Here, we propose an upper bound that stems from the Gibbs inequality, or
the log--sum inequality. 

The probability distribution of $\by$ is
\begin{equation}
P(\by)=\sum_{\bx}\prod_{t=1}^n[ W(y_t|x_t)Q(x_t|x_{t-1})].
\end{equation}
This summation does not lend itself to a nice closed--form 
expression, but if the $t$--th factor depended only on $t$ (and not also
on $t-1$) life would have been easy and simple as the sum of products would have
boiled down to a product of sums. So this motivates the following use of the
log--sum inequality: For a given $\by$,
let's think of $\bx$ as the index $i$ of the log--sum inequality and
then 
\begin{equation}
b(\bx)=\prod_{t=1}^n[ W(y_t|x_t)Q(x_t|x_{t-1})].
\end{equation}
Let us now define
\begin{equation}
a(\bx)=\prod_{t=1}^n P_0(x_t,y_t),
\end{equation}
where $P_0$ is an arbitrary joint distribution over $\calX\times\calY$,
to be optimized eventually.
Thus, applying the log--sum inequality, we get:
\begin{eqnarray}
\ln P(\by)&=&\ln\left(\sum_{\bx}b(\bx)\right)\nonumber\\
&\ge&\ln\left(\sum_{\bx}a(\bx)\right)+\frac{\sum_{\bx}a(\bx)
\ln[b(\bx)/a(\bx)]}{\sum_{\bx}a(\bx)}\nonumber\\
&=&\ln\left(\sum_{\bx}\prod_{t=1}^n P_0(x_t,y_t)\right)+\nonumber\\
& &+\frac{\sum_{\bx}\left[\prod_{t=1}^n P_0(x_t,y_t)\right]\cdot\ln[
\prod_{t=1}^n[Q(x_t|x_{t-1})W(y_t|x_t)/P_0(x_t,y_t)]}{\sum_{\bx}\prod_{t=1}^n
P_0(x_t,y_t)}.
\end{eqnarray}
Now, let us denote $P_0(y)=\sum_{x\in\calX}P_0(x,y)$, which is
the marginal of $y$ under $P_0$. Then, the first term is
simply $\sum_{t=1}^n\ln P_0(y_t)$. As for the second term, we have:
\begin{eqnarray}
&&
\frac{\sum_{\bx}\left[\prod_{t=1}^n P_0(x_t,y_t)\right]\cdot\ln[
\prod_{t=1}^n[Q(x_t|x_{t-1})W(y_t|x_t)/P_0(x_t,y_t)]}{\sum_{\bx}\prod_{t=1}^n
P_0(x_t,y_t)}\nonumber\\
&=&\sum_{t=1}^n \sum_{\bx}\frac{\prod_{t=1}^n P_0(x_t,y_t)\ln
[Q(x_t|x_{t-1})W(y_t|x_t)/P_0(x_t,y_t)]}{\prod_{t=1}^n
P_0(y_t)}\nonumber\\
&=&\sum_{t=1}^n \frac{\prod_{t'\ne t-1,t}P_0(y_{t'})}{\prod_{t=1}^nP_0(y_t)}\cdot
\sum_{x_{t-1},x_t}
P_0(x_{t-1},y_{t-1})P_0(x_t,y_t)\cdot
\ln\left[\frac{Q(x_t|x_{t-1})W(y_t|x_t)}{P_0(x_t,y_t)}\right]\nonumber\\
&=&\sum_{t=1}^n \sum_{x_{t-1},x_t}\frac{P_0(x_{t-1},y_{t-1})P_0(x_t,y_t)}{
P_0(y_{t-1})P_0(y_t)}\cdot
\ln\left[\frac{Q(x_t|x_{t-1})W(y_t|x_t)}{P_0(x_t,y_t)}\right]\nonumber\\
&=&\sum_{t=1}^n \sum_{x_{t-1},x_t}P_0(x_{t-1}|y_{t-1})P_0(x_t|y_t)
\cdot\ln\left[\frac{Q(x_t|x_{t-1})W(y_t|x_t)}{P_0(x_t,y_t)}\right]\nonumber\\
&\dfn&\sum_{t=1}^n \bE_0\left\{
\ln\left[\frac{Q(X_t|X_{t-1})W(y_t|X_t)}{P_0(X_t,y_t)}
\right]\bigg|Y_{t-1}=y_{t-1},Y_t=y_t\right\}\nonumber
\end{eqnarray}
where $\bE_0$ denotes expectation w.r.t.\ the product measure of $P_0$.
Adding now the first term of the r.h.s.\ of the log--sum inequality,
$\sum_{t=1}^n\ln P_0(y_t)$, we end up with the lower bound:
\begin{equation}
\ln P(\by)\ge\sum_{t=1}^n
\bE_0\left\{\ln\left[\frac{Q(X_t|X_{t-1})W(y_t|X_t)}{P_0(X_t|y_t)}
\right]\bigg|Y_{t-1}=y_{t-1},Y_t=y_t\right\}\dfn
\sum_{t=1}^n\Delta(y_{t-1},y_t;P_0).
\end{equation}
At this stage, we can perform the optimization over $P_0$ for each $\by$
individually, and then derive the bound on the expectation of $\ln P(\by)$
to get a bound on the entropy. Note, however, that
$\sum_t\Delta(y_{t-1},y_t;P_0)$ depends on $\by$ only via its Markov
statistics, i.e., the relative frequencies of transitions $y\Longrightarrow y'$ for all
$y,y'\in\calY$. Thus, the optimum $P_0$ depends on $\by$ also via these
statistics. Now, the expectation of $\sum_t\Delta(y_{t-1},y_t;P_0)$ is going
to be dominated by the typical $\{\by\}$ for which these transition counts
converge to the respective joint probabilities of $\{Y_{t-1}=y,~Y_t=y\}$. 
So, it is expected that for large $n$, nothing will essentially be lost if
we first take the expectation over both sides of the log--sum inequality and
only then optimize over $P_0$. This would give, assuming stationarity:
\begin{equation}
H(Y^n)\le -n\cdot\max_{P_0}\bE\{\Delta(Y_0,Y_1;P_0)\}.
\end{equation}
where the expectation on the r.h.s.\ is now under the {\it real}
joint distribution of two consecutive samples of $\{Y_n\}$, i.e.,
\begin{equation}
P(y_0,y_1)=\sum_{x_0,x_1}\pi(x_0)Q(x_1|x_0)P(y_0|x_0)P(y_1|x_1),
\end{equation}
where $\pi(\cdot)$ is the stationary distribution of the underlying Markov
process $\{x_t\}$.


\newpage
\subsection{Dynamics, Evolution of Info Measures, and Simulation}

The material here is taken mainly from the books by Reif, Kittel, and 
F.~P.~Kelly, {\it Reversibility and Stochastic Networks}, 
(Chaps 1--3), J.~Wiley \& Sons, 1979.

\subsubsection{Markovian Dynamics, Global Balance and Detailed Balance}

So far we discussed only physical systems in equilibrium. For these systems,
the Boltzmann--Gibbs distribution is nothing but the stationary distribution of
the microstate $x$ at every given time instant $t$. However, this is merely
one part of the picture. What is missing is the temporal probabilistic behavior, or 
in other words, the laws that underly the
evolution of the microstate with time. These are dictated by
dynamical properties of the system, which constitute
the underlying physical laws in the microscopic level. It is customary then to model
the microstate at time $t$ as a random process
$\{X_t\}$, where $t$ may denote either discrete time or continuous time,
and among the various models, one of the most common ones is the Markov
model. In this section, we discuss a few of the properties of these processes
as well as the evolution of information measures, like entropy, divergence
(and more) associated with them.

We begin with an isolated system in continuous time, which
is not necessarily assumed to have reached (yet) equilibrium. Let us suppose that
$X_t$, the
microstate at time $t$, can take on values in a discrete set $\calX$. For
$r,s\in\calX$, let
\begin{equation}
W_{rs}=\lim_{\delta\to
0}\frac{\mbox{Pr}\{X_{t+\delta}=s|X_t=r\}}{\delta}~~~~r\ne s
\end{equation}
in other words, $\mbox{Pr}\{X_{t+\delta}=s|X_t=r\}=W_{rs}\cdot \delta
+o(\delta)$. Letting $P_r(t)=\mbox{Pr}\{X_t=r\}$, it is easy to see that
\begin{equation}
P_r(t+\mbox{d}t)=\sum_{s\ne r}P_s(t)W_{sr}\mbox{d}t+P_r(t)\left(1-\sum_{s\ne
r}W_{rs}\mbox{d}t\right),
\end{equation}
where the first sum describes the probabilities of all possibile transitions from other states to
state $r$ and the second term describes the probability of not leaving state
$r$. Subtracting $P_r(t)$ from both sides and dividing by $\mbox{d}t$, we
immediately obtain the following set of differential equations:
\begin{equation}
\frac{\mbox{d}P_r(t)}{\mbox{d}t}=\sum_{s}[P_s(t)W_{sr}-P_r(t)W_{rs}],~~~r\in\calX,
\end{equation}
where $W_{rr}$ is defined in an arbitrary manner, e.g., $W_{rr}=0$ for all
$r$.
These equations are called the {\it master equations}.\footnote{Note that the
master equations apply in discrete time too, provided that the derivative at
the l.h.s.\ is replaced by a simple difference, $P_r(t+1)-P_r(t)$, and
$\{W_{rs}\}$ designate one--step state transition probabilities.}
When the process
reaches stationarity, i.e., for all $r\in\calX$, $P_r(t)$ converge to some $P_r$ that is
time--invariant, then
\begin{equation}
\sum_s[P_sW_{sr}-P_rW_{rs}]=0,~~~\forall~r\in\calX .
\end{equation}
This is called {\it global balance} or {\it steady state}. When the system is
isolated (microcanonical ensemble), the steady--state distribution must be
uniform, i.e., $P_r=1/|\calX|$ for all $r\in\calX$.
From quantum mechanical considerations, as well as considerations
pertaining to time reversibility in the microscopic level,\footnote{Think,
for example, of an isolated system 
of moving particles, obeying the differential
equations $m\mbox{d}^2\br_i(t)/\mbox{d}t^2=\sum_{j\ne i}F(\br_j(t)-\br_i(t))$, $i=1,2,\ldots,n$, which
remain valid if the time variable $t$ is replaced by $-t$ since
$\mbox{d}^2\br_i(t)/\mbox{d}t^2=\mbox{d}^2\br_i(-t)/\mbox{d}(-t)^2$.}
it is customary to
assume $W_{rs}=W_{sr}$ for all pairs $\{r,s\}$. We then observe that, not
only, $\sum_s[P_sW_{sr}-P_rW_{rs}]=0$, but moreover, each individual term in
the sum vanishes, as
\begin{equation}
P_sW_{sr}-P_rW_{rs}=\frac{1}{|\calX|}(W_{sr}-W_{rs})=0.
\end{equation}
This property is called {\it detailed balance}, which is stronger than global
balance, and it means equilibrium, which is stronger than steady state. While
both steady--state and equilibrium refer to a situation of time--invariant
state probabilities $\{P_r\}$, a steady--state still allows cyclic flows of
probability. For example, a Markov process with cyclic deterministic
transitions $1\to 2\to 3\to 1\to 2\to 3\to \cdot\cdot\cdot$ is in steady state
provided that the probability distribution of the initial state is uniform
$(1/3,1/3,1/3)$, however, the cyclic flow among the states is in one
direction. On the other hand, in detailed balance ($W_{rs}=W_{sr}$ for an
isolated system),
which is equilibrium, there
is no net flow in any cycle of states. All the net cyclic
probability fluxes vanish, and
therefore, time reversal would not change the probability law, that is,
$\{X_{-t}\}$ has the same probability law as $\{X_t\}$. For example,
if $\{Y_t\}$ is a Bernoulli process, taking values equiprobably in
$\{-1,+1\}$, then $X_t$ defined 
recursively by 
\begin{equation}
X_{t+1}=(X_t+Y_t)\mbox{mod} K,
\end{equation}
has a symmetric state--transition probability matrix $W$, a uniform stationary
state distribtuion, and it satisfies detailed balance.

\subsubsection{Evolution of Information Measures}

Returning to the case where the process 
$\{X_t\}$ pertaining to our isolated system has not necessarily reached
equilibrium,
let us take a look at the entropy of the state 
\begin{equation}
H(X_t)=-\sum_rP_r(t)\log
P_r(t).
\end{equation}
We argue that $H(X_t)$ is monotonically non--decreasing, which is in
agreement with the second law (a.k.a.\ the H--Theorem). To this end, 
we next show that 
\begin{equation}
\frac{\mbox{d}H(X_t)}{\mbox{d}t}\ge 0,
\end{equation}
where for convenience, we denote
$\mbox{d}P_r(t)/\mbox{d}t$ by $\dot{P}_r(t)$.
\begin{eqnarray}
\frac{\mbox{d}H(X_t)}{\mbox{d}t}&=&-\sum_r[\dot{P}_r(t)\log
P_r(t)+\dot{P}_r(t)]\nonumber\\
&=&-\sum_r\dot{P}_r(t)\log
P_r(t)~~~~~~~~~~~~~~\sum_r\dot{P}_r(t)=0\nonumber\\
&=&-\sum_r\sum_sW_{sr}[P_s(t)-P_r(t)]\log
P_r(t))~~~~~~~~~~~W_{sr}=W_{rs}\nonumber\\
&=&-\frac{1}{2}\sum_{r,s}W_{sr}[P_s(t)-P_r(t)]\log P_r(t)-\nonumber\\
& &\frac{1}{2}\sum_{s,r}W_{sr}[P_r(t)-P_s(t)]\log
P_s(t)\nonumber\\
&=&\frac{1}{2}\sum_{r,s}W_{sr}[P_s(t)-P_r(t)]\cdot[\log P_s(t)-\log
P_r(t)]\nonumber\\
&\ge& 0.
\end{eqnarray}
where the last inequality is due to the increasing monotonicity of the logarithmic
function: the product $[P_s(t)-P_r(t)]\cdot[\log P_s(t)-\log P_r(t)]$ cannot be
negative for any pair $(r,s)$, as the two factors of this product are either both negative,
both zero, or
both positive. Thus, $H(X_t)$ cannot decrease with time.

This result has a discrete--time analogue: If a finite--state Markov process
has a symmetric transition probability matrix, and so, the stationary
state distribution is uniform, then $H(X_t)$ is a monotonically
non--decreasing sequence. 

A considerably more general result is the following: If $\{X_t\}$
is a Markov process with a given state transition probability matrix
$W=\{W_{rs}\}$ (not necessarily symmetric) and $\{P_r\}$ is a stationary state
distribution, then the function
\begin{equation}
U(t)=\sum_rP_r\cdot V\left(\frac{P_r(t)}{P_r}\right)
\end{equation}
is monotonically strictly increasing provided that 
$V(\cdot)$ is strictly concave.
To see why this is true, we use the fact that $P_s=\sum_rP_rW_{rs}$ and define
$\tilde{W}_{sr}=P_rW_{rs}/P_s$. Obviously, $\sum_r\tilde{W}_{sr}=1$ for all
$s$, and so,
\begin{equation}
\frac{P_r(t+1)}{P_r}=\sum_s\frac{P_s(t)W_{sr}}{P_r}=\sum_s\frac{\tilde{W}_{rs}P_s(t)}{P_s}
\end{equation}
and so, by the concavity of $V(\cdot)$:
\begin{eqnarray}
U(t+1)&=&\sum_rP_r\cdot V\left(\frac{P_r(t+1)}{P_r}\right)\nonumber\\
&=&\sum_rP_r\cdot V\left(\sum_s\tilde{W}_{rs}\frac{P_s(t)}{P_s}\right)\nonumber\\
&>&\sum_r\sum_sP_r\tilde{W}_{rs}\cdot V\left(\frac{P_s(t)}{P_s}\right)\nonumber\\
&=&\sum_r\sum_sP_sW_{sr}\cdot V\left(\frac{P_s(t)}{P_s}\right)\nonumber\\
&=&\sum_sP_s\cdot V\left(\frac{P_s(t)}{P_s}\right)=U(t).
\end{eqnarray}
Here we required nothing except the existence of a stationary distribution.
Of course in the above derivation $t+1$ can be
replaced by $t+\tau$ for any positive real $\tau$ with the appropriate
transition probabilities, so the monotonicity of $U(t)$ applies to
continuous--time Markov processes as well.

Now, a few interesting choices of the function $V$ may be considered:
\begin{itemize}
\item
For $V(x)=-x\ln x$, we have $U(t)=-D(P(t)\|P)$. This means that
the divergence between $\{P_r(t)\}$ and the steady state distribution $\{P_r\}$
is monotonically strictly decreasing, whose physical interpretation could be
the decrease of the free energy, since we have already seen that the free
energy is the physical counterpart of the divergence.
This is a more general rule, that
governs not only isolated systems, but any Markov process with a stationary
limiting distribution (e.g., any Markov process whose distibution converges
to that of the Boltzmann--Gibbs distribution).
Having said that, if we now particularize this result to the case where
$\{P_r\}$ is the uniform distribution (as in an isolated system), then
\begin{equation}
D(P(t)\|P)=\log|\calX|-H(X_t),
\end{equation}
which means that the decrease of divergence is equivalent to the increase in
entropy, as before. The difference, however, is that 
here it is more general as 
we only required a uniform steady--state distribution, not
necessarily detailed balance.\footnote{For the uniform distribution to be a
stationary distribution, it is sufficient (and necessary) that $W$ would be a
doubly stochastic matrix, namely, $\sum_rW_{rs}=\sum_rW_{sr}=1$. This
condition is, of course, weaker than detailed balance, which means that $W$ is
moreover symmetric.}
\item
Another interesting choice of $V$ is $V(x)=\ln
x$, which gives $U(t)=-D(P\|P(t))$. Thus, $D(P\|P(t))$ is also monotonically
decreasing. In fact, both this and the monotonicity result of the previous item, are
in turn, special cases of a more general result concerning the divergence (see
also the book by Cover and Thomas, Section 4.4). Let
$\{P_r(t)\}$ and $\{P_r'(t)\}$ be two time--varying state--distributions
pertaining to the same Markov chain, but induced by two different initial state
distributions, $\{P_r(0)\}$ and $\{P_r'(0)\}$. Then $D(P(t)\|P'(t))$ is
monotonically non--increasing. This happens because
\begin{eqnarray}
D(P(t)\|P'(t))&=&\sum_rP_r(t)\log\frac{P_r(t)}{P_r'(t)}\nonumber\\
&=&\sum_{r,s}P_r(t)P(X_{t+\tau}=s|X_t=r)
\log\frac{P_r(t)P(X_{t+\tau}=s|X_t=r)}{P_r'(t)P(X_{t+\tau}=s|X_t=r)}\nonumber\\
&=&\sum_{r,s}P(X_t=r,~X_{t+\tau}=s)
\log\frac{P(X_t=r,~X_{t+\tau}=s)}{P'(X_t=r,~X_{t+\tau}=s)}\nonumber\\
&\ge&D(P(t+\tau)\|P'(t+\tau))
\end{eqnarray}
where the last inequality follows from the data processing theorem of the
divergence: the divergence between two joint distributions of $(X_t,X_{t+\tau})$ 
is never smaller than the
divergence between corresponding marginal distributions of $X_{t+\tau}$.
\item
Yet another choice is $V(x)=x^s$, where $s\in[0,1]$ is a
parameter. This would yield the increasing monotonicity of 
$\sum_r P_r^{1-s}P_r^{s}(t)$, a metric that plays a role in 
the theory of asymptotic
exponents of error probabilities pertaining to the optimum 
likelihood ratio test between two probability distributions. In particular,
the choice $s=1/2$ yields balance between the two kinds of error and it is
intimately related to the Bhattacharyya distance. Thus, we obtained some sorts
of generalizations of the second law to information measures other than
entropy.
\end{itemize}

\begin{figure}[ht]
\hspace*{2cm}\input{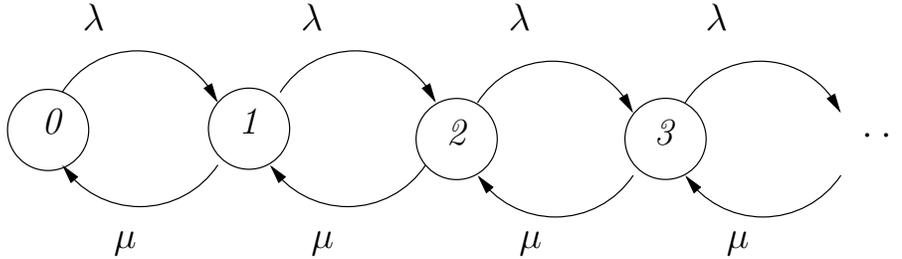}
\caption{\small State transition diagram of an M/M/1 queue.}
\label{mm1}
\end{figure}

For a general Markov process, whose steady state--distribution is not
necessarily uniform, the condition of detailed balance, which means
time--reversibility, reads
\begin{equation}
P_sW_{sr}=P_rW_{rs},
\end{equation}
both in discrete time and continuous time (with the corresponding meaning of
$\{W_{rs}\}$). The physical interpretation is that now our system is (a small)
part of a large isolated system, which obeys detailed balance w.r.t.\ the
uniform equilibrium distribution, as before.
A well known example of a process that obeys detailed balance in its more
general form
is an M/M/1 queue with an arrival rate $\lambda$ and service rate $\mu$
($\lambda < \mu$). Here, since all states are arranged along a line, with
bidirectional transitions between neighboring states only (see Fig.\
\ref{mm1}), there cannot be
any cyclic probability flux. The steady--state distibution is well--known
to be geometric
\begin{equation}
P_r=\left(1-\frac{\lambda}{\mu}\right)
\cdot\left(\frac{\lambda}{\mu}\right)^r, ~~~~~~r=0,1,2,\ldots,
\end{equation}
which indeed satisfies the detailed 
balance $P_r\lambda=P_{r+1}\mu$ for all $r$.
Thus, the Markov process $\{X_t\}$,
designating the number of customers in the queue at time $t$,
is time--reversible. 

It is interesting to point out that in order to check for the detailed balance property, one does
not necessarily have to know the equilibrium distribution $\{P_r\}$ as above.
Applying detailed balance to any $k$ pairs of states in a cycle, $(s_1,s_2)$,
$(s_2,s_3),\ldots, (s_k,s_1)$,
and multiplying the respective detailed balance equations, the steady state
probabilities cancel out and one easily obtains
\begin{equation}
W_{s_1s_2}W_{s_2s_3}\cdot\cdot\cdot W_{s_{k-1}s_k}W_{s_ks_1}=
W_{s_ks_{k-1}}W_{s_{k-1}s_{k-2}}\cdot\cdot\cdot W_{s_2s_1}W_{s_1s_k},
\end{equation}
so this is clearly a necessary condition for detailed balance. One can show
conversely, that if this equation applies to any finite cycle of states, then
the chain satisfies detailed balance, and so this is also a sufficient condition.
This is true both in discrete time and continuous time, with the corresponding
meanings of $\{W_{rs}\}$
(see Kelly's book, pp.\ 22--23).

\begin{figure}[ht]
\hspace*{2cm}\input{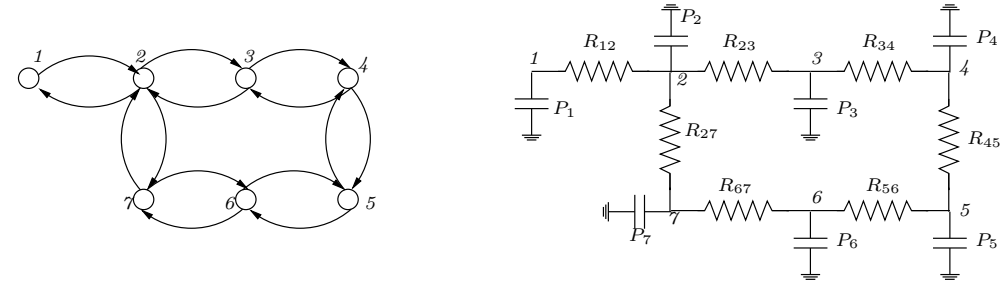}
\caption{\small State transition diagram of a Markov chain 
(left part) and the electric circuit that emulates the
dynamics of $\{P_r(t)
\}$ (right part).}
\label{circuit}
\end{figure}

In the case of detailed balance, there is another 
interpretation of the approach to equilibrium and the growth of $U(t)$.
We can write the master equations as follows:
\begin{equation}
\frac{\mbox{d}P_r(t)}{\mbox{d}t}=\sum_s\frac{1}{R_{sr}}\left(\frac{P_s(t)}{P_s}-\frac{P_r(t)}{P_r}\right)
\end{equation}
where $R_{sr}=(P_sW_{sr})^{-1}=(P_rW_{rs})^{-1}$. Imagine now an electrical
circuit where the indices $\{r\}$ designate the nodes. Nodes $r$ and $s$ are connected by
a wire with resistance $R_{sr}$ and every node $r$ is grounded via a capacitor
with capacitance $P_r$ (see Fig.\ \ref{circuit}). 
If $P_r(t)$ is the charge at node $r$ at time $t$,
then the master equations are the Kirchoff equations of the currents at each
node in the
circuit. Thus, the way in which probability spreads across the circuit is
analogous to the way charge spreads across the circuit and probability fluxes
are now analogous to electrical currents.
If we now choose
$V(x)=-\frac{1}{2}x^2$, then $-U(t)=\frac{1}{2}\sum_r\frac{P_r^2(t)}{P_r}$,
which means that the energy stored in the capacitors dissipates as heat in the
wires until the system reaches equilibrium, where all nodes have the same
potential, $P_r(t)/P_r=1$, and hence detailed balance corresponds to 
the situation where all individual currents vanish (not only their algebraic
sum).

We have seen, in the above examples, that various choices of the function $V$
yield various `metrics' between $\{P_r(t)\}$ and $\{P_r\}$, which are both
marginal distributions of a single symbol. What about joint distributions of
two or more symbols? Consider, for example, the function
\begin{equation}
J(t)=\sum_{r,s}P(X_0=r,~X_t=s)\cdot V\left(\frac{P(X_0=r)P(X_t=s)}{P(X_0=r,~X_t=s)}\right),
\end{equation}
where $V$ is concave as before. Here, by the same token, $J(t)$ is a `metric'
between the joint probability distribution $\{P(X_0=r,~X_t=s)\}$ and the
product of marginals $\{P(X_0=r)P(X_t=s)\}$, namely, it a measure of the
amount of statistical dependence between  $X_0$ and $X_t$. For $V(x)=\ln x$,
we have, of course, $J(t)=-I(X_0;X_t)$. Now, using
a similar chain of inequalities as before, we get the non--decreasing
monotonicity of $J(t)$ as follows:
\begin{eqnarray}
J(t)&=&\sum_{r,s,u}P(X_0=r,~X_t=s,~X_{t+\tau}=u)\cdot V\left(\frac{P(X_0=r)P(X_t=s)}{P(X_0=r,~X_t=s)}
\cdot\frac{P(X_{t+\tau}=u|X_t=s)}{P(X_{t+\tau}=u|X_t=s)}\right)\nonumber\\
&=&\sum_{r,u}P(X_0=r,~X_{t+\tau}=u)\sum_sP(X_t=s|X_0=r,~X_{t+\tau}=u)\times\nonumber\\
& &V\left(\frac{P(X_0=r)P(X_t=s,~X_{t+\tau}=u)}
{P(X_0=r,~X_t=s,~X_{t+\tau}=u)}\right)\nonumber\\
&\le&\sum_{r,u}P(X_0=r,~X_{t+\tau}=u)\times\nonumber\\
& &V\left(\sum_s P(X_t=s|X_0=r,~X_{t+\tau}=u)\cdot 
\frac{P(X_0=r)P(X_t=s,~X_{t+\tau}=u)}
{P(X_0=r,~X_t=s,~X_{t+\tau}=u)}\right)\nonumber\\
&=&\sum_{r,u}P(X_0=r,~X_{t+\tau}=u)\cdot V\left(\sum_s 
\frac{P(X_0=r)P(X_t=s,~X_{t+\tau}=u)}
{P(X_0=r,~X_{t+\tau}=u)}\right)\nonumber\\
&=&\sum_{r,u}P(X_0=r,~X_{t+\tau}=u)\cdot V\left(
\frac{P(X_0=r)P(X_{t+\tau}=u)}
{P(X_0=r,~X_{t+\tau}=u)}\right)=J(t+\tau).
\end{eqnarray}
This time, we assumed nothing beyond Markovity (not even homogeneity). This is
exactly the generalized data processing theorem of Ziv and Zakai (J.~Ziv and
M.~Zakai, ``On functionals satisfying a data-processing 
theorem,'' {\em IEEE Trans.~Inform.~Theory\/},
vol.~IT--19, no.~3, pp.~275--283, May 1973), which yields the ordinary data
processing theorem (of the mutual information) as a special case. 
Thus, we see
that the second law of thermodynamics is (at least indirectly) related to the data processing
theorem via the fact that they both stem from some more general principle
concerning monotonic evolution of `metrics' between probability distributions
defined using convex functions.
In a very
similar manner, one can easily show that the generalized conditional entropy
\begin{equation}
\sum_{r,s}P(X_0=r,~X_t=s)\cdot V\left(\frac{1}{P(X_0=r|X_t=s)}\right)
\end{equation}
is monotonically non--decreasing with $t$ for any concave $V$. 

\subsubsection{Monte Carlo Simulation}

Returning to the realm of Markov processes
with the detailed balance property,
suppose we want to simulate a 
physical system, namely, to sample from the Boltzmann--Gibbs
distribution 
\begin{equation}
P_r=\frac{e^{-\beta E_r}}{Z(\beta)}.
\end{equation}
In other words, we wish to
generate a discrete--time Markov process
$\{X_t\}$, possessing the detailed balance property, 
whose marginal converges to the Boltzmann--Gibbs distribution. 
This approach is called {\it dynamic Monte Carlo} or {\it Markov chain Monte Carlo}
(MCMC).
How should we select
the state transition probability matrix $W$ to this end? 
Substituting $P_r=e^{-\beta E_r}/Z(\beta)$ into the detailed balance
equation, we readily see that a necessary condition is
\begin{equation}
\frac{W_{rs}}{W_{sr}}=e^{-\beta(E_s-E_r)}.
\end{equation}
The {\it Metropolis algorithm} is one popular way to implement such a Markov
process in a rather efficient manner. It is based on the concept of factoring $W_{rs}$ as a
product $W_{rs}=C_{rs}A_{rs}$, where $C_{rs}$ is the conditional probability of selecting
$X_{t+1}=s$ as a {\it candidate} for the next state, and $A_{rs}$
designates the
probability of {\it acceptance}. In other words, we first choose a candidate
according to $C$, and then make a final decision whether we accept this
candidate or stay in state $r$. The Metropolis algorithm pics $C$ to
implement a uniform distribution among $n$ states `close' to $r$ (e.g.,
flipping one spin of a $n$--spin configuration). Thus,
$W_{rs}/W_{sr}=A_{rs}/A_{sr}$, and so, it remains to choose $A$ such that
\begin{equation}
\frac{A_{rs}}{A_{sr}}=e^{-\beta(E_s-E_r)}.
\end{equation}
The Metropolis algorithm defines
\begin{equation}
A_{rs}=\left\{\begin{array}{ll}
e^{-\beta(E_s-E_r)} & E_s > E_r\\
1 & \mbox{otherwise}\end{array}\right.
\end{equation}
In simple words, the algorithm works as follows:
Given that $X_t=r$,
first randomly select one candidate $s$ for $X_{t+1}$ among $n$ possible (neighboring)
states. If $E_s < E_r$ always accept $X_{t+1}=s$ as the next state. If $E_s\ge
E_r$, then
randomly draw a RV $Y\in\mbox{Unif}[0,1]$. If $Y <
e^{-\beta(E_s-E_r)}$, then again, accept $X_{t+1}=s$ as the next state.
Otherwise, stay in state $r$, i.e., $X_{t+1}=r$.
To see why this choice of $A$ works, observe that
\begin{equation}
\frac{A_{rs}}{A_{sr}}=\left\{\begin{array}{ll}
e^{-\beta(E_s-E_r)} & E_s > E_r\\
\frac{1}{e^{-\beta(E_r-E_s)}} & E_s \le E_r\end{array}\right.
= e^{-\beta(E_s-E_r)}.
\end{equation}
There are a few nice things about this algorithm: 
\begin{itemize}
\item Energy differences between neighboring states, $E_s-E_r$, are normally
easy to calculate. If $r$ and $s$ differ by a single component of the
microstate $\bx$, and the if the Hamiltonian structure consists of
short--range interactions only,
then most terms of the Hamiltonian are the same for $r$ and $s$, and only a
local calculation is required for evaluating the energy difference.
\item Calculation of $Z(\beta)$ is not required, and
\item Chances are that you don't get stuck in the same state for too long.
\end{itemize}
The drawback, however, is that aperiodicity is not guaranteed. This depends
on the Hamiltonian. 

The {\it heat bath} algorithm (a.k.a.\ {\it Glauber
dynamics}) alleviates this shortcoming
and although somewhat slower than Metropolis to equilibrate, it
guarantees all the good properties of a Markov chain: irreducibility,
aperiodicity, and convergence to stationarity. The only difference is that
instead of the above choice of $A_{rs}$, it is redefined as
\begin{eqnarray}
A_{rs}&=&\frac{1}{2}\left[1-\tanh\left(\frac{\beta(E_s-E_r)}{2}\right)\right]
\nonumber\\
&=&\frac{e^{-\beta(E_s-E_r)}}{1+e^{-\beta(E_s-E_r)}}\nonumber\\
&=&\frac{P_s}{P_s+P_r},
\end{eqnarray}
which is also easily shown to satisfy the detailed balance condition.
The heat bath algorithm generalizes
easily to sample from any distribution $P(\bx)$ whose configuration space is
of the form $\calX^n$. The algorithm can be described by the following
pseudocode:
\begin{enumerate}
\item Select $\bX_0$ uniformly at random across $\calX^n$.
\item {\bf For} $t=1$ {\bf to} $t=T$:
\item Draw an integer $i$ at random with uniform distribution across
$\{1,2,\ldots,n\}$.
\item For each $x\in\calX$, calculate
\begin{equation}
P(X^i=x|\bX^{\sim i}=\bx_t^{\sim i})=\frac{P(X^i=x,\bX^{\sim i}=\bx_t^{\sim
i})}{\sum_{x'\in\calX}P(X^i=x',\bX^{\sim i}=\bx_t^{\sim i})}.
\end{equation}
\item Set $x_{t+1}^j=x_t^j$ for all $j\ne i$ and $x_t^i=X^i$, 
where $X^i$ is drawn
according to 
$$P(X^i=x|\bX^{\sim i}=\bx_t^{\sim i}).$$
\item {\bf end}
\item {\bf Return} the sequence $\bX_t$, $t=1,2,\ldots,T$.
\end{enumerate}
It can be easily seen that the resulting Markov chain satisfies detailed
balance and that in the case of binary alphabet (spin array) it implements
the above expression of $A_{rs}$. One can also easily generalize the
Metropolis algorithm, in the same spirit, as $e^{-\beta(E_s-E_r)}$ is nothing
but the ratio $P_s/P_r$.


\end{document}